\PassOptionsToPackage{authoryear,round}{natbib}
\PassOptionsToPackage{table}{xcolor}

\documentclass[11pt,a4paper]{article}

\usepackage[a4paper,margin=1in,includefoot]{geometry}
\usepackage[T1]{fontenc}
\usepackage[utf8]{inputenc}
\usepackage{lmodern}

\usepackage{amsmath,amssymb}

\usepackage{graphicx}
\usepackage{xcolor}
\usepackage{float}
\usepackage{pdfpages}
\usepackage{adjustbox}
\usepackage{comment}

\usepackage{algorithm}
\usepackage{algorithmic}

\usepackage{multirow}
\usepackage{booktabs}
\usepackage{threeparttable}
\usepackage{enumitem}
\usepackage{tabularx,array}

\newcolumntype{L}[1]{>{\raggedright\arraybackslash}p{#1}}
\newcolumntype{C}{>{\centering\arraybackslash}X}

\renewcommand{\arraystretch}{1.06}

\usepackage{subcaption}
\usepackage{natbib}
\bibpunct{(}{)}{;}{a}{}{,}
\usepackage[hidelinks]{hyperref}
\usepackage{url}

\newcommand{\proglang}[1]{\textsf{#1}}

\newcommand{\email}[1]{\href{mailto:#1}{\texttt{#1}}}

\newcommand{\orcidlink}[1]{} 

\newcommand{\AJSAbstractText}{}
\newcommand{\AJSKeywordsText}{}

\newcommand{\Abstract}[1]{\gdef\AJSAbstractText{#1}}
\newcommand{\Keywords}[1]{\gdef\AJSKeywordsText{#1}}

\newcommand{\Plainauthor}[1]{}
\newcommand{\Plaintitle}[1]{}
\newcommand{\Shorttitle}[1]{}
\newcommand{\Pages}[1]{}

\definecolor{indianred1}{HTML}{FF6A6A}
\definecolor{indianred3}{HTML}{CD5555}
\definecolor{lightpink2}{HTML}{EEA2AD}

\definecolor{coral1}{HTML}{FF7256}
\definecolor{coral3}{HTML}{CD5B45}
\definecolor{darkorange}{HTML}{FF8C00}

\definecolor{gray70}{HTML}{B3B3B3}
\definecolor{gray40}{HTML}{666666}
\definecolor{gray24}{HTML}{3D3D3D}

\definecolor{nfortyA}{HTML}{C89414}
\definecolor{nfortyB}{HTML}{CCBC72}
\definecolor{nfortyC}{HTML}{8D8A0F}

\definecolor{nfiveA}{HTML}{C27BB0}
\definecolor{nfiveB}{HTML}{8B5F92}
\definecolor{nfiveC}{HTML}{B695C0}

\definecolor{steelblue4}{HTML}{36648B}
\definecolor{steelblue3}{HTML}{4F94CD}
\definecolor{lightblue}{HTML}{ADD8E6}

\newcommand{\chip}[1]{%
  \begingroup\setlength{\fboxsep}{0pt}%
  \colorbox{#1}{\rule{0pt}{0.9em}\rule{0.9em}{0pt}}%
  \endgroup
}

\newcommand{\Nchips}[1]{%
  \begingroup\count0=#1\relax
  \ifnum\count0=0
    \chip{white}\chip{white}\chip{white}\chip{white}\chip{white}\chip{white}%
  \else\ifnum\count0=20
    \chip{indianred3}\chip{indianred1}\chip{lightpink2}\chip{gray70}\chip{gray40}\chip{gray24}%
  \else\ifnum\count0=40
    \chip{nfortyA}\chip{nfortyB}\chip{nfortyC}\chip{gray70}\chip{gray40}\chip{gray24}%
  \else\ifnum\count0=80
    \chip{steelblue4}\chip{steelblue3}\chip{lightblue}\chip{gray70}\chip{gray40}\chip{gray24}%
  \else\ifnum\count0=200
    \chip{coral1}\chip{coral3}\chip{darkorange}\chip{gray70}\chip{gray40}\chip{gray24}%
  \else\ifnum\count0=500
    \chip{nfiveA}\chip{nfiveB}\chip{nfiveC}\chip{gray70}\chip{gray40}\chip{gray24}%
  \fi\fi\fi\fi\fi\fi
  \endgroup
}

\newcommand{\ParamBox}[8]{%
  \fbox{\parbox[t]{#1}{%
    n: #2\\[-1pt]
    $n_{\text{sim}}$: #3\\[-1pt]
    iter: #4\\[-1pt]
    $P_{\text{ext}}$: #5\\[-1pt]
    $P_{\text{miss}}$: #6\\[-1pt]
    $\rho$: #7\\[-1pt]
    Data: #8
  }}}

\title{Multiple Imputation Methods under Extreme Values}

\author{Enzo Porto Brasil\thanks{%
Department of Statistics, University of Brasília, Brazil.
Campus Universitário Darcy Ribeiro, Gleba A, Setor Norte, via L3 Norte, Federal District, Brazil, 70910-900.
Correspondence to: <\email{enzoportobrasil@gmail.com}>.%
}}

\Abstract{%
Missing data are ubiquitous in empirical databases, yet statistical analyses typically require complete data matrices. Multiple imputation offers a principled solution for filling these gaps. This study evaluates the performance of several multiple imputation methods, both in the presence and absence of extreme values, using the \proglang{MICE} package in R. Through Monte Carlo simulations, we generated incomplete data sets with three variables and assessed each imputation method within regression models. The results indicate that the linear regression based imputation method showed the best overall predictive performance (CV--MSE), whereas the sparse model approach was generally less efficient. Our findings underscore the relevance of extreme values when selecting an imputation strategy and highlight sample size, proportion of missingness, presence of extremes, and the type of fitted model as key determinants of performance. Despite its limitations, the study offers practical recommendations for researchers, stressing the need to examine the missingness mechanism and the occurrence of extreme values before choosing an imputation method.
}


\usepackage{fancyhdr}

\newcommand{\PreprintLine}{\textit{Preprint. \today}}

\fancypagestyle{firstpage}{%
  \fancyhf{}
  
  \fancyfoot[L]{\small \PreprintLine}
  \fancyfoot[C]{\small \thepage} 
  \fancyfoot[R]{}
}

\makeatletter
\renewcommand{\maketitle}{%
  \thispagestyle{firstpage}%
  \vspace*{-1.2em}%
  \hrule height 0.6pt%
  \vspace{0.9em}%
  \begin{center}
    {\Large\bfseries \@title\par}
    \vspace{1.0em}
    {\normalsize \@author\par}
  \end{center}

  \@thanks

  \vspace{0.9em}%
  \hrule height 0.6pt%
  \vspace{1.2em}%
}
\makeatother

\date{} 

\begin{document}

\maketitle

\begin{abstract}
\AJSAbstractText
\end{abstract}

\vspace{0.8em}

\vspace{2em}


\section{Introduction}
The unprecedented availability of large-scale data in recent years has fueled scientific and technological advances across diverse disciplines. However, the growing volume of information does not guarantee completeness or accuracy. Missing data remain a pervasive challenge, arising from a variety of causes such as nonresponse in surveys, sensor failures, data-entry errors, or loss of historical records. If unaddressed, missingness can distort parameter estimates, reduce statistical power, and bias inferential conclusions, particularly when traditional statistical techniques (which typically assume complete data) are applied without proper adaptation \citep{nunes2007, rubin1996}.

A variety of strategies exist to deal with incomplete data, ranging from ad hoc approaches such as mean substitution to more principled statistical frameworks. Among these, multiple imputation (MI) has emerged as a principled approach for handling missingness under the assumption of Missing Completely at Random (MCAR) and, more generally, Missing at Random (MAR) \citep{schafer1999multiple}. MI replaces each missing value with multiple plausible estimates drawn from a predictive distribution, generating several complete data sets. These are analyzed separately, and results are combined to incorporate the uncertainty associated with the imputation process. Compared to single imputation or case deletion, MI provides more valid statistical inference, preserving variability and reducing bias \citep{mcknight2007, kenward2007, van2006}.

Despite its advantages, the performance of multiple imputation can be compromised in the presence of extreme values, tail observations that, whether genuine or error–induced, exert high influence on model fitting and the predictive distributions used for imputation. \citet{kotz2000extreme} explain that extreme values may arise naturally in certain domains (e.g., finance, meteorology, astrophysical research) or may result from measurement errors. In either case, they can heavily influence model parameters, distort predictive distributions used for imputation, and lead to unstable or biased results \citep{ferrari2014missing, buuren2018, li2024comparison}. While the impact of missingness has been extensively studied, the interaction between missing data and extreme values remains comparatively underexplored. This gap is particularly relevant in applied research contexts where both problems occur simultaneously, and imputation models are often selected without considering their sensitivity to extreme values and other complex distributional characteristics, such as multimodality and skewness \citep{von2013should, templ2024robust}.

The present study provides a comprehensive and reproducible evaluation of widely used multiple imputation (MI) procedures under conditions with contaminated and clean data. A three-variable normal design \((y,x_1,x_2)\) is employed, MCAR missingness is induced only in \(x_2\), and casewise contamination replaces selected rows at \(\bar v \pm 3 s_v\) to generate vertical outliers and high-leverage points \citep{pukelsheim1994three, robert1995simulation}. Downstream models are aligned with the regime: ordinary least squares (OLS) is used for clean data and elastic net (EN) for contaminated data in order to stabilize estimation under leverage and collinearity \citep{zou2005regularization,ESL,glmnetJSS}. OLS serves as the efficiency benchmark under correct specification but is highly influence sensitive \citep{RousseeuwLeroy}.

The experimental design varies sample size, the proportion of missingness, the proportion of extreme values, the correlation between covariates, the number of imputations in \texttt{mice}, the number of Monte Carlo replicates, and the analysis model type. A single master seed with deterministic substreams, fixed cross-validation folds reused across imputations within each replicate, and a congenial predictor matrix for imputing \(x_2\) ensure paired comparisons and proper variance accounting \citep{buuren2018,van2011mice,arlot2010}. Performance is summarized primarily by out-of-sample CV--MSE through its mean, variance, and quantiles. Inferential validity in the clean regime is assessed by pooled bias, RMSE, and 95\% coverage for \((\beta_0,\beta_1,\beta_2)\) using Rubin’s rules \citep{rubin1987,white2011}, while under contamination coefficients are reported transparently after model selection.

The investigation addresses four questions in sequence: how leading MI methods compare in out-of-sample prediction with extremes present; what the consequences are for inference in the clean regime and how these metrics behave under contamination; how sample size and missingness level moderate performance and tail risk; and to what extent parametric MI yields tighter predictive dispersion while donor or machine learning (ML) procedures reduce slope bias under contamination.

Results indicate persistent contamination directionality in coefficients, with intercept inflation, a systematic tilt in \(\beta_1\), and shrinkage or sign pull in \(\beta_2\). Increasing sample size contracts variability rather than removing these shifts. A stable method trade-off is observed: parametric MI typically yields tighter predictive tails, whereas donor and ML procedures often temper slope bias as missingness increases. The evidence supports informed method selection when missingness and extreme values occur jointly. All computations were performed in R/RStudio 4.3.3 \citep{R}.

This paper is structured as follows. Section~\ref{material_methods} presents the materials and methods. Subsections~\ref{data_sets}--\ref{data_inputation} describe the data-generating process, the contamination mechanism, the statistical models, and the multiple imputation procedures. Subsections~\ref{sec_comparison}--\ref{sec:simulations} set out the comparison criteria, the evaluation protocol, and the reproducibility controls. Section~\ref{results} reports the main results. Section~\ref{conclusion} offers concluding remarks and implications. Appendix~\ref{appendix} provides transparency material and expanded results by sample size and scenario.


\section{Materials and methods} \label{material_methods}
This section records the elements needed to reproduce the study without revisiting the motivation. All scenarios share a paired design: a baseline clean data set and its contaminated counterpart; within each replicate, MCAR missingness is induced only in \(x_2\); the same fold partition is reused across imputations and methods; and the downstream model is fixed by regime (ordinary least squares for clean data and elastic net  for contaminated data). Remaining design factors and reporting conventions are stated in the Introduction and detailed in the subsections below.

\subsection{Data sets}\label{data_sets}
All data sets were generated from pseudo-random numbers drawn from normal distributions. Each data set comprises three continuous variables, denoted by \(y\), \(x_1\), and \(x_2\). Performance is evaluated under two baseline conditions: (i) data sets drawn from the specified normal model without contamination; and (ii) data sets with injected extreme values (contamination). For reproducibility, all simulation procedures were initialized with a fixed random seed (\texttt{241103414}), which has no methodological impact on the outcomes. Each variable is defined as follows.

\vspace*{2mm}
\begin{itemize}
  \item[$\cdot$] \textbf{$\mathbf{y}$}: a continuous response initially generated from a normal distribution with standard deviation \(1.5\) and linear predictor \(\mu(\mathbf{y_i})\) for the \(i\)-th observation,
  \begin{equation}
    \mathbf{y_i} \sim N\!\bigl(\mu(\mathbf{y_i}),\,1.5^{2}\bigr),
    \qquad
    \mu(\mathbf{y_i}) \;=\; 1 \;+\; 0.5\,\mathbf{x_{1i}} \;+\; 1.5\,\mathbf{x_{2i}}.
  \end{equation}
\end{itemize}

\noindent Predictors:
\begin{itemize}
  \item[$\cdot$] \textbf{$\mathbf{x_1}$}: a continuous predictor with \(\;x_1 \sim N(10,\,2^{2})\).
  \item[$\cdot$] \textbf{$\mathbf{x_2}$}: a continuous predictor generated conditionally on \(x_1\) to induce Pearson correlation \(\rho\):
  \begin{equation}
    \mathbf{x_{2i}} \;=\; 5 \;+\; \frac{\rho\,\sigma_{2}}{\sigma_{1}}\bigl(\mathbf{x_{1i}}-10\bigr)
    \;+\; \sigma_{2}\sqrt{1-\rho^{2}}\,\mathbf{z_i},
  \end{equation}
  where \(\sigma_{1}=2\), \(\sigma_{2}=1.5\), and \(\mathbf{z_i}\sim N(0,1)\) independent of \(x_1\).
\end{itemize}

\noindent Equivalently, \((x_1,x_2)\) follow a bivariate normal with mean vector \((10,5)\) and covariance matrix
\[
\Sigma \;=\;
\begin{pmatrix}
2^{2} & \rho\,\sigma_{1}\sigma_{2} \\
\rho\,\sigma_{1}\sigma_{2} & 1.5^{2}
\end{pmatrix}.
\]
Parameter \(\rho\) controls the induced correlation between \(x_1\) and \(x_2\) (larger \(|\rho|\) implies stronger linear association).

\vspace*{2mm}
Two reference data sets are considered for each scenario: one baseline (clean data, no contamination) and one contaminated with extreme values.

\subsubsection{Extreme values (contamination mechanism)}
\label{subsec:contamination}
Let $F_0$ be the baseline joint law of $(y,x_1,x_2)$ in Section~\ref{data_sets}. 
For a given contamination proportion $P_{\text{ext}}\in(0,1)$ and sample size $n$, set 
$k=\mathrm{round}(n \times P_{\text{ext}})$ and draw a set of indices 
$\mathcal{I}_{\text{ext}}\subset\{1,\ldots,n\}$ of size $k$ uniformly without replacement.
On the \emph{baseline} (pre-contamination) sample, compute for each $v\in\{y,x_1,x_2\}$ the sample summaries
\[
\bar v=\frac{1}{n}\sum_{i=1}^n v_i,
\qquad
s_v=\sqrt{\frac{1}{n-1}\sum_{i=1}^n (v_i-\bar v)^2}.
\]
Order $\mathcal{I}_{\text{ext}}=\{i_1,\ldots,i_k\}$ arbitrarily and set signs 
$\eta_{i_j}=+1$ for odd $j$ and $\eta_{i_j}=-1$ for even $j$. 
For each $i\in\mathcal{I}_{\text{ext}}$ and for every $v\in\{y,x_1,x_2\}$, perform the componentwise replacement
\[
v_i \;\leftarrow\; \bar v \;+\; 3\,\eta_i\, s_v,
\]
leaving all other observations unchanged. 
Equivalently, the contaminated sample follows 
$F=(1-P_{\text{ext}})F_0+P_{\text{ext}}\,H$, where $H=\tfrac12\delta_{\bar{\mathbf{m}}-3\mathbf{s}}+\tfrac12\delta_{\bar{\mathbf{m}}+3\mathbf{s}}$ 
acts componentwise with $\bar{\mathbf{m}}=(\bar y,\overline{x}_1,\overline{x}_2)$ and $\mathbf{s}=(s_y,s_{x_1},s_{x_2})$. 
This symmetric, casewise “three-sigma” replacement produces both vertical outliers and high-leverage points \citep{pukelsheim1994three,kotz2000extreme}. 
Alternative tail-based designs (e.g., truncated-normal beyond $\pm3s$) are possible \citep{robert1995simulation}, but all reported results use the $\pm3s$ scheme above.

\subsection{Statistical models}\label{statistical_models}
Imputation methods were compared via regression modelling for the response variable \(y\) with \(x_1\) and \(x_2\) as covariates, as detailed in Section \ref{sec_comparison}. 
For the reference data set without extreme values, an ordinary least squares (OLS) linear regression was fitted. For the data set containing extreme values, a sparse regression based on the elastic net (EN) penalty was employed to stabilize estimation under contamination and induced collinearity. The same modelling strategy (model form and tuning protocol) was applied after imputation: linear models for the baseline (“standard-value”) data and elastic-net models for the contaminated (“extreme-value”) data, so that differences in performance reflect the downstream effect of imputation rather than changes in the analysis model.

\emph{Comparability rationale.} The designs (with vs.\ without extreme values) keeps the estimand and the predictor set identical across analyses: the working model is linear in \((x_1,x_2)\) in both cases and is tuned by \(K\)-fold cross-validation with fixed folds reused across imputations within each replicate. The elastic net (EN) solves a penalized least-squares problem and reduces to unpenalized least squares when \(\lambda=0\); consequently, in clean, low-dimensional settings cross-validated EN typically selects a small \(\lambda\) and yields predictions close to OLS, while its \(\ell_2\) component stabilizes estimates when multicollinearity or contamination inflates variance \citep{zou2005regularization,ESL,glmnetJSS}. Ordinary least squares is well known to be highly sensitive to high-leverage points and outliers (unbounded influence; breakdown point \(1/n\)), so under extreme-value contamination EN was preferred for variance control, recognizing that standard EN is not robust in the strict sense because it retains squared loss; truly robust variants replace the loss by the type Huber/LAD losses combined with sparsity penalties \citep{RousseeuwLeroy,HuberRonchetti,YuYao2017,HuberLasso,LADLasso,Wang2013}.

\emph{Notation convention.} The same symbols for the population coefficients in both sections is considered: \(\boldsymbol{\beta}=(\beta_0,\beta_1,\beta_2)^\top\).
Their estimator is always written simply as \(\hat\beta\).
When disambiguation is needed, we do so in prose (e.g., “the OLS estimate \(\boldsymbol{\hat\beta}\)” vs. “the elastic–net estimate \(\boldsymbol{\hat\beta}\)”).
All coefficients are reported on the original data scale; penalized fits use internal standardization (slopes standardized, intercept unpenalized) with back-transformation on output for comparability.

\paragraph{Data without extreme values: linear regression.}\label{linear_no_extremes} When no extreme values were present, ordinary least squares (OLS) \citep{kutner2005} was fitted with \texttt{stats::lm()} \citep{R}:
\begin{equation}\label{linear_reg}
  y_i = \beta_0 + \beta_1 x_{1i} + \beta_2 x_{2i} + \epsilon_i,\qquad
  \epsilon_i \overset{\text{iid}}{\sim} N(0,\sigma^2),\ i=1,\ldots,n.
\end{equation}

Let \(\mathbf{y}=(y_1,\ldots,y_n)^\top\), \(\mathbf{x}_1=(x_{11},\ldots,x_{1n})^\top\), \(\mathbf{x}_2=(x_{21},\ldots,x_{2n})^\top\), and \(\mathbf{X}=[\,\mathbf{1},\,\mathbf{x}_1,\,\mathbf{x}_2\,]\in\mathbb{R}^{n\times 3}\). The OLS estimator of \(\boldsymbol{\beta}=(\beta_0,\beta_1,\beta_2)^\top\) is

\begin{equation}\label{ols_problem}
\hat{\boldsymbol{\beta}}
= \arg\min_{\boldsymbol{\beta}}\left\{\frac{1}{2n}\sum_{i=1}^{n}\bigl(y_i-\beta_{0}-\beta_{1}x_{1i}-\beta_{2}x_{2i}\bigr)^{2}
= (\mathbf{X}^\top \mathbf{X})^{-1}\mathbf{X}^\top \mathbf{y}\right\},
\end{equation}

whenever \(\mathbf{X}\) has full column rank; in practice \texttt{lm()} computes \(\hat{\boldsymbol{\beta}}\) via pivoted QR rather than forming \((\mathbf{X}^\top \mathbf{X})^{-1}\). Predictions are \(\hat{\mathbf{y}}=\mathbf{X}\hat{\boldsymbol{\beta}}\), i.e. $\hat{y}_{i} = \hat{\beta}_{0} + \hat{\beta}_{1}x_{1i} + \hat{\beta}_{2}x_{2i}.$

The residual variance is \(\hat\sigma^2 = \tfrac{1}{n-3}\sum_{i}(y_i-\hat y_i)^2\).
A predefined correlation between \(x_1\) and \(x_2\) was introduced as specified in Section~\ref{data_sets}.


\paragraph{Data with extreme values: sparse regression.}
\label{enet_extremes} The linear specification in \eqref{linear_reg} and the squared-error loss in \eqref{ols_problem} are retained, but for the datasets with extreme values present \(\boldsymbol{\beta}\) was estimated via elastic-net regularization \citep{zou2005regularization, bertsimas2020sparse, chang2021sparse} using \texttt{glmnet::glmnet()} \citep{glmnet}.
In this case, predictors are centered and scaled internally before penalization; the intercept is left unpenalized; all reported coefficients are back-transformed. The estimate \(\hat\beta=(\hat\beta_0,\hat\beta_1,\hat\beta_2)^\top\) is obtained by 

\begin{equation}\label{enet_problem}
\hat{\boldsymbol{\beta}}
= \arg\min_{\boldsymbol{\beta}}\left\{
\frac{1}{2n}\sum_{i=1}^{n}\!\bigl(y_i-\beta_0-\beta_1 x_{1i}-\beta_2 x_{2i}\bigr)^{2}
\;+\;
\lambda\!\left[\alpha\,(|\beta_1|+|\beta_2|)+\frac{1-\alpha}{2}\,(\beta_1^2+\beta_2^2)\right]\right\},
\end{equation}

where \(\lambda\ge 0\) controls overall shrinkage and \(\alpha\in[0,1]\) mixes \(\ell_1\) and \(\ell_2\) penalties (\(\alpha{=}1\) lasso; \(\alpha{=}0\) ridge).
Note that setting \(\lambda{=}0\) recovers the OLS objective in \eqref{ols_problem}.

In the simulations, \(\alpha\) is fixed at \(0.5\), and \(\lambda\) is selected by \(K\)-fold cross-validation within each completed data set, reusing the same fold partition across imputations within a replicate; the criterion is the mean squared error on held-out folds. Predictions have the same form as in the linear case $\hat y_i=\hat\beta_0+\hat\beta_1 x_{1i}+\hat\beta_2 x_{2i}$, with \(\mathbb{\hat\beta}\) obtained from \eqref{enet_problem}.

\noindent\textit{Remark.} Using elastic net regularization improves stability under collinearity and can reduce predictive variance with contaminated data; however, it does not provide outlier resistance in the strict sense used in robust statistics. When outliers are a concern, fitting with robust loss functions (e.g., Huber or Tukey losses) offers a complementary strategy \citep{hoerl1970ridge, tibshirani1996regression}.

\subsection{Data imputation.}\label{data_inputation} Methods for handling missing data can be grouped into two broad classes: \emph{single imputation}, in which each missing value is replaced only once and the procedure is typically straightforward to implement, and \emph{multiple imputation}, which relies on iterative schemes and therefore entails greater computational cost \citep{rubin1996,nunes2010}. Depending on the research context, both approaches may be applied within a single analysis, and efficient implementations are available in R packages such as \texttt{MICE}, \texttt{Hmisc}, and \texttt{mlr} \citep{aregImpute, mlr, mice}.

The primary aim of this study is to highlight potential differences among several multiple imputation (MI) methods under various scenarios, assessed via Monte Carlo simulations. Each scenario involves three variables ($y$, $x_1$, and $x_2$), where missingness is induced only in $x_2$ as illustrated in Table  \ref{tab:imputation_side_by_side}. Although a comprehensive investigation of all conceivable combinations of variable counts and types is ideal when selecting the best imputation method, such breadth demands substantial computational resources.

\definecolor{missred}{HTML}{FFCCC9}
\definecolor{extblue}{HTML}{CCE5FF}

\begin{table}[H]
\centering
\caption{Side-by-side illustrative data sets (10 rows each). Left: 3 missing values (red), no extremes induced. Right: 3 missing values (red) and selected extreme values (blue)}
\label{tab:imputation_side_by_side}
\setlength{\tabcolsep}{6pt}
\renewcommand{\arraystretch}{1.1}

\begin{minipage}[t]{0.48\linewidth}
\centering
(a) No extreme values\\[2mm]
\begin{tabular}{cccc}
\hline
\textbf{i} & \textbf{y} & $\mathbf{x_1}$ & $\mathbf{x_2}$ \\ \hline
1  & $y_{1}$   & $x_{1,1}$   & $x_{2,1}$ \\
2  & $y_{2}$   & $x_{1,2}$   & $x_{2,2}$ \\
3  & $y_{3}$   & $x_{1,3}$   & \cellcolor{missred}$\mathrm{NA}_{2,3}$ \\
4  & $y_{4}$   & $x_{1,4}$   & $x_{2,4}$ \\
5  & $y_{5}$   & $x_{1,5}$   & $x_{2,5}$ \\
6  & $y_{6}$   & $x_{1,6}$   & \cellcolor{missred}$\mathrm{NA}_{2,6}$ \\
7  & $y_{7}$   & $x_{1,7}$   & $x_{2,7}$ \\
8  & $y_{8}$   & $x_{1,8}$   & $x_{2,8}$ \\
9  & $y_{9}$   & $x_{1,9}$   & $x_{2,9}$ \\
10 & $y_{10}$  & $x_{1,10}$  & \cellcolor{missred}$\mathrm{NA}_{2,10}$ \\ \hline
\end{tabular}
\end{minipage}
\hfill
\begin{minipage}[t]{0.48\linewidth}
\centering
(b) With extreme values\\[2mm]
\begin{tabular}{cccc}
\hline
\textbf{i} & \textbf{y} & $\mathbf{x_1}$ & $\mathbf{x_2}$ \\ \hline
1  & \cellcolor{extblue}$y_{1}^{\star}$   & \cellcolor{extblue}$x_{1,1}^{\star}$   & \cellcolor{extblue}$x_{2,1}^{\star}$ \\
2  & \cellcolor{extblue}$y_{2}^{\star}$   & \cellcolor{extblue}$x_{1,2}^{\star}$   & \cellcolor{extblue}$x_{2,2}^{\star}$ \\
3  & $y_{3}$   & $x_{1,3}$   & \cellcolor{missred}$\mathrm{NA}_{2,3}$ \\
4  & $y_{4}$   & $x_{1,4}$   & $x_{2,4}$ \\
5  & $y_{5}$   & $x_{1,5}$   & $x_{2,5}$ \\
6  & $y_{6}$   & $x_{1,6}$   & \cellcolor{missred}$\mathrm{NA}_{2,6}$ \\
7  & $y_{7}$   & $x_{1,7}$   & $x_{2,7}$ \\
8  & $y_{8}$   & $x_{1,8}$   & $x_{2,8}$ \\
9  & \cellcolor{extblue}$y_{9}^{\star}$   & \cellcolor{extblue}$x_{1,9}^{\star}$   & \cellcolor{extblue}$x_{2,9}^{\star}$ \\
10 & $y_{10}$  & $x_{1,10}$  & \cellcolor{missred}$\mathrm{NA}_{2,10}$ \\ \hline
\end{tabular}
\end{minipage}
\end{table}

\subsubsection{Multiple Imputation}
Multiple imputation (MI) creates $M$ completed data sets by stochastically replacing missing values with draws from an imputation model fitted to the observed data. Each completed data set is analyzed with the same completed data method, and results are combined using Rubin’s rules to obtain point estimates and standard errors. Under Missing at Random (MAR) and with a properly specified imputation model/method, the pooled estimators are consistent and the standard errors reflect both within (and between) imputation variability. In practice, MI is commonly implemented via joint modelling (often with MCMC) or fully conditional specification, with options such as predictive mean matching to preserve distributional features \citep{rubin1996,schafer2002,nunes2007,vinha2016,buuren2018}. Despite the inherent complexity, \citet{buuren2018} considers multiple imputation the most effective strategy for incomplete data.

\citet{harrell2015} note that regression-based imputation adds a random residual to each prediction based on the model to preserve the conditional variance of the original variable. Each repetition yields a completed data set analysed with standard methods; the final estimates are the averages across imputations.

The number of imputations $M$ typically depends on the fraction of missing information $P_{\text{miss}}$. \citet{white2011} recommend $M = 100\times P_{\text{miss}}$, while \citet{buuren2018} and \citet{royston2004} advise a minimum of 20 imputations. \citet{bodner2008} gives rule-of-thumb values of $M$ for $P_{\text{miss}}$ ranging from 0.05 to 0.90, with interpolation for intermediate values. Ultimately, $M$ should reflect data complexity and the underlying missing-data mechanism.

The \texttt{mice} package, proposed by \citet{mice}, provides both parametric and nonparametric options for multiple imputation. In this study, scenarios were designed to include methods from both classes. Parametric approaches rely on explicit distributional assumptions, such as the normal linear model used by \texttt{method="norm.predict"} for continuous variables. Nonparametric approaches avoid such assumptions and include: (i) tree-based methods such as classification and regression trees (\texttt{cart}) \citep{steinberg2009cart}, predictive mean matching (\texttt{pmm}) \citep{allison2015imputation}, and random forests (\texttt{rf}) \citep{breiman2001random}; and (ii) simple donor or resampling approaches such as \texttt{sample} (hot-deck style). Tree-based imputers approximate conditional distributions without specifying a parametric form and draw replacements from donor sets defined by terminal nodes, accommodating nonlinearities and interactions without parametric assumptions.

\noindent\textit{Notes.}
The MCAR assumption remains the working premise for unbiasedness of pooled estimates.
When the goal is descriptive comparison of imputation strategies, distributional fidelity can be assessed via metrics and graphics (e.g. MSE distributions, boxplots, and descriptive statistics; and other coefficients results), as implemented in the simulation study \citep{rubin1987, buuren2018}.

\subsubsection{Imputation methods}\label{imputation_methods}
The simulation study compares six imputation methods, denoted by \{\text{T1},\dots, \text{T6}\} and differentiated by colour in figures and tables. These methods, also available in the \texttt{mice::mice()}, represent a mix of parametric and nonparametric approaches:

\begin{itemize}
    \item T1: \textbf{\texttt{norm.predict}} -- imputation by linear regression, being a prediction and parametric method, which imputes the value according to the model, also known as regression imputation. \citet{mice} and \citet{little2019statistical} also explain that \texttt{norm.predict} should be avoided for formal data analysis because it ignores imputation uncertainty and can artificially strengthen relationships between variables. Even with richer models, these issues persist. Stochastic methods, such as \texttt{pmm} or \texttt{norm}, are generally preferred, although deterministic prediction may still provide reasonable mean estimates under plausible normality assumptions;
    
    \item T2: \textbf{\texttt{lasso.select.norm}} -- parametric normal regression preceded by Lasso variable selection for sparsity. Formally imputes univariate missing data using Bayesian linear regression following a preprocessing lasso variable selection step. The method used on \texttt{mice} function is based on the Indirect Use of Regularized Regression (IURR) proposed by \citet{deng2016multiple} and \citet{zhao2016multiple};
    
    \item T3: \textbf{\texttt{norm.boot}}-- parametric normal regression with bootstrap resampling \citep{wu1986jackknife} to incorporate parameter uncertainty in imputations, so imputes univariate missing data using predictive values with bootstrap. On \texttt{mice}, the method draws a bootstrap sample, calculates regression weights and imputes with normal residuals \citep{mice};
    
    \item T4: \textbf{\texttt{pmm}} -- predictive mean matching; a donor-based method that imputes each missing value by selecting an observed value from cases with predicted means closest to that of the incomplete case. The method name was originally introduced by \citet{little1988missing}, and its implementation in \texttt{mice} follows the approach described by \citet{van2011mice}. Further methodological details and tuning guidelines are provided by \citet{morris2014tuning};
    
    \item T5: \textbf{\texttt{rf}} -- imputation of univariate missing data using random forests \citep{breiman2001random}, being a nonparametric ensemble of regression trees capturing nonlinearities and interactions. The method used on \texttt{mice} function calls \texttt{randomForest::randomForest()} which implements Breiman's random forest algorithm \citep{randomForest}. More details and simulation approach about alternative implementation of this method can be found on \citet{shah2014comparison} and \citet{salman2024random}; and
    
    \item T6: \textbf{\texttt{midastouch}} -- imputation occurs by predictive mean matching with distance aided donor selection, where donor selection is influenced by predictive distance weights. The method implemented on \texttt{mice} is based on \citet{rubin1987} and \citet{siddique2008multiple}. More details can be found on \citet{van2006fully} and \citet{gaffert2016towards}.

\end{itemize}

These techniques were chosen to reflect diverse modelling philosophies: fully parametric regression models (T1, T2, T3), donor methods with weaker distributional assumptions (T4, T6), and a nonparametric machine-learning approach (T5). This diversity enables evaluation of how model structure, predictor selection, and resampling strategies affect imputation performance across scenarios with and without extreme values. Further details about the imputation methods used and other techniques can be found in \citet{mice}.


\subsection{Comparison of multiple imputation methods}\label{sec_comparison}
The quality of an imputation strategy is assessed through its impact on \emph{downstream} analyses rather than by directly comparing imputed entries to their unobserved counterparts. As stressed by \citet{buuren2018}, imputation is not prediction, and diagnostics that treat it as such tend to be overstated or misleading. Accordingly, the comparison focuses on predictive and inferential performance under controlled simulation designs. 

\textbf{(i) Out-of-sample prediction error:} for each simulation replicate and completed data set, predictive performance is summarised by $K$-fold cross-validated mean squared error (CV--MSE), with fold assignments fixed within a replicate and reused across imputations and methods. This reduces variability from the fold assignments and ensures a paired, one-to-one comparison across methods \citep{arlot2010}. In the clean-data regime, ordinary least squares (OLS) is used; under contamination, elastic net (EN) with $\alpha=0.5$ and $\lambda$ chosen by \texttt{cv.glmnet} is used for prediction \citep{zou2005regularization,friedman2010}. For each MI method, the CV--MSE is averaged across the $M$ imputations within a replicate and then summarised across replicates by mean, variance, and the $2.5\%$, $50\%$, and $97.5\%$ quantiles. Lower quantiles with smaller dispersion indicate better and more stable performance.

\textbf{(ii) Inferential criteria and congeniality:} beyond prediction, the clean-data regime evaluates whether multiple imputation restores nominal inference for the true linear model: pooled bias, RMSE, and $95\%$ coverage for $(\beta_0,\beta_1,\beta_2)$ are reported via Rubin’s rules \citep{rubin1987,white2011,little2019statistical}. The imputation model for $x_2$ includes $(y,x_1)$ and excludes $x_2$ in the predictor matrix, promoting congeniality with the analysis model \citep{buuren2018,van2011mice}. In the contaminated regime, coefficients are summarised via post-selection pooled OLS after EN, using a pre-specified selection rule (retain variables selected in at least $50\%$ of imputations; union fallback otherwise). These summaries are conditional on selection and therefore optimistic for coverage by construction; they are reported transparently rather than as formal guarantees from inference that accounts for selection \citep{berk2013,taylor2015}.

\textbf{(iii) Visual summaries and calibration:} to complement tabular summaries, density plots and boxplots of replicate-level CV--MSE provide distributional shape and tail behaviour. Quantile--quantile (QQ) curves of predicted versus true $Y$ offer a calibration view (median across imputations per replicate, then aggregated across replicates), with the $45^\circ$ line as a reference.

\textbf{(iv) Fairness and reproducibility controls:} all methods are evaluated under a single master random seed, which initializes the generator once. Random draws are organised into independent, deterministic \emph{substreams} by task (fold assignment, MCAR masks, contamination, imputations), so that changing one task does not shift the others.  Based on \citet{arlot2010}, cross-validation folds are fixed within each replicate and reused across all imputations and methods, ensuring paired comparisons and reducing fold induced by noise. For contaminated data, elastic net is tuned using these fixed folds. 
The order of methods (T1--T6) is constant across figures and tables. As in \citet{varma2006,cawley2010}, it is acknowledge that selecting $\lambda$ by minimising CV error introduces mild optimism relative to an external test fold or nested CV.

\subsection{Simulation design (protocol and evaluation)}\label{sec_simulation}
\label{sec:simulations}

This subsection records only the Monte Carlo \emph{protocol} and the \emph{evaluation} pipeline. 
All generative details (baseline model, contamination scheme) are defined in Section \ref{data_sets} (Data sets), and the downstream analysis models are specified in Section \ref{statistical_models} (Statistical models). 
Imputation methods (subsection \ref{imputation_methods}) and their predictors are not repeated here.

\paragraph{Protocol per scenario.}
For each scenario defined by $\{n, P_{\text{miss}}, P_{\text{ext}}, \rho, M, n_{\text{sim}}, \text{MI}, \text{Branch}\}$:

\begin{table}[htbp]
\centering
\setlength{\fboxsep}{6pt}
\fbox{%
\begin{minipage}{0.875\textwidth}
\small
\centering
\caption{Design factors and analysis branches for the Monte Carlo study. Scenarios are given by the Cartesian product of the factors below; clean data are analysed with OLS and contaminated data (with extremes) with elastic net (EN)}
\label{tab:factors}
\begin{threeparttable}
\begin{tabularx}{\textwidth}{@{} l| l X @{}}
\toprule
\textbf{Factor} & \textbf{Levels} & \textbf{Definition} \\
\midrule
$n$ & $\{20,40,80,200,500\}$ &
Number of observations in each scenario \\

$P_{\text{miss}}$ & $\{0.05,0.10,0.25,0.30\}$ &
proportion of missings \\

$P_{\text{ext}}$ & $\{0.03,0.04,0.05,0.10,0.15,0.30\}$ &
Proportion of extreme values \\

$\rho$ & $\{0,0.6\}$ &
Correlation between $x_1$ and $x_2$ \\

$M$ & $\{5,10\}$ &
Number of imputations per scenario (iterations) \\

$n_{\text{sim}}$ & $\{50,300,1000, 3000\}$ &
Number of Monte Carlo replicates \\

MI & \{T1, T2, T3, T4, T5, T6\} &
Multiple imputation methods\\

Branch & \{clean/OLS, cont./EN\} &
Data type and respective statistical model \\
\bottomrule
\end{tabularx}

\begin{tablenotes}[flushleft]\scriptsize
\item $^*$Legend: scenarios combine all factor levels unless noted otherwise in subsections reporting stratified results. The number of imputations per scenario was also reported with ($iter$) symbol through this paper.
\end{tablenotes}
\end{threeparttable}
\end{minipage}%
}
\end{table}

\begin{enumerate}
\item Generate a single baseline (clean) data set; form a paired contaminated copy according to the contamination mechanism already stated there.
\item For each replicate $r=1,\ldots,n_{\text{sim}}$, independently draw an MCAR mask of size $k_{\text{miss}}=\mathrm{round}(n \times P_{\text{miss}})$ on $x_2$ for the clean copy and another MCAR mask for the contaminated copy. Fix a $K$-fold partition (\(K{=}5\)) for replicate $r$; reuse these fold identifiers across all imputations and methods within $r$.
\item For each data type $d\in\{\text{clean},\text{contaminated}\}$ and each MI method $t\in\{T1,\dots,T6\}$:
  \begin{enumerate}
  \item Run \texttt{mice} with $M$ imputations and the \emph{congenial} predictor matrix for $x_2$ (uses $y$ and $x_1$; excludes $x_2$ itself), as specified in Section~\ref{data_inputation};
  \item Fit the \emph{same} analysis model used for complete data: OLS for clean data; elastic net (EN) with $\alpha=0.5$ for contaminated data (Section~\ref{enet_extremes}). For EN, select $\lambda$ via cross--validation using the fixed folds of replicate $r$;
  \item Compute the replicate-level predictive error as the average $K$-fold CV--MSE across the $m$ completed data sets (fold identifiers fixed within replicate $r$ and reused across methods);
  \item (Clean regime only) Pool coefficient estimates over imputations by Rubin’s rules and record bias, RMSE, and $95\%$ coverage for $(\beta_0,\beta_1,\beta_2)$; and
  \item (Contaminated regime only) Record \emph{post-selection} pooled OLS summaries after EN under a fixed rule: retain a covariate if selected in at least $50\%$ of imputations (union fallback if none).
  \end{enumerate}
\end{enumerate}

\paragraph{Evaluation and summaries.}
The primary comparison criterion is the out-of-sample predictive mean squared error (CV--MSE). For each MI method $t$, CV--MSE is first averaged across the $M$ imputations within a replicate and then summarised across replicates by the mean, variance, and the $2.5\%$, $50\%$, and $97.5\%$ quantiles. 
These statistics correspond exactly to the ``Pred.~MSE (out-of-sample)'' columns in the tables and to the density/boxplot panels in the figures. 
Distributional calibration is conveyed by quantile--quantile curves of predicted versus true $Y$, computed by taking, for each replicate, the median across imputations, then aggregated over replicates. In the clean-data regime, those metrics assess inferential validity under congeniality \citep{rubin1987,white2011,buuren2018}. In the contaminated regime, coefficient summaries are reported transparently as conditional on selection, following the fixed rule described in Section~\ref{sec_comparison} \citep{berk2013,taylor2015}.

\paragraph{What is \emph{not} varied here.}
Missingness is induced only in $x_2$ under MCAR. This is intentional and sufficient for the aim, an algorithmic \emph{comparison} of MI procedures, because fully conditional specification updates one univariate conditional at a time; the single incomplete variable setting isolates each method’s core behaviour while preserving clean Rubin pooling under MCAR \citep{buuren2018,van2011mice,white2011,little2019statistical}. Extending to MAR/MNAR or to multiple incomplete variables would require re-specifying the conditional models, checking compatibility, and conducting sensitivity analyses, with a substantial computational burden relative to the already large grid (across $n, P_{\text{miss}}, P_{\text{ext}}, \rho, M, n_{\text{sim}}$, MI and Branch). Also, exploring alternative values for $\alpha$ or robust losses would introduce an additional tuning dimension (e.g., nested CV) and a different analysis regime \citep{RousseeuwLeroy,HuberRonchetti}. Finally, no parametric models based on the Extreme Value Theory (e.g., GEV or GPD distributions) is imposed for the generation of extremes; doing so could introduce extra choices and diagnostics, and reduce the generality of the insights \citep{fisher1928limiting,von1936distribution,gnedenko1943,jenkinson1955frequency}.

\newpage
Algorithm 1 outlines the Monte Carlo protocol, using a single master seed with deterministic substreams for folds, masks, imputation, and model tuning.

\begin{algorithm}[H]
\caption{Monte Carlo protocol with a single master seed and deterministic substreams}
\label{alg:simulations_2}
\begin{algorithmic}[1]
\STATE Initialize a pseudo-random generator with a single master seed $S=241103414$ and use deterministic substreams for folds, masks, imputation, and tuning.
\FOR{each scenario $\{n, P_{\text{miss}}, P_{\text{ext}}, \rho, M, n_{\text{sim}}, \text{MI}, \text{Branch}\}$}
  \FOR{$r=1,\dots,n_{\text{sim}}$}
    \STATE \textbf{Data generation (substream A).} Draw a clean baseline sample $\{(y_i,x_{1i},x_{2i})\}_{i=1}^n \sim F_0$ (Sec.~\ref{data_sets}); create its contaminated twin via the $\pm 3s$ mechanism (Sec.~\ref{subsec:contamination}).
    \STATE \textbf{Fold identifiers (substream B).} Draw a $K$-fold map ($K=5$) $f_r:\{1,\ldots,n\}\to\{1,\ldots,K\}$; pass $f_r$ to all cross-validation calls within replicate $r$.
    \STATE \textbf{MCAR masks (substreams C,D).} Independently sample $k_{\text{miss}}=\mathrm{round}(n \times P_{\text{miss}})$ indices for the clean and contaminated copies; set $x_2$ to \texttt{NA} at those indices.
    \FOR{data type $d\in\{\text{clean},\text{contaminated}\}$}
      \FOR{method $t\in\{\text{T1},\dots,\text{T6}\}$}
        \STATE \textbf{Imputation (substream E$_{r,t,d}$).} Run \texttt{mice::mice()} with $M$ imputations and a congenial predictor matrix for $x_2$ using $(y,x_1)$ only; obtain completed sets $\{\mathcal{D}^{(m)}_{r,t,d}\}_{m=1}^M$.
        \STATE \textbf{Model and prediction error.}
        \IF{$d=\text{clean}$}
          \STATE Fit OLS on each $\mathcal{D}^{(m)}_{r,t,d}$; compute CV--MSE using $f_r$; set $e^{(m)}_{r,t,d}=$ CV--MSE.
        \ELSIF{$d=\text{contaminated}$}
          \STATE Fit elastic net ($\alpha=0.5$) with $f_r$ as \texttt{foldid}; record CV--MSE at $\lambda_{\min}$ as $e^{(m)}_{r,t,d}$.
        \ENDIF
        \STATE Define the replicate-level error $\bar e_{r,t,d}=\frac{1}{M}\sum_{m=1}^M e^{(m)}_{r,t,d}$.
        \STATE \textbf{Coefficient summaries.}
        \IF{$d=\text{clean}$}
          \STATE Pool OLS across $M$ imputations via Rubin's rules; record bias, RMSE, and 95\% coverage for $(\beta_0,\beta_1,\beta_2)$.
        \ELSE
          \STATE Let $S^\star=\{j\in\{x_1,x_2\}:\Pr(j\ \text{selected})\ge \texttt{support\_thresh}\}$ across the $M$ EN fits; if $S^\star=\varnothing$, use the union of selected variables (fallback; if still empty, use $\{x_1,x_2\}$). Pool OLS on predictors $S^\star$.
        \ENDIF
      \ENDFOR
    \ENDFOR
  \ENDFOR
\STATE \textbf{Across-replicate summaries (CV--MSE).} For each $(t,d)$, summarise the replicate-level CV--MSE values $\{\bar e_{r,t,d}\}_{r=1}^{n_{\text{sim}}}$ by the mean, variance, and $2.5\%, 50\%, 97.5\%$ quantiles; plot the CV--MSE density and boxplots, and additionally produce QQ calibration curves of predicted vs.\ true $\mathbf{y}$.

\ENDFOR
\end{algorithmic}
\end{algorithm}

\newpage
\section{Results}\label{results}
This section reports a representative set of scenarios that illustrate the paper’s main conclusions. Complete simulation tables for all scenarios settings and replicates are provided in Appendix~\ref{appendix}.

\subsection{Complete-data benchmarks: structure and baseline fits} \label{results_example}
\begin{figure}[H]
  \centering
  \includegraphics[width=\textwidth]{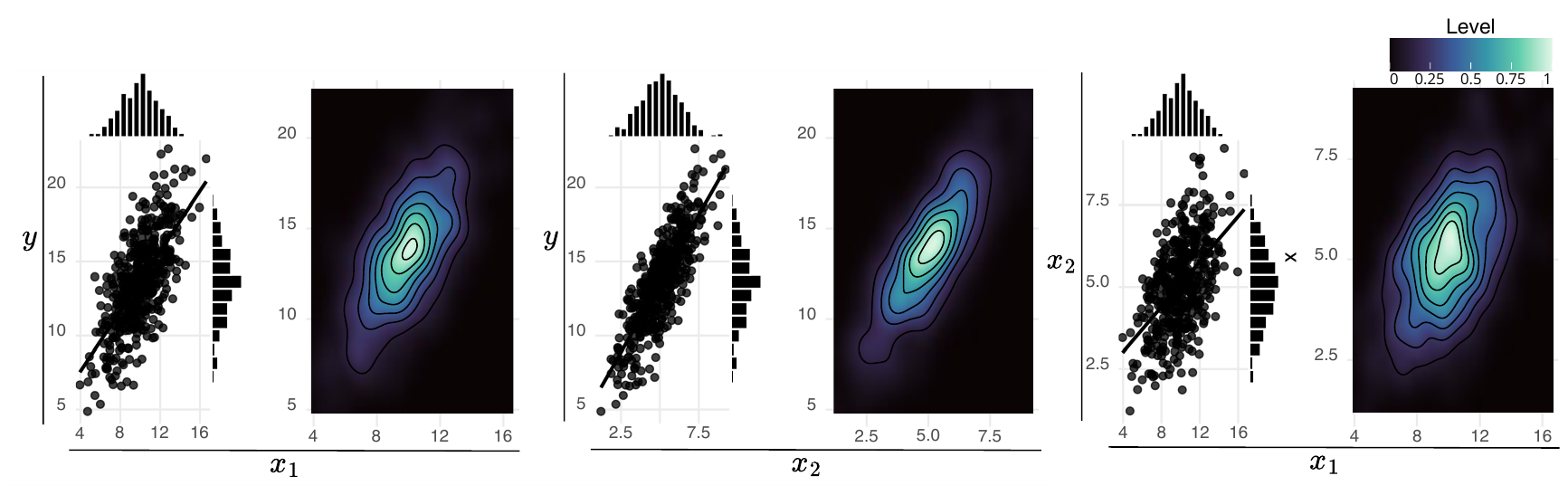}
  \caption{Relationships (correlation plot, marginal histograms and empirical densities) for \textbf{clean data} sets with \(n=500\), \(P_{\text{ext}}=0.10\), and \(\rho=0.6\).
  Axes are \(y, x_1, x_2\). Colorbars (`Level') indicate relative bivariate density per row}
  \label{fig:clean_example}
\end{figure}

\begin{figure}[H]
  \centering
  \includegraphics[width=\textwidth]{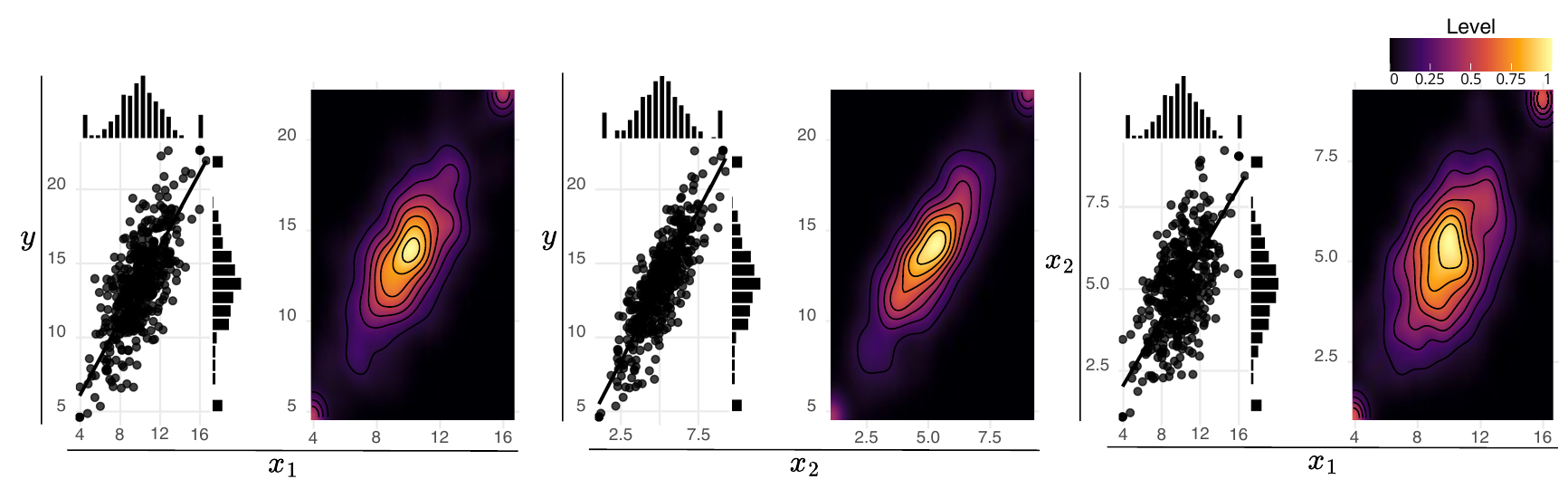}
  \caption{Relationships (correlation plot, marginal histograms and empirical densities) for \textbf{contaminated data} (with extreme values) sets with \(n=500\), \(P_{\text{ext}}=0.10\), and \(\rho=0.6\).
  Axes are \(y, x_1, x_2\). Colorbars (`Level') indicate relative bivariate density per row}
  \label{fig:cont_example}
\end{figure}

To fix ideas about the the complete datasets and the corresponding model adjustments, consider the case with \(n=500\) observations, contamination proportion of extreme values \(P_{\text{ext}}=0.10\) (therefore 50 values replaced in each variable), and correlation \(\rho=0.6\) between \(x_1\) and \(x_2\).
Figures \ref{fig:clean_example} and \ref{fig:cont_example} presents pairwise relationships among \((y, x_1, x_2)\) for two data regimes: the first row corresponds to the clean data (blue palette), and the second row to the contaminated data (yellow/red palette). Axes are labeled \(y, x_1, x_2\); each row includes its own colorbar (`Level') indicating relative bivariate density.

Three patterns stand out. (i) In the clean regime, scatterplots and linear smooth fits reveal near-linear relations with tight dispersion; also the density contours are approximately elliptical, consistent with \(\rho\approx0.6\) and the bivariate normal generator. (ii) Under contamination, dispersion increases in all pairwise views, level sets spread, and regions with high density appear farther from the central mass, indicating leverage effects from extremes in the tail. (iii) The fitted linear trend lines maintain similar directions across regimes, but prediction uncertainty increases substantially when extremes are present.

Table~\ref{tab1_example1} reports estimates and predictive error for reference models fitted to the complete data (no missingness): ordinary least squares (OLS) for clean data and elastic net (EN) for contaminated data. The linear model attains lower error on clean data, whereas EN is slightly more stable when contamination inflates variance and introduces high-leverage points. These benchmarks for this scenario serve as baselines for the imputation experiments that follow.

\begin{table}[H]
\centering
\caption{Estimated values for parameters and performance measure (MSE) for the linear (OLS) and sparse (EN) regression models fitted to the complete data ($n = 500$, $P_\text{ext}=0.10$, $\rho = 0.6$)}
\label{tab1_example1}
\begin{tabular}{llrrrr}
\toprule
      Data &                   Model &  $\hat{\beta}_0$ &  $\hat{\beta}_1$ &  $\hat{\beta}_2$ &   MSE \\
\midrule
Clean & OLS &   1.170 &   0.517 &   1.440 & 2.190 \\
Contaminated & EN &   1.190 &   0.532 &   1.410 & 2.020 \\
\bottomrule
\end{tabular}
\end{table}

\subsection{Simulation study: imputations under missingness and extreme values}\label{main_results}

Results are reported for six multiple imputation methods (T1--T6) under two analysis regimes: \emph{clean} data fitted by ordinary least squares (OLS) and \emph{contaminated} data with extreme values fitted by elastic net (EN). Comparisons are organised by sample-size regimes: small ($n\in\{20,40\}$), moderate ($n\in\{80,200\}$), and large ($n=500$). And by missingness blocks: \emph{low} (5\%, 10\%) and \emph{high} (25\%, 30\%). Inferential performance is summarised by bias, RMSE, and 95\% coverage for $(\beta_0,\beta_1,\beta_2)$; predictive performance is summarised by out-of-sample CV--MSE via its mean $\overline{X}$, variance $\sigma^2$, and median $Q_{50}$. Tail quantiles $Q_{2.5}$ and $Q_{97.5}$ are also reported in Appendix \ref{appendix}.

Method groupings used for general interpretation are: parametric (T1--T3), donor-based (T4, T6), and non-parametric/ML (T5). Three recurring patterns preview the detailed results: (i) under contamination, coefficients display a stable directionality (intercept inflation, $\beta_1$ tilt, $\beta_2$ shrinkage); (ii) increasing $n$ chiefly contracts variability rather than removing those shifts; and (iii) for prediction, parametric MI (T1--T3) typically yields tighter CV--MSE tails, whereas donor/ML procedures (T4--T6) often reduce slope bias at the expense of heavier dispersion. Subsequent subsections present the small, moderate, and large samples panels in this order, with the Appendix supplying the full numeric detail.

\begin{figure}[H]
  \centering
  \adjustbox{width=\textwidth, max height=\textheight, center}{%
    \begin{minipage}{.48\linewidth}
      \includegraphics[page=1 ,width=\linewidth]{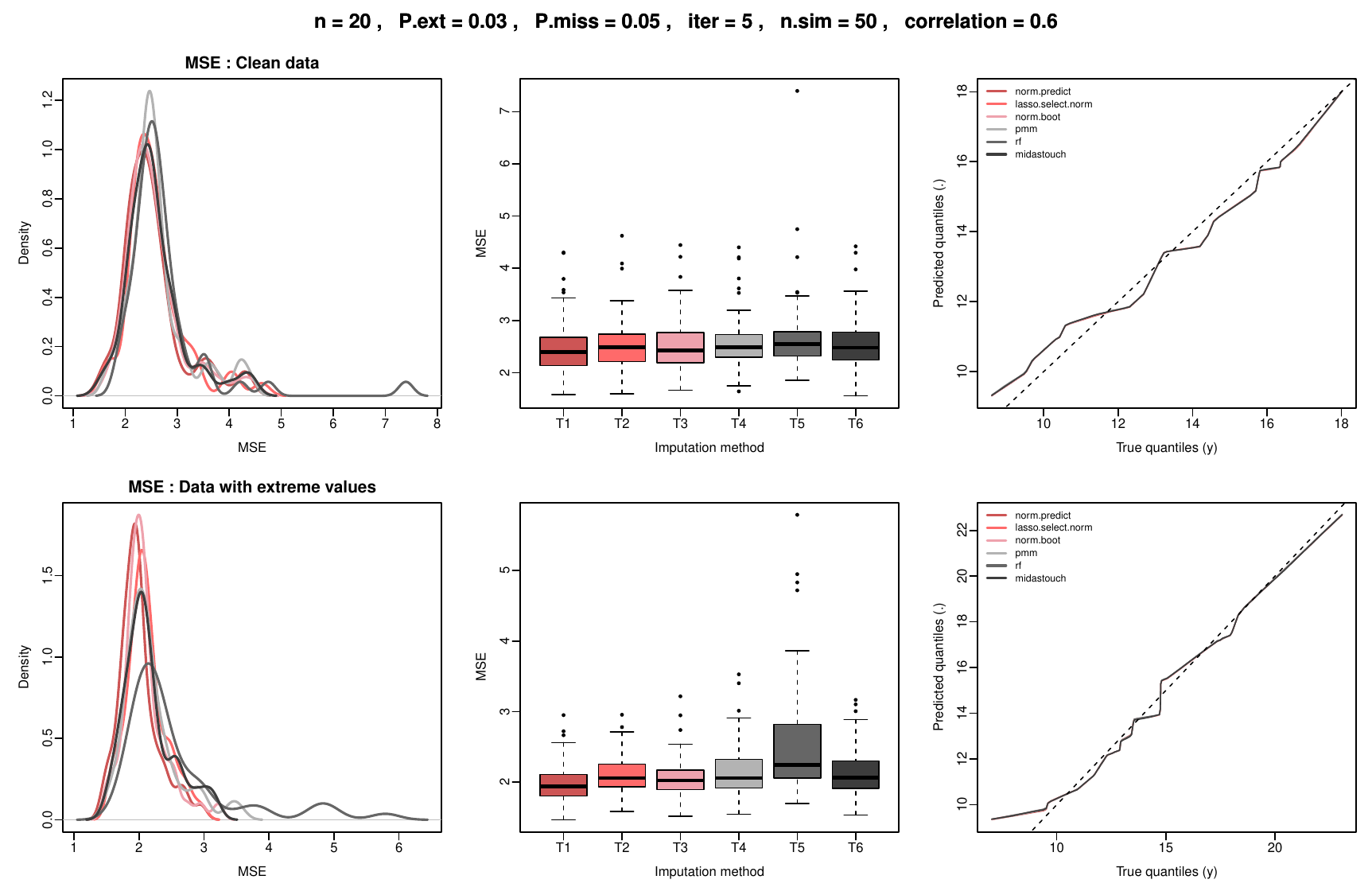}\\[20pt]
      \includegraphics[page=3 ,width=\linewidth]{figures/n_20_40_graficos.1_mesclado.pdf}\\[20pt]
      \includegraphics[page=20,width=\linewidth]{figures/n_20_40_graficos.1_mesclado.pdf}\\[20pt]
      \includegraphics[page=15,width=\linewidth]{figures/n_20_40_graficos.1_mesclado.pdf}
    \end{minipage}\hfill
    \begin{minipage}{.48\linewidth}
      \includegraphics[page=2 ,width=\linewidth]{figures/n_20_40_graficos.1_mesclado.pdf}\\[20pt]
      \includegraphics[page=19,width=\linewidth]{figures/n_20_40_graficos.1_mesclado.pdf}\\[20pt]
      \includegraphics[page=4 ,width=\linewidth]{figures/n_20_40_graficos.1_mesclado.pdf}\\[20pt]
      \includegraphics[page=29,width=\linewidth]{figures/n_20_40_graficos.1_mesclado.pdf}
    \end{minipage}%
  }
      \caption{Predictive MSE densities (clean vs contaminated with extremes data), MSE boxplots, and QQ-plots (predicted vs true quantiles of $y$) across six MI methods (T1--T6). For $\boldsymbol{n=20}$, ordered by $P_{\text{ext}}$, and $P_{\text{miss}}$ (panel 1 of 4). Clean data are analyzed with OLS and contaminated data with elastic net. Each subpanel shows the design values ($n$, $P_{\text{ext}}$, $P_{\text{miss}}$, iter, n.sim, and $\rho$).}
      \label{fig:n_20_40_graficos.1-mesclado-1}%
\end{figure}

\begin{figure}[H]
  \centering
  \adjustbox{width=\textwidth, max height=\textheight, center}{%
    \begin{minipage}{.48\linewidth}
      \includegraphics[page=16,width=\linewidth]{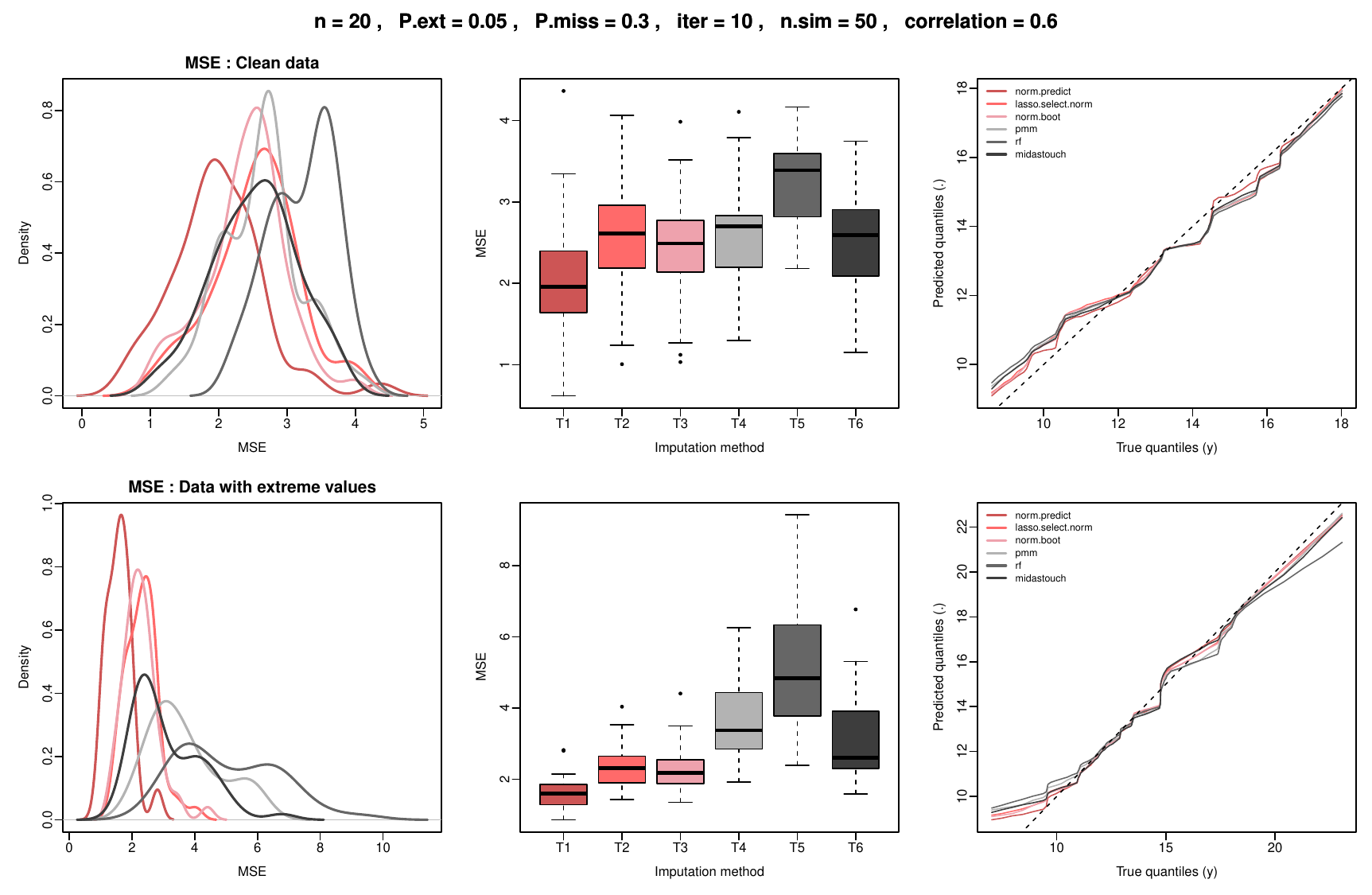}\\[20pt]
      \includegraphics[page=31,width=\linewidth]{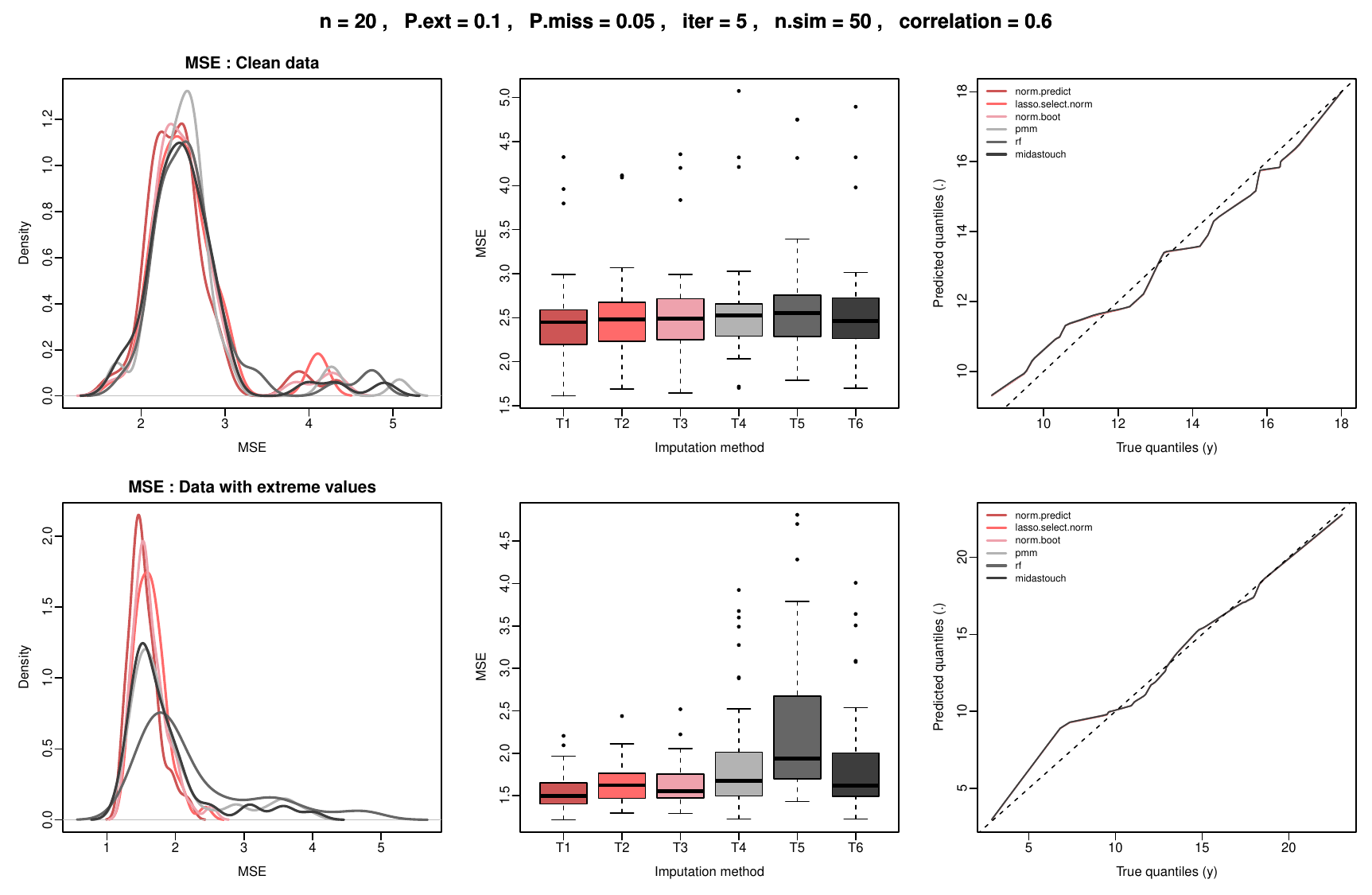}\\[20pt]
      \includegraphics[page=32,width=\linewidth]{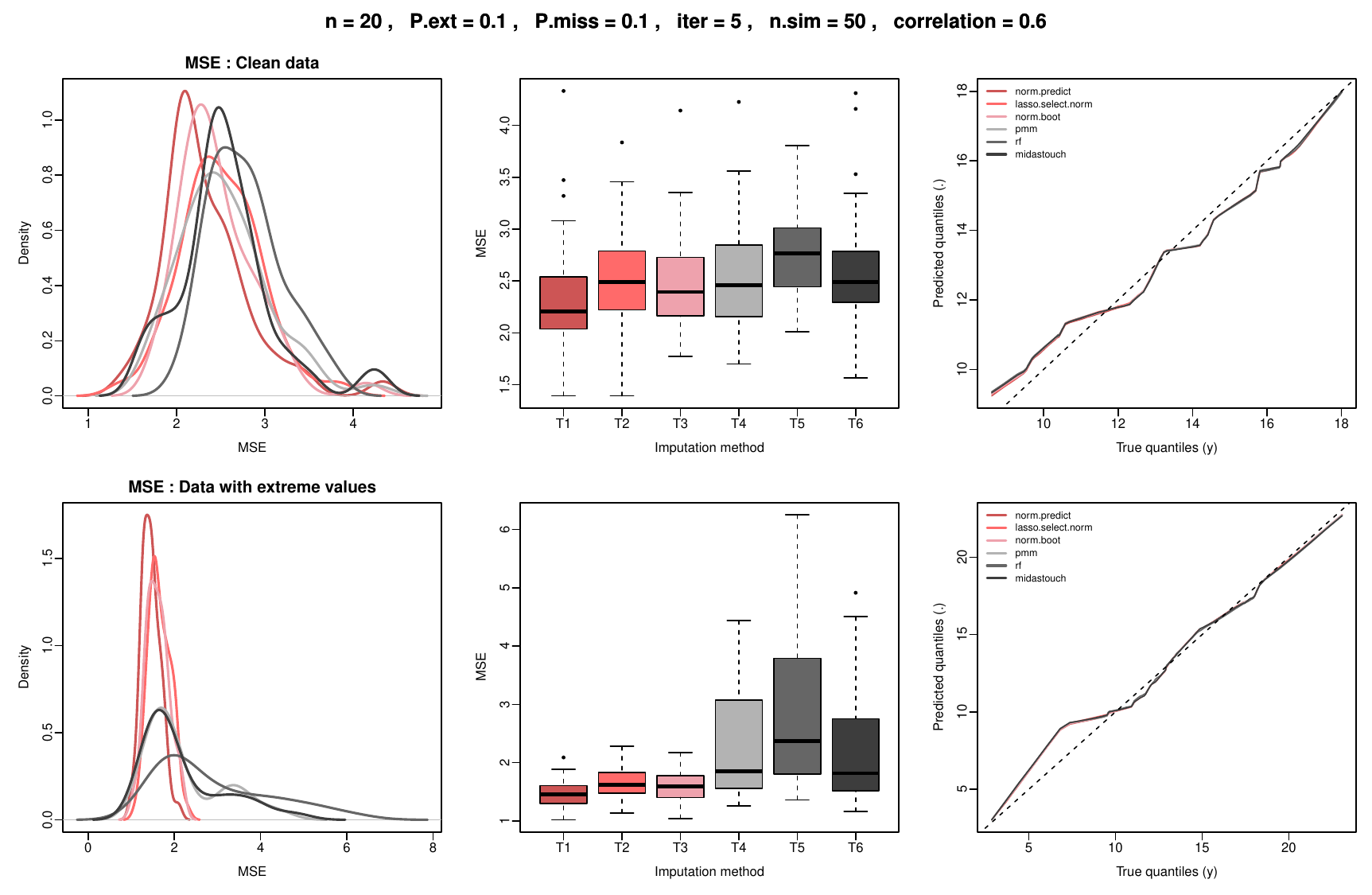}\\[20pt]
      \includegraphics[page=10,width=\linewidth]{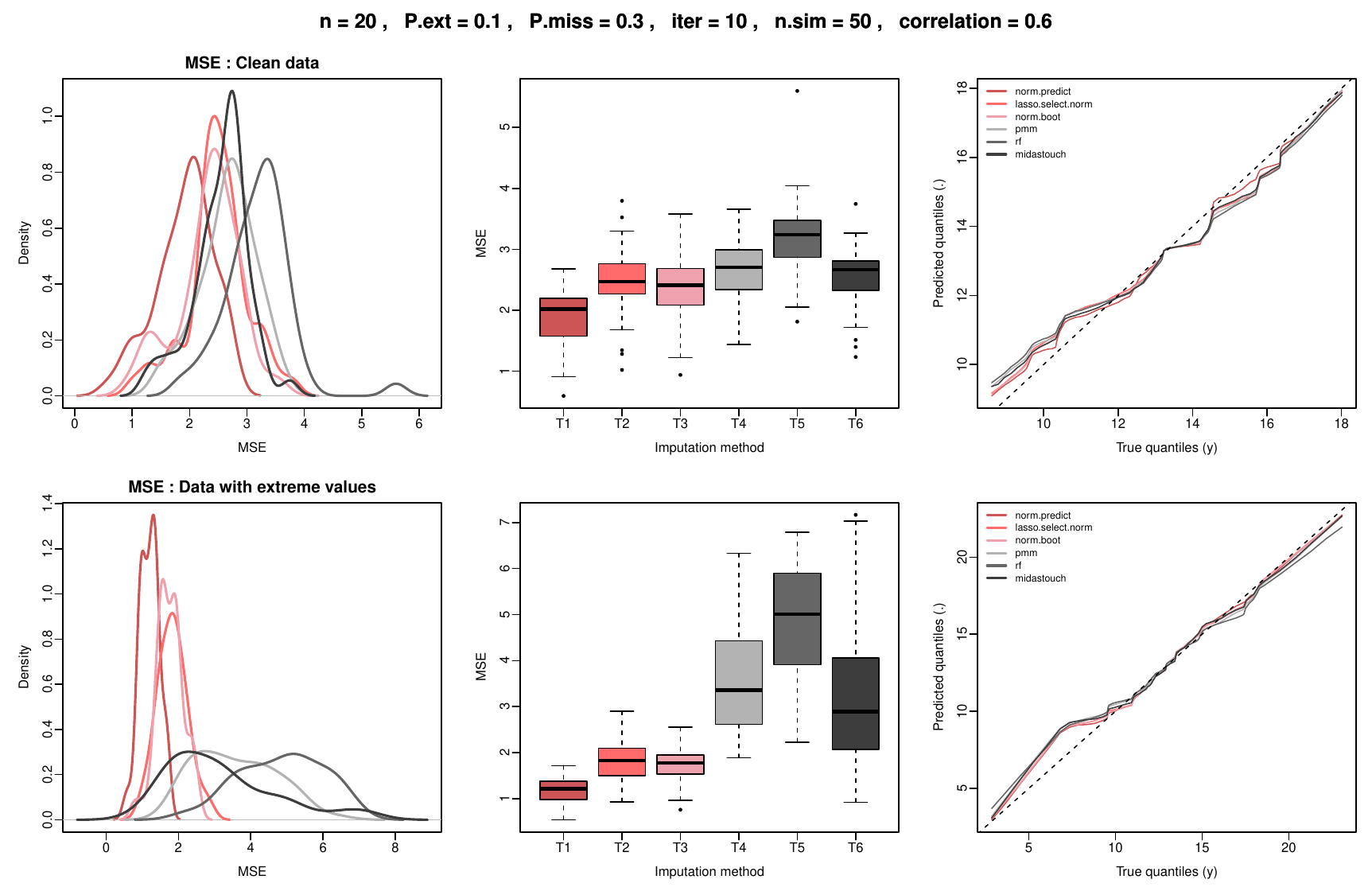}
    \end{minipage}\hfill
    \begin{minipage}{.48\linewidth}
      \includegraphics[page=30,width=\linewidth]{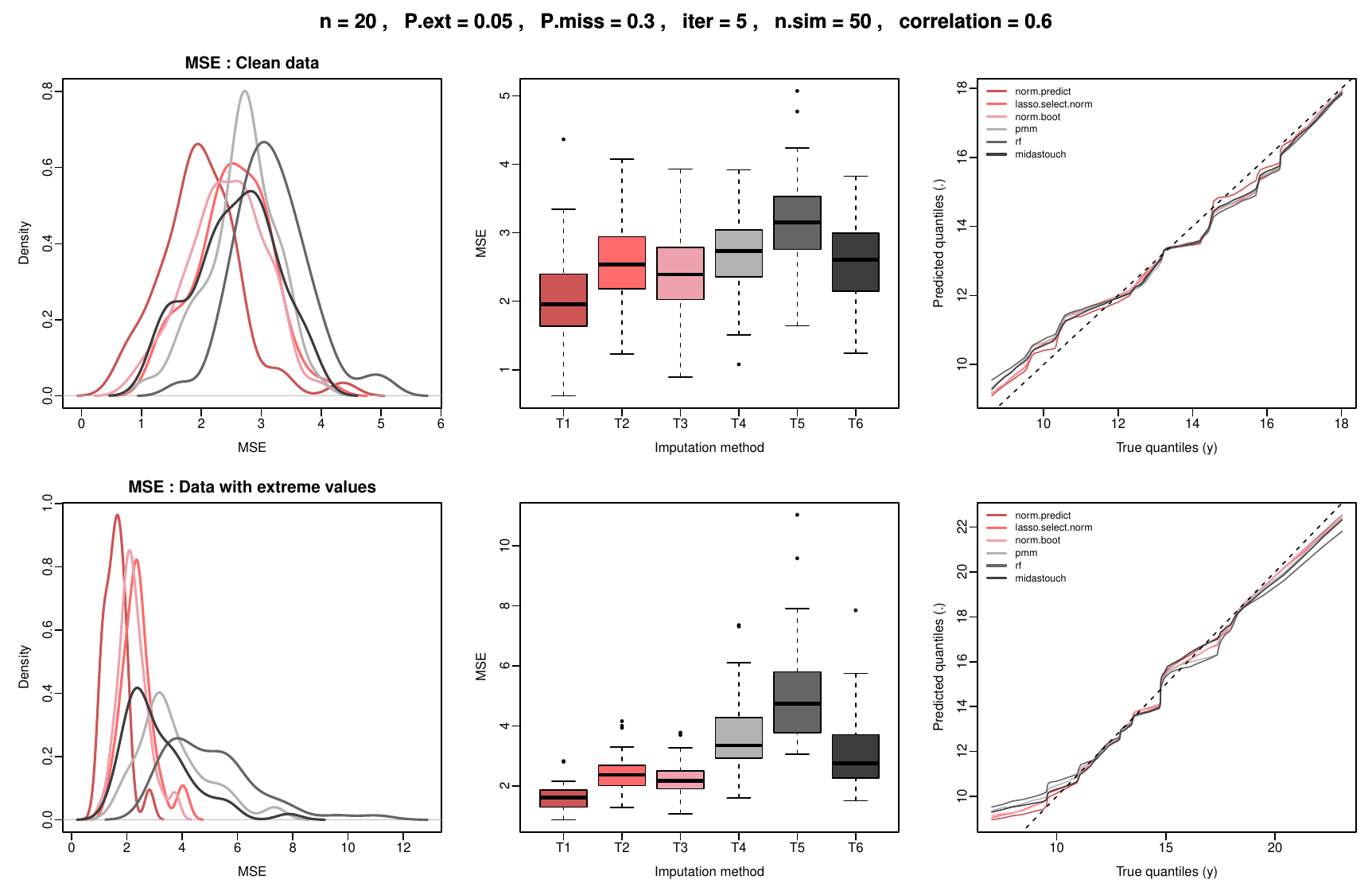}\\[20pt]
      \includegraphics[page=9 ,width=\linewidth]{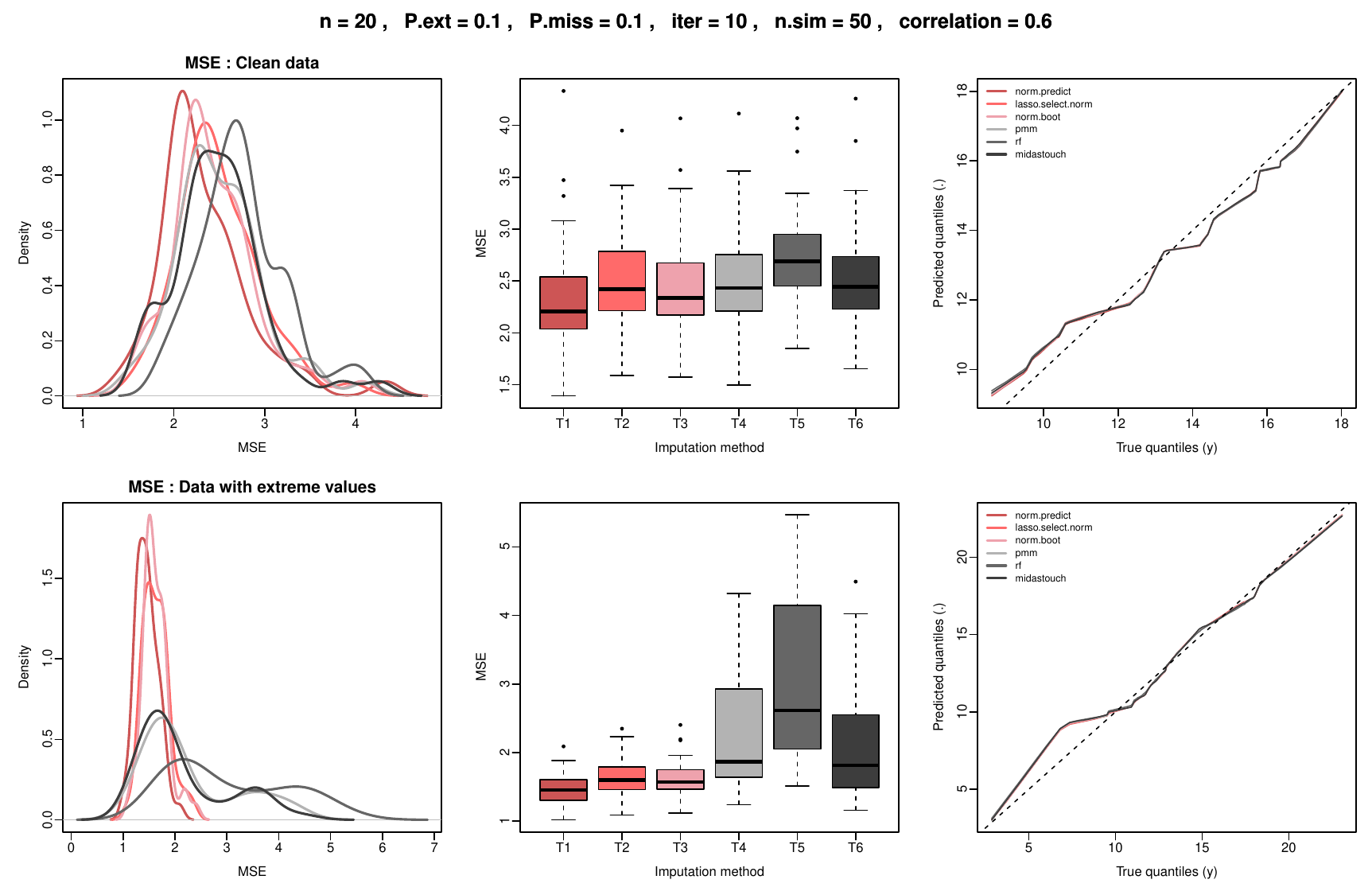}\\[20pt]
      \includegraphics[page=8 ,width=\linewidth]{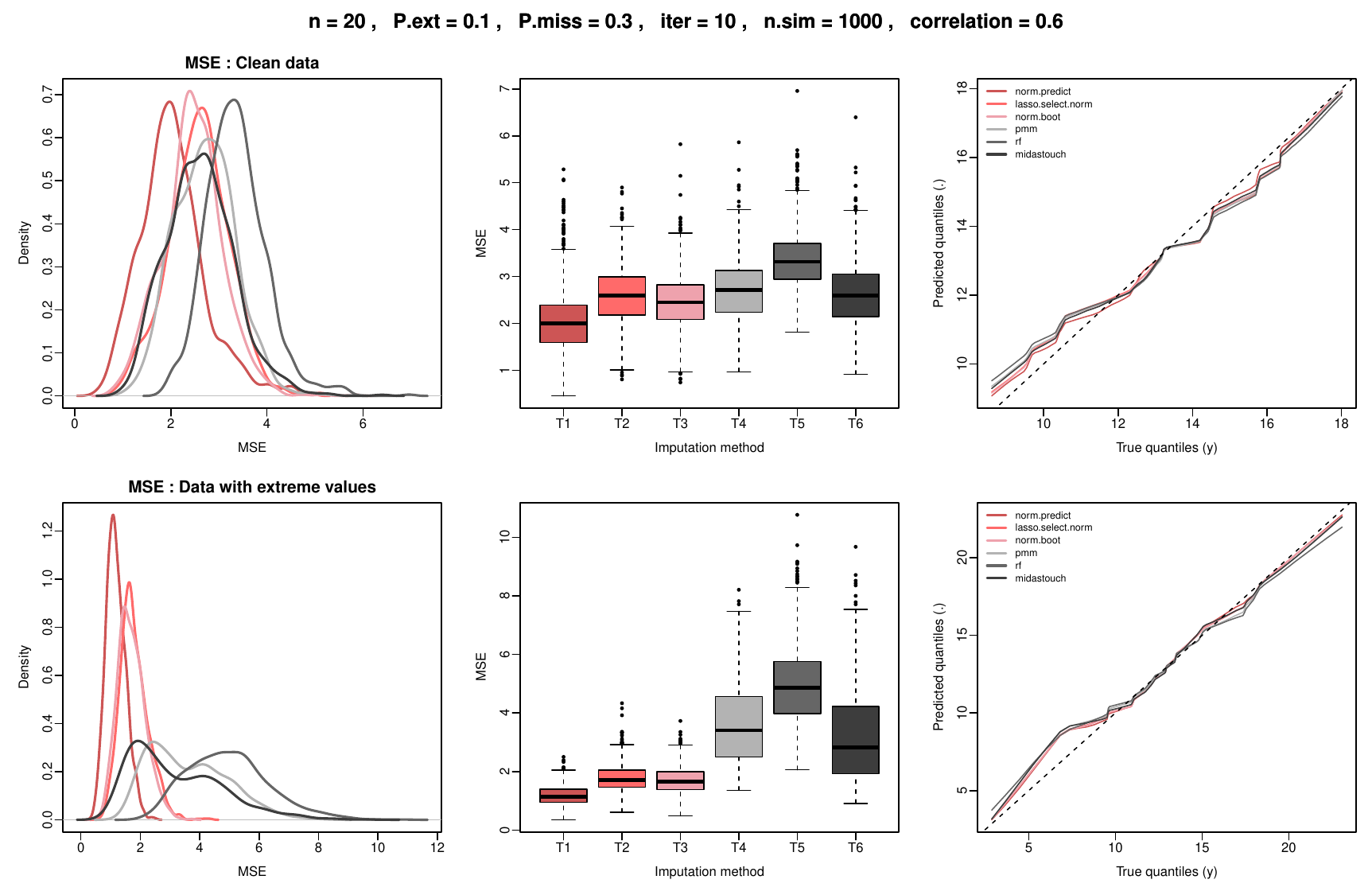}\\[20pt]
      \includegraphics[page=25,width=\linewidth]{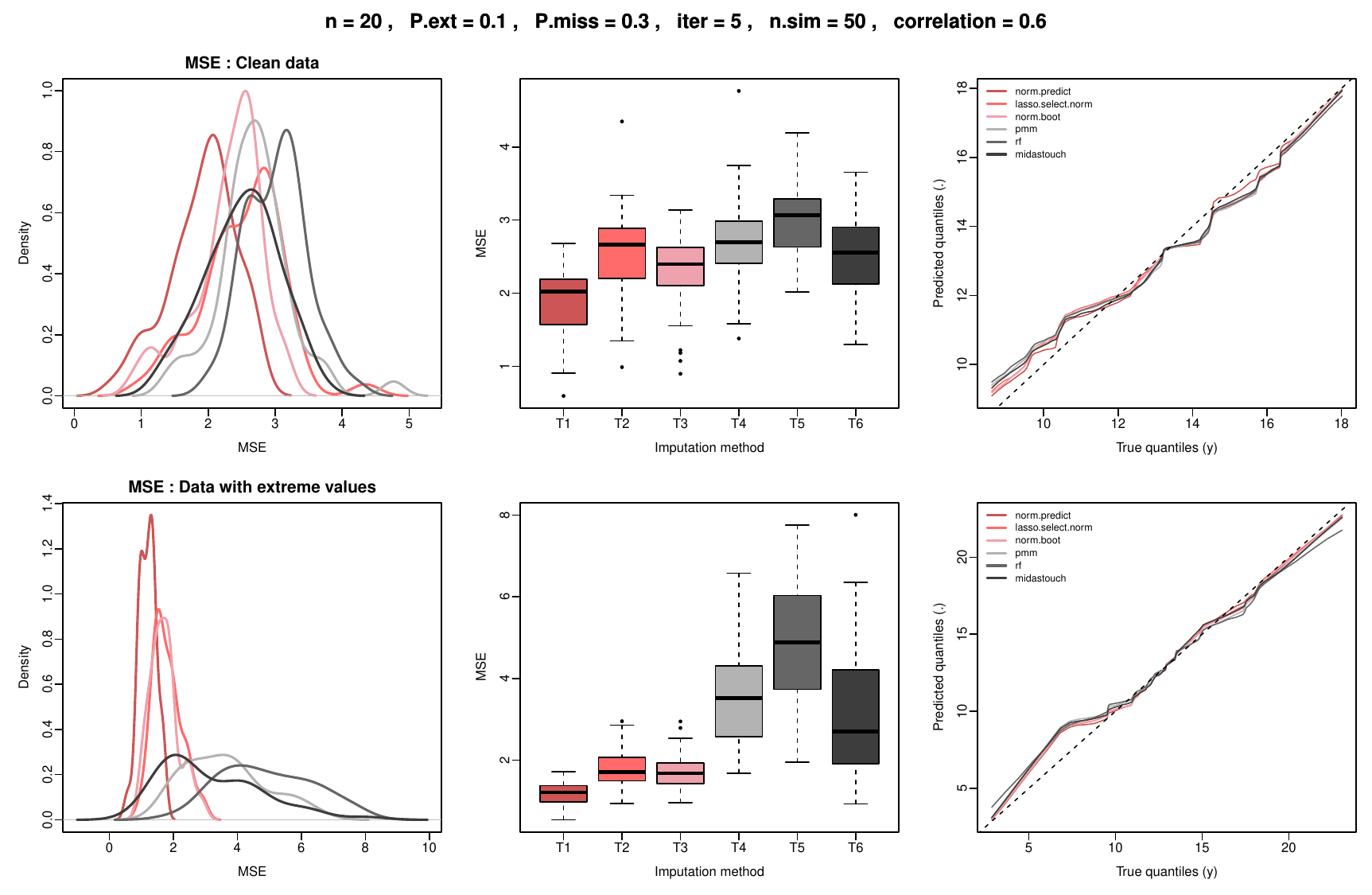}
    \end{minipage}%
  }
    \caption{Predictive MSE densities (clean vs contaminated with extremes data), MSE boxplots, and QQ-plots (predicted vs true quantiles of $y$) across six MI methods (T1--T6). For $\boldsymbol{n=20}$, ordered by $P_{\text{ext}}$, and $P_{\text{miss}}$ (panel 2 of 4). Clean data are analyzed with OLS and contaminated data with elastic net. Each subpanel shows the design values ($n$, $P_{\text{ext}}$, $P_{\text{miss}}$, iter, n.sim, and $\rho$).}
    \label{fig:n_20_40_graficos.1-mesclado-2}%
\end{figure}

\begin{figure}[H]
  \centering
  \adjustbox{width=\textwidth, max height=\textheight, center}{%
    \begin{minipage}{.48\linewidth}
      \includegraphics[page=13,width=\linewidth]{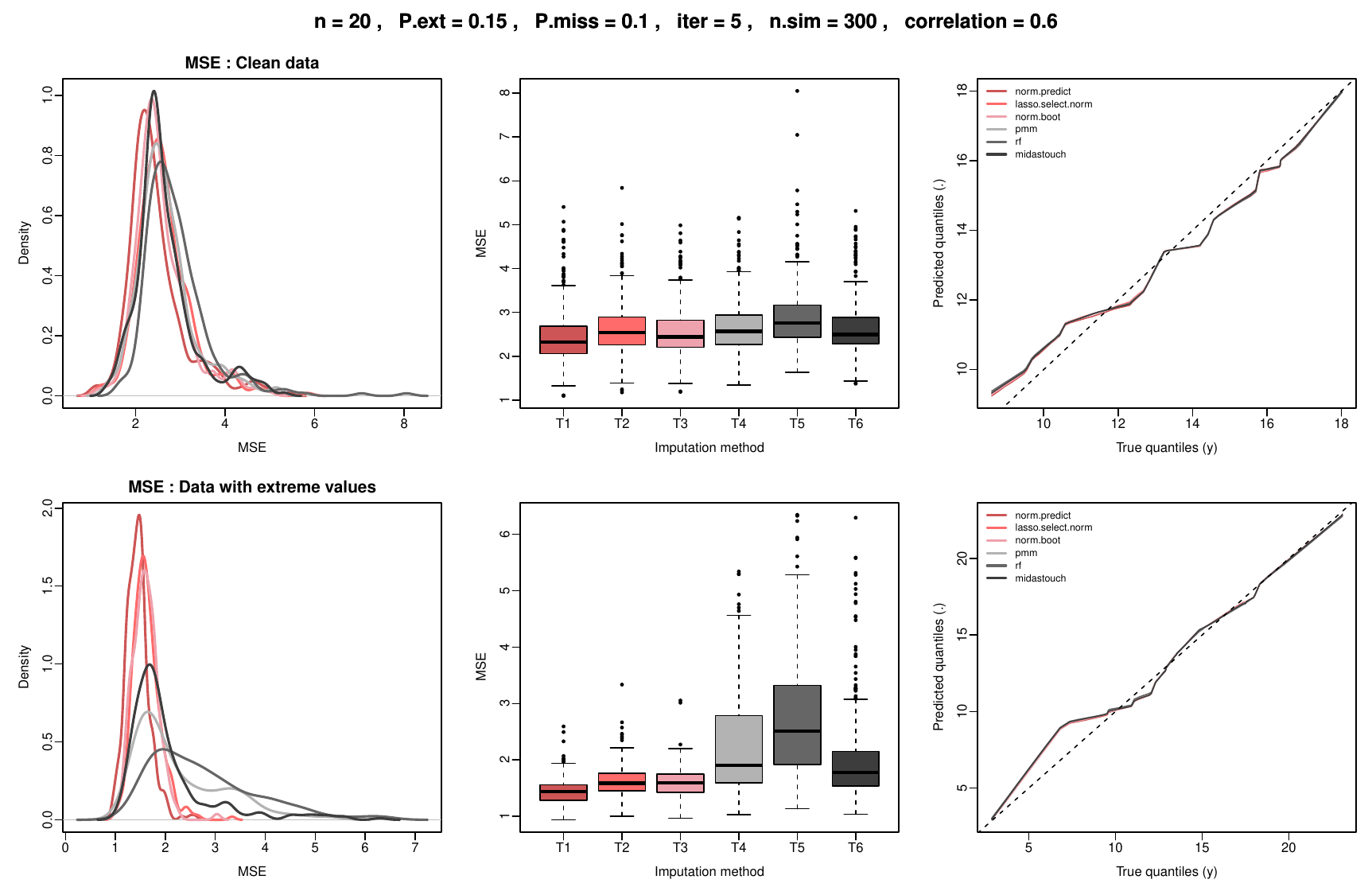}\\[20pt]
      \includegraphics[page=7 ,width=\linewidth]{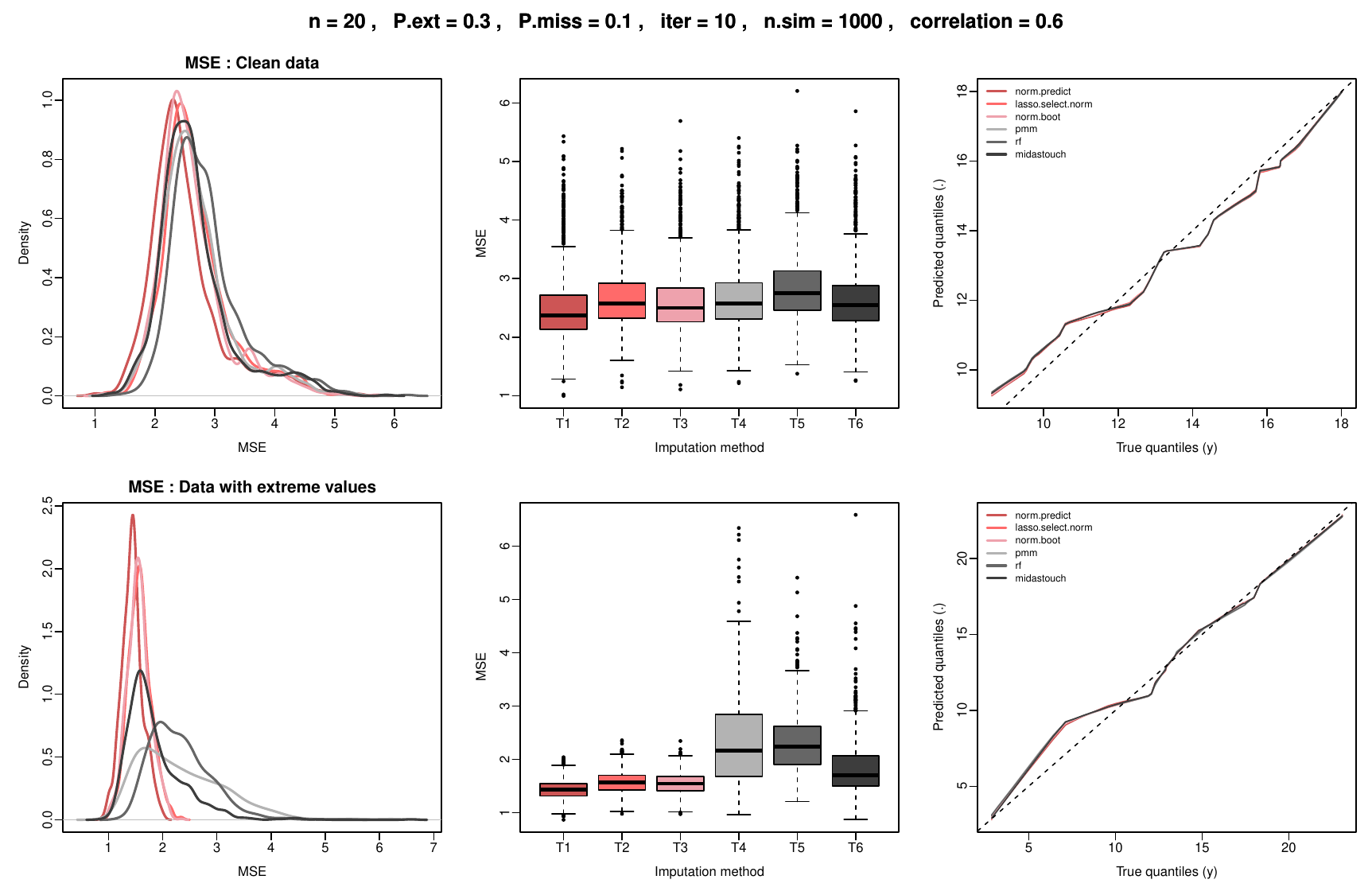}\\[20pt]
      \includegraphics[page=27,width=\linewidth]{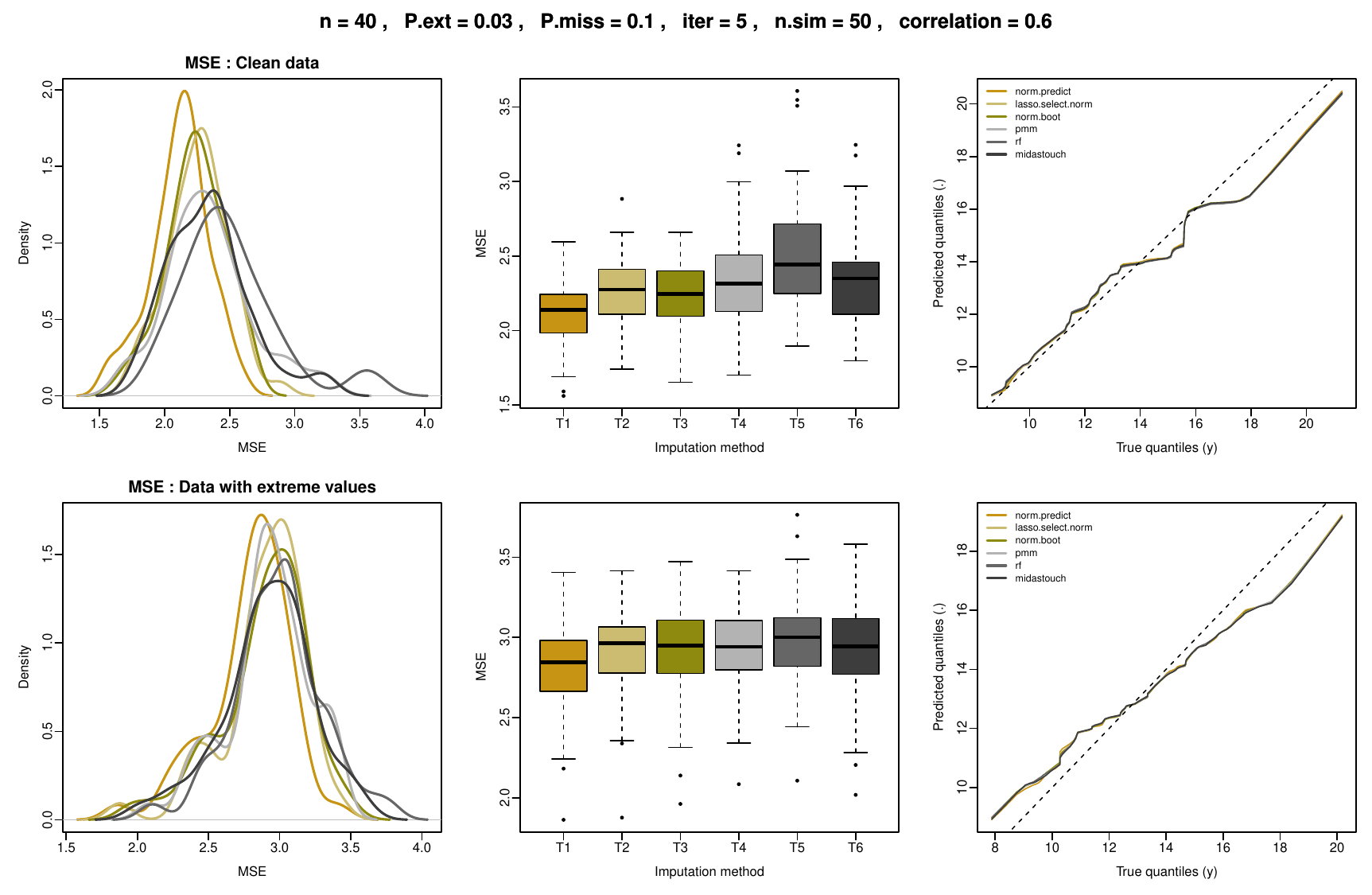}\\[20pt]
      \includegraphics[page=21,width=\linewidth]{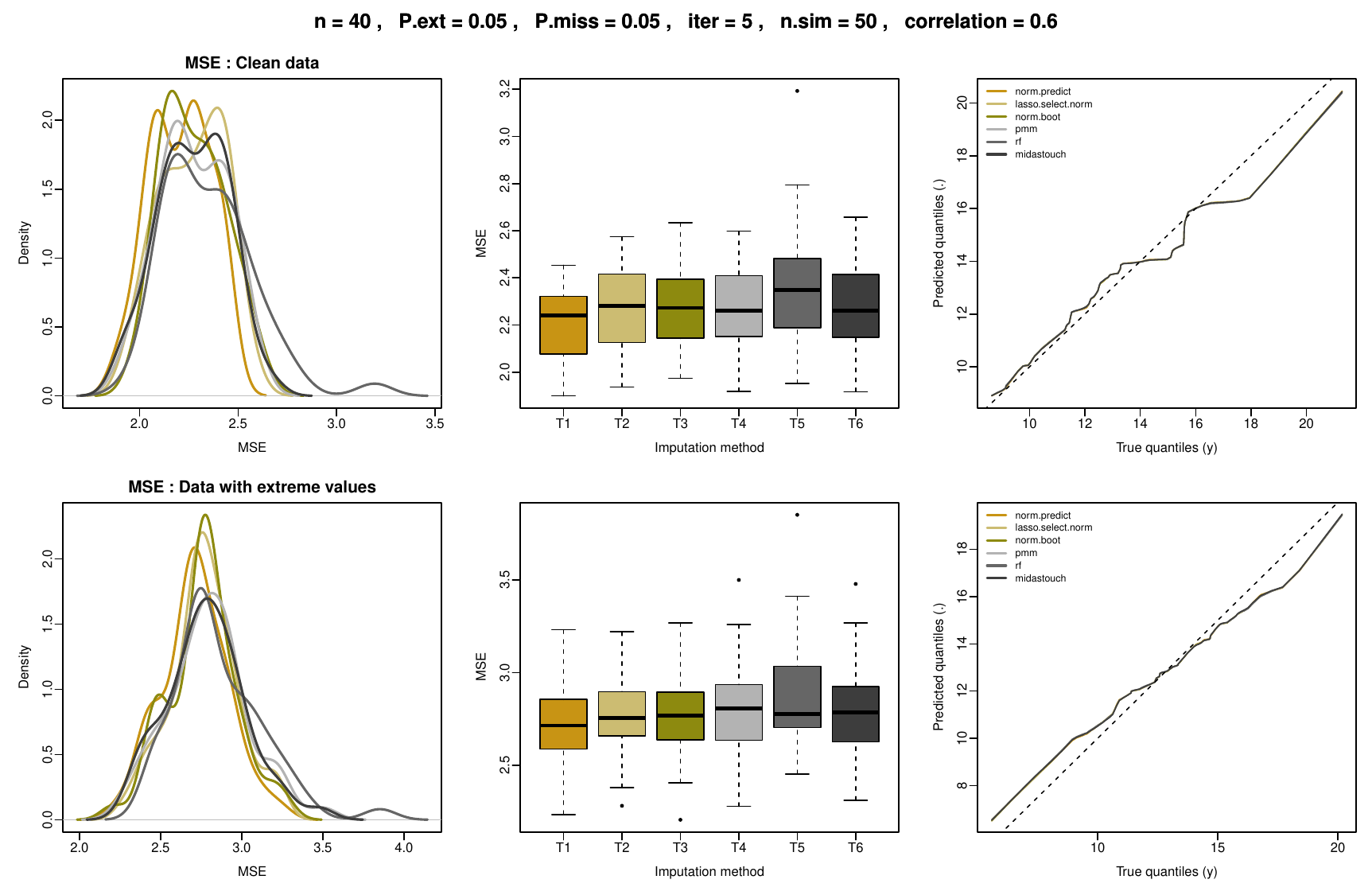}
    \end{minipage}\hfill
    \begin{minipage}{.48\linewidth}
      \includegraphics[page=14,width=\linewidth]{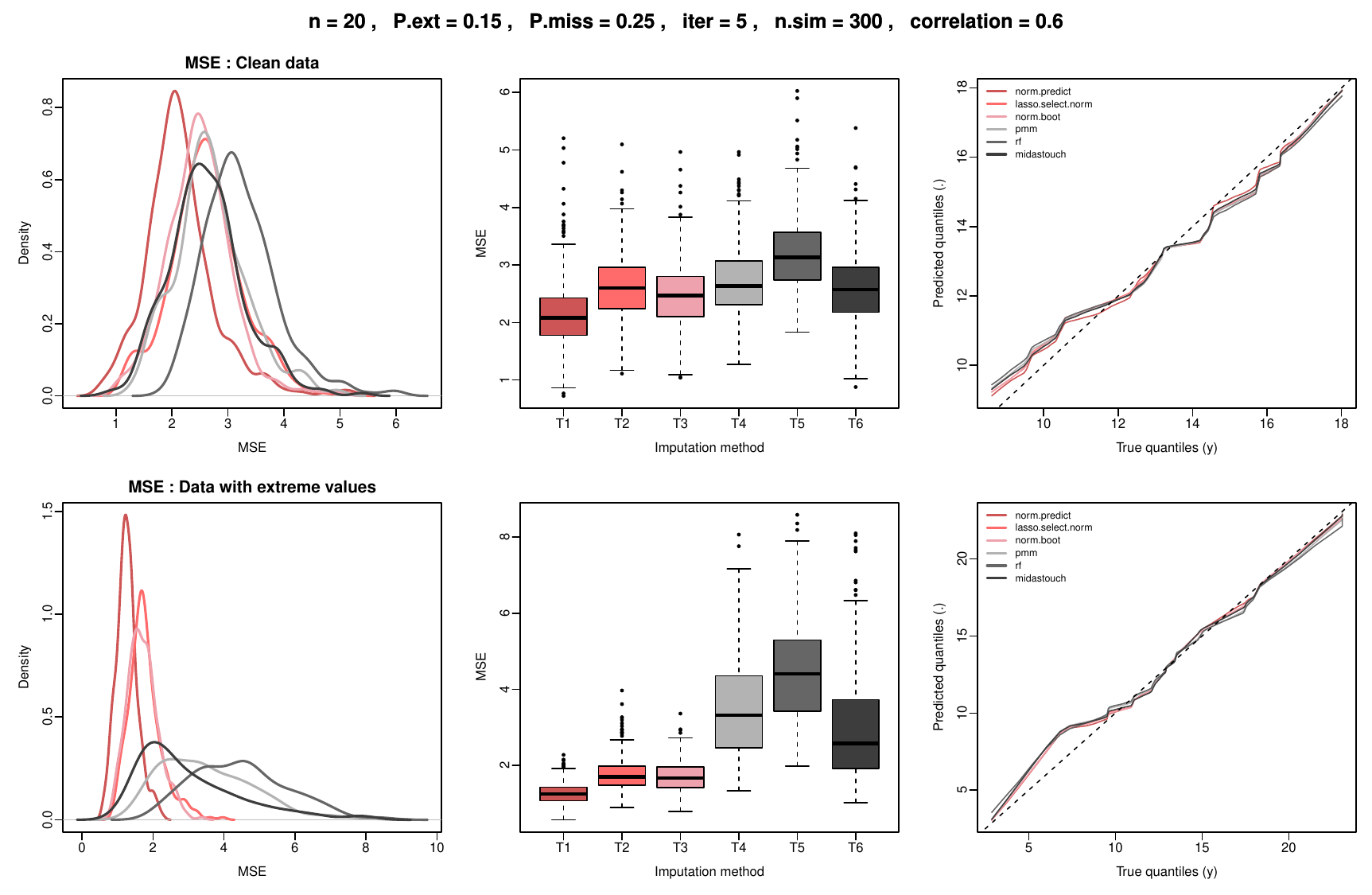}\\[20pt]
      \includegraphics[page=26,width=\linewidth]{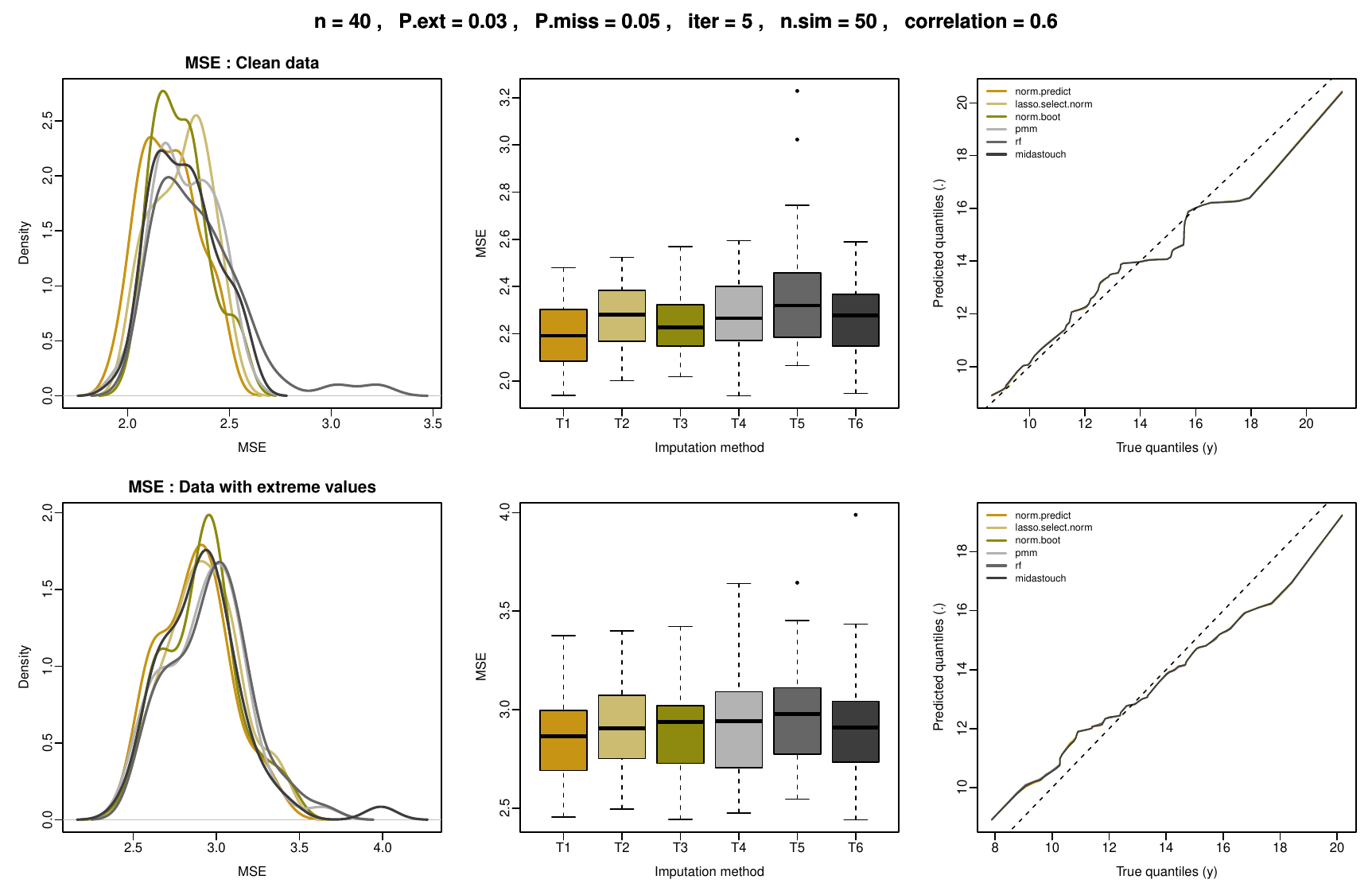}\\[20pt]
      \includegraphics[page=28,width=\linewidth]{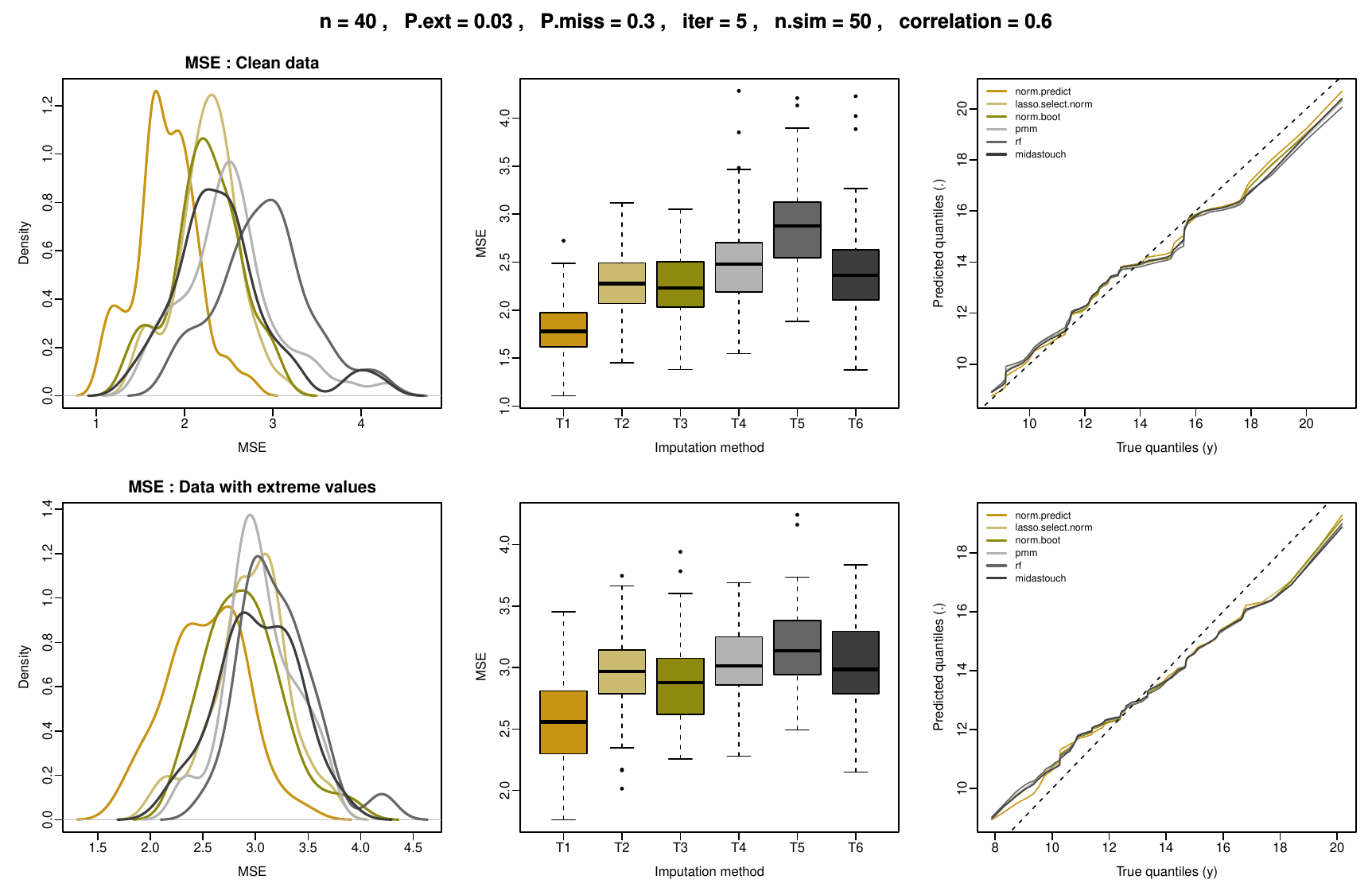}\\[20pt]
      \includegraphics[page=11,width=\linewidth]{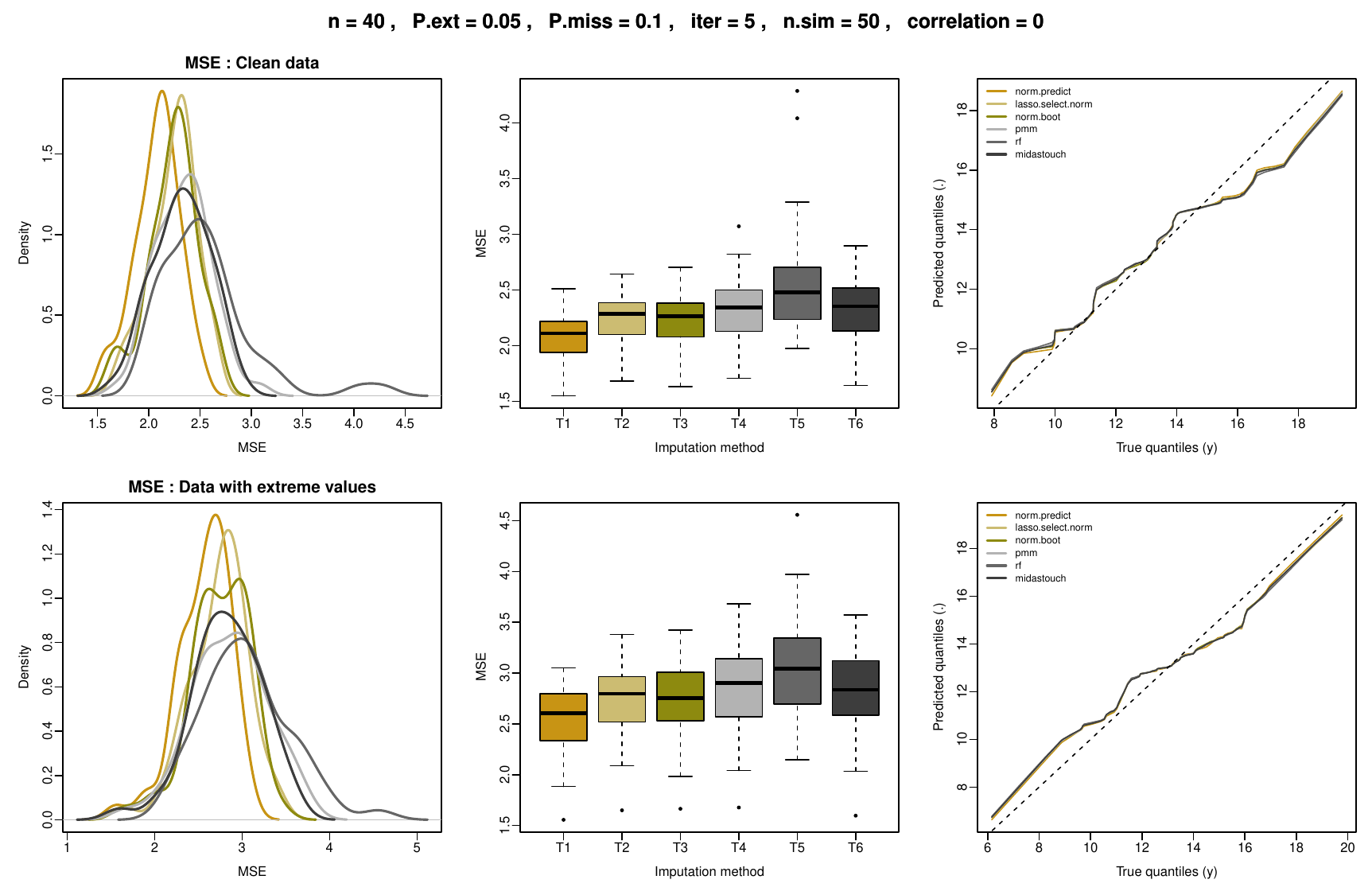}
    \end{minipage}%
  }
    \caption{Predictive MSE densities (clean vs contaminated with extremes data), MSE boxplots, and QQ-plots (predicted vs true quantiles of $y$) across six MI methods (T1--T6). For $\boldsymbol{n=20}$ and $\boldsymbol{n=40}$, ordered by $n$, $P_{\text{ext}}$, and $P_{\text{miss}}$ (panel 3 of 4). Clean data are analyzed with OLS and contaminated data with elastic net. Each subpanel shows the design values ($n$, $P_{\text{ext}}$, $P_{\text{miss}}$, iter, n.sim, and $\rho$).}
    \label{fig:n_20_40_graficos.1-mesclado-3}%
\end{figure}

\begin{figure}[H]
  \centering
  \adjustbox{width=\textwidth, max height=\textheight, center}{%
    \begin{minipage}{.48\linewidth}
      \includegraphics[page=22,width=\linewidth]{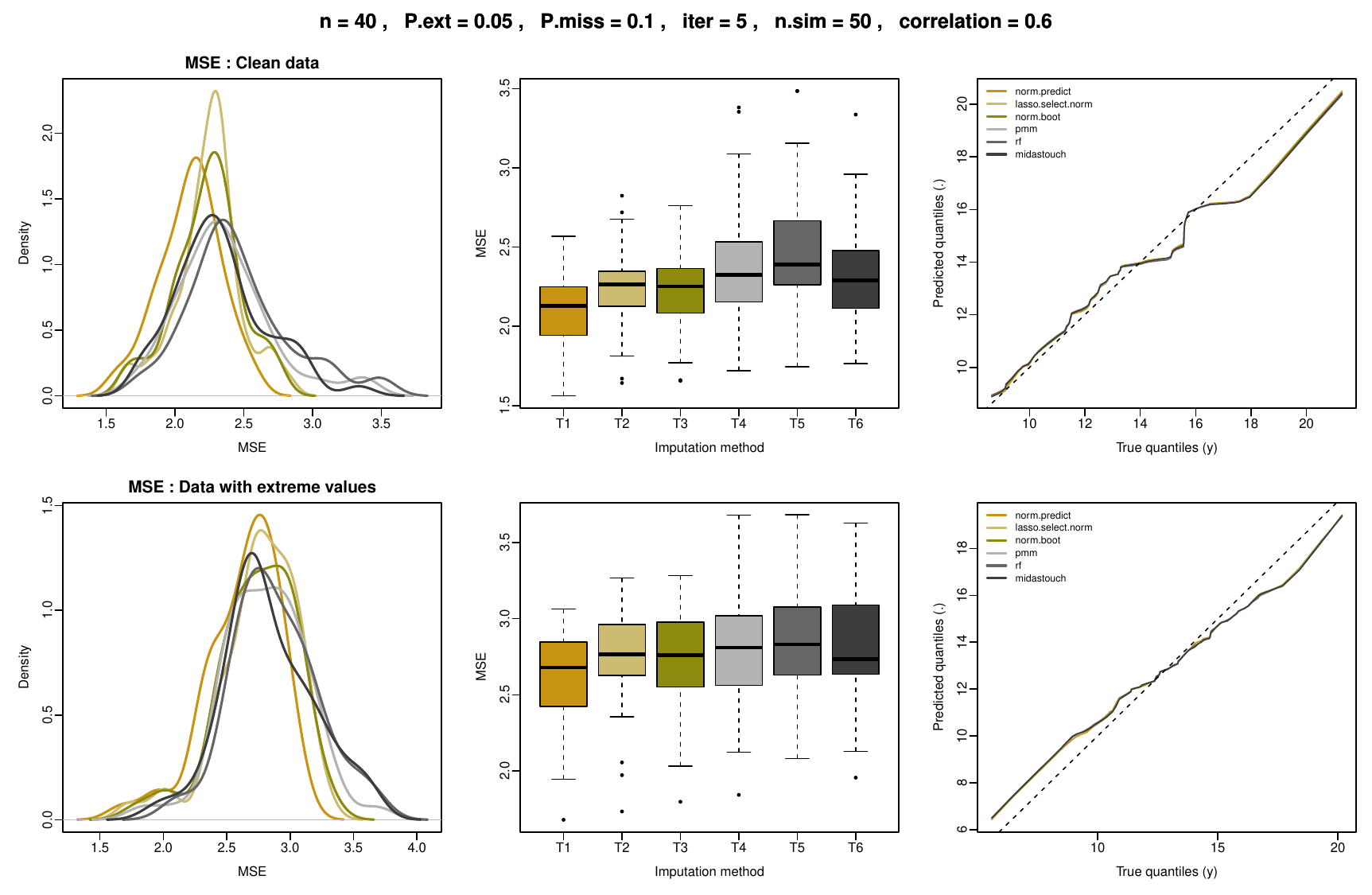}\\[20pt]
      \includegraphics[page=23,width=\linewidth]{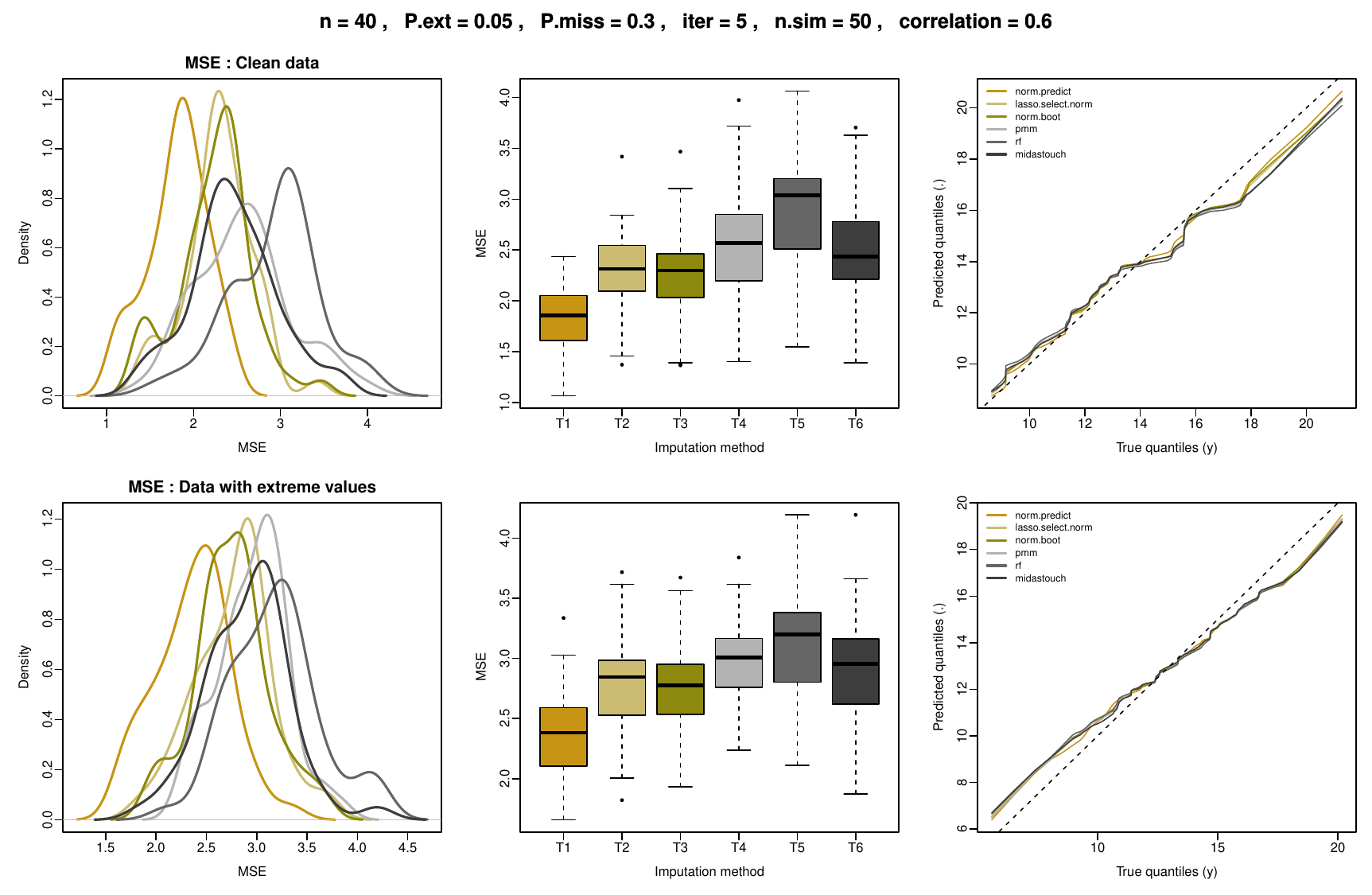}\\[20pt]
      \includegraphics[page=5 ,width=\linewidth]{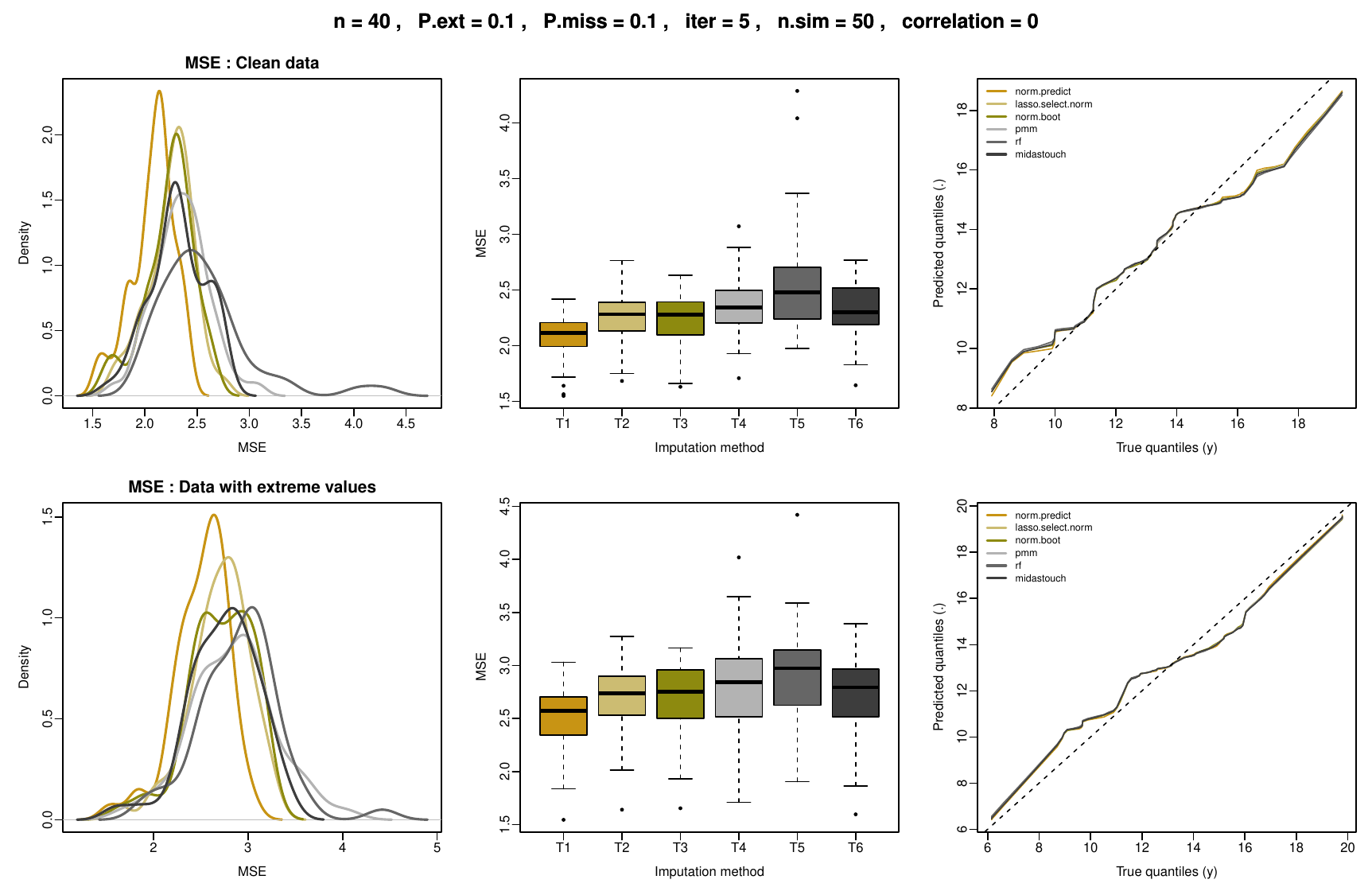}\\[20pt]
      \includegraphics[page=6 ,width=\linewidth]{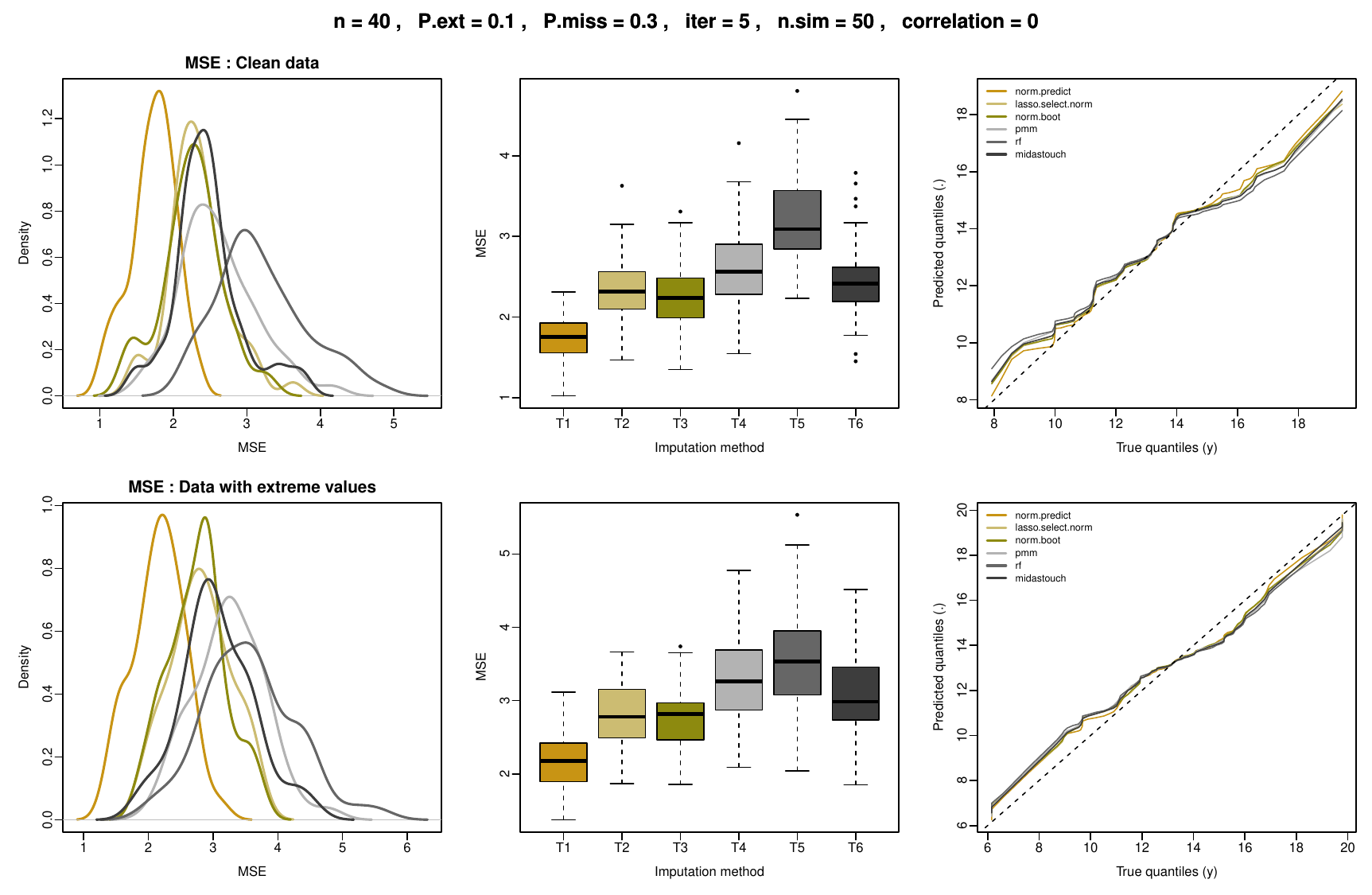}
    \end{minipage}\hfill
    \begin{minipage}{.48\linewidth}
      \includegraphics[page=12,width=\linewidth]{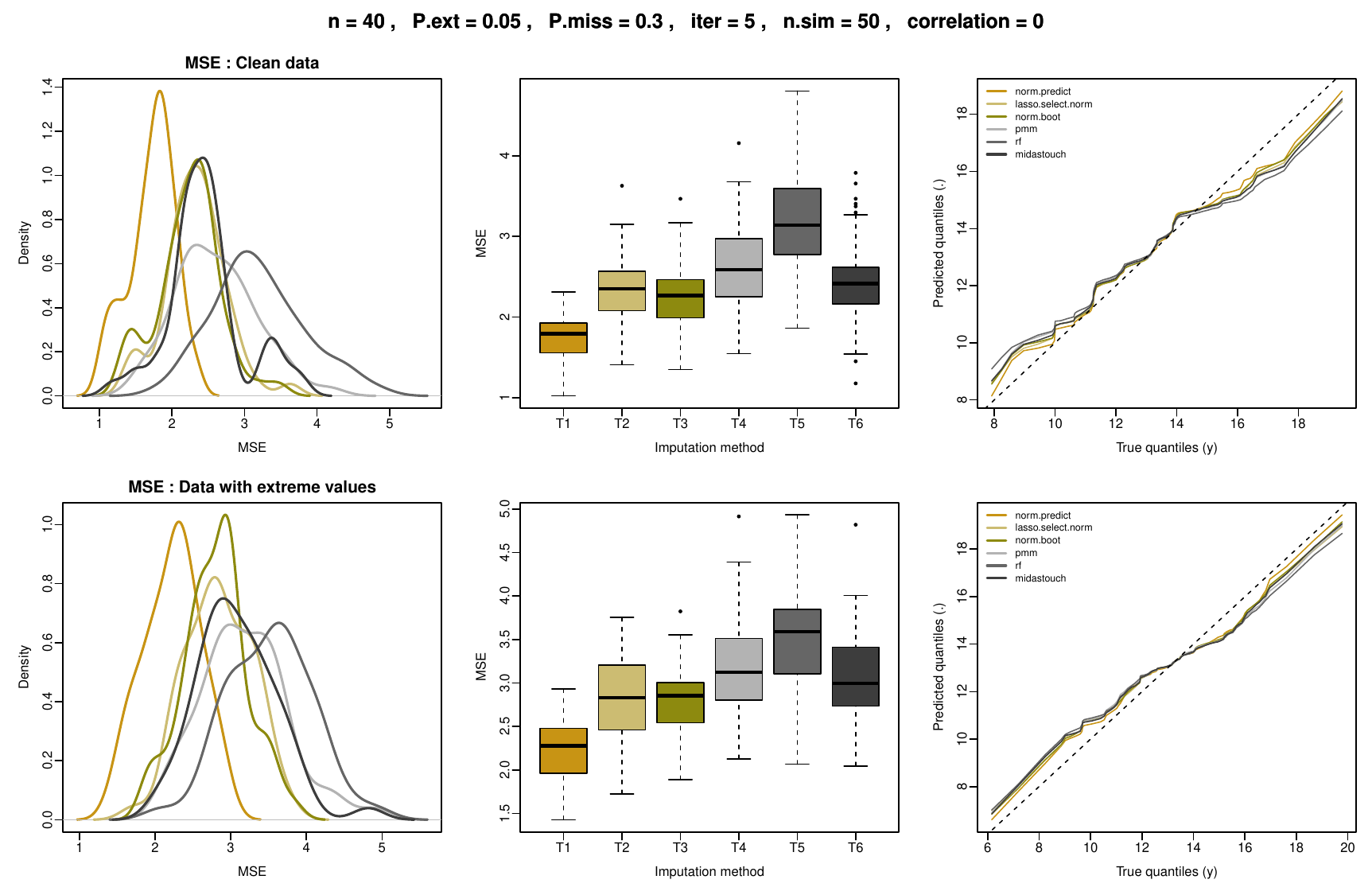}\\[20pt]
      \includegraphics[page=24,width=\linewidth]{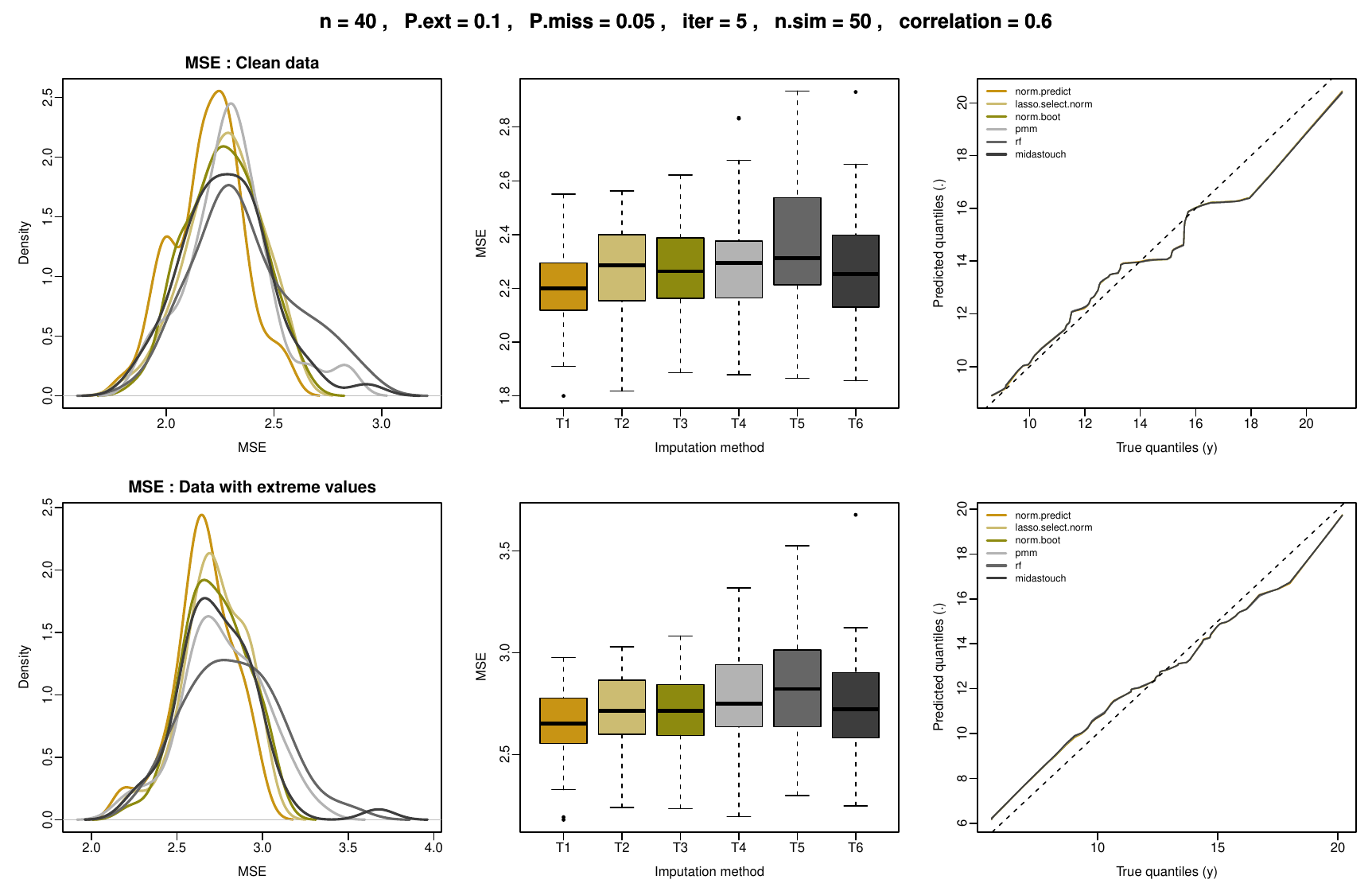}\\[20pt]
      \includegraphics[page=17,width=\linewidth]{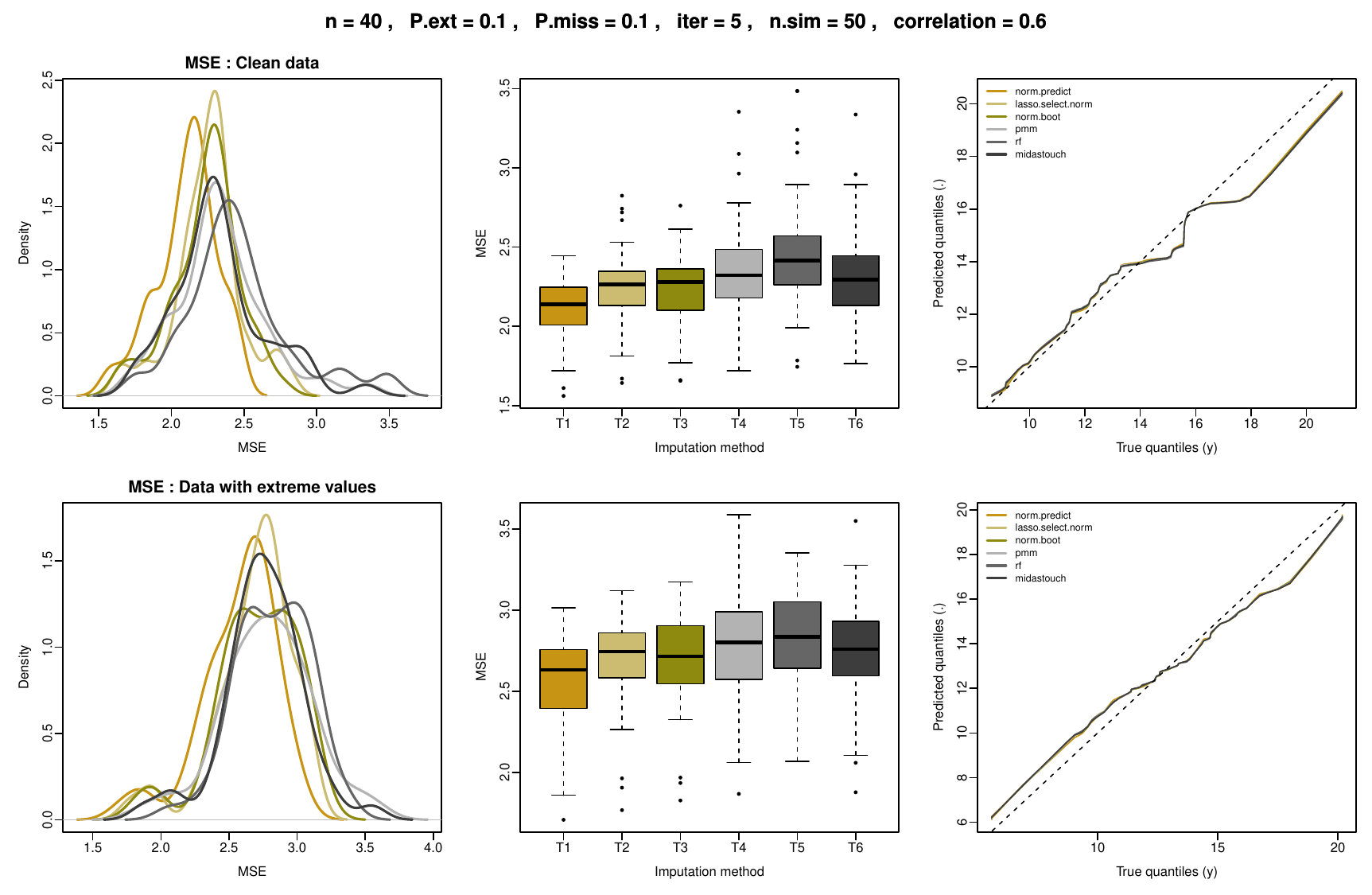}\\[20pt]
      \includegraphics[page=18,width=\linewidth]{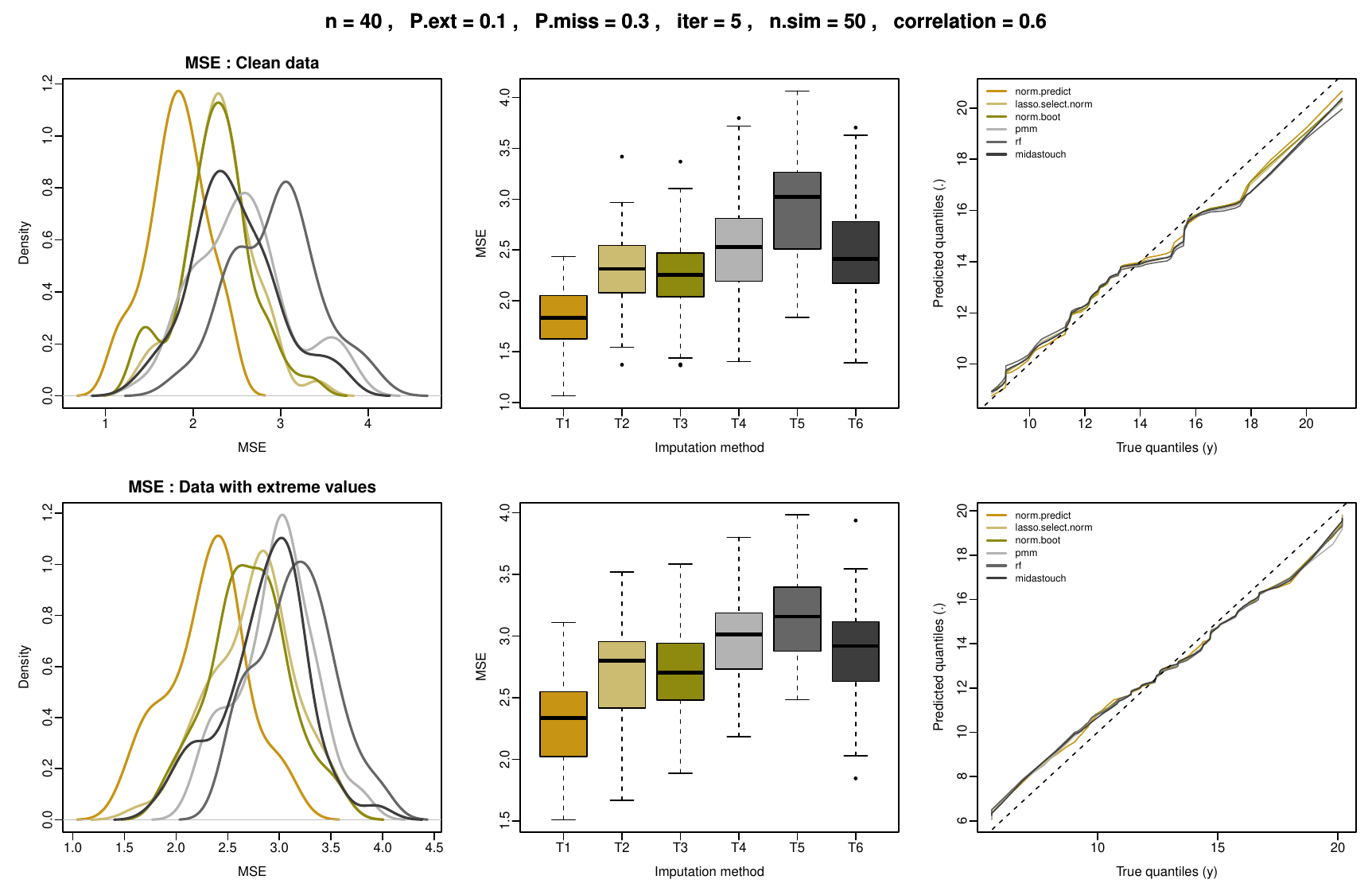}
    \end{minipage}%
  }
    \caption{Predictive MSE densities (clean vs contaminated with extremes data), MSE boxplots, and QQ-plots (predicted vs true quantiles of $y$) across six MI methods (T1--T6). For $\boldsymbol{n=40}$, ordered by $P_{\text{ext}}$, and $P_{\text{miss}}$ (panel 2 of 4). Clean data are analyzed with OLS and contaminated data with elastic net. Each subpanel shows the design values ($n$, $P_{\text{ext}}$, $P_{\text{miss}}$, iter, n.sim, and $\rho$).}
    \label{fig:n_20_40_graficos.1-mesclado-4}%
\end{figure}

\paragraph{Small sample sizes ($\boldsymbol{n=20}$ and $\boldsymbol{n=40}$).} See Tables \ref{tab:resumo_20_40_1} and \ref{tab:resumo_20_40_2}, and Figures \ref{fig:n_20_40_graficos.1-mesclado-1}--\ref{fig:n_20_40_graficos.1-mesclado-4}. Also, for further details see the Tables \ref{tab:n_20_40_graficos.1}--\ref{tab:n_20_40_graficos.8} on appendix regarding the simulation approach presented.

\emph{Coefficients (level and bias).}
Under \emph{clean} data with OLS and low missingness ($P_{\text{miss}}\le 0.10$), both T1--T3 and T4--T6 exhibit small slope biases and RMSE with near-nominal coverage at $n{=}20$; group differences are minor in this regime (Table~\ref{tab:resumo_20_40_1}). As $P_{\text{miss}}$ increases to $\{0.25,0.30\}$, slope RMSEs inflate for all methods; the donor/ML group (T4--T6) often shows slightly larger dispersion than T1--T3 in several cells, while combined summaries (T1--T6) sit between the two groups. Under \emph{contamination} with extreme values (EN), an intercept level shift is present at both $n{=}20$ and $n{=}40$. At $n{=}20$ and $P_{\text{miss}}\le 0.10$, T1--T3 show larger positive $\beta_0$ bias (e.g., $[0.888,1.662]$) than the donor/ML group T4--T6 (e.g., $[0.081,1.382]$), while $\beta_1$ is typically more negative under T1--T3 (e.g., $[-0.356,-0.225]$) than under T4--T6 (e.g., $[-0.264,-0.047]$). Hence, parametric MI tends to stronger systematic shifts under contamination, whereas donor/ML MI tempers coefficient bias. The sign of $\beta_2$ follows the design: commonly positive at $n{=}20$ and negative at $n{=}40$, with both groups moving in the same qualitative direction; distortions escalate with $P_{\text{miss}}$. Aggregated results (T1--T6) reflect an intermediate pattern between T1--T3 and T4--T6.

\emph{Coverage.}
Coverage is most fragile under contamination. At $n{=}20$, EN maintains non-trivial coverage across groups for $P_{\text{miss}}\le 0.10$ but degrades as missingness grows; reductions in bias for T4--T6 sometimes translate into equal or better coverage than T1--T3 in the high-$P_{\text{miss}}$ cells (Table~\ref{tab:resumo_20_40_1}). At $n{=}40$, intercept coverage can collapse under extremes for all groups (with several near-zero entries), suggesting that group differences are secondary to the overall interval unreliability in contaminated small-$n$ settings.

\emph{Predictive error (CV--MSE) and tail risk.}
Under OLS (clean), at $n{=}40$ and low $P_{\text{miss}}$ the parametric group T1--T3 concentrates at lower $\overline{X}$ and $\sigma^2$ than the donor/ML group: for example, T1--T3 typically show $\overline{X}\!\approx\!2.08$--$2.28$ and $\sigma^2\!\approx\!0.02$--$0.06$, whereas T4--T6 shift to higher means/variances (e.g., T4 around $\overline{X}=[2.27,2.36]$, $\sigma^2=[0.02,0.13]$; T5 around $[2.35,2.56]$, $[0.05,0.21]$). At $n{=}20$, this separation is clearer for T5, which exhibits the largest dispersion among clean scenarios. 

Under EN (contaminated), both the \emph{level} and \emph{dispersion} of CV--MSE increase, with a pronounced group contrast as $P_{\text{miss}}$ rises. For $n{=}20$ and $P_{\text{miss}}\in\{0.25,0.30\}$, T4--T6 occupy the upper tail of $\overline{X}$ and $\sigma^2$ (e.g., T4: $\overline{X}=[3.258,3.703]$, $\sigma^2=[1.237,1.734]$; T5: $[4.243,5.068]$, $[1.351,2.908]$; T6: $[2.847,3.197]$, $[1.182,2.461]$), whereas T1--T3 remain markedly tighter (e.g., T1: $[1.180,1.635]$, $[0.072,0.165]$; T2/T3 similarly low). At $n{=}40$, the ordering persists with attenuated dispersion. Combined summaries (T1--T6) widen as donor/ML methods contribute heavier tails (Table~\ref{tab:resumo_20_40_2}). Figures \ref{fig:n_20_40_graficos.1-mesclado-1}--\ref{fig:n_20_40_graficos.1-mesclado-4} corroborate these contrasts via right-shifted, more dispersed MSE densities and QQ-curves that sag below the identity at high quantiles under extremes.

\emph{Method sensitivity at small $n$.}
All MI methods was impacted by extreme values. Relative to T1--T3, the donor/ML group T4--T6 tends to reduce coefficients bias (and sometimes preserve coverage) at $n{=}20$, but at the cost of bit larger predictive tail risk (higher $\sigma^2$, especially for T5). The aggregate T1--T6 reflects this bias of the variance tail trade-off: medians remain moderate, while upper tails expand as missingness increases. At $n{=}40$, the gap narrows but the ordering remains: T1--T3 exhibit tighter tails; T4--T6 carry higher dispersion; T1--T6 lies in between (Tables \ref{tab:resumo_20_40_1} and \ref{tab:resumo_20_40_2}).

\emph{Implications.}
Intercept shifts and slope distortions are primary contamination signals across groups. CV--MSE dispersion ($\sigma^2$) is informative to discriminate parametric versus donor/ML behaviour at small $n$. For moderate-to-high missingness ($\geq 25\%$) with extremes, procedures with tighter tails (T1--T3) deliver more stable out-of-sample error, whereas donor/ML methods (T4--T6) may yield smaller coefficient bias at the expense of heavier predictive tails; aggregate behaviour (T1--T6) naturally reflects this compromise.

\begin{table}[H]
\centering
\begin{threeparttable}
\caption{Bias, RMSE, and 95\% coverage for $(\beta_0,\beta_1,\beta_2)$ under clean data with OLS and data with extreme values using EN for $\boldsymbol{n=20}$ (left) and $\boldsymbol{n=40}$ (right). Entries are reported as min--max within each missingness block (low: $P_{\text{miss}}\in\{0.05,0.10\}$; and high: $P_{\text{miss}}\in\{0.25,0.30\}$). Ranges further aggregate over all design combinations $P_{\text{ext}}\in\{0.03,0.05,0.10,0.15,0.30\}$, $n_{\text{sim}}$ $\in\{50,300,1000\}$, $\text{iter}\in\{5,10\}$, and $\rho\in\{0,0.6\}$. MI methods are grouped as parametric (T1--T3), donor/ML (T4--T6), and the combined set (T1--T6)}
\label{tab:resumo_20_40_1}

\begingroup
\scriptsize
\setlength{\tabcolsep}{2pt}
\renewcommand{\arraystretch}{1.06}
\begin{tabular*}{\textwidth}{@{\extracolsep{\fill}} l l c c c c c c @{}}
\toprule
\multicolumn{2}{c}{} & \multicolumn{3}{c}{\textbf{n = 20}} & \multicolumn{3}{c}{\textbf{n = 40}} \\
\cmidrule(lr){3-5} \cmidrule(lr){6-8}
MI & Metric & $\beta_0$ & $\beta_1$ & $\beta_2$ & $\beta_0$ & $\beta_1$ & $\beta_2$ \\
\midrule
\multicolumn{8}{l}{\textbf{Linear regression (OLS):} low missing-data proportion (5\% and 10\%)}\\ 
T1--T3 & Bias & [-0.424 , -0.097] & [-0.043 , -0.007] & [0.038 , 0.158] & [1.940 , 2.503] & [-0.203 , -0.131] & [-0.226 , -0.121] \\
T1--T3 & RMSE & [0.315 , 0.539] & [0.023 , 0.097] & [0.059 , 0.246] & [1.963 , 2.526] & [0.135 , 0.205] & [0.127 , 0.228] \\
T1--T3 & Coverage & [1.000 , 1.000] & [0.990 , 1.000] & [0.980 , 1.000] & [0.680 , 0.980] & [0.620 , 1.000] & [1.000 , 1.000] \\
T4--T6 & Bias & [-0.286 , -0.023] & [-0.025 , 0.002] & [0.012 , 0.086] & [1.886 , 2.574] & [-0.193 , -0.102] & [-0.266 , -0.167] \\
T4--T6 & RMSE & [0.290 , 0.478] & [0.026 , 0.097] & [0.050 , 0.205] & [1.931 , 2.619] & [0.121 , 0.197] & [0.173 , 0.276] \\
T4--T6 & Coverage & [1.000 , 1.000] & [0.993 , 1.000] & [0.990 , 1.000] & [0.720 , 0.980] & [0.880 , 1.000] & [1.000 , 1.000] \\
T1--T6 & Bias & [-0.424 , -0.023] & [-0.043 , 0.002] & [0.012 , 0.158] & [1.886 , 2.574] & [-0.203 , -0.102] & [-0.266 , -0.121] \\
T1--T6 & RMSE & [0.290 , 0.539] & [0.023 , 0.097] & [0.050 , 0.246] & [1.931 , 2.619] & [0.121 , 0.205] & [0.127 , 0.276] \\
T1--T6 & Coverage & [1.000 , 1.000] & [0.990 , 1.000] & [0.980 , 1.000] & [0.680 , 0.980] & [0.620 , 1.000] & [1.000 , 1.000] \\
\addlinespace[3pt]
\cmidrule(lr){1-8}
\multicolumn{8}{l}{\textbf{Linear regression (OLS):} high missing-data proportion (25\% and 30\%)}\\ 
T1--T3 & Bias & [-0.875 , 0.294] & [-0.113 , -0.051] & [0.085 , 0.331] & [1.858 , 2.766] & [-0.232 , -0.141] & [-0.228 , 0.008] \\
T1--T3 & RMSE & [0.726 , 1.151] & [0.145 , 0.214] & [0.303 , 0.470] & [1.927 , 2.830] & [0.154 , 0.241] & [0.065 , 0.245] \\
T1--T3 & Coverage & [0.980 , 1.000] & [0.903 , 0.987] & [0.799 , 0.973] & [0.640 , 0.820] & [0.320 , 1.000] & [1.000 , 1.000] \\
T4--T6 & Bias & [-0.406 , 0.191] & [-0.033 , 0.003] & [-0.016 , 0.117] & [1.863 , 3.087] & [-0.206 , -0.074] & [-0.365 , -0.161] \\
T4--T6 & RMSE & [0.643 , 0.837] & [0.116 , 0.164] & [0.238 , 0.353] & [1.946 , 3.163] & [0.110 , 0.217] & [0.182 , 0.383] \\
T4--T6 & Coverage & [1.000 , 1.000] & [0.980 , 1.000] & [0.940 , 1.000] & [0.720 , 0.940] & [0.860 , 1.000] & [1.000 , 1.000] \\
T1--T6 & Bias & [-0.875 , 0.294] & [-0.113 , 0.003] & [-0.016 , 0.331] & [1.858 , 3.087] & [-0.232 , -0.074] & [-0.365 , 0.008] \\
T1--T6 & RMSE & [0.643 , 1.151] & [0.116 , 0.214] & [0.238 , 0.470] & [1.927 , 3.163] & [0.110 , 0.241] & [0.065 , 0.383] \\
T1--T6 & Coverage & [0.980 , 1.000] & [0.903 , 1.000] & [0.799 , 1.000] & [0.640 , 0.940] & [0.320 , 1.000] & [1.000 , 1.000] \\
\addlinespace[3pt]
\cmidrule(lr){1-8}
\multicolumn{8}{l}{\textbf{Sparse regression (EN)}: low missing-data proportion (5\% and 10\%)} \\ 
T1--T3 & Bias & [0.888 , 1.662] & [-0.356 , -0.225] & [0.197 , 0.399] & [2.235 , 3.652] & [-0.239 , -0.082] & [-0.433 , -0.245] \\
T1--T3 & RMSE & [0.932 , 1.815] & [0.229 , 0.361] & [0.201 , 0.405] & [2.243 , 3.669] & [0.092 , 0.241] & [0.248 , 0.438] \\
T1--T3 & Coverage & [0.900 , 1.000] & [0.860 , 1.000] & [0.902 , 1.000] & [0.000 , 0.340] & [0.040 , 1.000] & [0.460 , 1.000] \\
T4--T6 & Bias & [0.081 , 1.382] & [-0.264 , -0.047] & [-0.007 , 0.237] & [2.197 , 3.581] & [-0.221 , -0.061] & [-0.454 , -0.298] \\
T4--T6 & RMSE & [0.558 , 1.582] & [0.149 , 0.275] & [0.115 , 0.283] & [2.212 , 3.617] & [0.079 , 0.226] & [0.303 , 0.460] \\
T4--T6 & Coverage & [0.950 , 1.000] & [0.940 , 1.000] & [0.950 , 1.000] & [0.000 , 0.320] & [0.460 , 1.000] & [0.480 , 1.000] \\
T1--T6 & Bias & [0.081 , 1.662] & [-0.356 , -0.047] & [-0.007 , 0.399] & [2.197 , 3.652] & [-0.239 , -0.061] & [-0.454 , -0.245] \\
T1--T6 & RMSE & [0.558 , 1.815] & [0.149 , 0.361] & [0.115 , 0.405] & [2.212 , 3.669] & [0.079 , 0.241] & [0.248 , 0.460] \\
T1--T6 & Coverage & [0.900 , 1.000] & [0.860 , 1.000] & [0.902 , 1.000] & [0.000 , 0.340] & [0.040 , 1.000] & [0.460 , 1.000] \\
\addlinespace[3pt]
\cmidrule(lr){1-8}
\multicolumn{8}{l}{\textbf{Sparse regression (EN)}: high missing-data proportion (25\% and 30\%)} \\ 
T1--T3 & Bias & [0.983 , 1.809] & [-0.396 , -0.241] & [0.211 , 0.423] & [2.037 , 3.599] & [-0.262 , -0.068] & [-0.448 , -0.090] \\
T1--T3 & RMSE & [1.158 , 2.124] & [0.261 , 0.429] & [0.242 , 0.459] & [2.095 , 3.661] & [0.102 , 0.272] & [0.151 , 0.465] \\
T1--T3 & Coverage & [0.740 , 1.000] & [0.428 , 1.000] & [0.660 , 1.000] & [0.060 , 0.640] & [0.100 , 1.000] & [0.720 , 1.000] \\
T4--T6 & Bias & [-0.612 , 0.867] & [-0.197 , 0.102] & [-0.228 , 0.191] & [1.833 , 3.456] & [-0.193 , 0.006] & [-0.520 , -0.314] \\
T4--T6 & RMSE & [1.229 , 1.707] & [0.144 , 0.259] & [0.124 , 0.276] & [1.903 , 3.573] & [0.079 , 0.213] & [0.329 , 0.534] \\
T4--T6 & Coverage & [0.978 , 1.000] & [0.980 , 1.000] & [0.980 , 1.000] & [0.220 , 0.720] & [0.720 , 1.000] & [0.600 , 0.900] \\
T1--T6 & Bias & [-0.612 , 1.809] & [-0.396 , 0.102] & [-0.228 , 0.423] & [1.833 , 3.599] & [-0.262 , 0.006] & [-0.520 , -0.090] \\
T1--T6 & RMSE & [1.158 , 2.124] & [0.144 , 0.429] & [0.124 , 0.459] & [1.903 , 3.661] & [0.079 , 0.272] & [0.151 , 0.534] \\
T1--T6 & Coverage & [0.740 , 1.000] & [0.428 , 1.000] & [0.660 , 1.000] & [0.060 , 0.720] & [0.100 , 1.000] & [0.600 , 1.000] \\
\addlinespace[3pt]
\bottomrule
\end{tabular*}
\endgroup

\begin{tablenotes}[flushleft]
\scriptsize
\setlength{\leftskip}{8em} 
\item $^*$Legend: MI methods (T1--T6): T1 \texttt{norm.predict}; T2 \texttt{lasso.select.norm};
T3 \texttt{norm.boot}; T4 \texttt{pmm}; T5 \texttt{rf}; T6 \texttt{midastouch}.
Coverage is the proportion of 95\% intervals containing the true coefficient.
\end{tablenotes}

\end{threeparttable}
\end{table}

\begin{table}[H]
\centering
\caption{Out-of-sample predictive error summaries (CV–MSE) by MI method for $\boldsymbol{n=20}$ (left) and $\boldsymbol{n=40}$ (right), stratified by model (OLS for clean data; elastic net (EN) for contaminated data) and by missingness blocks (low: $P_{\text{miss}}\in\{0.05,0.10\}$; and high: $P_{\text{miss}}\in\{0.25,0.30\}$). For each method it is reported the min-max of the MSE: mean $\overline{X}$, variance $\sigma^2$, and median $Q_{50}$ across the corresponding design cells. Ranges further aggregate over all design combinations $P_{\text{ext}}\in\{0.03,0.05,0.10,0.15,0.30\}$, $n_{\text{sim}}\in\{50,300,1000\}$, $\text{iter}\in\{5,10\}$, and $\rho\in\{0,0.6\}$}
\label{tab:resumo_20_40_2}
\begingroup
\scriptsize
\setlength{\tabcolsep}{3pt}
\renewcommand{\arraystretch}{1.06}
\begin{tabular*}{\textwidth}{@{\extracolsep{\fill}} l c c c c c c @{}}
\toprule
\multicolumn{1}{c}{} & \multicolumn{3}{c}{\textbf{n = 20}} & \multicolumn{3}{c}{\textbf{n = 40}} \\
\cmidrule(lr){2-4} \cmidrule(lr){5-7}
MI & $\overline{X}$ & $\sigma^2$ & $Q_{50}$ & $\overline{X}$ & $\sigma^2$ & $Q_{50}$ \\
\midrule
\multicolumn{7}{l}{\textbf{Linear regression (OLS):} low missing-data proportion (5\% and 10\%)} \\
T1 & [2.283 , 2.524] & [0.236 , 0.461] & [2.206 , 2.450] & [2.077 , 2.207] & [0.020 , 0.050] & [2.110 , 2.241] \\
T2 & [2.441 , 2.690] & [0.200 , 0.413] & [2.407 , 2.576] & [2.242 , 2.271] & [0.019 , 0.057] & [2.265 , 2.285] \\
T3 & [2.399 , 2.620] & [0.196 , 0.369] & [2.336 , 2.500] & [2.236 , 2.275] & [0.018 , 0.061] & [2.228 , 2.280] \\
T4 & [2.473 , 2.693] & [0.240 , 0.429] & [2.433 , 2.578] & [2.273 , 2.357] & [0.023 , 0.125] & [2.260 , 2.346] \\
T5 & [2.611 , 2.909] & [0.173 , 0.736] & [2.553 , 2.766] & [2.351 , 2.560] & [0.053 , 0.213] & [2.312 , 2.479] \\
T6 & [2.432 , 2.672] & [0.226 , 0.465] & [2.399 , 2.547] & [2.273 , 2.337] & [0.024 , 0.112] & [2.253 , 2.354] \\
\addlinespace[4pt]
\cmidrule(lr){1-7}
\multicolumn{7}{l}{\textbf{Linear regression (OLS):} high missing-data proportion (25\% and 30\%)} \\
T1 & [1.912 , 2.166] & [0.252 , 0.496] & [1.956 , 2.081] & [1.725 , 1.819] & [0.093 , 0.121] & [1.757 , 1.858] \\
T2 & [2.484 , 2.628] & [0.292 , 0.419] & [2.470 , 2.666] & [2.256 , 2.334] & [0.123 , 0.176] & [2.275 , 2.351] \\
T3 & [2.296 , 2.494] & [0.265 , 0.415] & [2.390 , 2.491] & [2.236 , 2.261] & [0.158 , 0.193] & [2.230 , 2.297] \\
T4 & [2.635 , 2.712] & [0.246 , 0.428] & [2.631 , 2.737] & [2.516 , 2.605] & [0.274 , 0.312] & [2.480 , 2.589] \\
T5 & [3.015 , 3.355] & [0.204 , 0.524] & [3.070 , 3.390] & [2.890 , 3.201] & [0.248 , 0.403] & [2.878 , 3.140] \\
T6 & [2.506 , 2.616] & [0.235 , 0.501] & [2.530 , 2.666] & [2.434 , 2.470] & [0.221 , 0.340] & [2.364 , 2.435] \\
\addlinespace[4pt]
\cmidrule(lr){1-7}
\multicolumn{7}{l}{\textbf{Sparse regression (EN)}: low missing-data proportion (5\% and 10\%)} \\
T1 & [1.441 , 1.995] & [0.038 , 0.095] & [1.438 , 1.943] & [2.505 , 2.867] & [0.032 , 0.091] & [2.574 , 2.867] \\
T2 & [1.577 , 2.118] & [0.046 , 0.162] & [1.566 , 2.072] & [2.687 , 2.923] & [0.033 , 0.113] & [2.716 , 2.964] \\
T3 & [1.558 , 2.083] & [0.045 , 0.142] & [1.549 , 2.056] & [2.704 , 2.911] & [0.035 , 0.116] & [2.715 , 2.948] \\
T4 & [1.906 , 2.553] & [0.175 , 1.002] & [1.675 , 2.274] & [2.771 , 2.925] & [0.057 , 0.199] & [2.750 , 2.944] \\
T5 & [2.275 , 3.103] & [0.302 , 1.721] & [1.935 , 2.740] & [2.822 , 3.045] & [0.059 , 0.246] & [2.776 , 3.042] \\
T6 & [1.846 , 2.380] & [0.150 , 0.877] & [1.618 , 2.180] & [2.735 , 2.914] & [0.056 , 0.160] & [2.722 , 2.943] \\
\addlinespace[4pt]
\cmidrule(lr){1-7}
\multicolumn{7}{l}{\textbf{Sparse regression (EN)}: high missing-data proportion (25\% and 30\%)} \\
T1 & [1.180 , 1.635] & [0.072 , 0.165] & [1.146 , 1.616] & [2.162 , 2.529] & [0.136 , 0.155] & [2.177 , 2.558] \\
T2 & [1.780 , 2.403] & [0.174 , 0.361] & [1.698 , 2.361] & [2.734 , 2.948] & [0.134 , 0.217] & [2.779 , 2.969] \\
T3 & [1.703 , 2.270] & [0.139 , 0.317] & [1.655 , 2.179] & [2.710 , 2.896] & [0.137 , 0.201] & [2.704 , 2.878] \\
T4 & [3.258 , 3.703] & [1.237 , 1.734] & [2.974 , 3.524] & [2.961 , 3.252] & [0.108 , 0.334] & [3.009 , 3.264] \\
T5 & [4.243 , 5.068] & [1.351 , 2.908] & [3.928 , 5.008] & [3.153 , 3.560] & [0.123 , 0.505] & [3.136 , 3.590] \\
T6 & [2.847 , 3.197] & [1.182 , 2.461] & [2.572 , 2.891] & [2.857 , 3.079] & [0.144 , 0.322] & [2.919 , 2.997] \\
\addlinespace[4pt]
\bottomrule
\end{tabular*}
\endgroup
\end{table}



\begin{figure}[H]
  \centering
  \adjustbox{width=\textwidth, max height=\textheight, center}{%
    \begin{minipage}{.48\linewidth}
      \includegraphics[page=1 ,width=\linewidth]{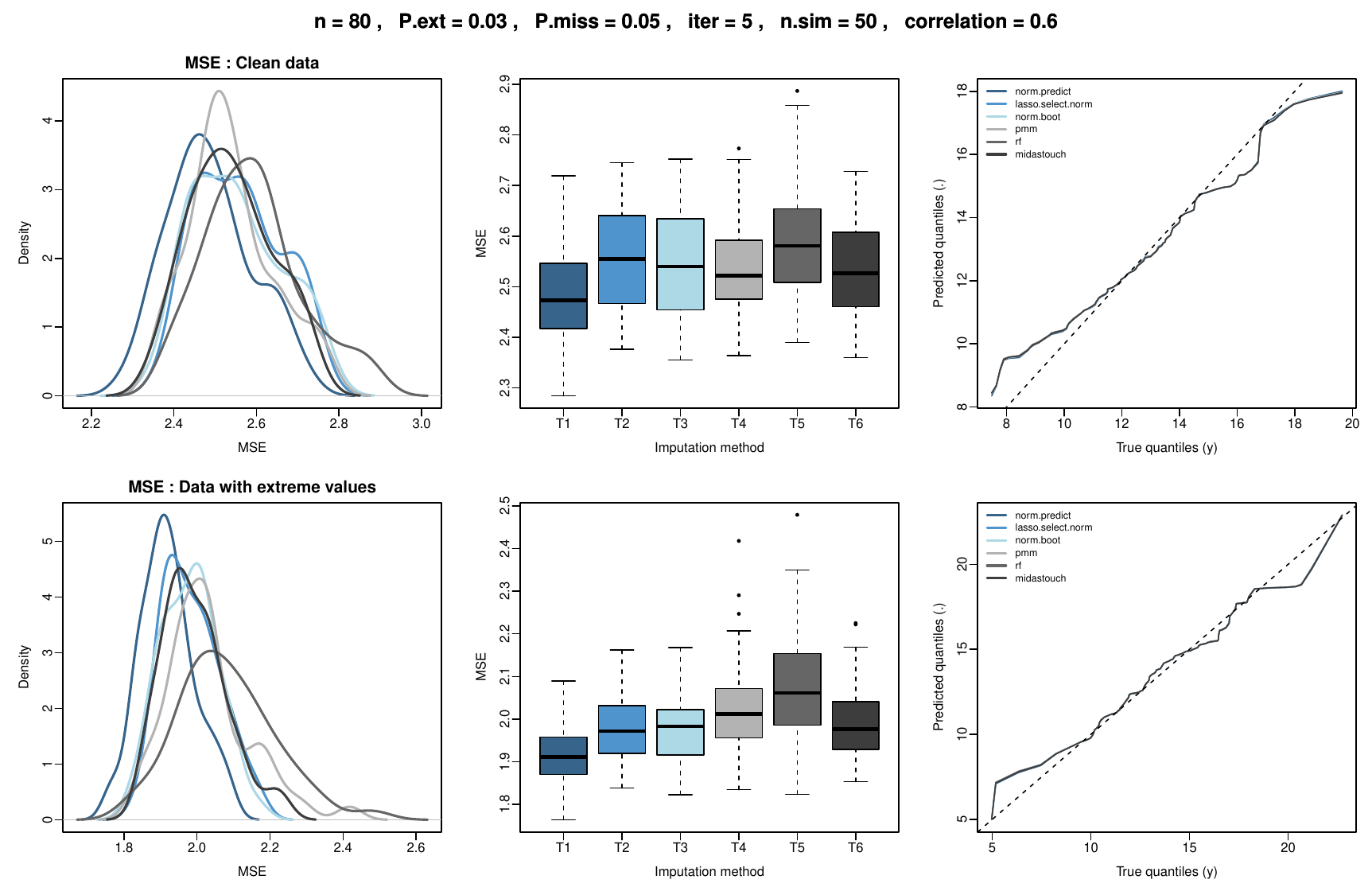}\\[20pt]
      \includegraphics[page=3 ,width=\linewidth]{figures/n_80_200_graficos.1_mesclado.pdf}\\[20pt]
      \includegraphics[page=20,width=\linewidth]{figures/n_80_200_graficos.1_mesclado.pdf}\\[20pt]
      \includegraphics[page=15,width=\linewidth]{figures/n_80_200_graficos.1_mesclado.pdf}
    \end{minipage}\hfill
    \begin{minipage}{.48\linewidth}
      \includegraphics[page=2 ,width=\linewidth]{figures/n_80_200_graficos.1_mesclado.pdf}\\[20pt]
      \includegraphics[page=19,width=\linewidth]{figures/n_80_200_graficos.1_mesclado.pdf}\\[20pt]
      \includegraphics[page=4 ,width=\linewidth]{figures/n_80_200_graficos.1_mesclado.pdf}\\[20pt]
      \includegraphics[page=29,width=\linewidth]{figures/n_80_200_graficos.1_mesclado.pdf}
    \end{minipage}%
  }
      \caption{Predictive MSE densities (clean vs contaminated with extremes data), MSE boxplots, and QQ-plots (predicted vs true quantiles of $y$) across six MI methods (T1--T6). For $\boldsymbol{n=80}$, ordered by $P_{\text{ext}}$, and $P_{\text{miss}}$ (panel 1 of 4). Clean data are analyzed with OLS and contaminated data with elastic net. Each subpanel shows the design values ($n$, $P_{\text{ext}}$, $P_{\text{miss}}$, iter, n.sim, and $\rho$).}
      \label{fig:n_80_200_graficos.1-mesclado-1}
\end{figure}

\begin{figure}[H]
  \centering
  \adjustbox{width=\textwidth, max height=\textheight, center}{%
    \begin{minipage}{.48\linewidth}
      \includegraphics[page=16,width=\linewidth]{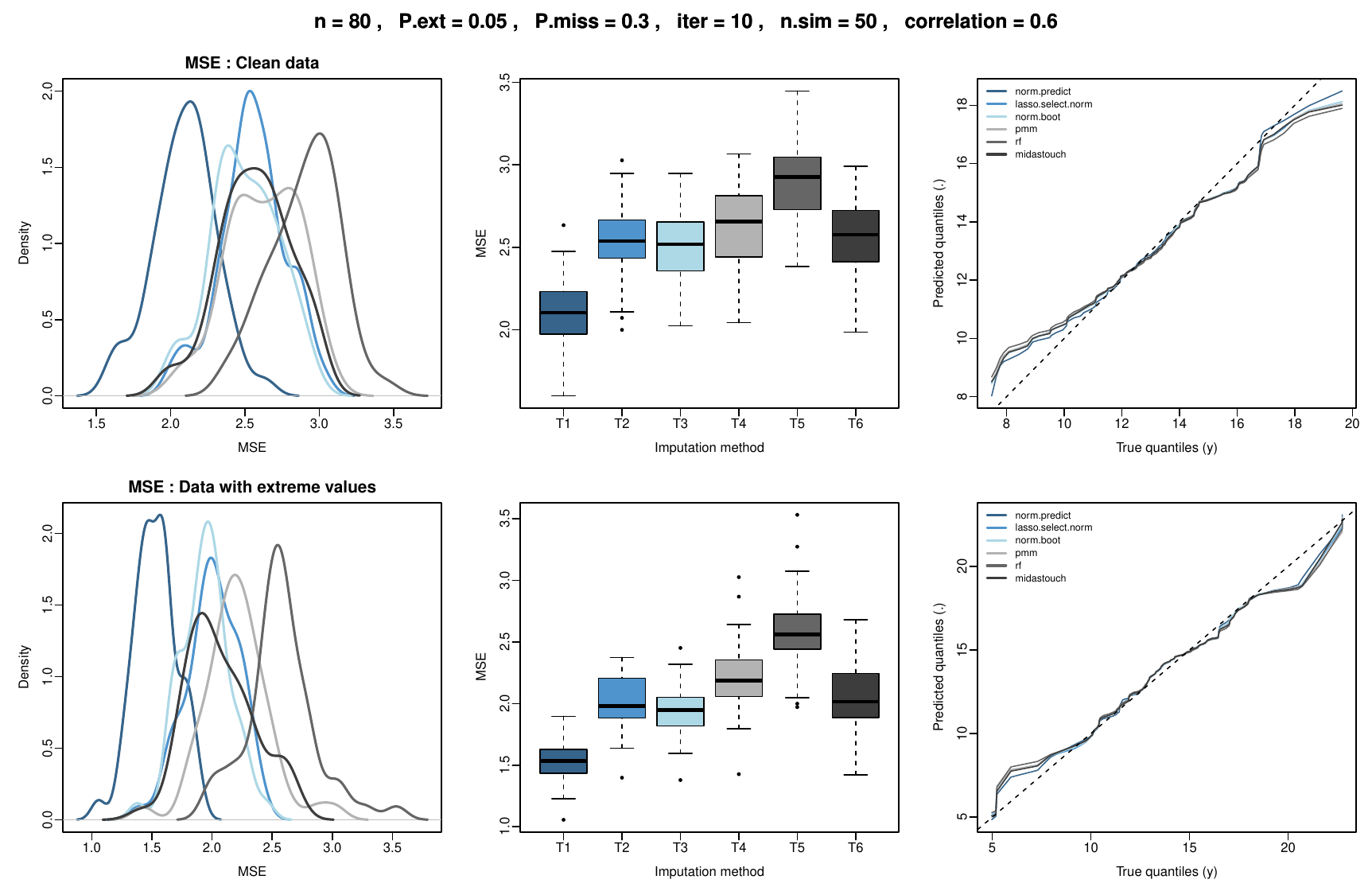}\\[20pt]
      \includegraphics[page=31,width=\linewidth]{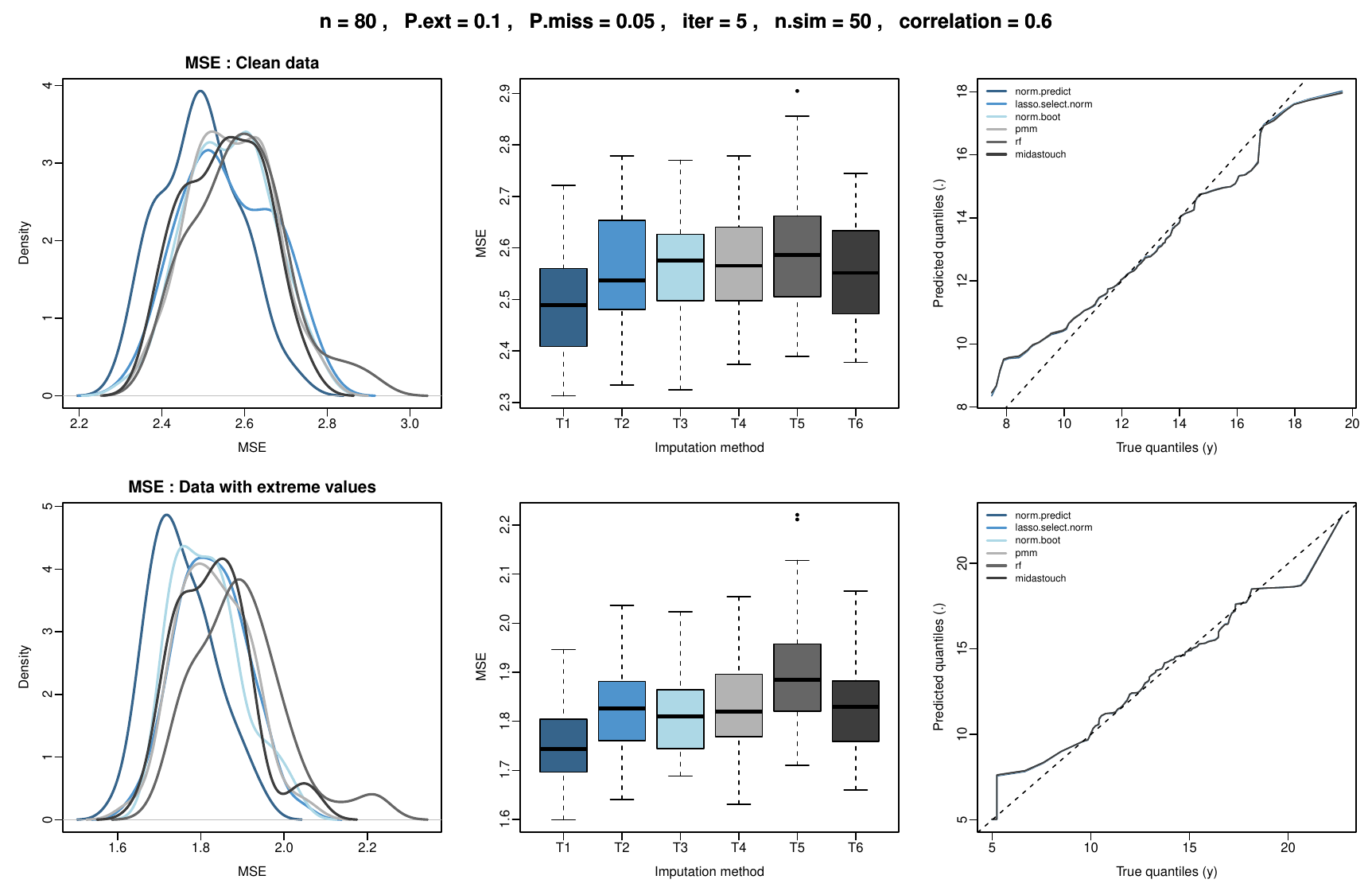}\\[20pt]
      \includegraphics[page=32,width=\linewidth]{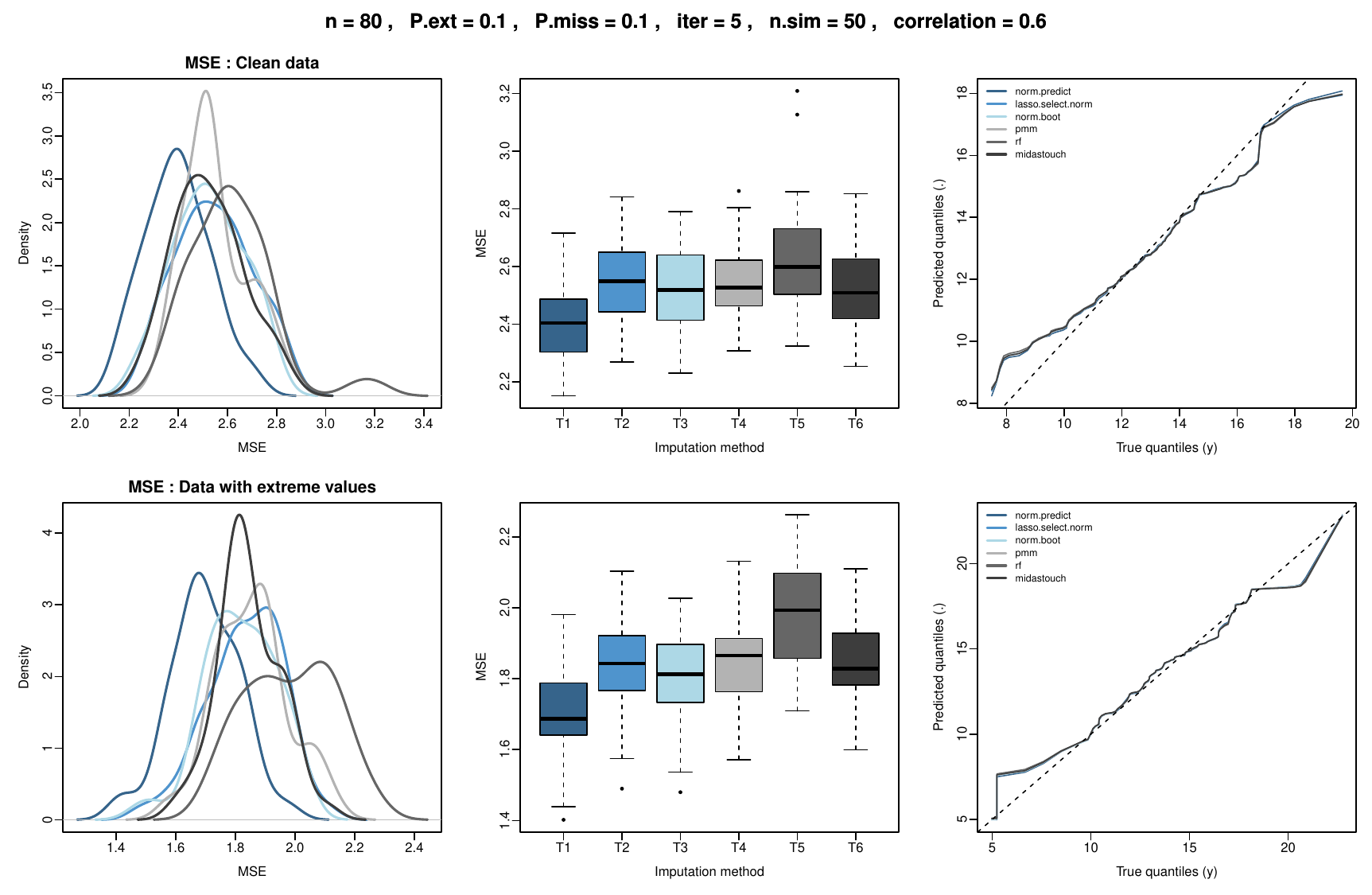}\\[20pt]
      \includegraphics[page=10,width=\linewidth]{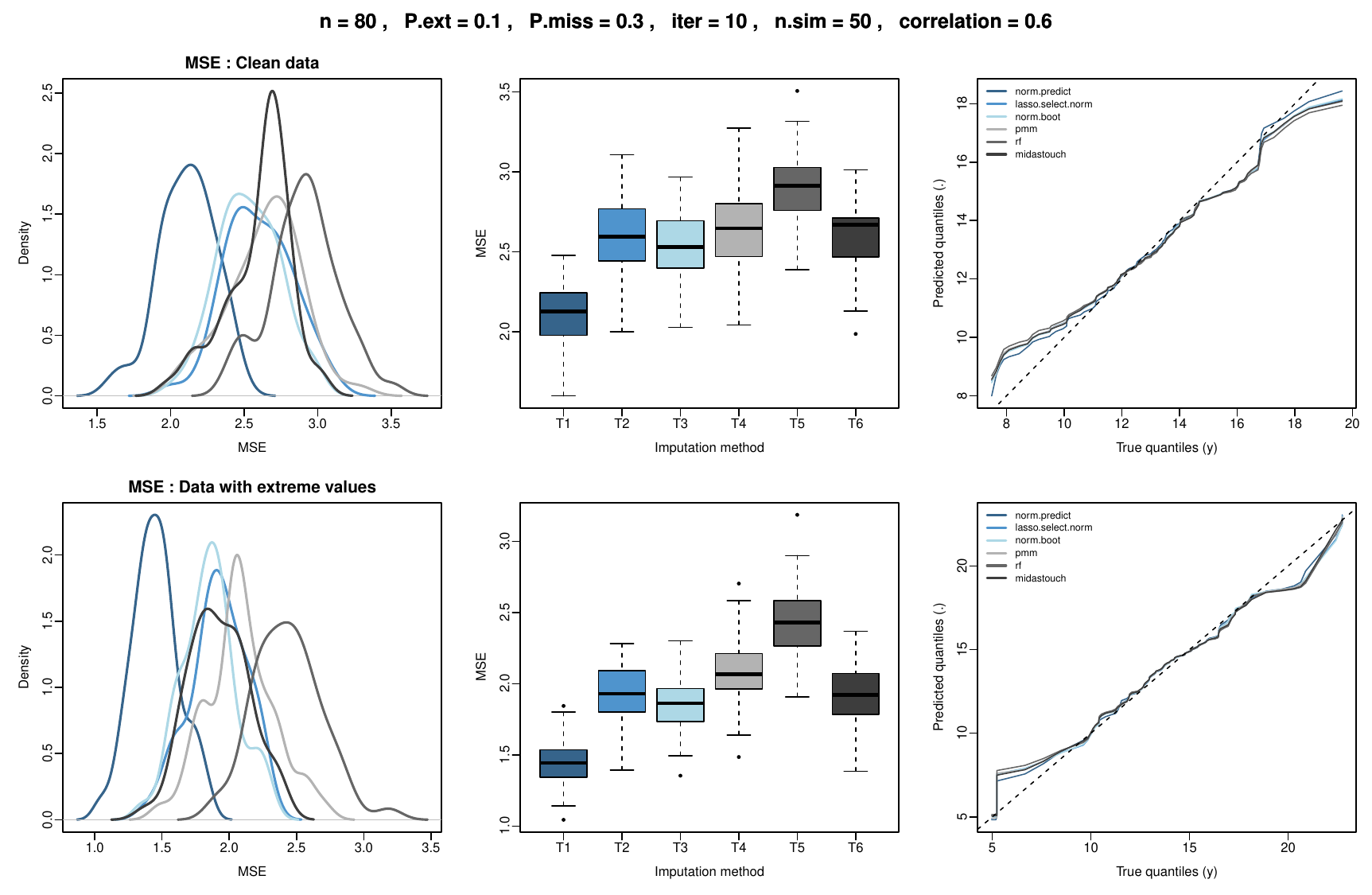}
    \end{minipage}\hfill
    \begin{minipage}{.48\linewidth}
      \includegraphics[page=30,width=\linewidth]{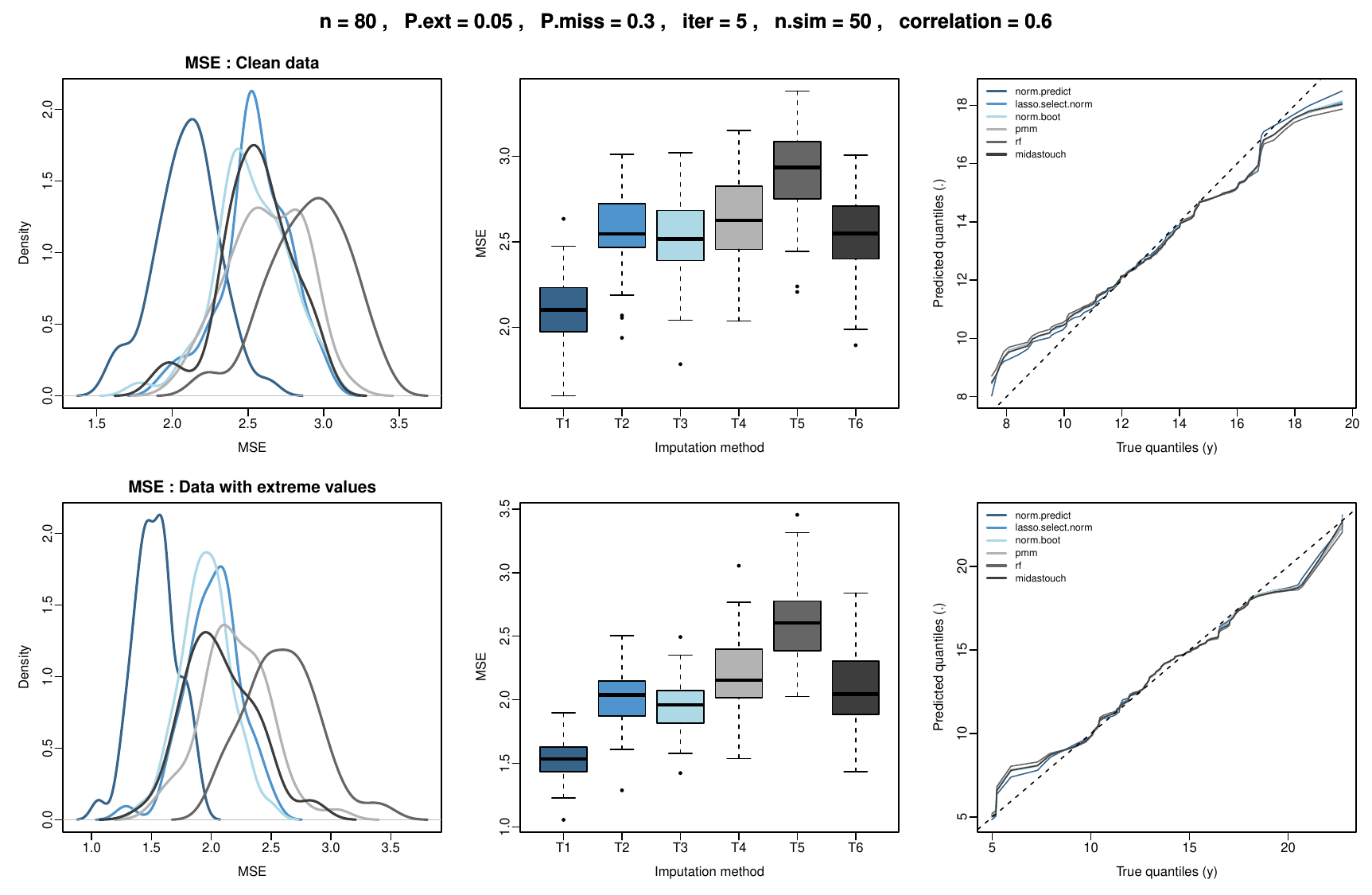}\\[20pt]
      \includegraphics[page=9 ,width=\linewidth]{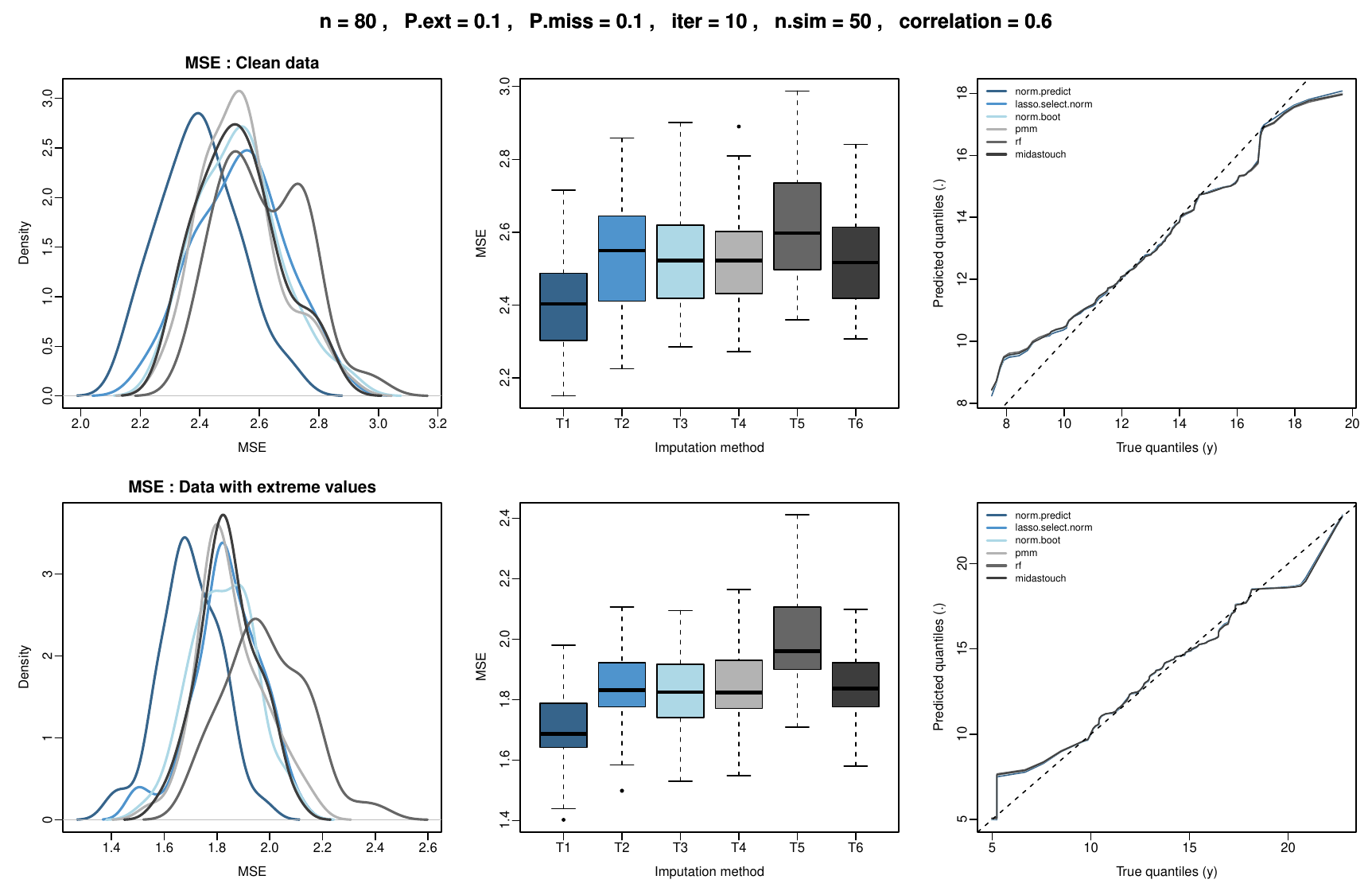}\\[20pt]
      \includegraphics[page=8 ,width=\linewidth]{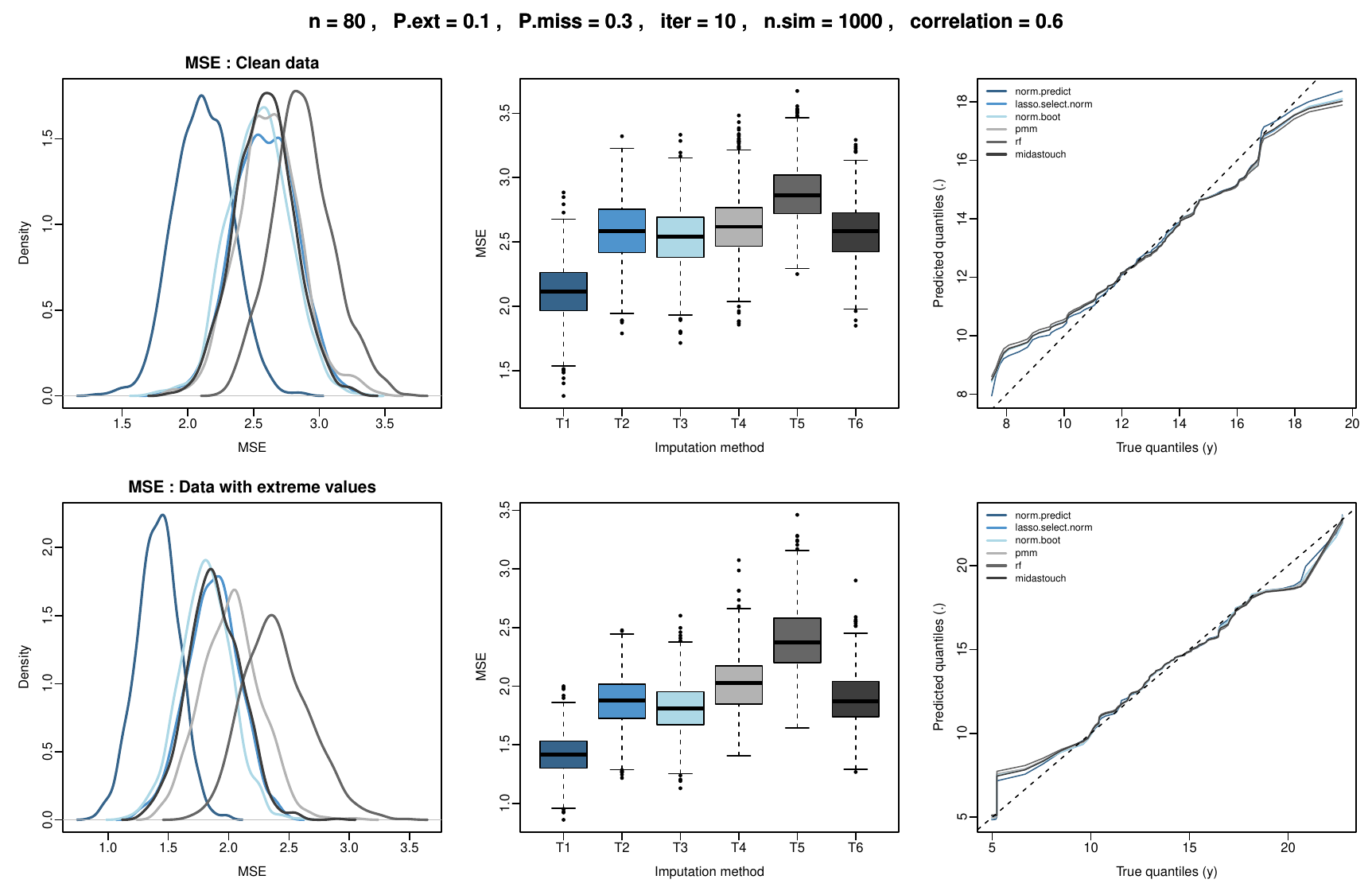}\\[20pt]
      \includegraphics[page=25,width=\linewidth]{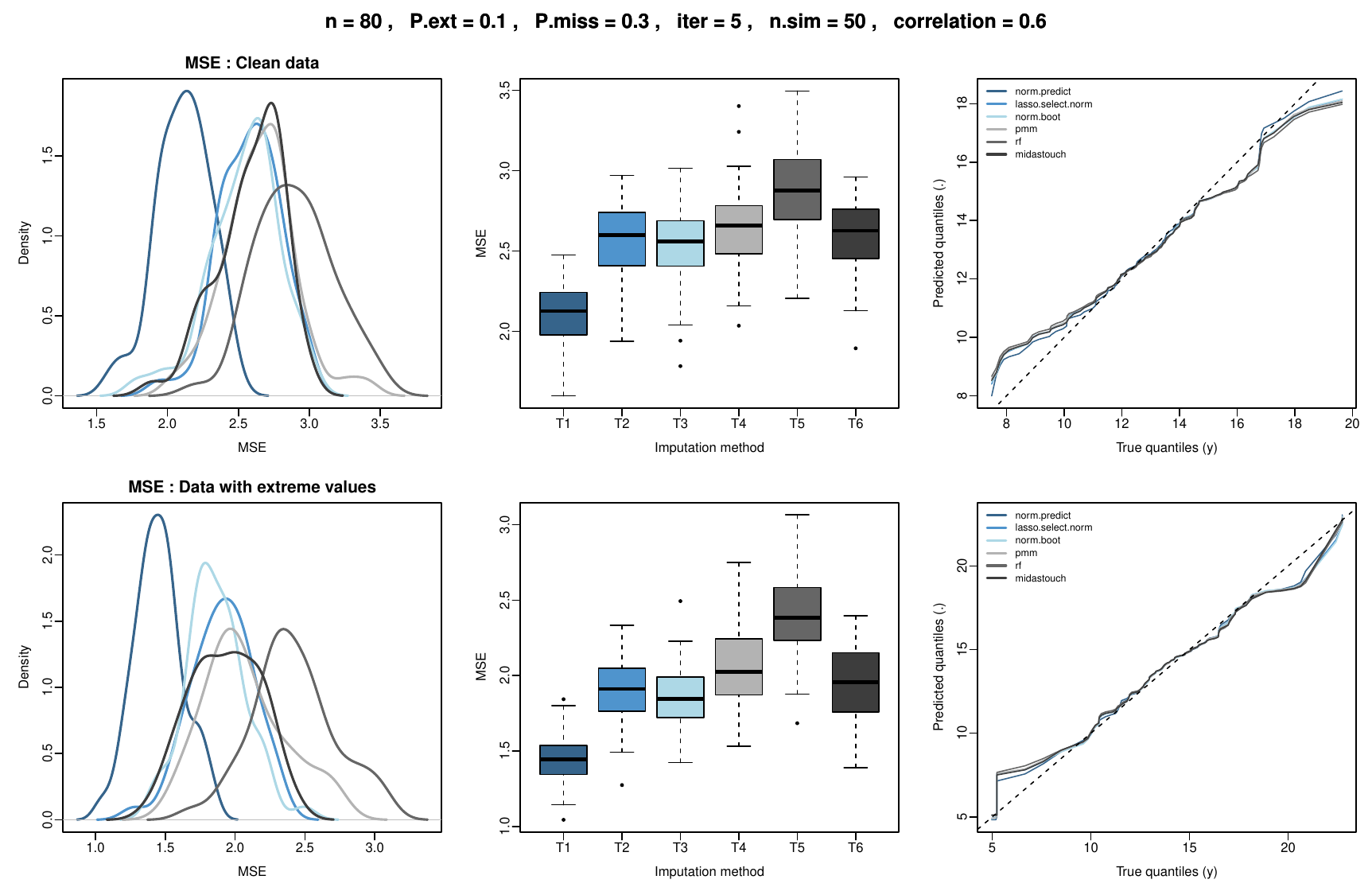}
    \end{minipage}%
  }
      \caption{Predictive MSE densities (clean vs contaminated with extremes data), MSE boxplots, and QQ-plots (predicted vs true quantiles of $y$) across six MI methods (T1--T6). For $\boldsymbol{n=80}$, ordered by $P_{\text{ext}}$, and $P_{\text{miss}}$ (panel 2 of 4). Clean data are analyzed with OLS and contaminated data with elastic net. Each subpanel shows the design values ($n$, $P_{\text{ext}}$, $P_{\text{miss}}$, iter, n.sim, and $\rho$).}
      \label{fig:n_80_200_graficos.1-mesclado-2}%
\end{figure}

\begin{figure}[H]
  \centering
  \adjustbox{width=\textwidth, max height=\textheight, center}{%
    \begin{minipage}{.48\linewidth}
      \includegraphics[page=13,width=\linewidth]{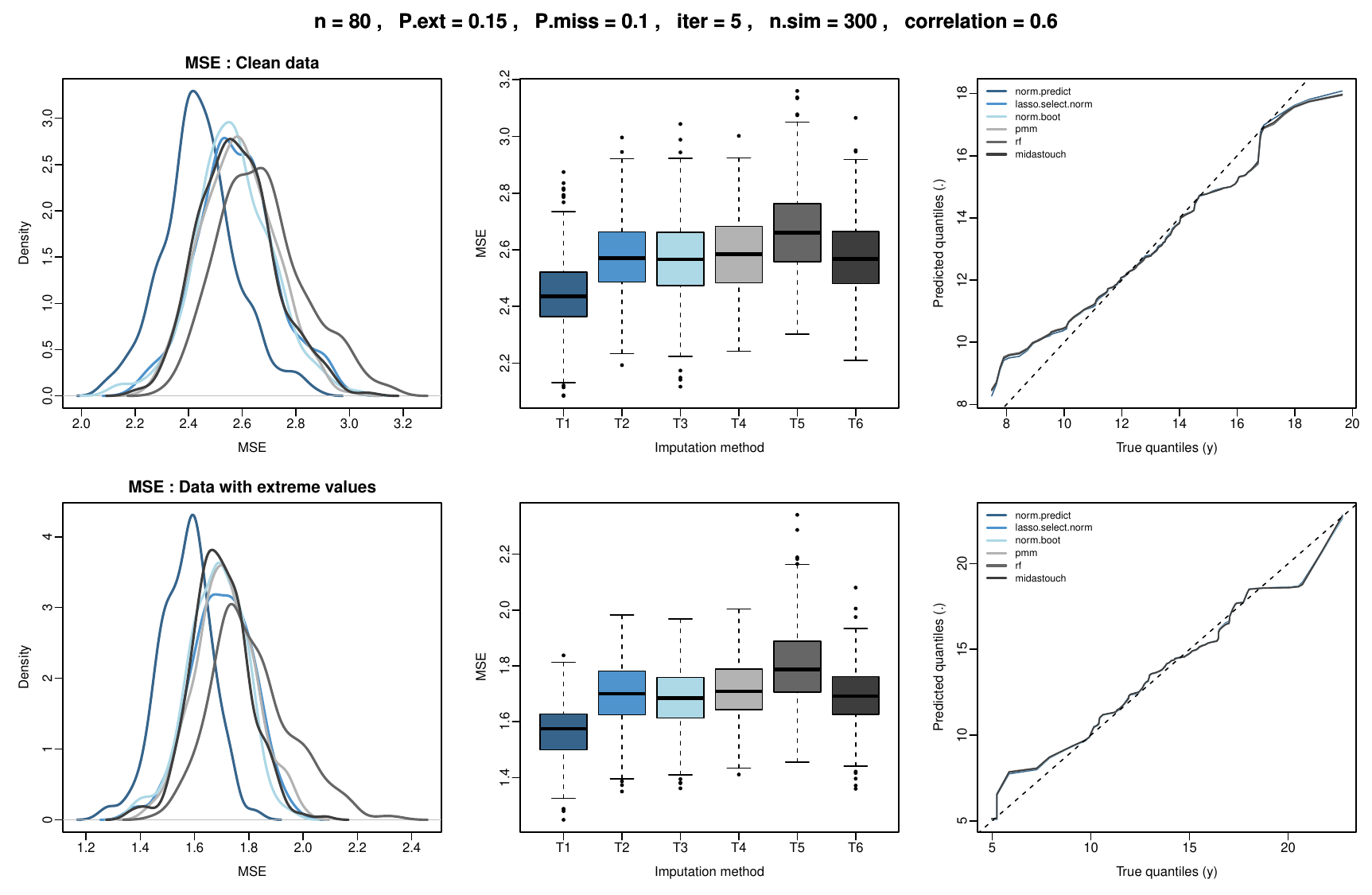}\\[20pt]
      \includegraphics[page=7 ,width=\linewidth]{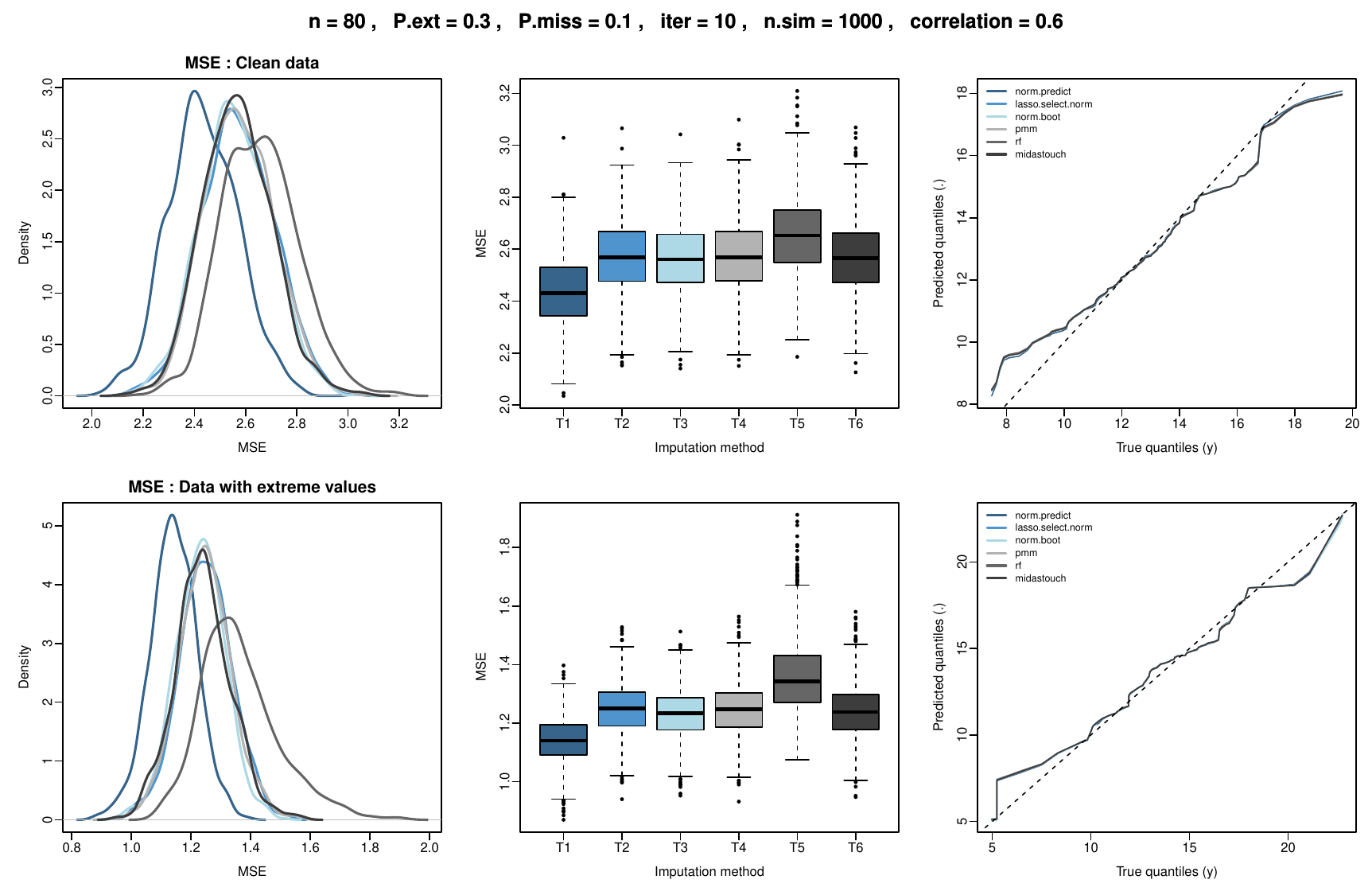}\\[20pt]
      \includegraphics[page=27,width=\linewidth]{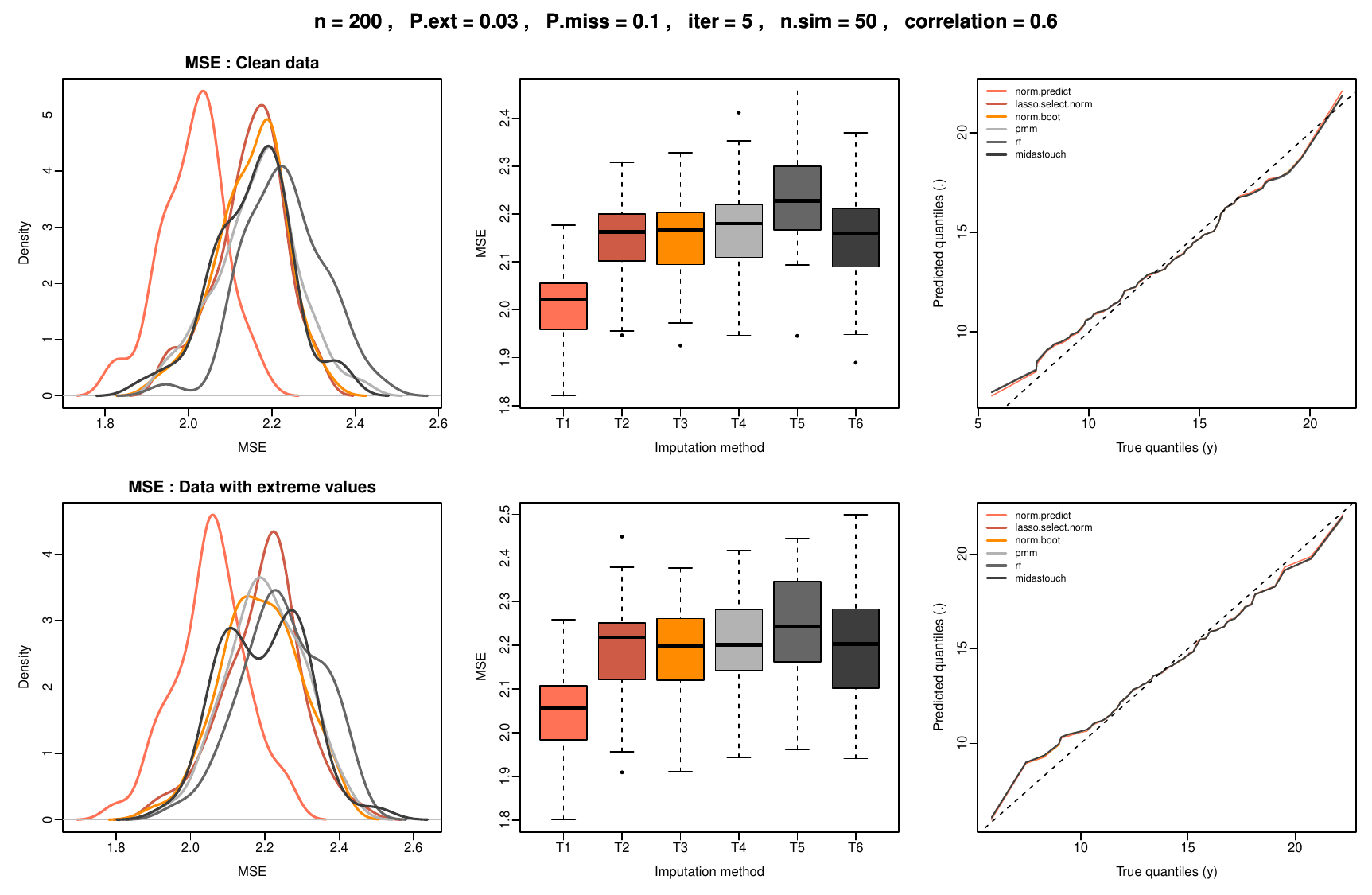}\\[20pt]
      \includegraphics[page=21,width=\linewidth]{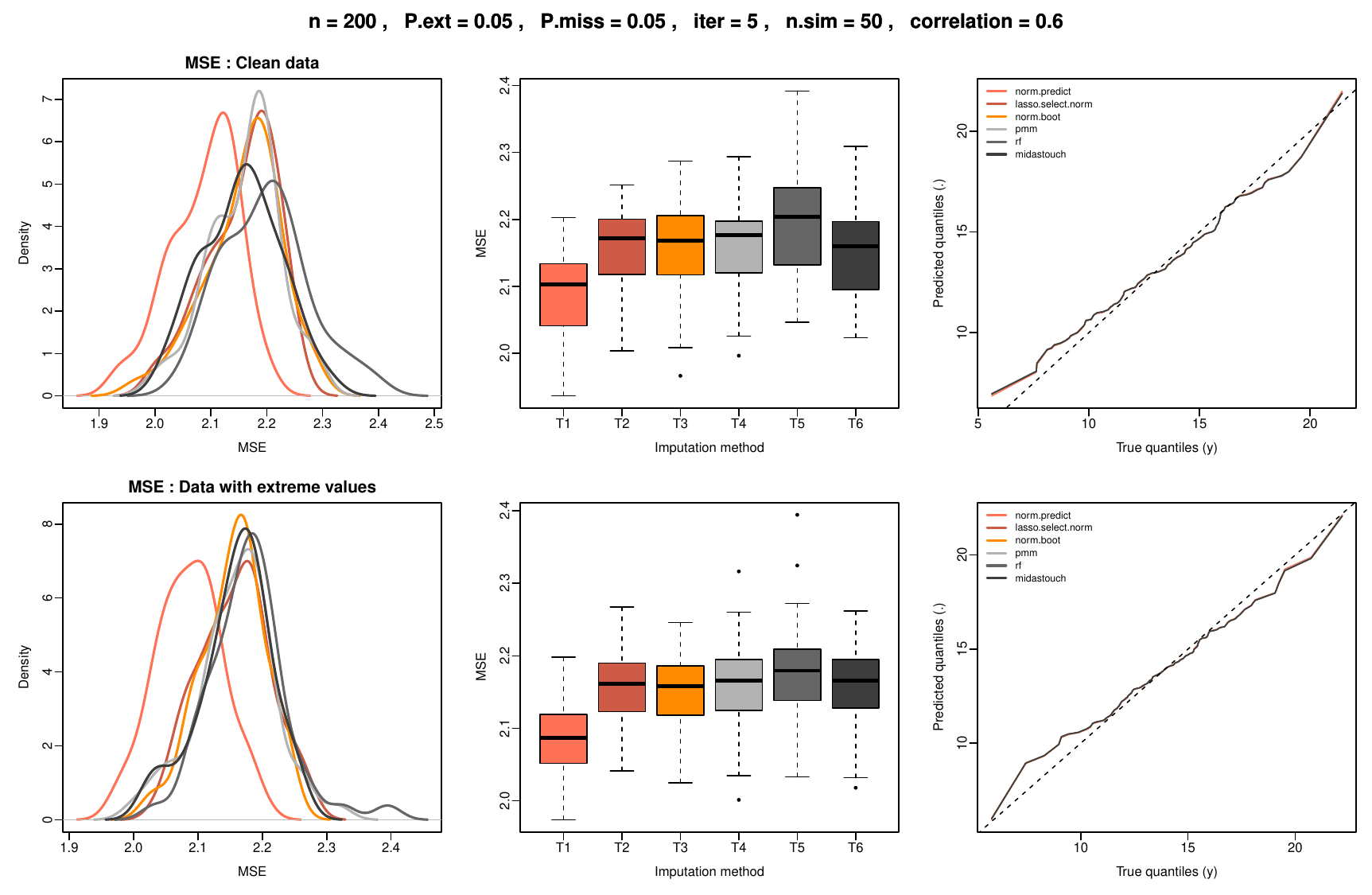}
    \end{minipage}\hfill
    \begin{minipage}{.48\linewidth}
      \includegraphics[page=14,width=\linewidth]{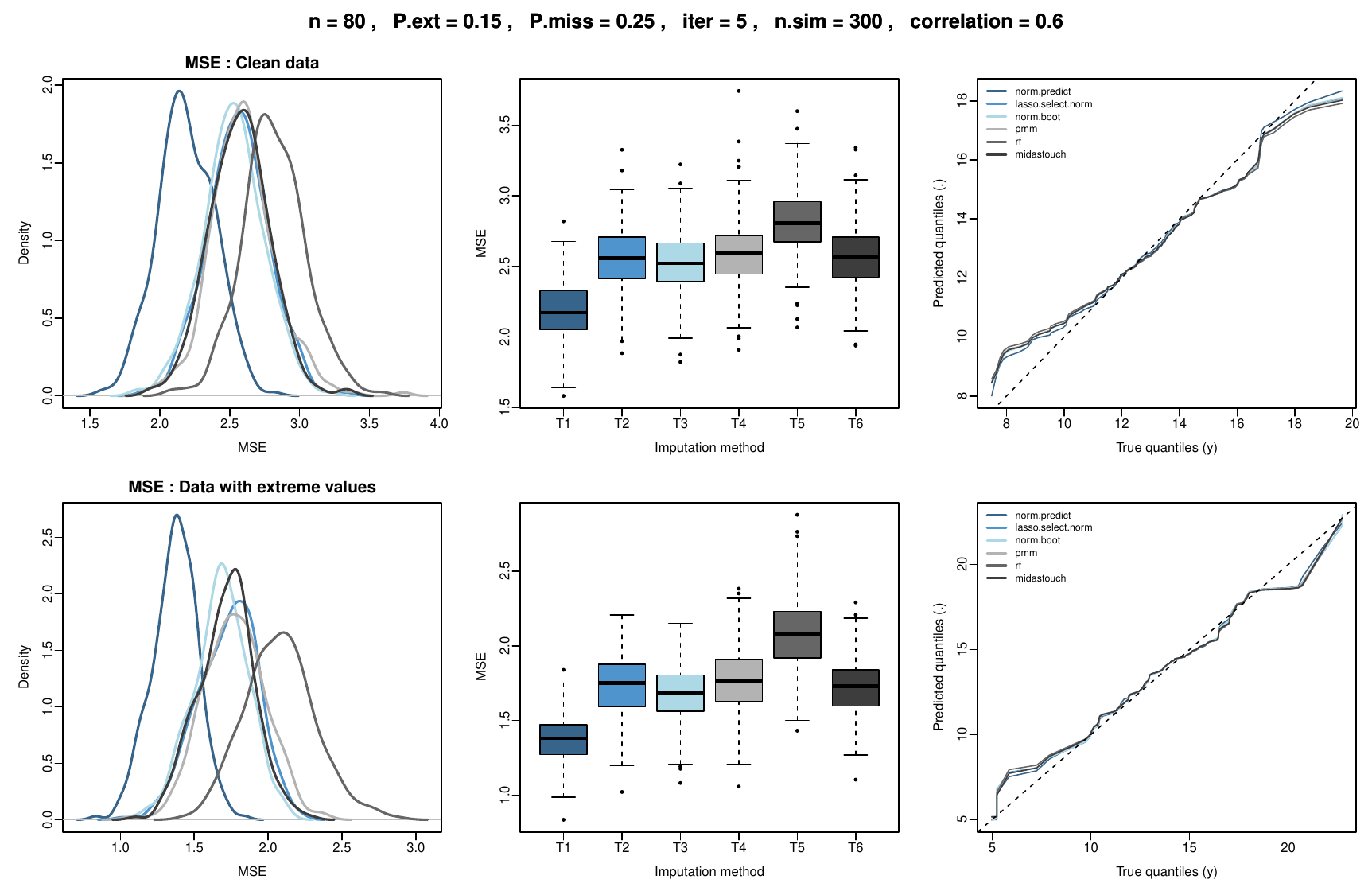}\\[20pt]
      \includegraphics[page=26,width=\linewidth]{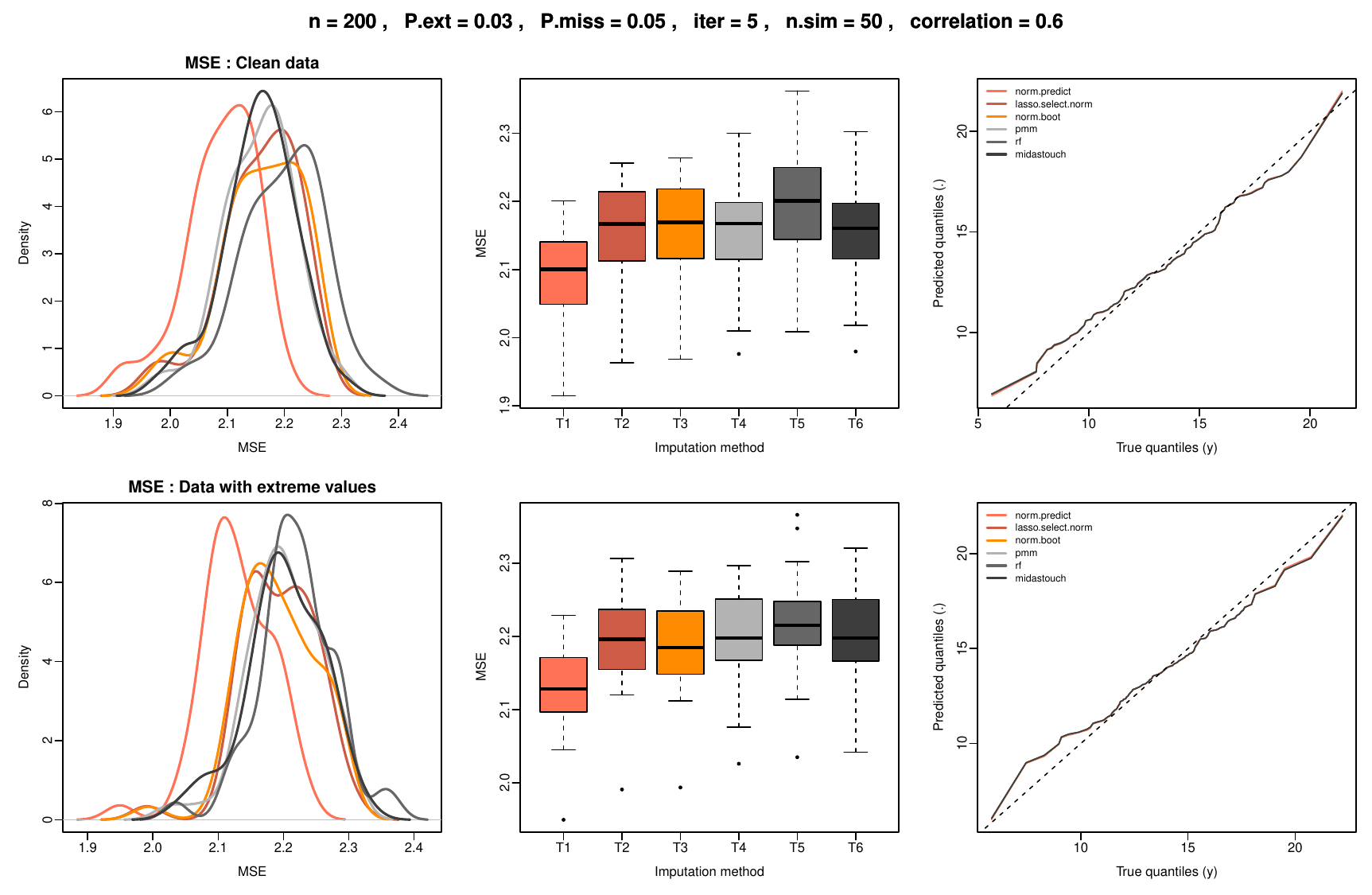}\\[20pt]
      \includegraphics[page=28,width=\linewidth]{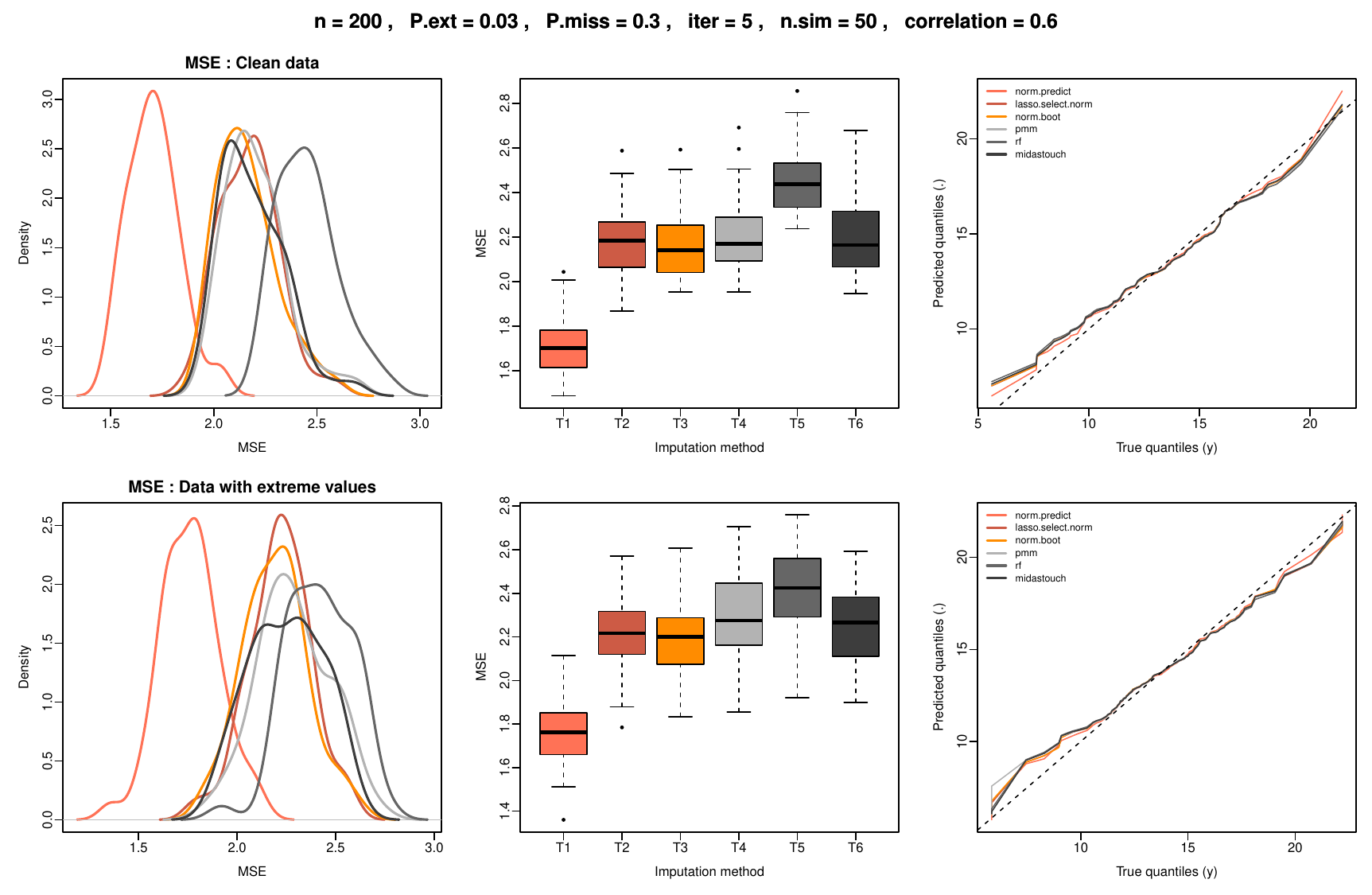}\\[20pt]
      \includegraphics[page=11,width=\linewidth]{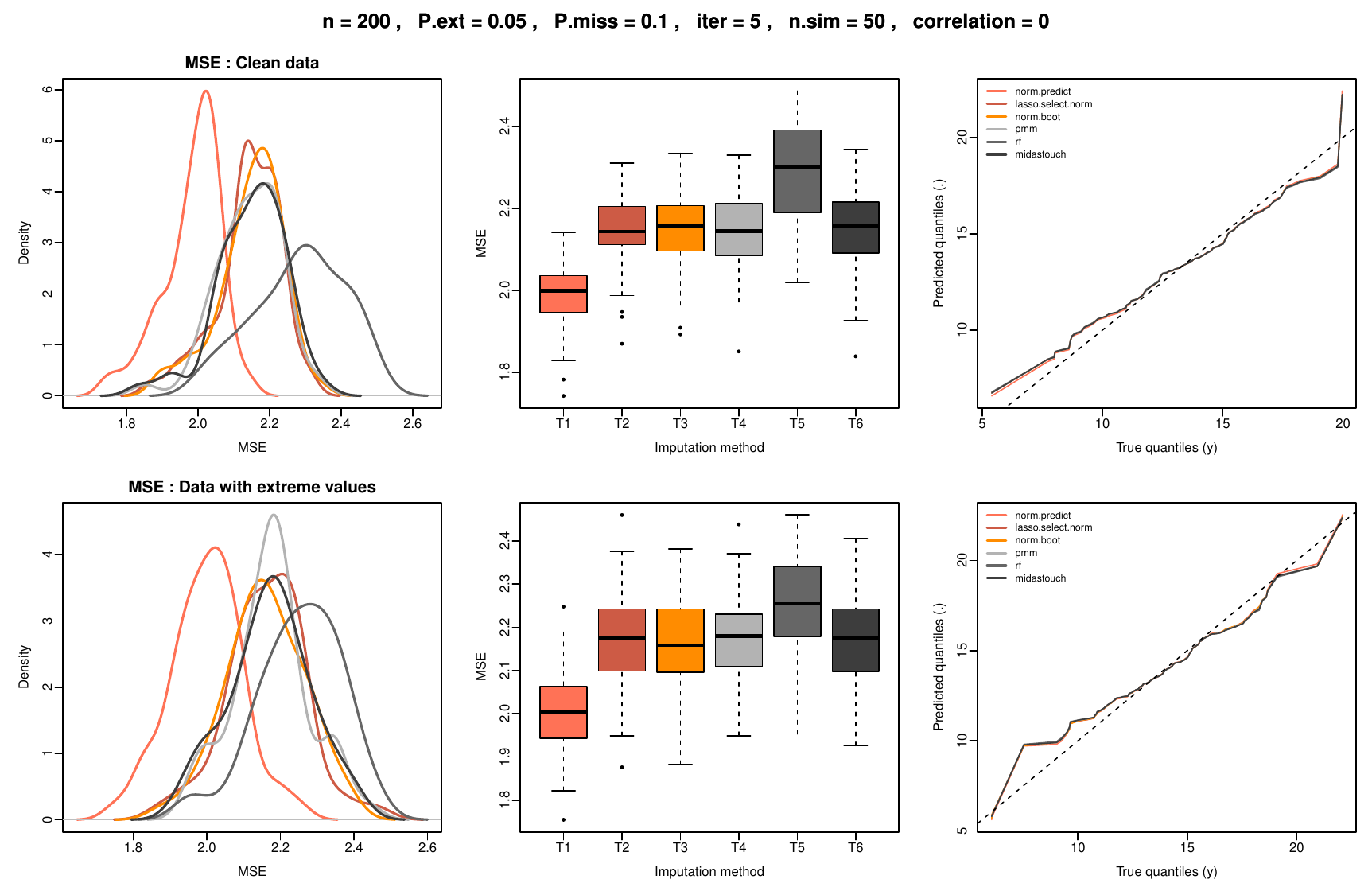}
    \end{minipage}%
  }
      \caption{Predictive MSE densities (clean vs contaminated with extremes data), MSE boxplots, and QQ-plots (predicted vs true quantiles of $y$) across six MI methods (T1--T6). For $\boldsymbol{n=80}$ and $\boldsymbol{n=200}$, ordered by $n$, $P_{\text{ext}}$, and $P_{\text{miss}}$ (panel 3 of 4). Clean data are analyzed with OLS and contaminated data with elastic net. Each subpanel shows the design values ($n$, $P_{\text{ext}}$, $P_{\text{miss}}$, iter, n.sim, and $\rho$).}
      \label{fig:n_80_200_graficos.1-mesclado-3}%
\end{figure}

\begin{figure}[H]
  \centering
  \adjustbox{width=\textwidth, max height=\textheight, center}{%
    \begin{minipage}{.48\linewidth}
      \includegraphics[page=22,width=\linewidth]{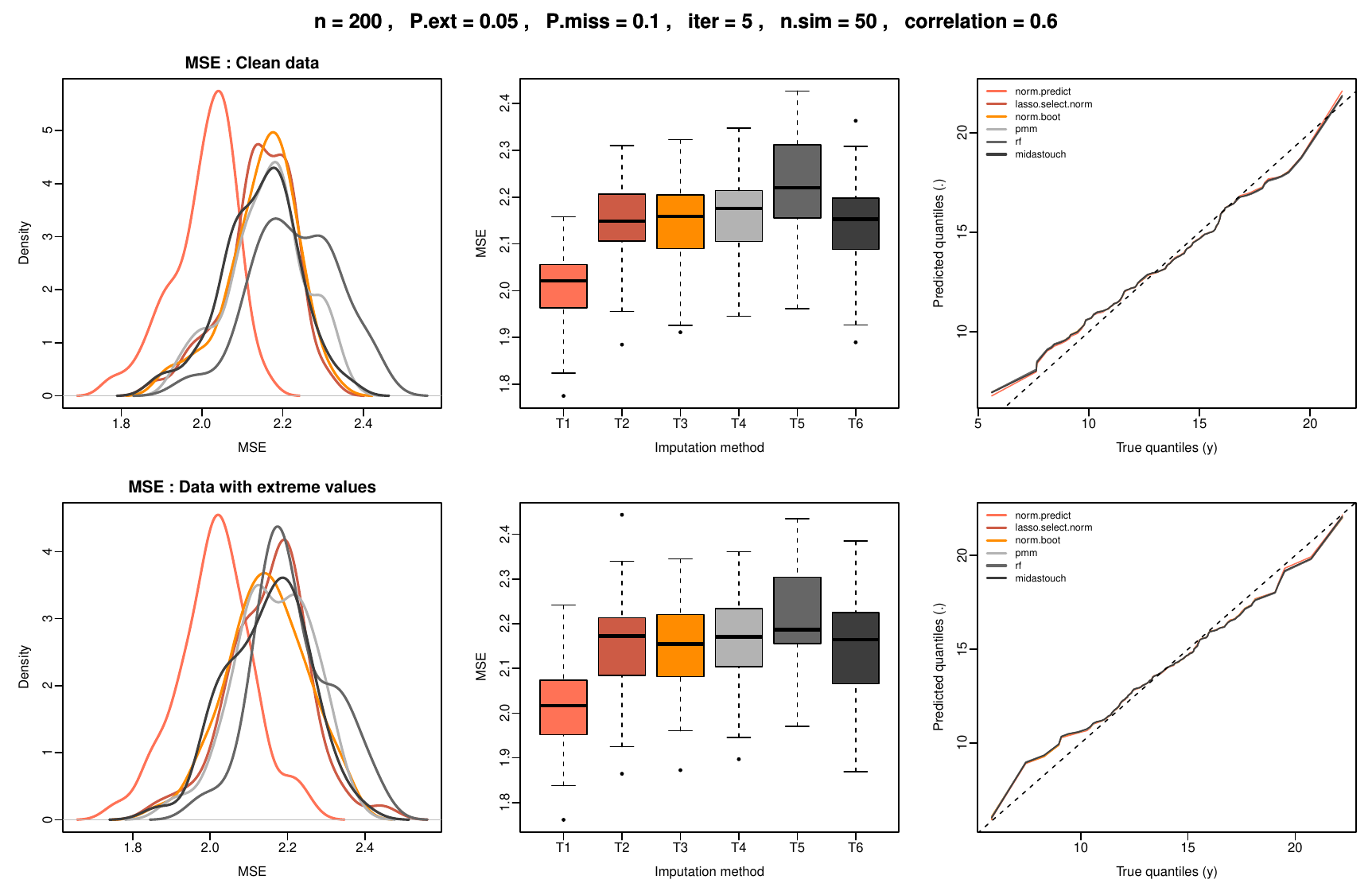}\\[20pt]
      \includegraphics[page=23,width=\linewidth]{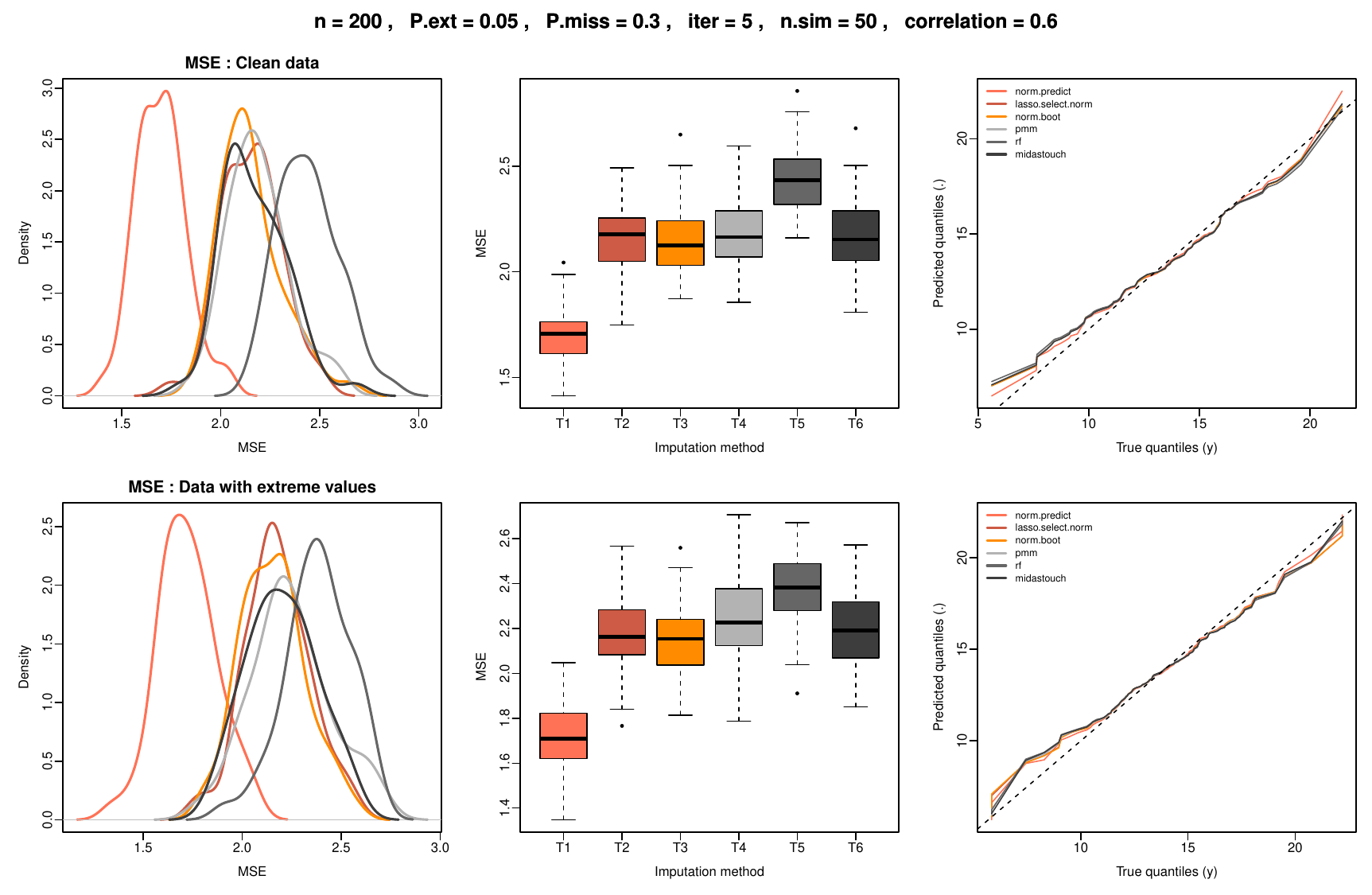}\\[20pt]
      \includegraphics[page=5 ,width=\linewidth]{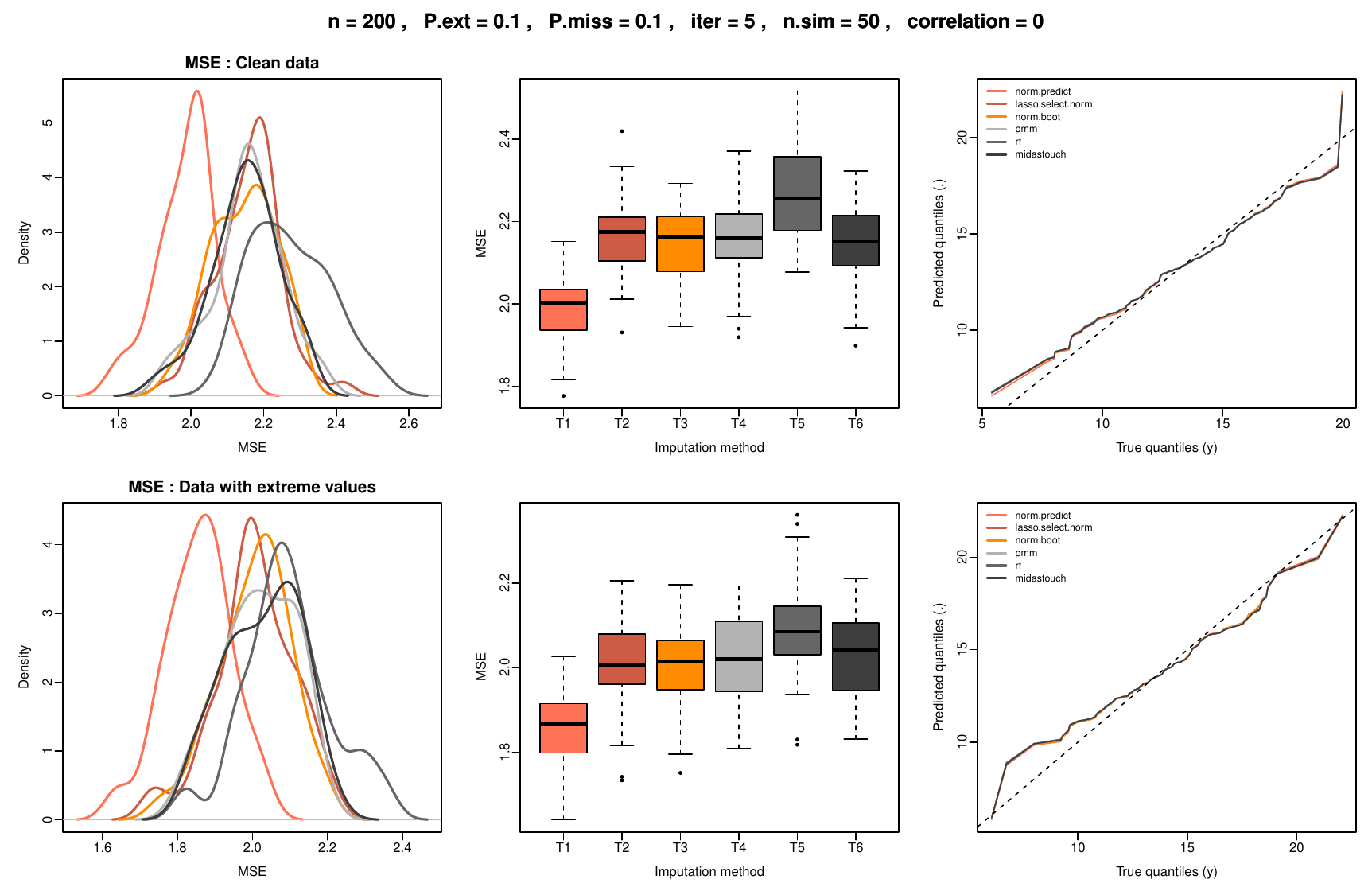}\\[20pt]
      \includegraphics[page=6 ,width=\linewidth]{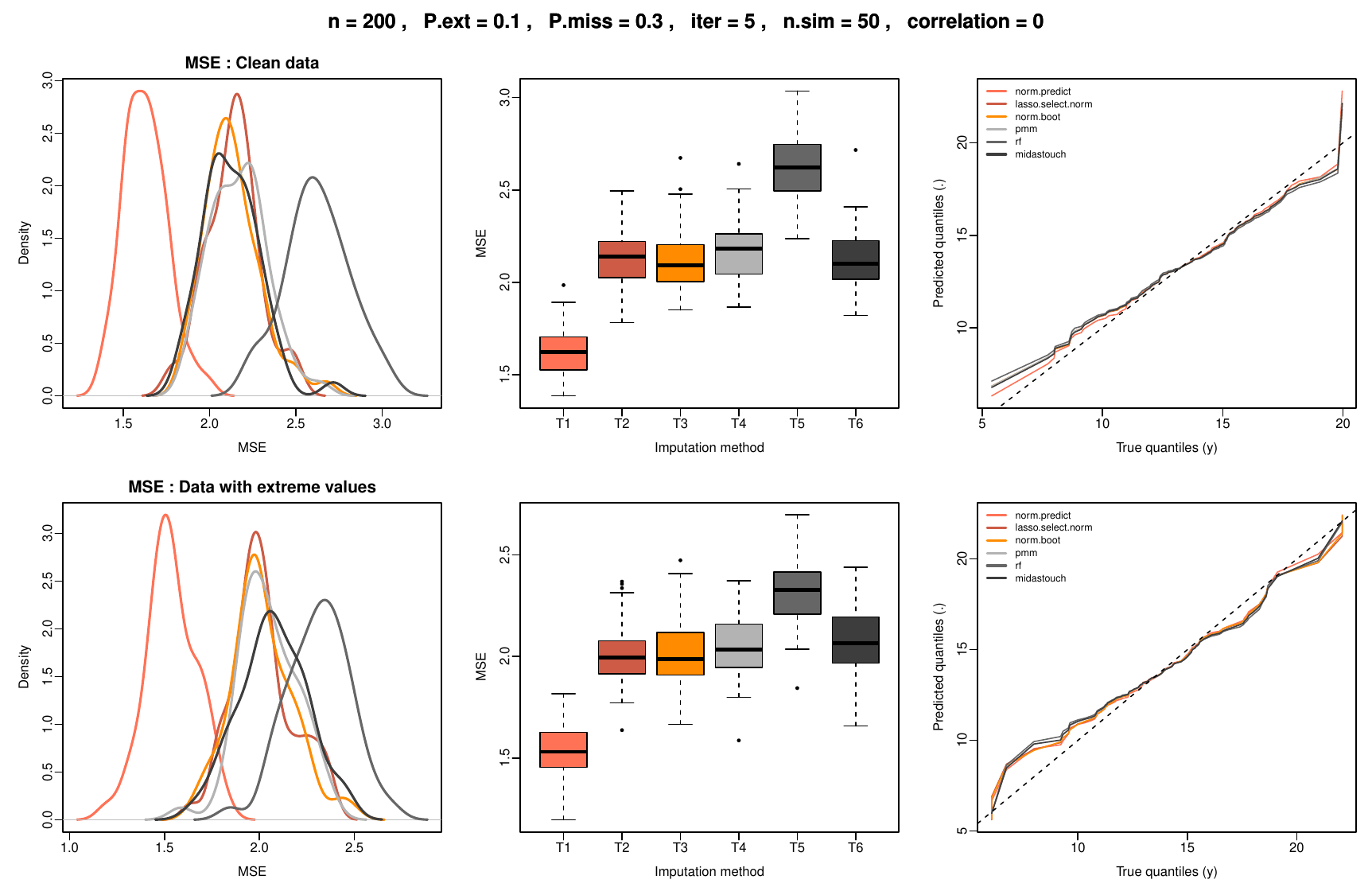}
    \end{minipage}\hfill
    \begin{minipage}{.48\linewidth}
      \includegraphics[page=12,width=\linewidth]{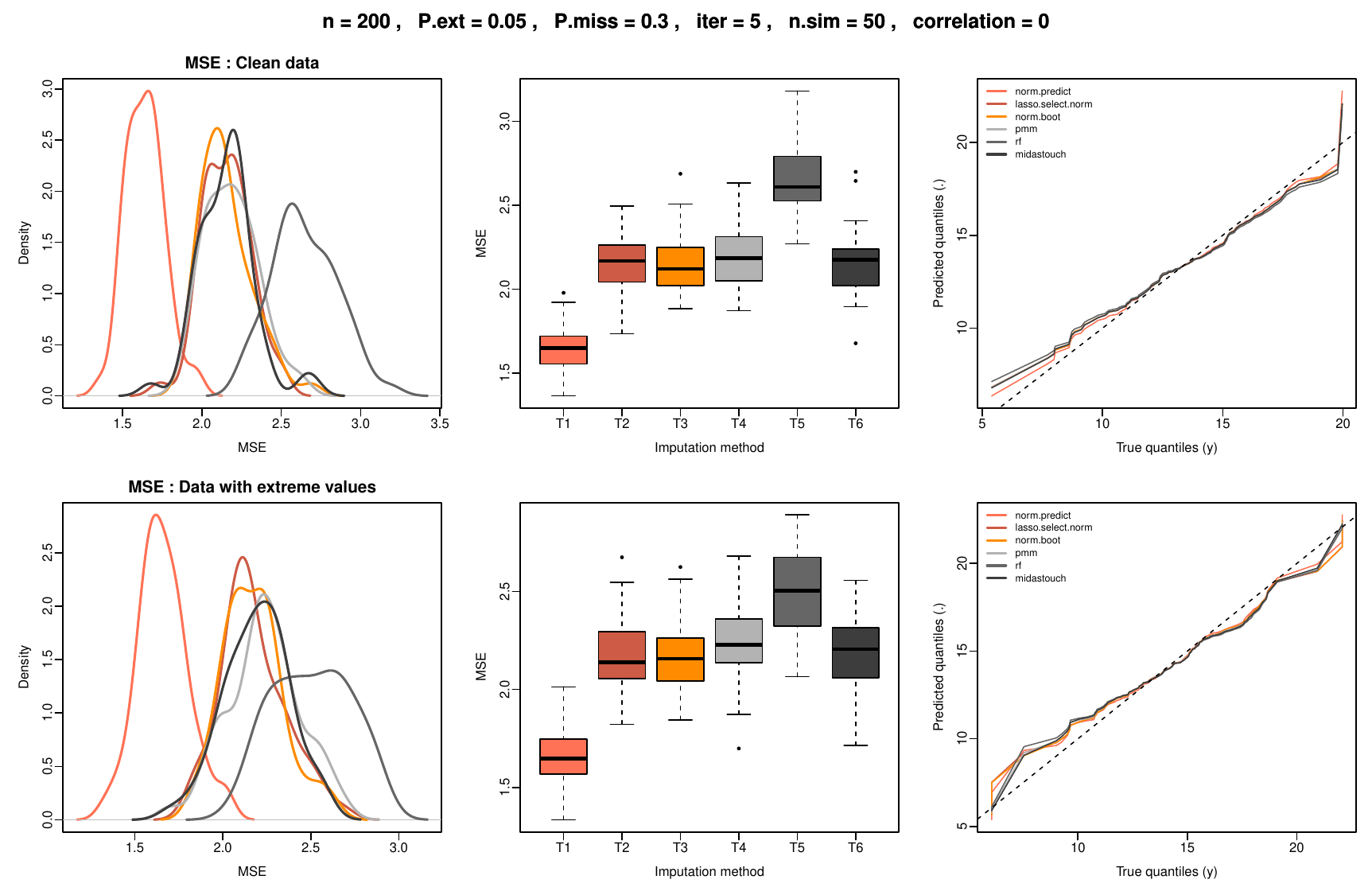}\\[20pt]
      \includegraphics[page=24,width=\linewidth]{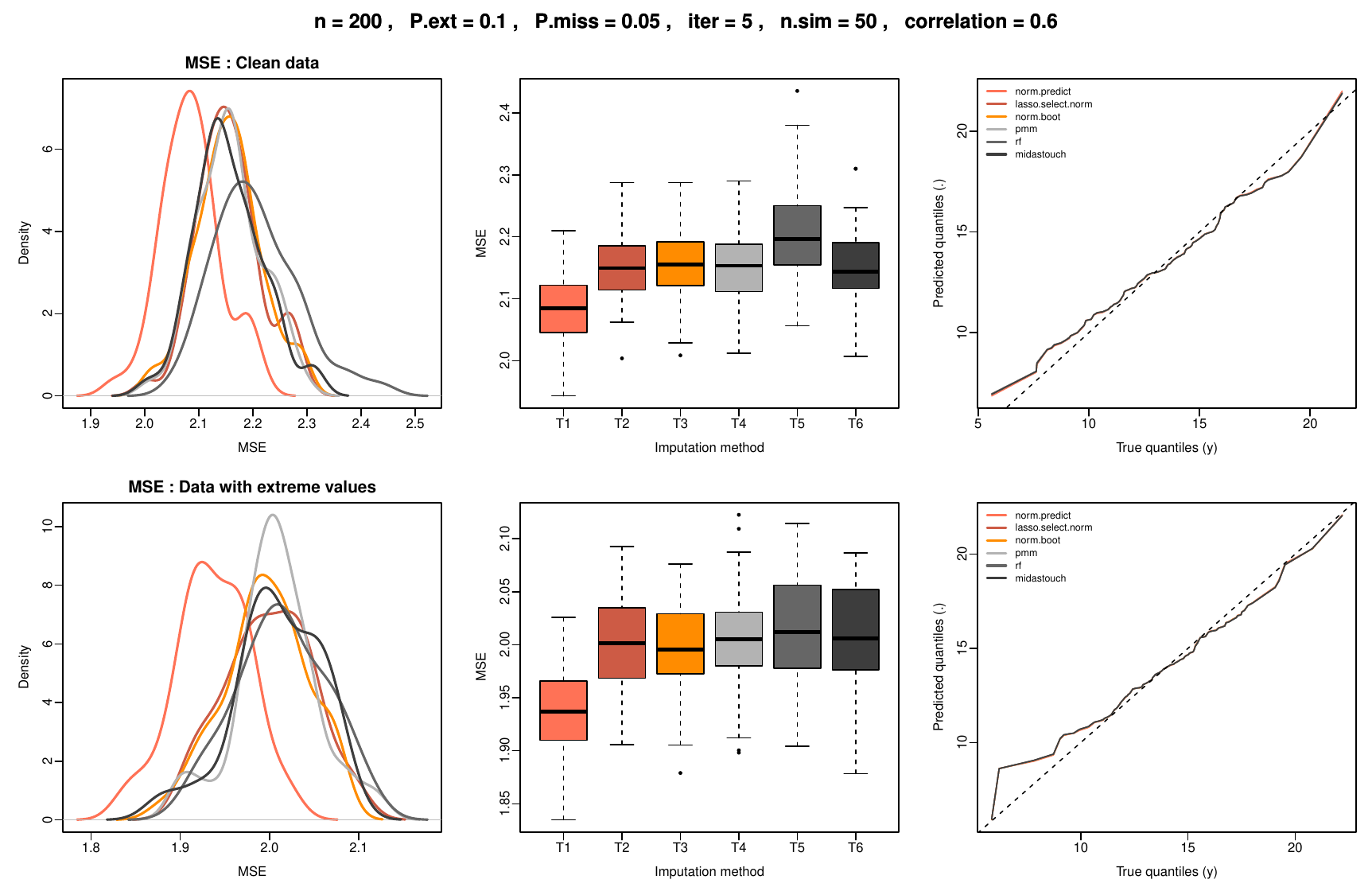}\\[20pt]
      \includegraphics[page=17,width=\linewidth]{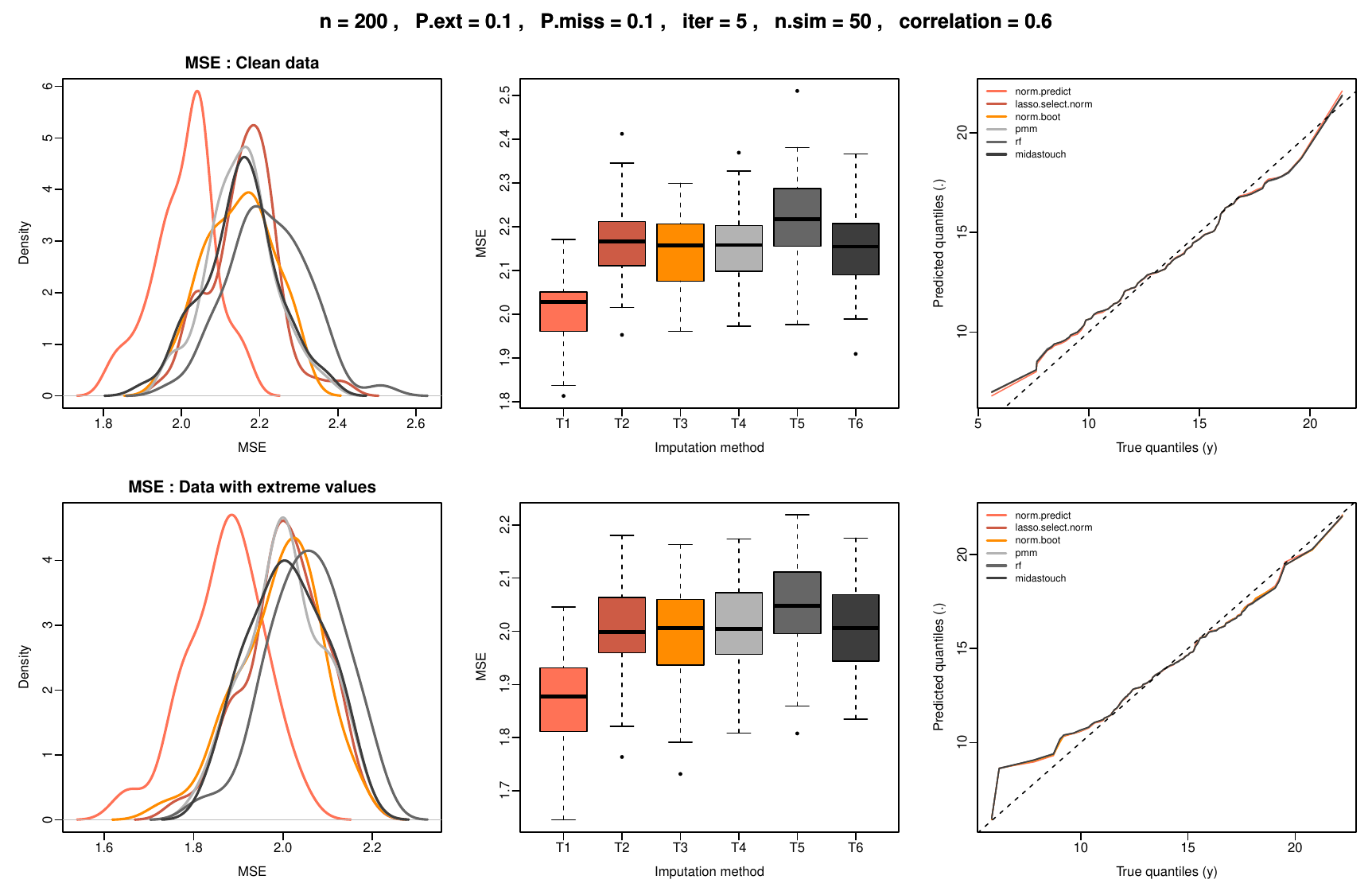}\\[20pt]
      \includegraphics[page=18,width=\linewidth]{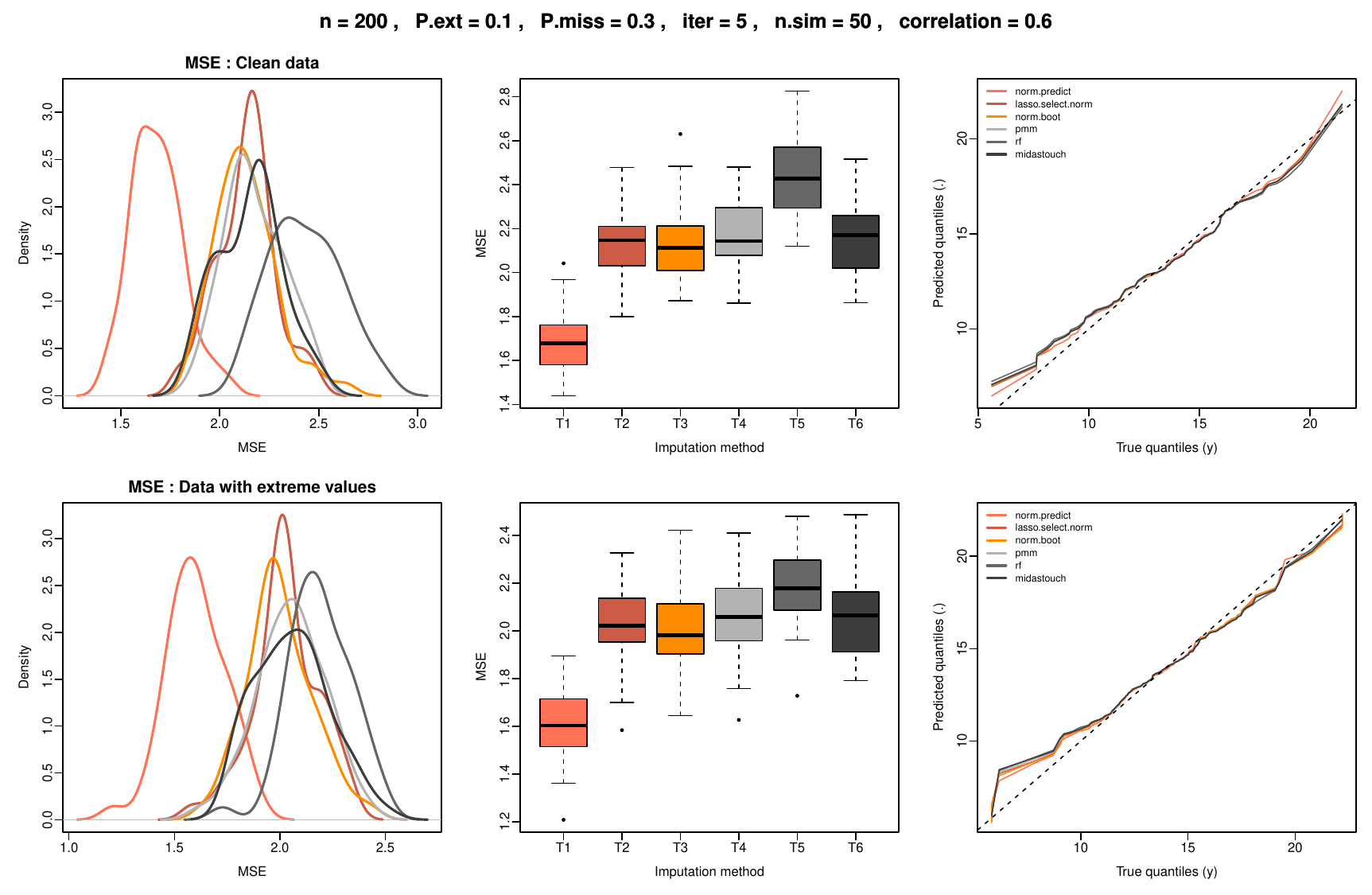}
    \end{minipage}%
  }
      \captionsetup{type=figure}%
      \caption{Predictive MSE densities (clean vs contaminated with extremes data), MSE boxplots, and QQ-plots (predicted vs true quantiles of $y$) across six MI methods (T1--T6). For $\boldsymbol{n=200}$, ordered by $P_{\text{ext}}$, and $P_{\text{miss}}$ (panel 4 of 4). Clean data are analyzed with OLS and contaminated data with elastic net. Each subpanel shows the design values ($n$, $P_{\text{ext}}$, $P_{\text{miss}}$, iter, n.sim, and $\rho$).}
      \label{fig:n_80_200_graficos.1-mesclado-4}%
\end{figure}

\paragraph{Moderate sample sizes ($\boldsymbol{n=80}$ and $\boldsymbol{n=200}$).}
See Tables~\ref{tab:resumo_80_200_1} (bias/RMSE/coverage) and \ref{tab:resumo_80_200_2} (CV--MSE), and Figures~\ref{fig:n_80_200_graficos.1-mesclado-1}--\ref{fig:n_80_200_graficos.1-mesclado-4}. For further details see the Tables \ref{tab:n_80_200_graficos.1}--\ref{tab:n_80_200_graficos.8} on appendix.

\emph{Coefficients (level and bias).}
Under \emph{clean} OLS at $n=80$ with low missingness, slopes are already well-behaved for T1--T3 ($\beta_1$ bias $[0.021,0.037]$, RMSE $[0.030,0.046]$; $\beta_2$ bias $[-0.199,-0.115]$, RMSE $[0.124,0.203]$), with T4--T6 comparable though slightly more dispersed ($\beta_2$ RMSE up to $0.218$). At $n=200$, slope biases contract further toward $0$ (e.g., T1--T3 $\beta_1$ bias $[-0.018,0.003]$, RMSE $[0.011,0.022]$; $\beta_2$ bias $[0.002,0.067]$, RMSE $[0.015,0.071]$), and the donor/ML group converges similarly (T4--T6 $\beta_1$ RMSE $[0.010,0.022]$, $\beta_2$ RMSE $[0.014,0.028]$). With higher missingness ($\geq 25\%$), inflation of RMSE is visible at $n=80$ for both groups (e.g., T4--T6 $\beta_2$ RMSE $[0.209,0.280]$), but the escalation is substantially muted at $n=200$ (T4--T6 $\beta_2$ RMSE $[0.037,0.064]$). Aggregate ranges (T1--T6) sit between the two groups in all panels. Under \emph{contamination} with extreme values (EN), the three canonical distortions persist intercept level shift, $\beta_1$ tilt, and $\beta_2$ shrinkage—and exhibit clear $n$-effects. At $n=80$ and low $P_{\text{miss}}$, T1--T3 show small-to-moderate $\beta_0$ bias ([$-0.285,0.047$]) and positive $\beta_1$ bias ($[0.083,0.136]$), while $\beta_2$ is pulled negative ([$-0.227,-0.144$]); T4--T6 behave similarly with slightly stronger negative pull on $\beta_2$ ([$-0.252,-0.196$]). At $n=200$ the \emph{signs} remain (intercept inflation, positive $\beta_1$, negative $\beta_2$), but dispersion contracts sharply: e.g., T1--T3 $\beta_1$ RMSE drops to $[0.010,0.091]$ and $\beta_2$ RMSE to $[0.017,0.145]$; T4--T6 show analogous tightening. Hence, increasing $n$ reduces volatility but does not eliminate the structural shifts induced by extremes.

\emph{Coverage.}
With OLS on clean data, coverage is essentially nominal at $n=80$ for low missingness across groups ($\approx 1.00$ for all coefficients), and remains high at $n=200$. Under higher missingness, coverage erosion concentrates on $\beta_2$: at $n=80$ T4--T6 reach $0.84$--$0.96$; at $n=200$ the \emph{parametric} block can under-cover markedly in some cells (as low as $0.10$), despite small biases an instance where RMSE and coverage diverge due to variance underestimation. Under contamination (EN), intercept coverage remains unreliable—even at $n=200$ several cells are near zero—while slope coverages improve relative to $n=20$ and $n=40$ yet remain sub-nominal in multiple panels. Donor/ML T4--T6 methods occasionally preserves coverage relative to T1--T3 (especially for $\beta_2$ under higher $P_{\text{miss}}$), consistent with its slightly reduced coefficient biases.

\emph{Predictive error (CV--MSE) and tail risk.}
Under OLS (clean), sample size has strong stabilising effects on both \emph{level} and \emph{dispersion}: for $P_{\text{miss}}\le 0.10$, $n=80$ exhibits small variances (e.g., T1 $\sigma^2=[0.009,0.019]$, T5 up to $0.030$), while at $n=200$ variances are nearly negligible across methods (often $\le 0.01$). Donor/ML remains right-shifted relative to T1--T3 (e.g., at high $P_{\text{miss}}$, $n=80$: T5 $\overline{X}=[2.811,2.905]$, $\sigma^2=[0.049,0.074]$ vs.\ T2 $\overline{X}=[2.551,2.600]$, $\sigma^2=[0.046,0.056]$), but the gap narrows at $n=200$ (the only difference lies in T1, which yielded lower values compared to the other methods).

Under EN (contaminated), both the mean level ($\overline{X}$) and the dispersion ($\sigma^2$) exceed their OLS counterparts, indicating higher predictive variability; figures also show right-skewed densities. For $n=80$ at high $P_{\text{miss}}$, donor/ML methods present larger dispersion (e.g., T5 $\sigma^2\in[0.058,0.116]$, T4 up to $0.088$), whereas parametric T1--T3 remain tighter (e.g., T1 $\sigma^2\in[0.023,0.033]$). At $n=200$, dispersions compress across methods (e.g., T5 $\sigma^2\in[0.022,0.049]$; T1 $\sigma^2\in[0.017,0.022]$), preserving the ordering: T1--T3 exhibit lower spread than T4--T6, while absolute gaps shrink.

\emph{Cross-$n$ synthesis (vs.\ small $n=20,40$).}
(i) Under clean OLS, moving from $n=20,40$ to $n=80,200$ suppresses slope bias toward zero and collapses variance; residual coverage failures at high $P_{\text{miss}}$ target $\beta_2$ and are method-dependent.  
(ii) Under contamination (EN), the \emph{direction} of distortions is stable across $n$ (intercept up, $\beta_1$ up, $\beta_2$ down); larger $n$ chiefly reduces dispersion and upper-tail risk rather than removing bias.  
(iii) Methodologically, the T1--T3 vs.\ T4--T6 trade-off observed at $n=20,40$ persists: donor/ML often shows \emph{smaller coefficient bias} (helping coverage in some cells) but \emph{heavier predictive tails}; parametric MI keeps \emph{tighter} CV--MSE spreads. The aggregate (T1--T6) lies between, with its tail behaviour driven by the donor/ML contribution.

\emph{Practical guidance.}
For moderate $n$ and clean data, method choice is less consequential at low $P_{\text{miss}}$; with $P_{\text{miss}}\gtrsim 0.25$, monitor $\beta_2$ coverage parametric MI (T1--T3) may under cover even with small RMSEs. Under contamination, prefer MI procedures with tighter tails if the goal is stable prediction (T1--T3), while acknowledging that donor/ML (T4--T6) can yield smaller coefficient biases but with higher tail risk. At $n=200$, absolute differences are smaller but remain decision-relevant when upper-tail control is critical.

\begin{table}[H]
\centering
\begin{threeparttable}
\caption{Bias, RMSE, and 95\% coverage for $(\beta_0,\beta_1,\beta_2)$ under clean data with OLS and data with extreme values using EN for $\boldsymbol{n=80}$ (left) and $\boldsymbol{n=200}$ (right). Entries are reported as min--max within each missingness block (low: $P_{\text{miss}}\in\{0.05,0.10\}$; and high: $P_{\text{miss}}\in\{0.25,0.30\}$). Ranges further aggregate over all design combinations $P_{\text{ext}}\in\{0.03,0.05,0.10,0.15,0.30\}$, $n_{\text{sim}}$ $\in\{50,300,1000\}$, $\text{iter}\in\{5,10\}$, and $\rho\in\{0,0.6\}$. MI methods are grouped as parametric (T1--T3), donor/ML (T4--T6), and the combined set (T1--T6)}
\label{tab:resumo_80_200_1}

\begingroup
\scriptsize
\setlength{\tabcolsep}{2pt}
\renewcommand{\arraystretch}{1.06}
\begin{tabular*}{\textwidth}{@{\extracolsep{\fill}} l l c c c c c c @{}}
\toprule
\multicolumn{2}{c}{} & \multicolumn{3}{c}{\textbf{n = 80}} & \multicolumn{3}{c}{\textbf{n = 200}} \\
\cmidrule(lr){3-5} \cmidrule(lr){6-8}
MI & Metric & $\beta_0$ & $\beta_1$ & $\beta_2$ & $\beta_0$ & $\beta_1$ & $\beta_2$ \\
\midrule
\multicolumn{8}{l}{\textbf{Linear regression (OLS):} low missing-data proportion (5\% and 10\%)} \\ 
T1--T3 & Bias & [0.147 , 0.425] & [0.021 , 0.037] & [-0.199 , -0.115] & [-0.268 , 0.034] & [-0.018 , 0.003] & [0.002 , 0.067] \\
T1--T3 & RMSE & [0.301 , 0.487] & [0.030 , 0.046] & [0.124 , 0.203] & [0.128 , 0.350] & [0.011 , 0.022] & [0.015 , 0.071] \\
T1--T3 & Coverage & [1.000 , 1.000] & [1.000 , 1.000] & [1.000 , 1.000] & [1.000 , 1.000] & [1.000 , 1.000] & [1.000 , 1.000] \\
T4--T6 & Bias & [0.359 , 0.475] & [0.026 , 0.037] & [-0.213 , -0.183] & [-0.047 , 0.144] & [-0.009 , 0.001] & [-0.010 , 0.010] \\
T4--T6 & RMSE & [0.411 , 0.533] & [0.037 , 0.045] & [0.189 , 0.218] & [0.127 , 0.295] & [0.010 , 0.022] & [0.014 , 0.028] \\
T4--T6 & Coverage & [1.000 , 1.000] & [1.000 , 1.000] & [0.980 , 1.000] & [1.000 , 1.000] & [1.000 , 1.000] & [1.000 , 1.000] \\
T1--T6 & Bias & [0.147 , 0.475] & [0.021 , 0.037] & [-0.213 , -0.115] & [-0.268 , 0.144] & [-0.018 , 0.003] & [-0.010 , 0.067] \\
T1--T6 & RMSE & [0.301 , 0.533] & [0.030 , 0.046] & [0.124 , 0.218] & [0.127 , 0.350] & [0.010 , 0.022] & [0.014 , 0.071] \\
T1--T6 & Coverage & [1.000 , 1.000] & [1.000 , 1.000] & [0.980 , 1.000] & [1.000 , 1.000] & [1.000 , 1.000] & [1.000 , 1.000] \\
\addlinespace[3pt]
\cmidrule(lr){1-8}
\multicolumn{8}{l}{\textbf{Linear regression (OLS):} high missing-data proportion (25\% and 30\%)} \\ 
T1--T3 & Bias & [-0.368 , 0.444] & [-0.007 , 0.039] & [-0.207 , 0.047] & [-0.872 , -0.004] & [-0.045 , 0.018] & [-0.001 , 0.196] \\
T1--T3 & RMSE & [0.485 , 0.683] & [0.041 , 0.059] & [0.075 , 0.230] & [0.336 , 0.990] & [0.029 , 0.055] & [0.035 , 0.201] \\
T1--T3 & Coverage & [0.980 , 1.000] & [0.993 , 1.000] & [0.900 , 1.000] & [0.680 , 1.000] & [0.940 , 1.000] & [0.100 , 1.000] \\
T4--T6 & Bias & [0.354 , 0.632] & [0.032 , 0.049] & [-0.259 , -0.188] & [-0.175 , 0.473] & [-0.020 , 0.015] & [-0.051 , 0.015] \\
T4--T6 & RMSE & [0.527 , 0.796] & [0.050 , 0.062] & [0.209 , 0.280] & [0.385 , 0.656] & [0.033 , 0.042] & [0.037 , 0.064] \\
T4--T6 & Coverage & [1.000 , 1.000] & [1.000 , 1.000] & [0.840 , 0.960] & [1.000 , 1.000] & [1.000 , 1.000] & [1.000 , 1.000] \\
T1--T6 & Bias & [-0.368 , 0.632] & [-0.007 , 0.049] & [-0.259 , 0.047] & [-0.872 , 0.473] & [-0.045 , 0.018] & [-0.051 , 0.196] \\
T1--T6 & RMSE & [0.485 , 0.796] & [0.041 , 0.062] & [0.075 , 0.280] & [0.336 , 0.990] & [0.029 , 0.055] & [0.035 , 0.201] \\
T1--T6 & Coverage & [0.980 , 1.000] & [0.993 , 1.000] & [0.840 , 1.000] & [0.680 , 1.000] & [0.940 , 1.000] & [0.100 , 1.000] \\
\addlinespace[3pt]
\cmidrule(lr){1-8}
\multicolumn{8}{l}{\textbf{Sparse regression (EN)}: low missing-data proportion (5\% and 10\%)} \\ 
T1--T3 & Bias & [-0.285 , 0.047] & [0.083 , 0.136] & [-0.227 , -0.144] & [0.663 , 1.738] & [-0.090 , 0.011] & [-0.144 , 0.007] \\
T1--T3 & RMSE & [0.113 , 0.346] & [0.086 , 0.139] & [0.145 , 0.228] & [0.666 , 1.743] & [0.010 , 0.091] & [0.017 , 0.145] \\
T1--T3 & Coverage & [1.000 , 1.000] & [0.980 , 1.000] & [0.188 , 1.000] & [0.000 , 1.000] & [0.560 , 1.000] & [0.100 , 1.000] \\
T4--T6 & Bias & [-0.352 , -0.010] & [0.109 , 0.155] & [-0.252 , -0.196] & [0.682 , 1.830] & [-0.090 , 0.011] & [-0.151 , -0.058] \\
T4--T6 & RMSE & [0.100 , 0.437] & [0.112 , 0.158] & [0.197 , 0.255] & [0.684 , 1.833] & [0.013 , 0.090] & [0.061 , 0.152] \\
T4--T6 & Coverage & [1.000 , 1.000] & [0.920 , 1.000] & [0.170 , 1.000] & [0.000 , 1.000] & [0.800 , 1.000] & [0.040 , 1.000] \\
T1--T6 & Bias & [-0.352 , 0.047] & [0.083 , 0.155] & [-0.252 , -0.144] & [0.663 , 1.830] & [-0.090 , 0.011] & [-0.151 , 0.007] \\
T1--T6 & RMSE & [0.100 , 0.437] & [0.086 , 0.158] & [0.145 , 0.255] & [0.666 , 1.833] & [0.010 , 0.091] & [0.017 , 0.152] \\
T1--T6 & Coverage & [1.000 , 1.000] & [0.920 , 1.000] & [0.170 , 1.000] & [0.000 , 1.000] & [0.560 , 1.000] & [0.040 , 1.000] \\
\addlinespace[3pt]
\cmidrule(lr){1-8}
\multicolumn{8}{l}{\textbf{Sparse regression (EN)}: high missing-data proportion (25\% and 30\%)} \\ 
T1--T3 & Bias & [-0.506 , 0.039] & [0.037 , 0.138] & [-0.221 , -0.014] & [0.604 , 1.717] & [-0.122 , 0.013] & [-0.148 , 0.151] \\
T1--T3 & RMSE & [0.261 , 0.707] & [0.057 , 0.148] & [0.055 , 0.228] & [0.665 , 1.733] & [0.027 , 0.125] & [0.037 , 0.155] \\
T1--T3 & Coverage & [0.980 , 1.000] & [0.880 , 1.000] & [0.673 , 1.000] & [0.000 , 0.900] & [0.000 , 1.000] & [0.380 , 1.000] \\
T4--T6 & Bias & [-0.707 , -0.016] & [0.108 , 0.215] & [-0.334 , -0.199] & [0.715 , 1.976] & [-0.106 , 0.018] & [-0.175 , -0.066] \\
T4--T6 & RMSE & [0.233 , 0.939] & [0.116 , 0.228] & [0.207 , 0.342] & [0.751 , 1.983] & [0.024 , 0.111] & [0.076 , 0.177] \\
T4--T6 & Coverage & [0.920 , 1.000] & [0.660 , 0.967] & [0.380 , 0.920] & [0.000 , 0.940] & [0.500 , 1.000] & [0.180 , 1.000] \\
T1--T6 & Bias & [-0.707 , 0.039] & [0.037 , 0.215] & [-0.334 , -0.014] & [0.604 , 1.976] & [-0.122 , 0.018] & [-0.175 , 0.151] \\
T1--T6 & RMSE & [0.233 , 0.939] & [0.057 , 0.228] & [0.055 , 0.342] & [0.665 , 1.983] & [0.024 , 0.125] & [0.037 , 0.177] \\
T1--T6 & Coverage & [0.920 , 1.000] & [0.660 , 1.000] & [0.380 , 1.000] & [0.000 , 0.940] & [0.000 , 1.000] & [0.180 , 1.000] \\
\addlinespace[3pt]
\bottomrule
\end{tabular*}
\endgroup

\begin{tablenotes}[flushleft]
\scriptsize
\setlength{\leftskip}{8em} 
\item $^*$Legend: MI methods (T1--T6): T1 \texttt{norm.predict}; T2 \texttt{lasso.select.norm};
T3 \texttt{norm.boot}; T4 \texttt{pmm}; T5 \texttt{rf}; T6 \texttt{midastouch}.
Coverage is the proportion of 95\% intervals containing the true coefficient.
\end{tablenotes}

\end{threeparttable}
\end{table}

\begin{table}[H]
\centering
\caption{Out-of-sample predictive error summaries (CV–MSE) by MI method for $\boldsymbol{n=80}$ (left) and $\boldsymbol{n=200}$ (right), stratified by model (OLS for clean data; elastic net (EN) for contaminated data) and by missingness blocks (low: $P_{\text{miss}}\in\{0.05,0.10\}$; and high: $P_{\text{miss}}\in\{0.25,0.30\}$). For each method it is reported the min-max of the MSE: mean $\overline{X}$, variance $\sigma^2$, and median $Q_{50}$ across the corresponding design cells. Ranges further aggregate over all design combinations $P_{\text{ext}}\in\{0.03,0.05,0.10,0.15,0.30\}$, $n_{\text{sim}}\in\{50,300,1000\}$, $\text{iter}\in\{5,10\}$, and $\rho\in\{0,0.6\}$}
\label{tab:resumo_80_200_2}
\begingroup
\scriptsize
\setlength{\tabcolsep}{3pt}
\renewcommand{\arraystretch}{1.06}
\begin{tabular*}{\textwidth}{@{\extracolsep{\fill}} l c c c c c c @{}}
\toprule
\multicolumn{1}{c}{} & \multicolumn{3}{c}{\textbf{n = 80}} & \multicolumn{3}{c}{\textbf{n = 200}} \\
\cmidrule(lr){2-4} \cmidrule(lr){5-7}
MI & $\overline{X}$ & $\sigma^2$ & $Q_{50}$ & $\overline{X}$ & $\sigma^2$ & $Q_{50}$ \\
\midrule
\multicolumn{7}{l}{\textbf{Linear regression (OLS):} low missing-data proportion (5\% and 10\%)} \\
T1 & [2.397 , 2.493] & [0.009 , 0.019] & [2.403 , 2.490] & [1.983 , 2.089] & [0.003 , 0.007] & [2.000 , 2.103] \\
T2 & [2.536 , 2.580] & [0.010 , 0.023] & [2.537 , 2.583] & [2.141 , 2.159] & [0.004 , 0.008] & [2.144 , 2.175] \\
T3 & [2.523 , 2.568] & [0.010 , 0.021] & [2.518 , 2.576] & [2.146 , 2.161] & [0.004 , 0.009] & [2.155 , 2.169] \\
T4 & [2.534 , 2.585] & [0.009 , 0.019] & [2.522 , 2.584] & [2.142 , 2.164] & [0.003 , 0.010] & [2.144 , 2.180] \\
T5 & [2.589 , 2.673] & [0.013 , 0.030] & [2.569 , 2.660] & [2.198 , 2.284] & [0.005 , 0.016] & [2.196 , 2.302] \\
T6 & [2.522 , 2.580] & [0.009 , 0.020] & [2.509 , 2.574] & [2.144 , 2.156] & [0.004 , 0.009] & [2.144 , 2.160] \\
\addlinespace[4pt]
\cmidrule(lr){1-7}
\multicolumn{7}{l}{\textbf{Linear regression (OLS):} high missing-data proportion (25\% and 30\%)} \\
T1 & [2.093 , 2.189] & [0.036 , 0.047] & [2.095 , 2.191] & [1.628 , 1.707] & [0.015 , 0.017] & [1.624 , 1.707] \\
T2 & [2.551 , 2.600] & [0.046 , 0.056] & [2.537 , 2.601] & [2.131 , 2.170] & [0.020 , 0.024] & [2.140 , 2.185] \\
T3 & [2.499 , 2.540] & [0.045 , 0.064] & [2.514 , 2.561] & [2.132 , 2.162] & [0.021 , 0.027] & [2.092 , 2.140] \\
T4 & [2.598 , 2.652] & [0.049 , 0.065] & [2.582 , 2.659] & [2.168 , 2.203] & [0.023 , 0.029] & [2.143 , 2.184] \\
T5 & [2.811 , 2.905] & [0.049 , 0.074] & [2.807 , 2.941] & [2.433 , 2.648] & [0.021 , 0.042] & [2.427 , 2.621] \\
T6 & [2.552 , 2.600] & [0.043 , 0.059] & [2.547 , 2.669] & [2.125 , 2.189] & [0.024 , 0.032] & [2.102 , 2.176] \\
\addlinespace[4pt]
\cmidrule(lr){1-7}
\multicolumn{7}{l}{\textbf{Sparse regression (EN)}: low missing-data proportion (5\% and 10\%)} \\
T1 & [1.141 , 1.919] & [0.006 , 0.013] & [1.140 , 1.912] & [1.856 , 2.129] & [0.002 , 0.010] & [1.867 , 2.128] \\
T2 & [1.249 , 2.002] & [0.006 , 0.017] & [1.250 , 2.002] & [1.999 , 2.194] & [0.002 , 0.011] & [1.998 , 2.219] \\
T3 & [1.231 , 1.979] & [0.006 , 0.015] & [1.234 , 1.983] & [1.991 , 2.192] & [0.002 , 0.011] & [1.995 , 2.198] \\
T4 & [1.246 , 2.058] & [0.007 , 0.024] & [1.247 , 2.041] & [2.004 , 2.203] & [0.002 , 0.011] & [2.005 , 2.201] \\
T5 & [1.363 , 2.206] & [0.012 , 0.034] & [1.343 , 2.192] & [2.014 , 2.252] & [0.003 , 0.014] & [2.012 , 2.254] \\
T6 & [1.240 , 2.035] & [0.008 , 0.015] & [1.238 , 2.008] & [2.008 , 2.202] & [0.002 , 0.013] & [2.006 , 2.203] \\
\addlinespace[4pt]
\cmidrule(lr){1-7}
\multicolumn{7}{l}{\textbf{Sparse regression (EN)}: high missing-data proportion (25\% and 30\%)} \\
T1 & [1.369 , 1.578] & [0.023 , 0.033] & [1.381 , 1.586] & [1.536 , 1.761] & [0.017 , 0.022] & [1.531 , 1.763] \\
T2 & [1.726 , 2.047] & [0.040 , 0.059] & [1.751 , 2.038] & [2.017 , 2.217] & [0.024 , 0.034] & [1.994 , 2.216] \\
T3 & [1.679 , 2.010] & [0.035 , 0.048] & [1.689 , 2.007] & [2.001 , 2.193] & [0.025 , 0.029] & [1.981 , 2.201] \\
T4 & [1.772 , 2.237] & [0.044 , 0.088] & [1.767 , 2.240] & [2.047 , 2.287] & [0.024 , 0.043] & [2.034 , 2.275] \\
T5 & [2.080 , 2.639] & [0.058 , 0.116] & [2.075 , 2.680] & [2.190 , 2.506] & [0.022 , 0.049] & [2.180 , 2.504] \\
T6 & [1.716 , 2.149] & [0.036 , 0.099] & [1.732 , 2.137] & [2.059 , 2.251] & [0.030 , 0.034] & [2.065 , 2.265] \\
\addlinespace[4pt]
\bottomrule
\end{tabular*}
\endgroup
\end{table}

\begin{figure}[H]
  \centering
  \adjustbox{width=\textwidth, max height=\textheight, center}{%
    \begin{minipage}{.48\linewidth}
      \includegraphics[page=9 ,width=\linewidth]{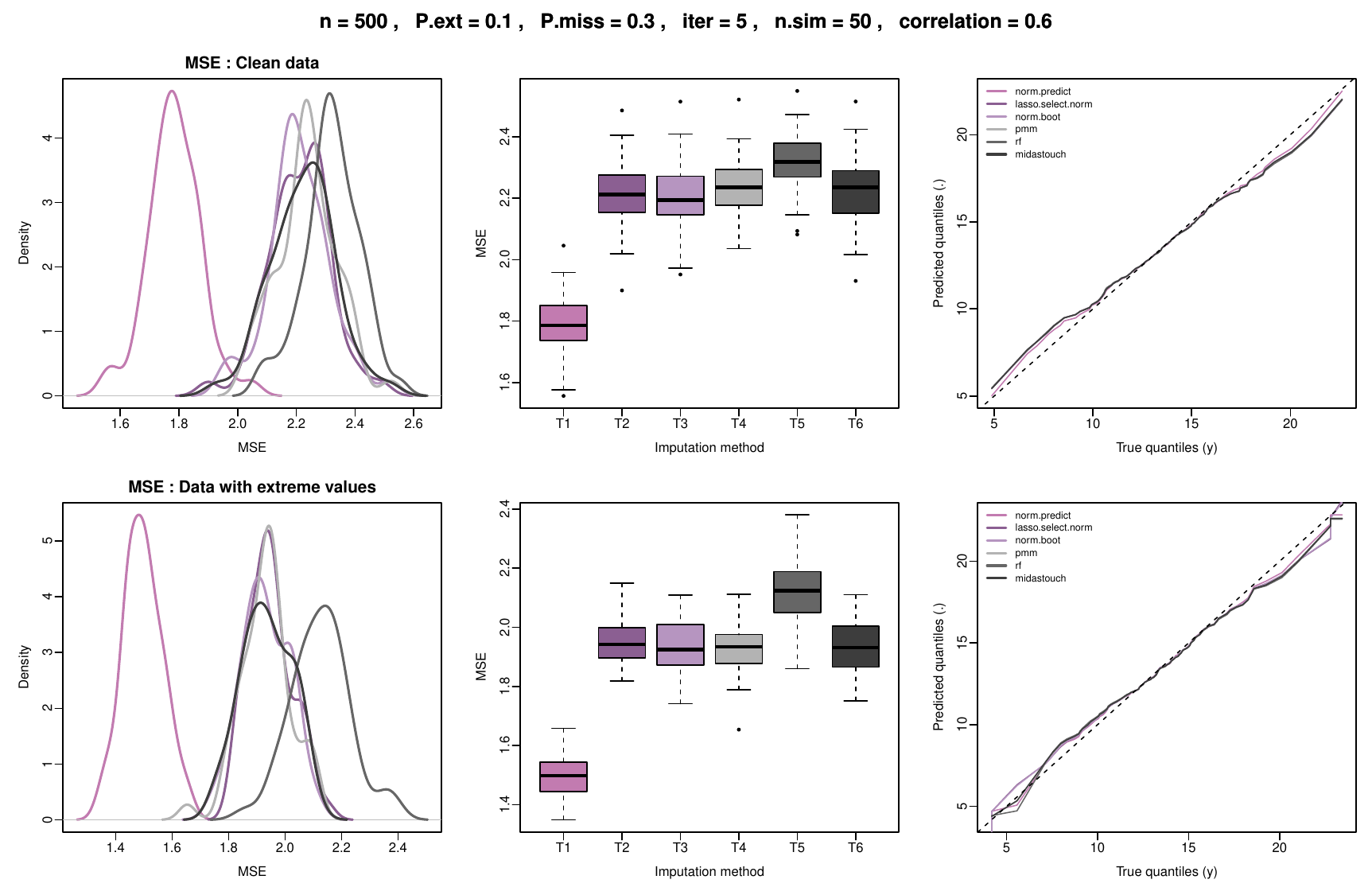}\\[20pt]
      \includegraphics[page=11,width=\linewidth]{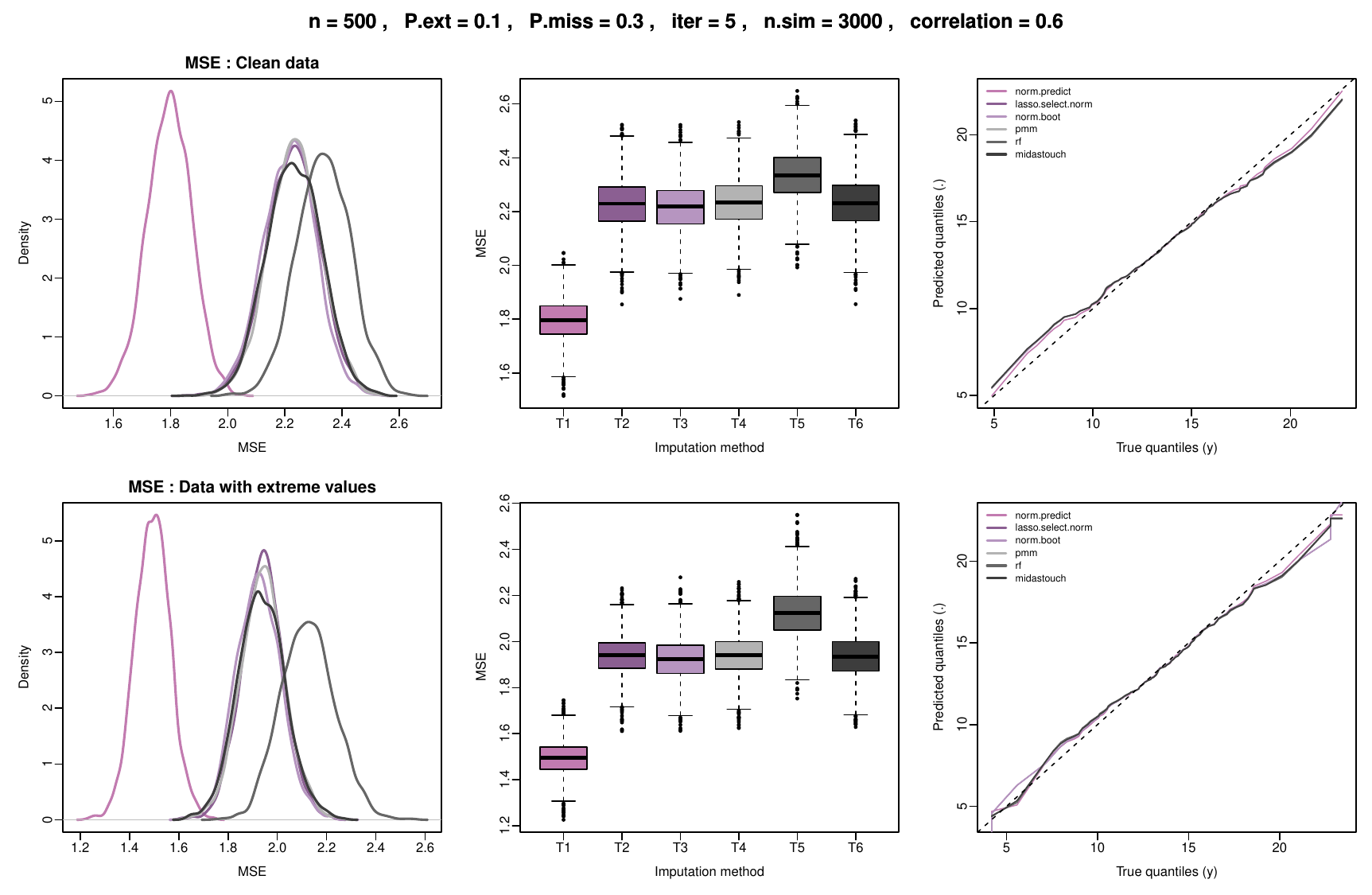}
    \end{minipage}\hfill
    \begin{minipage}{.48\linewidth}
      \includegraphics[page=10,width=\linewidth]{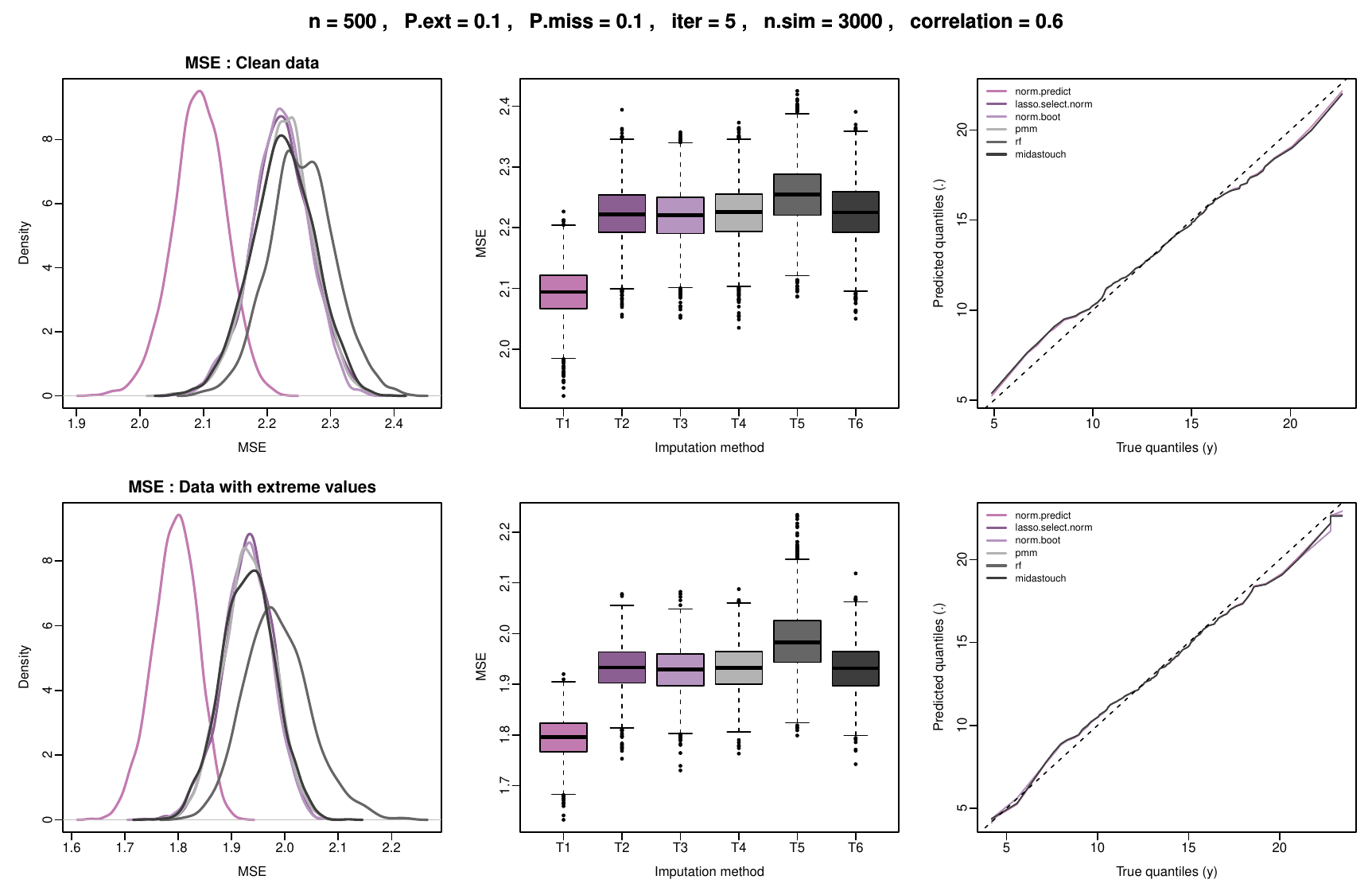}\\[20pt]
      \includegraphics[page=12,width=\linewidth]{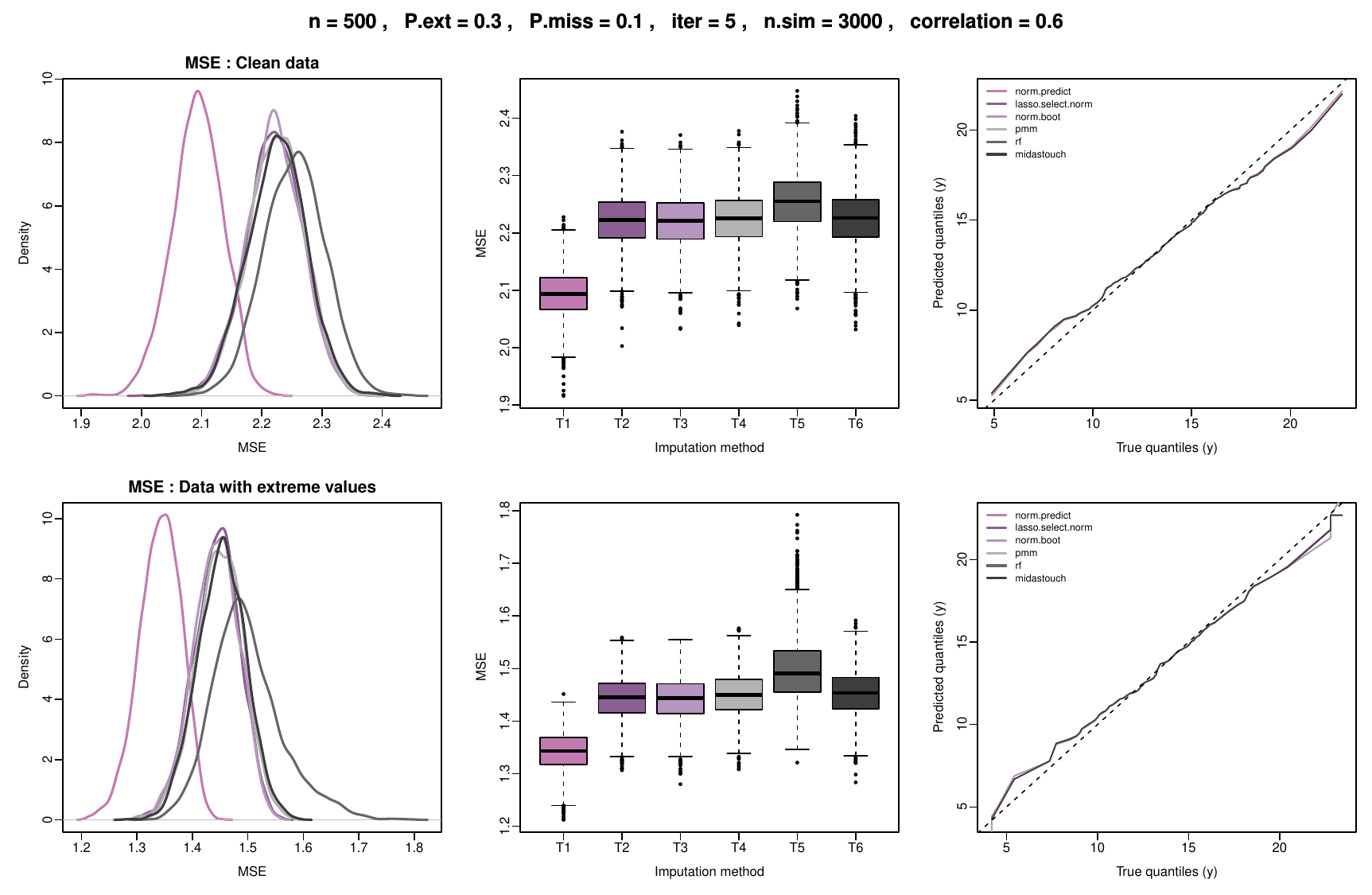}
    \end{minipage}%
  }
      \caption{Predictive MSE densities (clean vs contaminated with extremes data), MSE boxplots, and QQ-plots (predicted vs true quantiles of $y$) across six MI methods (T1--T6). For $\boldsymbol{n=500}$, ordered by $P_{\text{ext}}$, and $P_{\text{miss}}$ (panel 1 of 2). Clean data are analyzed with OLS and contaminated data with elastic net. Each subpanel shows the design values ($n$, $P_{\text{ext}}$, $P_{\text{miss}}$, iter, $n_{\text{sim}}$, and $\rho$)}
      \label{fig:n_500_graficos.1_mesclado-2}%
\end{figure}
\begin{figure}[H]
  \centering
  \adjustbox{width=\textwidth, max height=\textheight, center}{%
    \begin{minipage}{.48\linewidth}
      \includegraphics[page=1 ,width=\linewidth]{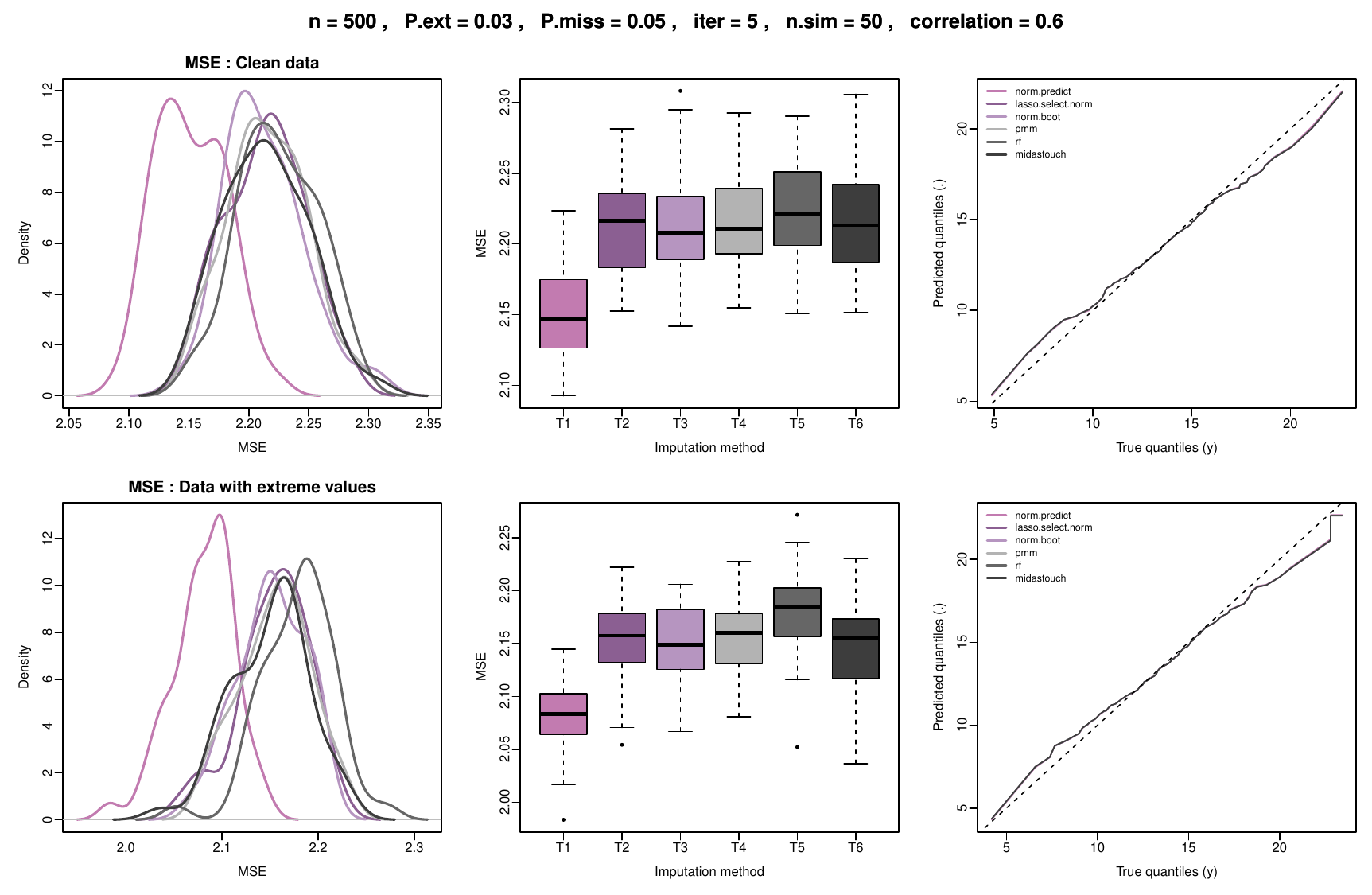}\\[20pt]
      \includegraphics[page=3 ,width=\linewidth]{figures/n_500_graficos.1_mesclado.pdf}\\[20pt]
      \includegraphics[page=5 ,width=\linewidth]{figures/n_500_graficos.1_mesclado.pdf}\\[20pt]
      \includegraphics[page=7 ,width=\linewidth]{figures/n_500_graficos.1_mesclado.pdf}
    \end{minipage}\hfill
    \begin{minipage}{.48\linewidth}
      \includegraphics[page=2 ,width=\linewidth]{figures/n_500_graficos.1_mesclado.pdf}\\[20pt]
      \includegraphics[page=4 ,width=\linewidth]{figures/n_500_graficos.1_mesclado.pdf}\\[20pt]
      \includegraphics[page=6 ,width=\linewidth]{figures/n_500_graficos.1_mesclado.pdf}\\[20pt]
      \includegraphics[page=8 ,width=\linewidth]{figures/n_500_graficos.1_mesclado.pdf}
    \end{minipage}%
  }
      \caption{Predictive MSE densities (clean vs contaminated with extremes data), MSE boxplots, and QQ-plots (predicted vs true quantiles of $y$) across six MI methods (T1--T6). For $\boldsymbol{n=500}$, ordered by $P_{\text{ext}}$, and $P_{\text{miss}}$ (panel 2 of 2). Clean data are analyzed with OLS and contaminated data with elastic net. Each subpanel shows the design values ($n$, $P_{\text{ext}}$, $P_{\text{miss}}$, iter, $n_{\text{sim}}$, and $\rho$)}
      \label{fig:n_500_graficos.1_mesclado-1}%
\end{figure}

\paragraph{Relatively large sample size ($\boldsymbol{n=500}$).}
See Tables~\ref{tab:resumo_500_1}--\ref{tab:resumo_500_2} and Figures~\ref{fig:n_500_graficos.1_mesclado-1}--\ref{fig:n_500_graficos.1_mesclado-2}. For further details, see Tables~\ref{tab:n_500_graficos.1}--\ref{tab:n_500_graficos.3} in the appendix.

\emph{Coefficients (level and bias).}
With \emph{clean} data (OLS) and low missingness ($\le 10\%$), slopes are essentially unbiased for all method blocks at $n=500$. For T1--T3, $\beta_1$ bias lies in $[-0.007,\,0.018]$ with RMSE $[0.008,\,0.020]$, and $\beta_2$ bias in $[-0.058,\,0.013]$ with RMSE $[0.017,$ $\,0.059]$. T4--T6 are comparable, showing slightly larger slope RMSEs (e.g., $\beta_2$ RMSE $[0.055,$ $\,0.066]$) and a mild negative $\beta_2$ bias $[-0.064,$ $\,-0.053]$. 
At higher missingness ($\geq 25\%$), OLS slope RMSEs increase moderately for both blocks (e.g., T1--T3: $\beta_1$ RMSE $[0.024,\,0.067]$, $\beta_2$ RMSE $[0.060,\,0.176]$; T4--T6: $[0.025,\,0.033]$ and $[0.060,\,0.088]$) with small, stable biases in T4--T6 (e.g., $\beta_2$ bias $[-0.083,\,-0.055]$). Under \emph{contamination} with extreme values (EN), the large-$n$ panels retain the same directional pattern: intercept inflation, negative shift in $\beta_1$, and positive shift in $\beta_2$. For $P_{\text{miss}}\le 0.10$, T1--T3 show $\beta_0$ bias $[0.076,$ $\,0.270]$, $\beta_1$ bias $[-0.108,\,-0.060]$, and $\beta_2$ bias $[0.080,$ $\,0.162]$; T4--T6 display similar directions with slightly smaller slope biases (e.g., $\beta_1$ $[-0.083,$ $\,-0.056]$, $\beta_2$ $[0.076,\,0.109]$). When $P_{\text{miss}} \geq 0.25$, T1--T3 intensify the $\beta_1$ tilt ($[-0.168,\,-0.066]$) and broaden $\beta_2$ bias ($[0.076,\,0.284]$), whereas T4--T6 keep comparatively smaller slope biases (e.g., $\beta_2$ $[0.046,\,0.112]$) with a somewhat larger intercept bias ($\beta_0$ up to $0.367$). These differences remain modest in magnitude given the sample size.

\emph{Coverage.}
With OLS on clean data, coverage is uniformly nominal for $P_{\text{miss}}\le 0.10$ across all methods. At $P_{\text{miss}}\in\{0.25,0.30\}$, coverage degradation concentrates in the parametric block for the slopes, most notably $\beta_2$ (T1--T3: $\beta_1$ coverage $[0.580,\,1.000]$, $\beta_2$ $[0.000,\,1.000]$), whereas the donor/ML block retains near-nominal behaviour for all coefficients (T4--T6: $\beta_1$ $[0.980,$ $\,1.000]$, $\beta_2$ $[0.960,\,1.000]$). Under EN, intercept coverage is generally high (T1--T3: $[0.912,$ $\,1.000]$; T4--T6: $[0.920,\,1.000]$), while slope coverage, especially for $\beta_1$, can be sub-nominal even at $n=500$ (e.g., T1--T3: $\beta_1$ coverage $[0.000,\,0.840]$ at high missigness). Donor/ML methods often improve $\beta_2$ coverage relative to T1--T3 as missingness rises (e.g., $[0.420,\,1.000]$ vs.\ $[0.000,\,0.980]$), in line with their smaller $\beta_2$ biases under contamination.

\emph{Predictive error (CV--MSE) and dispersion.}
Predictive distributions are highly concentrated at $n=500$. Under OLS for $P_{\text{miss}}\le 0.10$, method means $\overline{X}$ lie in tight bands (e.g., T1: $[2.081,\,2.153]$; T5: $[2.224,\,2.255]$) with variances $\sigma^2$ near $10^{-3}$. At higher missingness, dispersion increases slightly yet remains small (e.g., T1: $\sigma^2=[0.005,\,0.008]$; T5: $[0.006,\,0.010]$).  
Under EN, both level and variance remain controlled despite contamination. For low $P_{\text{miss}}$, T1--T3 attain the lowest or near-lowest $\overline{X}$ (e.g., T1: $[1.342,\,2.080]$) with $\sigma^2\le 0.004$, whereas T5 tends to be right-shifted (e.g., $[1.499,\,2.203]$). For $P_{\text{miss}}\geq 0.25$, the ordering persists (T1--T3 with lower level and dispersion than T4--T6): T1 shows $\overline{X}=[1.494,\,1.694]$ with $\sigma^2=[0.005,\,0.006]$; T2, $\overline{X}=[1.939,\,2.172]$ with $\sigma^2=[0.006,\,0.009]$; T5, $\overline{X}=[2.124,\,2.354]$ with $\sigma^2=[0.008,\,0.015]$. Medians $Q_{50}$ (and also $Q_{2.5}$ and $Q_{97.5}$) mirror these rankings.

\emph{Method sensitivity and trade-offs at large $n$.}
The contrast between parametric (T1--T3) and donor/ML (T4--T6) observed at smaller $n$ persists but is attenuated:  
(i) for \emph{prediction}, T1--T3 retain smaller CV--MSE level and dispersion, including under contamination;  
(ii) for \emph{inference} at high missingness under OLS, T4--T6 provide more reliable slope coverage, linked to smaller $\beta_2$ biases;  
(iii) the combined block (T1--T6) remains intermediate, with its tail behaviour largely driven by T5.

\emph{Cross-$n$ perspective.}
Relative to the small-sample panels $n=20,40$ and the moderate cases $n=80,200$, the $n=500$ results show:  
(a) progressive contraction of RMSE and CV--MSE dispersion from $n=20\!\rightarrow\!40\!\rightarrow\!80\!\rightarrow\!200\!\rightarrow\!500$ in both clean and contaminated regimes;  
(b) persistence of the contamination \emph{directionality} across all $n$, characterized by intercept inflation, negative shift in $\beta_1$, and positive shift in $\beta_2$; variability contracts as $n$ increases rather than the biases vanishing. 
(c) the same method ordering across sample sizes: parametric T1--T3 exhibit lower predictive dispersion and lower CV--MSE level than donor/ML T4--T6, most visibly at small $n$ (e.g., $n=20,40$) and still present, though with smaller absolute gaps, at $n=80,200$ and $n=500$;  
(d) for coverage, severe fragility under contamination at $n=20,40$ (including near-zero intercept coverage in several cells) improves with $n=80,200$ and further with $n=500$, yet $\beta_1$ coverage remains the most vulnerable under EN at higher $P_{\text{miss}}$;  
(e) at higher missingness in clean data, the donor/ML block maintains near-nominal slope coverage from moderate $n$ onward, while parametric methods tend to under-cover slopes, particularly $\beta_2$, a contrast that persists (though attenuated) at $n=500$.

\begin{table}[H]
\centering
\caption{Out-of-sample predictive error summaries (CV–MSE) by MI method for $\boldsymbol{n = 500}$, stratified by model (OLS for clean data; elastic net (EN) for contaminated data) and by missingness blocks (low: $P_{\text{miss}}\in\{0.05,0.10\}$; and high: $P_{\text{miss}}\in\{0.25,0.30\}$). For each method it is reported the min-max of the MSE: mean $\overline{X}$, variance $\sigma^2$, and median $Q_{50}$ across the corresponding design cells. Ranges further aggregate over all design combinations $P_{\text{ext}}\in\{0.03,0.05,0.10,0.15,0.30\}$, $n_{\text{sim}}\in\{50,300,1000\}$, $\text{iter}\in\{5,10\}$, and $\rho\in\{0,0.6\}$}
\label{tab:resumo_500_1}
\begingroup
\scriptsize
\setlength{\tabcolsep}{3pt}
\renewcommand{\arraystretch}{1.06}
\begin{tabular*}{\textwidth}{@{\extracolsep{\fill}} l l ccc ccc @{}}
\toprule
& & \multicolumn{3}{c}{low missingness (5\% and 10\%)} & \multicolumn{3}{c}{high missingness (25\% and 30\%)} \\
\cmidrule(lr){3-5}\cmidrule(lr){6-8}
\textbf{MI} & \textbf{Model} & $\overline{X}$ & $\sigma^2$ & $Q_{50}$ & $\overline{X}$ & $\sigma^2$ & $Q_{50}$ \\
\midrule
T1 & OLS & [2.081 , 2.153] & [0.001 , 0.002] & [2.079 , 2.154] & [1.784 , 1.797] & [0.005 , 0.008] & [1.786 , 1.796] \\
T1 & EN  & [1.342 , 2.080] & [0.001 , 0.003] & [1.343 , 2.083] & [1.494 , 1.694] & [0.005 , 0.006] & [1.495 , 1.694] \\
\addlinespace[2pt]
T2 & OLS & [2.208 , 2.222] & [0.001 , 0.003] & [2.202 , 2.222] & [2.211 , 2.230] & [0.007 , 0.011] & [2.213 , 2.229] \\
T2 & EN  & [1.443 , 2.154] & [0.001 , 0.003] & [1.446 , 2.158] & [1.939 , 2.172] & [0.006 , 0.009] & [1.940 , 2.165] \\
\addlinespace[2pt]
T3 & OLS & [2.209 , 2.220] & [0.001 , 0.003] & [2.206 , 2.221] & [2.208 , 2.221] & [0.007 , 0.012] & [2.194 , 2.220] \\
T3 & EN  & [1.442 , 2.151] & [0.001 , 0.003] & [1.444 , 2.149] & [1.924 , 2.153] & [0.007 , 0.009] & [1.923 , 2.143] \\
\addlinespace[2pt]
T4 & OLS & [2.211 , 2.224] & [0.001 , 0.003] & [2.204 , 2.226] & [2.221 , 2.235] & [0.005 , 0.011] & [2.226 , 2.246] \\
T4 & EN  & [1.450 , 2.155] & [0.001 , 0.004] & [1.450 , 2.160] & [1.931 , 2.160] & [0.007 , 0.010] & [1.934 , 2.151] \\
\addlinespace[2pt]
T5 & OLS & [2.224 , 2.255] & [0.001 , 0.003] & [2.221 , 2.255] & [2.323 , 2.335] & [0.006 , 0.010] & [2.314 , 2.340] \\
T5 & EN  & [1.499 , 2.203] & [0.001 , 0.004] & [1.491 , 2.213] & [2.124 , 2.354] & [0.008 , 0.015] & [2.122 , 2.340] \\
\addlinespace[2pt]
T6 & OLS & [2.207 , 2.225] & [0.001 , 0.003] & [2.211 , 2.226] & [2.221 , 2.237] & [0.006 , 0.012] & [2.232 , 2.235] \\
T6 & EN  & [1.453 , 2.150] & [0.001 , 0.004] & [1.454 , 2.156] & [1.936 , 2.147] & [0.008 , 0.009] & [1.931 , 2.149] \\
\bottomrule
\end{tabular*}
\endgroup
\end{table}

\begin{table}[H]
\centering
\begin{threeparttable}
\captionsetup{width=\textwidth,justification=justified}
\caption{Bias, RMSE, and 95\% coverage for $(\beta_0,\beta_1,\beta_2)$ under clean data with OLS and data with extreme values using EN for $\boldsymbol{n=20}$ (left) and $\boldsymbol{n=40}$ (right). Entries are reported as min--max within each missingness block (low: $P_{\text{miss}}\in\{0.05,0.10\}$; and high: $P_{\text{miss}}\in\{0.25,0.30\}$). Ranges further aggregate over all design combinations $P_{\text{ext}}\in\{0.03,0.05,0.10,0.15,0.30\}$, $n_{\text{sim}}\in\{50,300,1000\}$, $\text{iter}\in\{5,10\}$, and $\rho\in\{0,0.6\}$. MI methods are grouped as parametric (T1--T3), donor/ML (T4--T6), and the combined set (T1--T6). Coverage is the proportion of 95\% intervals containing the true coefficient}
\label{tab:resumo_500_2}

\begingroup
\scriptsize
\begin{tabular*}{0.7\textwidth}{@{\extracolsep{\fill}} l l c c c @{}}
\toprule
\multicolumn{2}{c}{} & \multicolumn{3}{c}{\textbf{n = 500}} \\
\cmidrule(lr){3-5}
Case & Metrics & $\beta_0$ & $\beta_1$ & $\beta_2$ \\
\midrule

\multicolumn{5}{@{}l@{}}{\textbf{Linear regression (OLS)} \quad \textit{\( P_{\text{miss}} \in \{0.05, 0.10\} \)}} \\ 
T1--T3 & Bias & [0.057 , 0.177] & [-0.007 , 0.018] & [-0.058 , 0.013] \\
T1--T3 & RMSE & [0.109 , 0.206] & [0.008 , 0.020] & [0.017 , 0.059] \\
T1--T3 & Coverage & [1.000 , 1.000] & [1.000 , 1.000] & [1.000 , 1.000] \\
T4--T6 & Bias & [0.158 , 0.188] & [0.016 , 0.019] & [-0.064 , -0.053] \\
T4--T6 & RMSE & [0.179 , 0.215] & [0.018 , 0.022] & [0.055 , 0.066] \\
T4--T6 & Coverage & [1.000 , 1.000] & [1.000 , 1.000] & [1.000 , 1.000] \\
T1--T6 & Bias & [0.057 , 0.188] & [-0.007 , 0.019] & [-0.064 , 0.013] \\
T1--T6 & RMSE & [0.109 , 0.215] & [0.008 , 0.022] & [0.017 , 0.066] \\
T1--T6 & Coverage & [1.000 , 1.000] & [1.000 , 1.000] & [1.000 , 1.000] \\

\addlinespace[2pt]
\cmidrule(lr){1-5}

\multicolumn{5}{@{}l@{}}{\textbf{Linear regression (OLS)} \quad \textit{\( P_{\text{miss}} \in \{0.25, 0.30\} \)}} \\ 
T1--T3 & Bias & [-0.200 , 0.206] & [-0.064 , 0.017] & [-0.061 , 0.174] \\
T1--T3 & RMSE & [0.251 , 0.294] & [0.024 , 0.067] & [0.060 , 0.176] \\
T1--T3 & Coverage & [0.960 , 1.000] & [0.580 , 1.000] & [0.000 , 1.000] \\
T4--T6 & Bias & [0.138 , 0.227] & [0.017 , 0.026] & [-0.083 , -0.055] \\
T4--T6 & RMSE & [0.222 , 0.299] & [0.025 , 0.033] & [0.060 , 0.088] \\
T4--T6 & Coverage & [0.997 , 1.000] & [0.980 , 1.000] & [0.960 , 1.000] \\
T1--T6 & Bias & [-0.200 , 0.227] & [-0.064 , 0.026] & [-0.083 , 0.174] \\
T1--T6 & RMSE & [0.222 , 0.299] & [0.024 , 0.067] & [0.060 , 0.176] \\
T1--T6 & Coverage & [0.960 , 1.000] & [0.580 , 1.000] & [0.000 , 1.000] \\

\addlinespace[2pt]
\cmidrule(lr){1-5}

\multicolumn{5}{@{}l@{}}{\textbf{Sparse regression (elastic net)} \quad \textit{\( P_{\text{miss}} \in \{0.05, 0.10\} \)}} \\ 
T1--T3 & Bias & [0.076 , 0.270] & [-0.108 , -0.060] & [0.080 , 0.162] \\
T1--T3 & RMSE & [0.087 , 0.285] & [0.061 , 0.108] & [0.081 , 0.163] \\
T1--T3 & Coverage & [1.000 , 1.000] & [0.000 , 0.860] & [0.000 , 1.000] \\
T4--T6 & Bias & [0.082 , 0.289] & [-0.083 , -0.056] & [0.076 , 0.109] \\
T4--T6 & RMSE & [0.092 , 0.300] & [0.057 , 0.084] & [0.076 , 0.110] \\
T4--T6 & Coverage & [0.997 , 1.000] & [0.000 , 0.900] & [0.300 , 1.000] \\
T1--T6 & Bias & [0.076 , 0.289] & [-0.108 , -0.056] & [0.076 , 0.162] \\
T1--T6 & RMSE & [0.087 , 0.300] & [0.057 , 0.108] & [0.076 , 0.163] \\
T1--T6 & Coverage & [0.997 , 1.000] & [0.000 , 0.900] & [0.000 , 1.000] \\

\addlinespace[2pt]
\cmidrule(lr){1-5}

\multicolumn{5}{@{}l@{}}{\textbf{Sparse regression (elastic net)} \quad \textit{\( P_{\text{miss}} \in \{0.25, 0.30\} \)}} \\ 
T1--T3 & Bias & [0.007 , 0.294] & [-0.168 , -0.066] & [0.076 , 0.284] \\
T1--T3 & RMSE & [0.150 , 0.334] & [0.068 , 0.169] & [0.079 , 0.285] \\
T1--T3 & Coverage & [0.912 , 1.000] & [0.000 , 0.840] & [0.000 , 0.980] \\
T4--T6 & Bias & [0.206 , 0.367] & [-0.086 , -0.055] & [0.046 , 0.112] \\
T4--T6 & RMSE & [0.236 , 0.385] & [0.058 , 0.088] & [0.050 , 0.115] \\
T4--T6 & Coverage & [0.920 , 1.000] & [0.272 , 0.940] & [0.420 , 1.000] \\
T1--T6 & Bias & [0.007 , 0.367] & [-0.168 , -0.055] & [0.046 , 0.284] \\
T1--T6 & RMSE & [0.150 , 0.385] & [0.058 , 0.169] & [0.050 , 0.285] \\
T1--T6 & Coverage & [0.912 , 1.000] & [0.000 , 0.940] & [0.000 , 1.000] \\

\bottomrule
\end{tabular*}
\endgroup

\begin{tablenotes}[flushleft]
\scriptsize
\setlength{\leftskip}{1em} 
\item $^*$Legend: MI methods (T1--T6): T1 \texttt{norm.predict}; T2 \texttt{lasso.select.norm};
T3 \texttt{norm.boot}; T4 \texttt{pmm}; T5 \texttt{rf}; T6 \texttt{midastouch}.
Coverage is the proportion of 95\% intervals containing the true coefficient.
\end{tablenotes}

\end{threeparttable}
\end{table}

\newpage

\section{Conclusion}\label{conclusion}

This study compared six multiple imputation methods implemented in \texttt{mice}: parametric regression-based approaches (T1: \texttt{norm.predict}, T2: \texttt{lasso.select.norm}, T3: \texttt{norm.boot}), donor based methods (T4: \texttt{pmm}, T6: \texttt{midastouch}), and a nonparametric machine learning method (T5: \texttt{rf}). Performance was evaluated under clean data analysed with OLS and under casewise extreme contamination analysed with elastic net, across multiple missingness levels.

Across the design, three patterns were stable. First, under clean data and low missingness, slope bias was negligible and coverage was close to nominal as sample size increased, with uncertainty shrinking markedly with $n$. Second, under clean data and high missingness, the main differences were inferential: donor based and flexible methods (T4--T6) more consistently maintained coverage close to nominal for slopes, whereas the parametric block (T1--T3) exhibited below nominal coverage for slopes, most notably for $\beta_2$, even when bias and RMSE were small. Third, under extreme contamination, coefficient estimates exhibited systematic distortions relative to the clean data benchmarks. Increasing $n$ mainly reduced dispersion and tail risk, but it did not eliminate these shifts induced by contamination.

A clear trade-off emerged between prediction and inference. For prediction, the parametric block (T1--T3) generally achieved lower and more concentrated CV--MSE, including in contaminated regimes, while T5 (\texttt{rf}) tended to produce larger and more dispersed errors, especially in small samples and higher missingness. For inference under higher missingness in clean data, donor based methods (T4 and T6) and the flexible method (T5) more often delivered reliable slope coverage and smaller slope distortions. Overall, multiple imputation is a modelling choice rather than a neutral, purely technical preprocessing step. The imputer's modelling assumptions can materially affect both predictive performance and inferential validity, and can meaningfully shape downstream conclusions, especially in the presence of extreme values and missingness.


\bibliographystyle{plainnat}
\bibliography{bibliography}

\vspace*{1cm}
\noindent \textbf{Acknowledgments:}\\
I thank Dr.\ Eduardo Yoshio Nakano for insightful discussions and suggestions on data imputation, modelling, and the design of the Monte Carlo simulations. I also thank Dr.\ Cira Etheowalda Guevara Otiniano for discussions on the theory of extreme events and probability (both from the Department of Statistics, University of Brasília), and Dr.\ Fidel Ernesto Castro Morales (Department of Statistics, Federal University of Rio Grande do Norte) for helpful discussions on extreme events and related computational aspects.

\newpage

\thispagestyle{plain} 
\appendix
\section{Overview \& table guide}\label{appendix}

This appendix provides transparency and a more detailed view of the results reported in the Results section, focusing on the different sample sizes and scenarios considered. The tables are organized as follows:
\begin{enumerate}
  \item Small sample sizes
  \item Moderate sample sizes
  \item Relatively large sample sizes
\end{enumerate}

Each table reports: the scenario description (\textit{setup}); the multiple imputation approach used (\textit{method}); and descriptive statistics (mean, median, quantiles, and variance) for each fitted regression model (linear or sparse). Readers interested in inspecting the actual numerical values and making more precise cross-scenario comparisons are encouraged to consult the relevant tables on the following pages.

\begin{table}[htbp]
\centering
\caption{Models coefficients bias, RMSE, 95\% coverage, and out-of-sample predictive MSE across MI methods for OLS (clean) and elastic net (EN, contaminated), with \(\boldsymbol{n=20}\), \(n_{\text{sim}}=50\), \(\text{iter}=5\), \(\rho=0.6\), contamination \(P_{\text{ext}}\in\{0.03,0.05\}\) and missingness \(P_{\text{miss}}\in\{0.05,0.10,0.30\}\)}

\label{tab:n_20_40_graficos.1}

\begingroup
\scriptsize
\begin{tabular*}{\textwidth}{@{\extracolsep{\fill}} l l
ccc 
ccc 
ccc 
ccccc 
@{}}
\toprule
& & \multicolumn{3}{c}{Bias} & \multicolumn{3}{c}{RMSE} & \multicolumn{3}{c}{Coverage} & \multicolumn{5}{c}{Pred.~MSE (out-of-sample)} \\
\cmidrule(lr){3-5}\cmidrule(lr){6-8}\cmidrule(lr){9-11}\cmidrule(lr){12-16}
Setup & MI &
$\beta_0$ & $\beta_1$ & $\beta_2$ &
$\beta_0$ & $\beta_1$ & $\beta_2$ &
$\beta_0$ & $\beta_1$ & $\beta_2$ &
$\overline{X}$ & $\sigma^2$ & $Q_{2.5}$ & $Q_{50}$ & $Q_{97.5}$
 \\
\midrule
\multicolumn{16}{@{}l@{}}{\textbf{Linear regression (OLS)}} \\
\multirow{6}{*}{\ParamBox{0.11\textwidth}{20}{50}{5}{0.03}{0.05}{0.6}{clean}} & T1 & -0.287 & -0.014 & 0.074 & 0.367 & 0.024 & 0.083 & 1.000 & 1.000 & 1.000 & 2.520 & 0.344 & 1.680 & 2.397 & 4.182 \\
& T2 & -0.176 & -0.007 & 0.039 & 0.323 & 0.024 & 0.059 & 1.000 & 1.000 & 1.000 & 2.581 & 0.323 & 1.728 & 2.485 & 4.070 \\
& T3 & -0.199 & -0.008 & 0.046 & 0.315 & 0.024 & 0.064 & 1.000 & 1.000 & 1.000 & 2.565 & 0.322 & 1.835 & 2.423 & 4.134 \\
& T4 & -0.154 & -0.004 & 0.031 & 0.319 & 0.026 & 0.061 & 1.000 & 1.000 & 1.000 & 2.624 & 0.358 & 1.780 & 2.487 & 4.206 \\
& T5 & -0.152 & -0.002 & 0.027 & 0.290 & 0.030 & 0.061 & 1.000 & 1.000 & 1.000 & 2.741 & 0.736 & 1.936 & 2.555 & 4.629 \\
& T6 & -0.241 & -0.006 & 0.050 & 0.349 & 0.033 & 0.061 & 1.000 & 1.000 & 1.000 & 2.576 & 0.339 & 1.743 & 2.483 & 4.226 \\
\addlinespace[2pt]
\multicolumn{16}{@{}l@{}}{\textbf{Sparse regression (EN)}} \\
\multirow{6}{*}{\ParamBox{0.11\textwidth}{20}{50}{5}{0.03}{0.05}{0.6}{{ext.}}} & T1 & 1.457 & -0.288 & 0.242 & 1.471 & 0.290 & 0.247 & 1.000 & 1.000 & 1.000 & 1.995 & 0.092 & 1.518 & 1.943 & 2.709 \\
& T2 & 1.452 & -0.272 & 0.210 & 1.472 & 0.274 & 0.215 & 1.000 & 1.000 & 1.000 & 2.111 & 0.092 & 1.600 & 2.060 & 2.765 \\
& T3 & 1.452 & -0.276 & 0.219 & 1.471 & 0.278 & 0.224 & 1.000 & 1.000 & 1.000 & 2.077 & 0.106 & 1.551 & 2.023 & 2.900 \\
& T4 & 1.347 & -0.264 & 0.218 & 1.443 & 0.273 & 0.225 & 1.000 & 1.000 & 1.000 & 2.170 & 0.175 & 1.575 & 2.059 & 3.314 \\
& T5 & 1.382 & -0.238 & 0.155 & 1.491 & 0.251 & 0.170 & 1.000 & 1.000 & 1.000 & 2.575 & 0.787 & 1.841 & 2.241 & 4.921 \\
& T6 & 1.366 & -0.264 & 0.212 & 1.453 & 0.271 & 0.219 & 1.000 & 1.000 & 1.000 & 2.147 & 0.150 & 1.595 & 2.063 & 3.081 \\
\addlinespace[2pt]
\cmidrule(lr){1-16}
\multicolumn{16}{@{}l@{}}{\textbf{Linear regression (OLS)}} \\
\multirow{6}{*}{\ParamBox{0.11\textwidth}{20}{50}{5}{0.03}{0.10}{0.6}{clean}} & T1 & -0.411 & -0.025 & 0.117 & 0.533 & 0.067 & 0.175 & 1.000 & 1.000 & 0.980 & 2.283 & 0.243 & 1.521 & 2.221 & 3.317 \\
& T2 & -0.118 & -0.017 & 0.046 & 0.417 & 0.063 & 0.138 & 1.000 & 1.000 & 1.000 & 2.441 & 0.203 & 1.660 & 2.407 & 3.290 \\
& T3 & -0.185 & -0.019 & 0.064 & 0.436 & 0.070 & 0.147 & 1.000 & 1.000 & 1.000 & 2.399 & 0.226 & 1.623 & 2.372 & 3.461 \\
& T4 & -0.198 & -0.002 & 0.032 & 0.427 & 0.058 & 0.132 & 1.000 & 1.000 & 1.000 & 2.487 & 0.333 & 1.627 & 2.450 & 4.147 \\
& T5 & -0.133 & -0.007 & 0.033 & 0.336 & 0.057 & 0.126 & 1.000 & 1.000 & 1.000 & 2.611 & 0.260 & 1.945 & 2.569 & 3.874 \\
& T6 & -0.280 & -0.000 & 0.044 & 0.434 & 0.062 & 0.129 & 1.000 & 1.000 & 1.000 & 2.432 & 0.226 & 1.639 & 2.399 & 3.436 \\
\addlinespace[2pt]
\multicolumn{16}{@{}l@{}}{\textbf{Sparse regression (EN)}} \\
\multirow{6}{*}{\ParamBox{0.11\textwidth}{20}{50}{5}{0.03}{0.10}{0.6}{{ext.}}} & T1 & 1.650 & -0.314 & 0.267 & 1.778 & 0.323 & 0.274 & 0.900 & 0.880 & 0.980 & 1.866 & 0.095 & 1.349 & 1.849 & 2.415 \\
& T2 & 1.635 & -0.287 & 0.213 & 1.776 & 0.297 & 0.223 & 0.960 & 0.980 & 1.000 & 2.118 & 0.162 & 1.487 & 2.058 & 2.964 \\
& T3 & 1.662 & -0.298 & 0.230 & 1.815 & 0.309 & 0.241 & 0.940 & 0.980 & 0.980 & 2.052 & 0.142 & 1.447 & 1.969 & 2.872 \\
& T4 & 1.189 & -0.238 & 0.201 & 1.582 & 0.275 & 0.221 & 1.000 & 1.000 & 1.000 & 2.553 & 1.002 & 1.623 & 2.274 & 4.937 \\
& T5 & 1.215 & -0.199 & 0.107 & 1.563 & 0.233 & 0.143 & 1.000 & 1.000 & 1.000 & 3.103 & 1.640 & 1.825 & 2.740 & 6.595 \\
& T6 & 1.240 & -0.246 & 0.207 & 1.573 & 0.272 & 0.218 & 0.980 & 0.940 & 1.000 & 2.380 & 0.594 & 1.510 & 2.172 & 4.495 \\
\addlinespace[2pt]
\cmidrule(lr){1-16}
\multicolumn{16}{@{}l@{}}{\textbf{Linear regression (OLS)}} \\
\multirow{6}{*}{\ParamBox{0.11\textwidth}{20}{50}{5}{0.03}{0.30}{0.6}{clean}} & T1 & -0.640 & -0.112 & 0.331 & 1.063 & 0.214 & 0.470 & 0.980 & 0.920 & 0.800 & 1.980 & 0.459 & 0.837 & 1.956 & 3.319 \\
& T2 & 0.277 & -0.084 & 0.098 & 1.003 & 0.187 & 0.358 & 1.000 & 0.960 & 0.920 & 2.526 & 0.389 & 1.349 & 2.536 & 3.600 \\
& T3 & 0.113 & -0.090 & 0.141 & 0.951 & 0.199 & 0.353 & 1.000 & 0.980 & 0.920 & 2.410 & 0.415 & 1.192 & 2.390 & 3.341 \\
& T4 & -0.317 & -0.012 & 0.089 & 0.731 & 0.144 & 0.322 & 1.000 & 1.000 & 0.960 & 2.669 & 0.335 & 1.539 & 2.737 & 3.572 \\
& T5 & 0.058 & -0.015 & 0.021 & 0.783 & 0.142 & 0.273 & 1.000 & 1.000 & 1.000 & 3.185 & 0.397 & 2.259 & 3.152 & 4.651 \\
& T6 & -0.235 & -0.026 & 0.094 & 0.663 & 0.151 & 0.351 & 1.000 & 0.980 & 0.940 & 2.561 & 0.479 & 1.273 & 2.610 & 3.761 \\
\addlinespace[2pt]
\multicolumn{16}{@{}l@{}}{\textbf{Sparse regression (EN)}} \\
\multirow{6}{*}{\ParamBox{0.11\textwidth}{20}{50}{5}{0.03}{0.30}{0.6}{{ext.}}} & T1 & 1.748 & -0.396 & 0.423 & 2.124 & 0.429 & 0.459 & 0.740 & 0.560 & 0.660 & 1.594 & 0.165 & 1.046 & 1.608 & 2.656 \\
& T2 & 1.577 & -0.291 & 0.236 & 1.921 & 0.318 & 0.274 & 0.960 & 0.980 & 1.000 & 2.403 & 0.361 & 1.517 & 2.361 & 3.984 \\
& T3 & 1.809 & -0.323 & 0.255 & 2.122 & 0.348 & 0.290 & 0.940 & 1.000 & 1.000 & 2.228 & 0.299 & 1.403 & 2.163 & 3.598 \\
& T4 & 0.271 & -0.106 & 0.123 & 1.665 & 0.222 & 0.187 & 0.980 & 0.980 & 1.000 & 3.688 & 1.703 & 1.980 & 3.345 & 7.045 \\
& T5 & 0.507 & -0.015 & -0.132 & 1.416 & 0.145 & 0.236 & 1.000 & 1.000 & 1.000 & 5.065 & 2.908 & 3.066 & 4.735 & 9.210 \\
& T6 & 0.675 & -0.177 & 0.191 & 1.707 & 0.256 & 0.248 & 0.980 & 0.980 & 0.980 & 3.126 & 1.581 & 1.641 & 2.752 & 5.728 \\
\addlinespace[2pt]
\cmidrule(lr){1-16}
\multicolumn{16}{@{}l@{}}{\textbf{Linear regression (OLS)}} \\
\multirow{6}{*}{\ParamBox{0.11\textwidth}{20}{50}{5}{0.05}{0.05}{0.6}{clean}} & T1 & -0.287 & -0.014 & 0.074 & 0.367 & 0.024 & 0.083 & 1.000 & 1.000 & 1.000 & 2.520 & 0.344 & 1.680 & 2.397 & 4.182 \\
& T2 & -0.176 & -0.007 & 0.039 & 0.323 & 0.024 & 0.059 & 1.000 & 1.000 & 1.000 & 2.581 & 0.323 & 1.728 & 2.485 & 4.070 \\
& T3 & -0.199 & -0.008 & 0.046 & 0.315 & 0.024 & 0.064 & 1.000 & 1.000 & 1.000 & 2.565 & 0.322 & 1.835 & 2.423 & 4.134 \\
& T4 & -0.154 & -0.004 & 0.031 & 0.319 & 0.026 & 0.061 & 1.000 & 1.000 & 1.000 & 2.624 & 0.358 & 1.780 & 2.487 & 4.206 \\
& T5 & -0.152 & -0.002 & 0.027 & 0.290 & 0.030 & 0.061 & 1.000 & 1.000 & 1.000 & 2.741 & 0.736 & 1.936 & 2.555 & 4.629 \\
& T6 & -0.241 & -0.006 & 0.050 & 0.349 & 0.033 & 0.061 & 1.000 & 1.000 & 1.000 & 2.576 & 0.339 & 1.743 & 2.483 & 4.226 \\
\addlinespace[2pt]
\multicolumn{16}{@{}l@{}}{\textbf{Sparse regression (EN)}} \\
\multirow{6}{*}{\ParamBox{0.11\textwidth}{20}{50}{5}{0.05}{0.05}{0.6}{{ext.}}} & T1 & 1.457 & -0.288 & 0.242 & 1.471 & 0.290 & 0.247 & 1.000 & 1.000 & 1.000 & 1.995 & 0.092 & 1.518 & 1.943 & 2.709 \\
& T2 & 1.452 & -0.272 & 0.210 & 1.472 & 0.274 & 0.215 & 1.000 & 1.000 & 1.000 & 2.111 & 0.092 & 1.600 & 2.060 & 2.765 \\
& T3 & 1.452 & -0.276 & 0.219 & 1.471 & 0.278 & 0.224 & 1.000 & 1.000 & 1.000 & 2.077 & 0.106 & 1.551 & 2.023 & 2.900 \\
& T4 & 1.347 & -0.264 & 0.218 & 1.443 & 0.273 & 0.225 & 1.000 & 1.000 & 1.000 & 2.170 & 0.175 & 1.575 & 2.059 & 3.314 \\
& T5 & 1.382 & -0.238 & 0.155 & 1.491 & 0.251 & 0.170 & 1.000 & 1.000 & 1.000 & 2.575 & 0.787 & 1.841 & 2.241 & 4.921 \\
& T6 & 1.366 & -0.264 & 0.212 & 1.453 & 0.271 & 0.219 & 1.000 & 1.000 & 1.000 & 2.147 & 0.150 & 1.595 & 2.063 & 3.081 \\
\bottomrule
\end{tabular*}
\makebox[\textwidth][r]{\Nchips{20}} 
\endgroup
\end{table}

\begin{table}[htbp]
\centering
\caption{Models coefficients bias, RMSE, 95\% coverage, and out-of-sample predictive MSE across MI methods for OLS (clean) and EN (contaminated), with \(\boldsymbol{n=20}\), \(n_{\text{sim}}=50\), \(\text{iter}=5\), \(\rho=0.6\), contamination \(P_{\text{ext}}\in\{0.05,0.10\}\) and missingness \(P_{\text{miss}}\in\{0.05,0.10,0.30\}\)}

\label{tab:n_20_40_graficos.2}

\begingroup
\scriptsize
\begin{tabular*}{\textwidth}{@{\extracolsep{\fill}} l l
ccc 
ccc 
ccc 
ccccc 
@{}}
\toprule
& & \multicolumn{3}{c}{Bias} & \multicolumn{3}{c}{RMSE} & \multicolumn{3}{c}{Coverage} & \multicolumn{5}{c}{Pred.~MSE (out-of-sample)} \\
\cmidrule(lr){3-5}\cmidrule(lr){6-8}\cmidrule(lr){9-11}\cmidrule(lr){12-16}
Setup & MI &
$\beta_0$ & $\beta_1$ & $\beta_2$ &
$\beta_0$ & $\beta_1$ & $\beta_2$ &
$\beta_0$ & $\beta_1$ & $\beta_2$ &
$\overline{X}$ & $\sigma^2$ & $Q_{2.5}$ & $Q_{50}$ & $Q_{97.5}$
 \\
\midrule
\multicolumn{16}{@{}l@{}}{\textbf{Linear regression (OLS)}} \\
\multirow{6}{*}{\ParamBox{0.11\textwidth}{20}{50}{5}{0.05}{0.10}{0.6}{clean}} & T1 & -0.411 & -0.025 & 0.117 & 0.533 & 0.067 & 0.175 & 1.000 & 1.000 & 0.980 & 2.283 & 0.243 & 1.521 & 2.221 & 3.317 \\
& T2 & -0.118 & -0.017 & 0.046 & 0.417 & 0.063 & 0.138 & 1.000 & 1.000 & 1.000 & 2.441 & 0.203 & 1.660 & 2.407 & 3.290 \\
& T3 & -0.185 & -0.019 & 0.064 & 0.436 & 0.070 & 0.147 & 1.000 & 1.000 & 1.000 & 2.399 & 0.226 & 1.623 & 2.372 & 3.461 \\
& T4 & -0.198 & -0.002 & 0.032 & 0.427 & 0.058 & 0.132 & 1.000 & 1.000 & 1.000 & 2.487 & 0.333 & 1.627 & 2.450 & 4.147 \\
& T5 & -0.133 & -0.007 & 0.033 & 0.336 & 0.057 & 0.126 & 1.000 & 1.000 & 1.000 & 2.611 & 0.260 & 1.945 & 2.569 & 3.874 \\
& T6 & -0.280 & -0.000 & 0.044 & 0.434 & 0.062 & 0.129 & 1.000 & 1.000 & 1.000 & 2.432 & 0.226 & 1.639 & 2.399 & 3.436 \\
\addlinespace[2pt]
\multicolumn{16}{@{}l@{}}{\textbf{Sparse regression (EN)}} \\
\multirow{6}{*}{\ParamBox{0.11\textwidth}{20}{50}{5}{0.05}{0.10}{0.6}{{ext.}}} & T1 & 1.650 & -0.314 & 0.267 & 1.778 & 0.323 & 0.274 & 0.900 & 0.880 & 0.980 & 1.866 & 0.095 & 1.349 & 1.849 & 2.415 \\
& T2 & 1.635 & -0.287 & 0.213 & 1.776 & 0.297 & 0.223 & 0.960 & 0.980 & 1.000 & 2.118 & 0.162 & 1.487 & 2.058 & 2.964 \\
& T3 & 1.662 & -0.298 & 0.230 & 1.815 & 0.309 & 0.241 & 0.940 & 0.980 & 0.980 & 2.052 & 0.142 & 1.447 & 1.969 & 2.872 \\
& T4 & 1.189 & -0.238 & 0.201 & 1.582 & 0.275 & 0.221 & 1.000 & 1.000 & 1.000 & 2.553 & 1.002 & 1.623 & 2.274 & 4.937 \\
& T5 & 1.215 & -0.199 & 0.107 & 1.563 & 0.233 & 0.143 & 1.000 & 1.000 & 1.000 & 3.103 & 1.640 & 1.825 & 2.740 & 6.595 \\
& T6 & 1.240 & -0.246 & 0.207 & 1.573 & 0.272 & 0.218 & 0.980 & 0.940 & 1.000 & 2.380 & 0.594 & 1.510 & 2.172 & 4.495 \\
\addlinespace[2pt]
\cmidrule(lr){1-16}
\multicolumn{16}{@{}l@{}}{\textbf{Linear regression (OLS)}} \\
\multirow{6}{*}{\ParamBox{0.11\textwidth}{20}{50}{5}{0.05}{0.30}{0.6}{clean}} & T1 & -0.640 & -0.112 & 0.331 & 1.063 & 0.214 & 0.470 & 0.980 & 0.920 & 0.800 & 1.980 & 0.459 & 0.837 & 1.956 & 3.319 \\
& T2 & 0.277 & -0.084 & 0.098 & 1.003 & 0.187 & 0.358 & 1.000 & 0.960 & 0.920 & 2.526 & 0.389 & 1.349 & 2.536 & 3.600 \\
& T3 & 0.113 & -0.090 & 0.141 & 0.951 & 0.199 & 0.353 & 1.000 & 0.980 & 0.920 & 2.410 & 0.415 & 1.192 & 2.390 & 3.341 \\
& T4 & -0.317 & -0.012 & 0.089 & 0.731 & 0.144 & 0.322 & 1.000 & 1.000 & 0.960 & 2.669 & 0.335 & 1.539 & 2.737 & 3.572 \\
& T5 & 0.058 & -0.015 & 0.021 & 0.783 & 0.142 & 0.273 & 1.000 & 1.000 & 1.000 & 3.185 & 0.397 & 2.259 & 3.152 & 4.651 \\
& T6 & -0.235 & -0.026 & 0.094 & 0.663 & 0.151 & 0.351 & 1.000 & 0.980 & 0.940 & 2.561 & 0.479 & 1.273 & 2.610 & 3.761 \\
\addlinespace[2pt]
\multicolumn{16}{@{}l@{}}{\textbf{Sparse regression (EN)}} \\
\multirow{6}{*}{\ParamBox{0.11\textwidth}{20}{50}{5}{0.05}{0.30}{0.6}{{ext.}}} & T1 & 1.748 & -0.396 & 0.423 & 2.124 & 0.429 & 0.459 & 0.740 & 0.560 & 0.660 & 1.594 & 0.165 & 1.046 & 1.608 & 2.656 \\
& T2 & 1.577 & -0.291 & 0.236 & 1.921 & 0.318 & 0.274 & 0.960 & 0.980 & 1.000 & 2.403 & 0.361 & 1.517 & 2.361 & 3.984 \\
& T3 & 1.809 & -0.323 & 0.255 & 2.122 & 0.348 & 0.290 & 0.940 & 1.000 & 1.000 & 2.228 & 0.299 & 1.403 & 2.163 & 3.598 \\
& T4 & 0.271 & -0.106 & 0.123 & 1.665 & 0.222 & 0.187 & 0.980 & 0.980 & 1.000 & 3.688 & 1.703 & 1.980 & 3.345 & 7.045 \\
& T5 & 0.507 & -0.015 & -0.132 & 1.416 & 0.145 & 0.236 & 1.000 & 1.000 & 1.000 & 5.065 & 2.908 & 3.066 & 4.735 & 9.210 \\
& T6 & 0.675 & -0.177 & 0.191 & 1.707 & 0.256 & 0.248 & 0.980 & 0.980 & 0.980 & 3.126 & 1.581 & 1.641 & 2.752 & 5.728 \\
\addlinespace[2pt]
\cmidrule(lr){1-16}
\multicolumn{16}{@{}l@{}}{\textbf{Linear regression (OLS)}} \\
\multirow{6}{*}{\ParamBox{0.11\textwidth}{20}{50}{5}{0.10}{0.05}{0.6}{clean}} & T1 & -0.292 & -0.014 & 0.074 & 0.369 & 0.023 & 0.083 & 1.000 & 1.000 & 1.000 & 2.480 & 0.236 & 1.847 & 2.450 & 3.925 \\
& T2 & -0.149 & -0.009 & 0.038 & 0.326 & 0.027 & 0.061 & 1.000 & 1.000 & 1.000 & 2.565 & 0.239 & 1.992 & 2.479 & 4.104 \\
& T3 & -0.197 & -0.009 & 0.048 & 0.331 & 0.025 & 0.064 & 1.000 & 1.000 & 1.000 & 2.552 & 0.239 & 1.943 & 2.490 & 4.119 \\
& T4 & -0.156 & -0.010 & 0.041 & 0.350 & 0.033 & 0.056 & 1.000 & 1.000 & 1.000 & 2.573 & 0.336 & 1.789 & 2.527 & 4.297 \\
& T5 & -0.144 & -0.006 & 0.032 & 0.294 & 0.028 & 0.057 & 1.000 & 1.000 & 1.000 & 2.644 & 0.353 & 1.957 & 2.553 & 4.651 \\
& T6 & -0.206 & -0.002 & 0.034 & 0.348 & 0.034 & 0.050 & 1.000 & 1.000 & 1.000 & 2.561 & 0.318 & 1.818 & 2.464 & 4.245 \\
\addlinespace[2pt]
\multicolumn{16}{@{}l@{}}{\textbf{Sparse regression (EN)}} \\
\multirow{6}{*}{\ParamBox{0.11\textwidth}{20}{50}{5}{0.10}{0.05}{0.6}{{ext.}}} & T1 & 0.939 & -0.241 & 0.220 & 0.949 & 0.242 & 0.223 & 1.000 & 1.000 & 1.000 & 1.544 & 0.047 & 1.256 & 1.498 & 2.065 \\
& T2 & 0.913 & -0.227 & 0.197 & 0.935 & 0.229 & 0.201 & 1.000 & 1.000 & 1.000 & 1.636 & 0.052 & 1.305 & 1.625 & 2.108 \\
& T3 & 0.912 & -0.229 & 0.200 & 0.932 & 0.231 & 0.204 & 1.000 & 1.000 & 1.000 & 1.625 & 0.062 & 1.309 & 1.553 & 2.185 \\
& T4 & 0.594 & -0.189 & 0.178 & 1.051 & 0.218 & 0.194 & 1.000 & 1.000 & 1.000 & 1.906 & 0.451 & 1.290 & 1.675 & 3.659 \\
& T5 & 0.561 & -0.164 & 0.128 & 1.060 & 0.196 & 0.145 & 1.000 & 1.000 & 1.000 & 2.275 & 0.765 & 1.473 & 1.935 & 4.607 \\
& T6 & 0.644 & -0.196 & 0.184 & 1.053 & 0.221 & 0.196 & 1.000 & 1.000 & 1.000 & 1.848 & 0.388 & 1.260 & 1.618 & 3.611 \\
\addlinespace[2pt]
\cmidrule(lr){1-16}
\multicolumn{16}{@{}l@{}}{\textbf{Linear regression (OLS)}} \\
\multirow{6}{*}{\ParamBox{0.11\textwidth}{20}{50}{5}{0.10}{0.10}{0.6}{clean}} & T1 & -0.420 & -0.025 & 0.120 & 0.539 & 0.067 & 0.177 & 1.000 & 1.000 & 0.980 & 2.331 & 0.265 & 1.604 & 2.206 & 3.439 \\
& T2 & -0.121 & -0.018 & 0.047 & 0.520 & 0.067 & 0.135 & 1.000 & 1.000 & 1.000 & 2.503 & 0.207 & 1.793 & 2.492 & 3.401 \\
& T3 & -0.190 & -0.016 & 0.059 & 0.391 & 0.069 & 0.149 & 1.000 & 1.000 & 1.000 & 2.460 & 0.196 & 1.819 & 2.395 & 3.313 \\
& T4 & -0.234 & -0.007 & 0.052 & 0.419 & 0.058 & 0.143 & 1.000 & 1.000 & 1.000 & 2.535 & 0.260 & 1.763 & 2.457 & 3.533 \\
& T5 & -0.023 & -0.011 & 0.019 & 0.346 & 0.063 & 0.141 & 1.000 & 1.000 & 1.000 & 2.776 & 0.173 & 2.179 & 2.766 & 3.638 \\
& T6 & -0.224 & -0.005 & 0.041 & 0.401 & 0.068 & 0.139 & 1.000 & 1.000 & 1.000 & 2.550 & 0.297 & 1.656 & 2.490 & 4.018 \\
\addlinespace[2pt]
\multicolumn{16}{@{}l@{}}{\textbf{Sparse regression (EN)}} \\
\multirow{6}{*}{\ParamBox{0.11\textwidth}{20}{50}{5}{0.10}{0.10}{0.6}{{ext.}}} & T1 & 1.081 & -0.262 & 0.242 & 1.156 & 0.269 & 0.251 & 1.000 & 0.860 & 1.000 & 1.457 & 0.047 & 1.083 & 1.452 & 1.866 \\
& T2 & 0.979 & -0.233 & 0.202 & 1.059 & 0.239 & 0.209 & 1.000 & 1.000 & 1.000 & 1.657 & 0.062 & 1.283 & 1.615 & 2.061 \\
& T3 & 1.037 & -0.243 & 0.210 & 1.145 & 0.252 & 0.221 & 1.000 & 0.960 & 1.000 & 1.604 & 0.069 & 1.139 & 1.589 & 2.144 \\
& T4 & 0.474 & -0.168 & 0.165 & 1.167 & 0.218 & 0.194 & 1.000 & 0.980 & 1.000 & 2.200 & 0.782 & 1.346 & 1.855 & 4.294 \\
& T5 & 0.415 & -0.123 & 0.076 & 1.254 & 0.198 & 0.155 & 1.000 & 1.000 & 1.000 & 2.902 & 1.721 & 1.471 & 2.370 & 5.749 \\
& T6 & 0.518 & -0.173 & 0.165 & 1.155 & 0.220 & 0.195 & 1.000 & 0.960 & 1.000 & 2.175 & 0.877 & 1.218 & 1.817 & 4.397 \\
\bottomrule
\end{tabular*}
\makebox[\textwidth][r]{\Nchips{20}} 
\endgroup
\end{table}

\begin{table}[htbp]
\centering
\caption{Models coefficients bias, RMSE, 95\% coverage, and out-of-sample predictive MSE across MI methods for OLS (clean) and EN (contaminated), mixing \(\boldsymbol{n=20}\) (block \(P_{\text{ext}}=0.10\), \(P_{\text{miss}}=0.30\)) and \(\boldsymbol{n=40}\) (blocks \(P_{\text{ext}}=0.03\), \(P_{\text{miss}}\in\{0.05,0.10,0.30\}\)); \(n_{\text{sim}}=50\), \(\text{iter}=5\), \(\rho=0.6\)}

\label{tab:n_20_40_graficos.3}

\begingroup
\scriptsize
\begin{tabular*}{\textwidth}{@{\extracolsep{\fill}} l l
ccc 
ccc 
ccc 
ccccc 
@{}}
\toprule
& & \multicolumn{3}{c}{Bias} & \multicolumn{3}{c}{RMSE} & \multicolumn{3}{c}{Coverage} & \multicolumn{5}{c}{Pred.~MSE (out-of-sample)} \\
\cmidrule(lr){3-5}\cmidrule(lr){6-8}\cmidrule(lr){9-11}\cmidrule(lr){12-16}
Setup & MI &
$\beta_0$ & $\beta_1$ & $\beta_2$ &
$\beta_0$ & $\beta_1$ & $\beta_2$ &
$\beta_0$ & $\beta_1$ & $\beta_2$ &
$\overline{X}$ & $\sigma^2$ & $Q_{2.5}$ & $Q_{50}$ & $Q_{97.5}$
 \\
\midrule
\multicolumn{16}{@{}l@{}}{\textbf{Linear regression (OLS)}} \\
\multirow{6}{*}{\ParamBox{0.11\textwidth}{20}{50}{5}{0.10}{0.30}{0.6}{clean}} & T1 & -0.617 & -0.113 & 0.329 & 1.031 & 0.213 & 0.469 & 0.980 & 0.920 & 0.800 & 1.912 & 0.252 & 0.917 & 2.023 & 2.649 \\
& T2 & 0.294 & -0.085 & 0.096 & 1.063 & 0.204 & 0.351 & 0.980 & 0.960 & 0.960 & 2.512 & 0.374 & 1.357 & 2.666 & 3.324 \\
& T3 & 0.039 & -0.077 & 0.133 & 0.861 & 0.195 & 0.362 & 1.000 & 0.960 & 0.920 & 2.296 & 0.265 & 1.099 & 2.399 & 3.108 \\
& T4 & -0.245 & -0.013 & 0.077 & 0.767 & 0.139 & 0.306 & 1.000 & 1.000 & 0.980 & 2.692 & 0.321 & 1.595 & 2.698 & 3.734 \\
& T5 & 0.063 & -0.020 & 0.028 & 0.674 & 0.143 & 0.298 & 1.000 & 1.000 & 1.000 & 3.015 & 0.204 & 2.222 & 3.070 & 3.837 \\
& T6 & -0.311 & -0.026 & 0.114 & 0.713 & 0.142 & 0.334 & 1.000 & 1.000 & 0.940 & 2.506 & 0.301 & 1.410 & 2.554 & 3.362 \\
\addlinespace[2pt]
\multicolumn{16}{@{}l@{}}{\textbf{Sparse regression (EN)}} \\
\multirow{6}{*}{\ParamBox{0.11\textwidth}{20}{50}{5}{0.10}{0.30}{0.6}{{ext.}}} & T1 & 1.294 & -0.338 & 0.361 & 1.519 & 0.362 & 0.394 & 0.780 & 0.540 & 0.660 & 1.202 & 0.072 & 0.696 & 1.210 & 1.692 \\
& T2 & 1.117 & -0.265 & 0.241 & 1.350 & 0.289 & 0.277 & 1.000 & 0.960 & 0.980 & 1.803 & 0.202 & 1.175 & 1.706 & 2.799 \\
& T3 & 1.178 & -0.267 & 0.233 & 1.428 & 0.296 & 0.278 & 0.960 & 0.960 & 0.960 & 1.723 & 0.212 & 1.026 & 1.678 & 2.730 \\
& T4 & -0.493 & -0.010 & 0.029 & 1.440 & 0.180 & 0.154 & 1.000 & 1.000 & 1.000 & 3.629 & 1.619 & 1.834 & 3.524 & 6.175 \\
& T5 & -0.320 & 0.085 & -0.217 & 1.311 & 0.175 & 0.276 & 1.000 & 1.000 & 1.000 & 4.919 & 2.066 & 2.735 & 4.882 & 7.577 \\
& T6 & -0.167 & -0.070 & 0.094 & 1.419 & 0.210 & 0.199 & 1.000 & 0.980 & 1.000 & 3.189 & 2.461 & 1.333 & 2.703 & 6.326 \\
\addlinespace[2pt]
\cmidrule(lr){1-16}
\multicolumn{16}{@{}l@{}}{\textbf{Linear regression (OLS)}} \\
\multirow{6}{*}{\ParamBox{0.11\textwidth}{40}{50}{5}{0.03}{0.05}{0.6}{clean}} & T1 & 2.021 & -0.147 & -0.189 & 2.028 & 0.149 & 0.191 & 0.900 & 1.000 & 1.000 & 2.205 & 0.020 & 1.969 & 2.192 & 2.455 \\
& T2 & 2.051 & -0.134 & -0.221 & 2.059 & 0.136 & 0.223 & 0.920 & 1.000 & 1.000 & 2.263 & 0.019 & 2.027 & 2.280 & 2.482 \\
& T3 & 2.058 & -0.135 & -0.221 & 2.066 & 0.137 & 0.223 & 0.960 & 1.000 & 1.000 & 2.259 & 0.018 & 2.076 & 2.228 & 2.548 \\
& T4 & 2.045 & -0.134 & -0.219 & 2.050 & 0.136 & 0.221 & 0.960 & 1.000 & 1.000 & 2.280 & 0.023 & 2.020 & 2.266 & 2.500 \\
& T5 & 2.104 & -0.131 & -0.239 & 2.116 & 0.133 & 0.242 & 0.960 & 1.000 & 1.000 & 2.351 & 0.053 & 2.094 & 2.320 & 2.960 \\
& T6 & 2.055 & -0.136 & -0.219 & 2.062 & 0.138 & 0.222 & 0.960 & 1.000 & 1.000 & 2.277 & 0.024 & 2.070 & 2.278 & 2.557 \\
\addlinespace[2pt]
\multicolumn{16}{@{}l@{}}{\textbf{Sparse regression (EN)}} \\
\multirow{6}{*}{\ParamBox{0.11\textwidth}{40}{50}{5}{0.03}{0.05}{0.6}{{ext.}}} & T1 & 2.733 & -0.117 & -0.386 & 2.738 & 0.121 & 0.388 & 0.140 & 1.000 & 1.000 & 2.867 & 0.045 & 2.541 & 2.867 & 3.271 \\
& T2 & 2.772 & -0.100 & -0.427 & 2.777 & 0.105 & 0.429 & 0.120 & 1.000 & 0.980 & 2.923 & 0.048 & 2.564 & 2.907 & 3.355 \\
& T3 & 2.790 & -0.104 & -0.424 & 2.795 & 0.108 & 0.426 & 0.060 & 1.000 & 1.000 & 2.911 & 0.050 & 2.592 & 2.939 & 3.404 \\
& T4 & 2.770 & -0.102 & -0.422 & 2.776 & 0.107 & 0.424 & 0.080 & 1.000 & 0.980 & 2.925 & 0.057 & 2.533 & 2.944 & 3.359 \\
& T5 & 2.783 & -0.097 & -0.436 & 2.791 & 0.104 & 0.439 & 0.120 & 1.000 & 0.960 & 2.962 & 0.059 & 2.576 & 2.978 & 3.440 \\
& T6 & 2.770 & -0.103 & -0.419 & 2.777 & 0.110 & 0.423 & 0.100 & 1.000 & 0.980 & 2.913 & 0.068 & 2.559 & 2.910 & 3.406 \\
\addlinespace[2pt]
\cmidrule(lr){1-16}
\multicolumn{16}{@{}l@{}}{\textbf{Linear regression (OLS)}} \\
\multirow{6}{*}{\ParamBox{0.11\textwidth}{40}{50}{5}{0.03}{0.10}{0.6}{clean}} & T1 & 1.948 & -0.158 & -0.150 & 1.972 & 0.162 & 0.157 & 0.820 & 1.000 & 1.000 & 2.117 & 0.050 & 1.613 & 2.139 & 2.496 \\
& T2 & 2.038 & -0.131 & -0.221 & 2.063 & 0.137 & 0.228 & 0.880 & 1.000 & 1.000 & 2.259 & 0.054 & 1.822 & 2.274 & 2.646 \\
& T3 & 2.025 & -0.136 & -0.209 & 2.052 & 0.140 & 0.216 & 0.840 & 1.000 & 1.000 & 2.242 & 0.053 & 1.784 & 2.244 & 2.639 \\
& T4 & 1.906 & -0.113 & -0.231 & 1.951 & 0.127 & 0.238 & 0.940 & 1.000 & 1.000 & 2.345 & 0.115 & 1.728 & 2.316 & 3.146 \\
& T5 & 1.973 & -0.102 & -0.266 & 2.019 & 0.121 & 0.276 & 0.940 & 1.000 & 1.000 & 2.496 & 0.144 & 1.947 & 2.442 & 3.538 \\
& T6 & 1.917 & -0.116 & -0.226 & 1.966 & 0.131 & 0.234 & 0.880 & 1.000 & 1.000 & 2.337 & 0.102 & 1.813 & 2.349 & 3.128 \\
\addlinespace[2pt]
\multicolumn{16}{@{}l@{}}{\textbf{Sparse regression (EN)}} \\
\multirow{6}{*}{\ParamBox{0.11\textwidth}{40}{50}{5}{0.03}{0.10}{0.6}{{ext.}}} & T1 & 2.625 & -0.124 & -0.354 & 2.640 & 0.131 & 0.359 & 0.340 & 0.980 & 1.000 & 2.788 & 0.083 & 2.195 & 2.844 & 3.172 \\
& T2 & 2.752 & -0.097 & -0.433 & 2.769 & 0.106 & 0.438 & 0.320 & 0.980 & 0.880 & 2.897 & 0.082 & 2.343 & 2.964 & 3.296 \\
& T3 & 2.765 & -0.099 & -0.431 & 2.778 & 0.109 & 0.437 & 0.340 & 1.000 & 0.900 & 2.892 & 0.097 & 2.178 & 2.948 & 3.368 \\
& T4 & 2.711 & -0.093 & -0.434 & 2.729 & 0.105 & 0.439 & 0.320 & 1.000 & 0.920 & 2.916 & 0.087 & 2.351 & 2.941 & 3.380 \\
& T5 & 2.711 & -0.085 & -0.447 & 2.733 & 0.102 & 0.452 & 0.320 & 1.000 & 0.880 & 2.984 & 0.098 & 2.444 & 3.001 & 3.598 \\
& T6 & 2.704 & -0.093 & -0.430 & 2.722 & 0.109 & 0.437 & 0.300 & 1.000 & 0.900 & 2.914 & 0.102 & 2.222 & 2.943 & 3.475 \\
\addlinespace[2pt]
\cmidrule(lr){1-16}
\multicolumn{16}{@{}l@{}}{\textbf{Linear regression (OLS)}} \\
\multirow{6}{*}{\ParamBox{0.11\textwidth}{40}{50}{5}{0.03}{0.30}{0.6}{clean}} & T1 & 1.869 & -0.227 & 0.003 & 1.941 & 0.237 & 0.079 & 0.640 & 0.640 & 1.000 & 1.791 & 0.121 & 1.135 & 1.780 & 2.478 \\
& T2 & 2.178 & -0.141 & -0.226 & 2.229 & 0.156 & 0.244 & 0.820 & 1.000 & 1.000 & 2.256 & 0.123 & 1.544 & 2.275 & 2.860 \\
& T3 & 2.132 & -0.144 & -0.213 & 2.194 & 0.160 & 0.234 & 0.760 & 0.960 & 1.000 & 2.236 & 0.158 & 1.423 & 2.230 & 2.945 \\
& T4 & 1.873 & -0.096 & -0.253 & 1.963 & 0.134 & 0.279 & 0.840 & 0.980 & 1.000 & 2.516 & 0.308 & 1.602 & 2.480 & 3.770 \\
& T5 & 2.185 & -0.075 & -0.358 & 2.256 & 0.116 & 0.378 & 0.820 & 1.000 & 1.000 & 2.890 & 0.286 & 1.922 & 2.878 & 4.080 \\
& T6 & 1.863 & -0.105 & -0.232 & 1.946 & 0.139 & 0.257 & 0.820 & 1.000 & 1.000 & 2.434 & 0.340 & 1.539 & 2.364 & 3.993 \\
\addlinespace[2pt]
\multicolumn{16}{@{}l@{}}{\textbf{Sparse regression (EN)}} \\
\multirow{6}{*}{\ParamBox{0.11\textwidth}{40}{50}{5}{0.03}{0.30}{0.6}{{ext.}}} & T1 & 2.209 & -0.178 & -0.160 & 2.328 & 0.199 & 0.199 & 0.640 & 0.880 & 1.000 & 2.529 & 0.138 & 1.819 & 2.558 & 3.201 \\
& T2 & 2.691 & -0.082 & -0.448 & 2.772 & 0.116 & 0.465 & 0.640 & 1.000 & 0.760 & 2.948 & 0.134 & 2.165 & 2.969 & 3.624 \\
& T3 & 2.693 & -0.090 & -0.433 & 2.749 & 0.124 & 0.457 & 0.580 & 0.960 & 0.860 & 2.896 & 0.137 & 2.299 & 2.878 & 3.743 \\
& T4 & 2.450 & -0.056 & -0.447 & 2.542 & 0.098 & 0.460 & 0.600 & 1.000 & 0.780 & 3.042 & 0.108 & 2.345 & 3.014 & 3.655 \\
& T5 & 2.636 & -0.048 & -0.503 & 2.709 & 0.090 & 0.515 & 0.680 & 1.000 & 0.740 & 3.199 & 0.123 & 2.643 & 3.136 & 4.065 \\
& T6 & 2.562 & -0.065 & -0.450 & 2.642 & 0.112 & 0.468 & 0.680 & 0.980 & 0.780 & 3.015 & 0.144 & 2.290 & 2.986 & 3.699 \\
\bottomrule
\end{tabular*}
\noindent\parbox{\textwidth}{\raggedleft
  \Nchips{20}\\[-0.05em]
  \Nchips{40}%
}
\endgroup
\end{table}

\begin{table}[htbp]
\centering
\caption{Models coefficients bias, RMSE, 95\% coverage, and out-of-sample predictive MSE across MI methods for OLS (clean) and EN (contaminated), with \(\boldsymbol{n=40}\), \(n_{\text{sim}}=50\), \(\text{iter}=5\), \(\rho=0.6\), contamination \(P_{\text{ext}}\in\{0.05,0.10\}\) and missingness \(P_{\text{miss}}\in\{0.05,0.10,0.30\}\)}
\label{tab:n_20_40_graficos.4}

\begingroup
\scriptsize
\begin{tabular*}{\textwidth}{@{\extracolsep{\fill}} l l
ccc 
ccc 
ccc 
ccccc 
@{}}
\toprule
& & \multicolumn{3}{c}{Bias} & \multicolumn{3}{c}{RMSE} & \multicolumn{3}{c}{Coverage} & \multicolumn{5}{c}{Pred.~MSE (out-of-sample)} \\
\cmidrule(lr){3-5}\cmidrule(lr){6-8}\cmidrule(lr){9-11}\cmidrule(lr){12-16}
Setup & MI &
$\beta_0$ & $\beta_1$ & $\beta_2$ &
$\beta_0$ & $\beta_1$ & $\beta_2$ &
$\beta_0$ & $\beta_1$ & $\beta_2$ &
$\overline{X}$ & $\sigma^2$ & $Q_{2.5}$ & $Q_{50}$ & $Q_{97.5}$
 \\
\midrule
\multicolumn{16}{@{}l@{}}{\textbf{Linear regression (OLS)}} \\
\multirow{6}{*}{\ParamBox{0.11\textwidth}{40}{50}{5}{0.05}{0.05}{0.6}{clean}} & T1 & 2.021 & -0.148 & -0.188 & 2.027 & 0.149 & 0.190 & 0.860 & 1.000 & 1.000 & 2.207 & 0.023 & 1.914 & 2.241 & 2.442 \\
& T2 & 2.051 & -0.134 & -0.220 & 2.058 & 0.136 & 0.222 & 0.960 & 1.000 & 1.000 & 2.271 & 0.027 & 1.967 & 2.280 & 2.527 \\
& T3 & 2.048 & -0.134 & -0.221 & 2.054 & 0.135 & 0.223 & 0.980 & 1.000 & 1.000 & 2.275 & 0.026 & 2.016 & 2.272 & 2.566 \\
& T4 & 2.046 & -0.135 & -0.218 & 2.052 & 0.136 & 0.220 & 0.960 & 1.000 & 1.000 & 2.273 & 0.030 & 1.952 & 2.260 & 2.564 \\
& T5 & 2.078 & -0.129 & -0.236 & 2.084 & 0.132 & 0.240 & 0.980 & 1.000 & 1.000 & 2.352 & 0.053 & 2.069 & 2.349 & 2.780 \\
& T6 & 2.052 & -0.136 & -0.218 & 2.057 & 0.137 & 0.221 & 0.980 & 1.000 & 1.000 & 2.273 & 0.030 & 1.930 & 2.262 & 2.561 \\
\addlinespace[2pt]
\multicolumn{16}{@{}l@{}}{\textbf{Sparse regression (EN)}} \\
\multirow{6}{*}{\ParamBox{0.11\textwidth}{40}{50}{5}{0.05}{0.05}{0.6}{{ext.}}} & T1 & 2.574 & -0.103 & -0.373 & 2.579 & 0.108 & 0.375 & 0.000 & 1.000 & 1.000 & 2.712 & 0.043 & 2.357 & 2.716 & 3.107 \\
& T2 & 2.600 & -0.086 & -0.412 & 2.604 & 0.092 & 0.414 & 0.020 & 1.000 & 0.960 & 2.769 & 0.043 & 2.385 & 2.757 & 3.207 \\
& T3 & 2.617 & -0.090 & -0.409 & 2.621 & 0.095 & 0.411 & 0.000 & 1.000 & 0.980 & 2.757 & 0.045 & 2.408 & 2.769 & 3.189 \\
& T4 & 2.559 & -0.083 & -0.411 & 2.569 & 0.094 & 0.415 & 0.140 & 1.000 & 0.920 & 2.796 & 0.063 & 2.342 & 2.809 & 3.255 \\
& T5 & 2.566 & -0.072 & -0.433 & 2.577 & 0.088 & 0.438 & 0.160 & 1.000 & 0.860 & 2.858 & 0.076 & 2.469 & 2.776 & 3.388 \\
& T6 & 2.556 & -0.082 & -0.411 & 2.566 & 0.095 & 0.415 & 0.160 & 1.000 & 0.940 & 2.782 & 0.060 & 2.366 & 2.785 & 3.255 \\
\addlinespace[2pt]
\cmidrule(lr){1-16}
\multicolumn{16}{@{}l@{}}{\textbf{Linear regression (OLS)}} \\
\multirow{6}{*}{\ParamBox{0.11\textwidth}{40}{50}{5}{0.05}{0.10}{0.6}{clean}} & T1 & 1.941 & -0.159 & -0.149 & 1.963 & 0.162 & 0.156 & 0.780 & 1.000 & 1.000 & 2.109 & 0.050 & 1.635 & 2.130 & 2.524 \\
& T2 & 2.054 & -0.134 & -0.220 & 2.076 & 0.138 & 0.225 & 0.780 & 1.000 & 1.000 & 2.246 & 0.057 & 1.702 & 2.265 & 2.710 \\
& T3 & 2.030 & -0.136 & -0.210 & 2.051 & 0.140 & 0.216 & 0.800 & 1.000 & 1.000 & 2.237 & 0.061 & 1.685 & 2.252 & 2.699 \\
& T4 & 1.921 & -0.114 & -0.231 & 1.960 & 0.126 & 0.239 & 0.940 & 1.000 & 1.000 & 2.357 & 0.125 & 1.776 & 2.325 & 3.294 \\
& T5 & 1.963 & -0.107 & -0.255 & 2.014 & 0.124 & 0.263 & 0.900 & 1.000 & 1.000 & 2.457 & 0.147 & 1.813 & 2.391 & 3.410 \\
& T6 & 1.904 & -0.115 & -0.225 & 1.949 & 0.131 & 0.233 & 0.940 & 1.000 & 1.000 & 2.331 & 0.112 & 1.781 & 2.288 & 2.944 \\
\addlinespace[2pt]
\multicolumn{16}{@{}l@{}}{\textbf{Sparse regression (EN)}} \\
\multirow{6}{*}{\ParamBox{0.11\textwidth}{40}{50}{5}{0.05}{0.10}{0.6}{{ext.}}} & T1 & 2.512 & -0.113 & -0.344 & 2.523 & 0.120 & 0.349 & 0.020 & 0.960 & 1.000 & 2.628 & 0.085 & 1.951 & 2.680 & 3.011 \\
& T2 & 2.616 & -0.085 & -0.420 & 2.625 & 0.095 & 0.425 & 0.100 & 0.980 & 0.820 & 2.750 & 0.096 & 1.991 & 2.767 & 3.178 \\
& T3 & 2.589 & -0.083 & -0.420 & 2.598 & 0.093 & 0.425 & 0.100 & 1.000 & 0.860 & 2.742 & 0.094 & 2.033 & 2.759 & 3.209 \\
& T4 & 2.508 & -0.071 & -0.427 & 2.529 & 0.089 & 0.432 & 0.280 & 1.000 & 0.920 & 2.805 & 0.105 & 2.192 & 2.811 & 3.323 \\
& T5 & 2.548 & -0.061 & -0.454 & 2.567 & 0.086 & 0.460 & 0.220 & 1.000 & 0.780 & 2.879 & 0.112 & 2.253 & 2.830 & 3.581 \\
& T6 & 2.514 & -0.070 & -0.429 & 2.534 & 0.094 & 0.437 & 0.220 & 1.000 & 0.820 & 2.830 & 0.128 & 2.170 & 2.736 & 3.565 \\
\addlinespace[2pt]
\cmidrule(lr){1-16}
\multicolumn{16}{@{}l@{}}{\textbf{Linear regression (OLS)}} \\
\multirow{6}{*}{\ParamBox{0.11\textwidth}{40}{50}{5}{0.05}{0.30}{0.6}{clean}} & T1 & 1.858 & -0.229 & 0.008 & 1.927 & 0.238 & 0.075 & 0.660 & 0.640 & 1.000 & 1.814 & 0.121 & 1.129 & 1.858 & 2.417 \\
& T2 & 2.181 & -0.141 & -0.228 & 2.235 & 0.154 & 0.245 & 0.740 & 1.000 & 1.000 & 2.292 & 0.154 & 1.476 & 2.314 & 2.838 \\
& T3 & 2.187 & -0.152 & -0.208 & 2.236 & 0.164 & 0.225 & 0.760 & 0.960 & 1.000 & 2.244 & 0.188 & 1.379 & 2.297 & 3.047 \\
& T4 & 1.874 & -0.092 & -0.264 & 1.955 & 0.128 & 0.290 & 0.940 & 0.980 & 1.000 & 2.572 & 0.307 & 1.738 & 2.570 & 3.682 \\
& T5 & 2.219 & -0.076 & -0.365 & 2.275 & 0.110 & 0.383 & 0.900 & 1.000 & 1.000 & 2.931 & 0.271 & 1.858 & 3.038 & 3.912 \\
& T6 & 1.903 & -0.110 & -0.231 & 1.978 & 0.138 & 0.255 & 0.860 & 0.980 & 1.000 & 2.464 & 0.259 & 1.456 & 2.435 & 3.575 \\
\addlinespace[2pt]
\multicolumn{16}{@{}l@{}}{\textbf{Sparse regression (EN)}} \\
\multirow{6}{*}{\ParamBox{0.11\textwidth}{40}{50}{5}{0.05}{0.30}{0.6}{{ext.}}} & T1 & 2.232 & -0.175 & -0.157 & 2.308 & 0.195 & 0.193 & 0.400 & 0.820 & 1.000 & 2.365 & 0.136 & 1.674 & 2.382 & 3.009 \\
& T2 & 2.522 & -0.068 & -0.431 & 2.567 & 0.102 & 0.447 & 0.500 & 1.000 & 0.800 & 2.771 & 0.151 & 2.023 & 2.845 & 3.559 \\
& T3 & 2.539 & -0.079 & -0.415 & 2.597 & 0.114 & 0.431 & 0.460 & 0.980 & 0.820 & 2.754 & 0.140 & 2.017 & 2.776 & 3.534 \\
& T4 & 2.203 & -0.026 & -0.450 & 2.268 & 0.079 & 0.461 & 0.600 & 1.000 & 0.740 & 2.961 & 0.121 & 2.334 & 3.009 & 3.616 \\
& T5 & 2.383 & -0.010 & -0.520 & 2.446 & 0.084 & 0.534 & 0.580 & 1.000 & 0.600 & 3.181 & 0.200 & 2.502 & 3.200 & 4.176 \\
& T6 & 2.298 & -0.035 & -0.451 & 2.359 & 0.092 & 0.471 & 0.540 & 1.000 & 0.780 & 2.909 & 0.171 & 2.163 & 2.953 & 3.636 \\
\addlinespace[2pt]
\cmidrule(lr){1-16}
\multicolumn{16}{@{}l@{}}{\textbf{Linear regression (OLS)}} \\
\multirow{6}{*}{\ParamBox{0.11\textwidth}{40}{50}{5}{0.10}{0.05}{0.6}{clean}} & T1 & 2.015 & -0.146 & -0.190 & 2.022 & 0.148 & 0.192 & 0.900 & 1.000 & 1.000 & 2.195 & 0.027 & 1.912 & 2.201 & 2.528 \\
& T2 & 2.076 & -0.134 & -0.226 & 2.084 & 0.136 & 0.228 & 0.880 & 1.000 & 1.000 & 2.270 & 0.029 & 1.940 & 2.285 & 2.554 \\
& T3 & 2.062 & -0.134 & -0.223 & 2.069 & 0.137 & 0.225 & 0.980 & 1.000 & 1.000 & 2.267 & 0.028 & 2.009 & 2.263 & 2.564 \\
& T4 & 2.024 & -0.128 & -0.227 & 2.035 & 0.132 & 0.230 & 0.940 & 1.000 & 1.000 & 2.294 & 0.041 & 1.925 & 2.294 & 2.797 \\
& T5 & 2.092 & -0.128 & -0.242 & 2.104 & 0.131 & 0.245 & 0.980 & 1.000 & 1.000 & 2.364 & 0.059 & 2.026 & 2.312 & 2.860 \\
& T6 & 2.041 & -0.133 & -0.221 & 2.054 & 0.137 & 0.224 & 0.940 & 1.000 & 1.000 & 2.278 & 0.042 & 1.918 & 2.253 & 2.661 \\
\addlinespace[2pt]
\multicolumn{16}{@{}l@{}}{\textbf{Sparse regression (EN)}} \\
\multirow{6}{*}{\ParamBox{0.11\textwidth}{40}{50}{5}{0.10}{0.05}{0.6}{{ext.}}} & T1 & 2.282 & -0.108 & -0.304 & 2.285 & 0.113 & 0.307 & 0.000 & 1.000 & 1.000 & 2.657 & 0.032 & 2.223 & 2.653 & 2.957 \\
& T2 & 2.316 & -0.091 & -0.344 & 2.320 & 0.098 & 0.347 & 0.020 & 1.000 & 1.000 & 2.718 & 0.033 & 2.295 & 2.716 & 2.988 \\
& T3 & 2.294 & -0.089 & -0.345 & 2.298 & 0.097 & 0.349 & 0.040 & 1.000 & 1.000 & 2.724 & 0.035 & 2.406 & 2.716 & 3.030 \\
& T4 & 2.224 & -0.077 & -0.354 & 2.234 & 0.092 & 0.360 & 0.240 & 1.000 & 1.000 & 2.771 & 0.058 & 2.264 & 2.750 & 3.181 \\
& T5 & 2.267 & -0.072 & -0.372 & 2.278 & 0.089 & 0.378 & 0.140 & 1.000 & 1.000 & 2.822 & 0.073 & 2.327 & 2.822 & 3.352 \\
& T6 & 2.283 & -0.087 & -0.345 & 2.290 & 0.100 & 0.351 & 0.060 & 1.000 & 0.980 & 2.735 & 0.056 & 2.297 & 2.722 & 3.110 \\
\bottomrule
\end{tabular*}
\makebox[\textwidth][r]{\Nchips{40}} 
\endgroup
\end{table}

\begin{table}[htbp]
\centering
\caption{Models coefficients bias, RMSE, 95\% coverage, and out-of-sample predictive MSE across MI methods for OLS (clean) and EN (contaminated). First block: \(\boldsymbol{n=40}\), \(n_{\text{sim}}=50\), \(\text{iter}=5\), \(\rho=0.6\), \(P_{\text{ext}}=0.10\), \(P_{\text{miss}}\in\{0.10,0.30\}\). Second block: \(\boldsymbol{n=20}\), \(n_{\text{sim}}=300\), \(\text{iter}=5\), \(\rho=0.6\), \(P_{\text{ext}}\in\{0.04\}\), \(P_{\text{miss}}\in\{0.10,0.25\}\)}
\label{tab:n_20_40_graficos.5}

\begingroup
\scriptsize
\begin{tabular*}{\textwidth}{@{\extracolsep{\fill}} l l
ccc 
ccc 
ccc 
ccccc 
@{}}
\toprule
& & \multicolumn{3}{c}{Bias} & \multicolumn{3}{c}{RMSE} & \multicolumn{3}{c}{Coverage} & \multicolumn{5}{c}{Pred.~MSE (out-of-sample)} \\
\cmidrule(lr){3-5}\cmidrule(lr){6-8}\cmidrule(lr){9-11}\cmidrule(lr){12-16}
Setup & MI &
$\beta_0$ & $\beta_1$ & $\beta_2$ &
$\beta_0$ & $\beta_1$ & $\beta_2$ &
$\beta_0$ & $\beta_1$ & $\beta_2$ &
$\overline{X}$ & $\sigma^2$ & $Q_{2.5}$ & $Q_{50}$ & $Q_{97.5}$
 \\
\midrule
\multicolumn{16}{@{}l@{}}{\textbf{Linear regression (OLS)}} \\
\multirow{6}{*}{\ParamBox{0.11\textwidth}{40}{50}{5}{0.10}{0.10}{0.6}{clean}} & T1 & 1.940 & -0.158 & -0.150 & 1.964 & 0.162 & 0.157 & 0.820 & 1.000 & 1.000 & 2.106 & 0.044 & 1.635 & 2.138 & 2.444 \\
& T2 & 2.062 & -0.134 & -0.221 & 2.089 & 0.140 & 0.227 & 0.760 & 1.000 & 1.000 & 2.255 & 0.057 & 1.702 & 2.265 & 2.737 \\
& T3 & 2.030 & -0.135 & -0.212 & 2.053 & 0.140 & 0.218 & 0.800 & 1.000 & 1.000 & 2.240 & 0.055 & 1.685 & 2.278 & 2.613 \\
& T4 & 1.910 & -0.114 & -0.227 & 1.950 & 0.127 & 0.234 & 0.920 & 1.000 & 1.000 & 2.339 & 0.101 & 1.773 & 2.322 & 3.060 \\
& T5 & 1.972 & -0.105 & -0.258 & 2.025 & 0.122 & 0.266 & 0.940 & 1.000 & 1.000 & 2.464 & 0.139 & 1.831 & 2.415 & 3.429 \\
& T6 & 1.886 & -0.113 & -0.226 & 1.931 & 0.129 & 0.234 & 0.940 & 1.000 & 1.000 & 2.332 & 0.103 & 1.781 & 2.296 & 2.944 \\
\addlinespace[2pt]
\multicolumn{16}{@{}l@{}}{\textbf{Sparse regression (EN)}} \\
\multirow{6}{*}{\ParamBox{0.11\textwidth}{40}{50}{5}{0.10}{0.10}{0.6}{{ext.}}} & T1 & 2.235 & -0.119 & -0.276 & 2.243 & 0.126 & 0.283 & 0.020 & 0.960 & 1.000 & 2.572 & 0.077 & 1.865 & 2.631 & 2.964 \\
& T2 & 2.312 & -0.088 & -0.352 & 2.319 & 0.098 & 0.359 & 0.040 & 0.980 & 1.000 & 2.687 & 0.084 & 1.920 & 2.745 & 3.090 \\
& T3 & 2.262 & -0.082 & -0.354 & 2.271 & 0.095 & 0.362 & 0.120 & 0.980 & 1.000 & 2.704 & 0.091 & 1.943 & 2.715 & 3.122 \\
& T4 & 2.197 & -0.069 & -0.368 & 2.212 & 0.090 & 0.377 & 0.180 & 1.000 & 1.000 & 2.777 & 0.114 & 2.073 & 2.801 & 3.429 \\
& T5 & 2.251 & -0.064 & -0.386 & 2.263 & 0.079 & 0.392 & 0.180 & 1.000 & 0.980 & 2.823 & 0.072 & 2.317 & 2.836 & 3.288 \\
& T6 & 2.260 & -0.078 & -0.361 & 2.268 & 0.092 & 0.369 & 0.140 & 1.000 & 0.980 & 2.749 & 0.087 & 2.070 & 2.760 & 3.258 \\
\addlinespace[2pt]
\cmidrule(lr){1-16}
\multicolumn{16}{@{}l@{}}{\textbf{Linear regression (OLS)}} \\
\multirow{6}{*}{\ParamBox{0.11\textwidth}{40}{50}{5}{0.10}{0.30}{0.6}{clean}} & T1 & 1.888 & -0.232 & 0.008 & 1.949 & 0.241 & 0.080 & 0.640 & 0.640 & 1.000 & 1.819 & 0.114 & 1.149 & 1.830 & 2.415 \\
& T2 & 2.199 & -0.144 & -0.227 & 2.256 & 0.159 & 0.245 & 0.700 & 1.000 & 1.000 & 2.301 & 0.150 & 1.547 & 2.314 & 2.956 \\
& T3 & 2.213 & -0.154 & -0.209 & 2.251 & 0.166 & 0.226 & 0.780 & 0.960 & 1.000 & 2.261 & 0.176 & 1.389 & 2.256 & 3.064 \\
& T4 & 1.885 & -0.095 & -0.258 & 1.964 & 0.134 & 0.288 & 0.920 & 0.960 & 1.000 & 2.565 & 0.307 & 1.738 & 2.531 & 3.704 \\
& T5 & 2.205 & -0.074 & -0.364 & 2.268 & 0.112 & 0.382 & 0.900 & 1.000 & 1.000 & 2.939 & 0.248 & 1.983 & 3.020 & 3.912 \\
& T6 & 1.899 & -0.112 & -0.227 & 1.974 & 0.144 & 0.255 & 0.840 & 0.960 & 1.000 & 2.470 & 0.251 & 1.611 & 2.414 & 3.589 \\
\addlinespace[2pt]
\multicolumn{16}{@{}l@{}}{\textbf{Sparse regression (EN)}} \\
\multirow{6}{*}{\ParamBox{0.11\textwidth}{40}{50}{5}{0.10}{0.30}{0.6}{{ext.}}} & T1 & 2.037 & -0.190 & -0.090 & 2.095 & 0.210 & 0.154 & 0.260 & 0.760 & 1.000 & 2.307 & 0.148 & 1.585 & 2.336 & 3.044 \\
& T2 & 2.223 & -0.068 & -0.370 & 2.260 & 0.107 & 0.392 & 0.440 & 0.980 & 0.880 & 2.734 & 0.163 & 2.046 & 2.799 & 3.484 \\
& T3 & 2.235 & -0.081 & -0.349 & 2.277 & 0.118 & 0.372 & 0.400 & 0.980 & 0.960 & 2.710 & 0.146 & 1.979 & 2.704 & 3.470 \\
& T4 & 1.833 & -0.009 & -0.411 & 1.903 & 0.094 & 0.428 & 0.720 & 1.000 & 0.800 & 2.964 & 0.136 & 2.296 & 3.012 & 3.672 \\
& T5 & 2.043 & 0.006 & -0.483 & 2.121 & 0.089 & 0.497 & 0.540 & 1.000 & 0.840 & 3.153 & 0.138 & 2.537 & 3.159 & 3.923 \\
& T6 & 2.048 & -0.037 & -0.395 & 2.101 & 0.108 & 0.425 & 0.460 & 0.980 & 0.900 & 2.857 & 0.174 & 2.036 & 2.919 & 3.535 \\
\addlinespace[2pt]
\cmidrule(lr){1-16}
\multicolumn{16}{@{}l@{}}{\textbf{Linear regression (OLS)}} \\
\multirow{6}{*}{\ParamBox{0.11\textwidth}{20}{300}{5}{0.04}{0.10}{0.6}{clean}} & T1 & -0.423 & -0.042 & 0.157 & 0.527 & 0.096 & 0.244 & 1.000 & 0.990 & 0.983 & 2.455 & 0.421 & 1.589 & 2.292 & 4.360 \\
& T2 & -0.142 & -0.035 & 0.089 & 0.441 & 0.092 & 0.210 & 1.000 & 0.990 & 0.990 & 2.633 & 0.413 & 1.644 & 2.510 & 4.449 \\
& T3 & -0.213 & -0.032 & 0.098 & 0.418 & 0.094 & 0.214 & 1.000 & 0.990 & 0.990 & 2.550 & 0.369 & 1.719 & 2.396 & 4.183 \\
& T4 & -0.221 & -0.020 & 0.079 & 0.478 & 0.090 & 0.203 & 1.000 & 0.993 & 0.990 & 2.645 & 0.401 & 1.651 & 2.543 & 4.366 \\
& T5 & -0.120 & -0.025 & 0.070 & 0.410 & 0.092 & 0.180 & 1.000 & 1.000 & 1.000 & 2.807 & 0.368 & 1.987 & 2.679 & 4.501 \\
& T6 & -0.266 & -0.020 & 0.084 & 0.476 & 0.090 & 0.194 & 1.000 & 0.997 & 0.990 & 2.603 & 0.379 & 1.633 & 2.504 & 4.387 \\
\addlinespace[2pt]
\multicolumn{16}{@{}l@{}}{\textbf{Sparse regression (EN)}} \\
\multirow{6}{*}{\ParamBox{0.11\textwidth}{20}{300}{5}{0.04}{0.10}{0.6}{{ext.}}} & T1 & 1.526 & -0.303 & 0.266 & 1.607 & 0.309 & 0.274 & 0.957 & 0.913 & 1.000 & 1.894 & 0.092 & 1.381 & 1.843 & 2.575 \\
& T2 & 1.516 & -0.277 & 0.215 & 1.591 & 0.283 & 0.225 & 0.997 & 0.983 & 1.000 & 2.099 & 0.137 & 1.559 & 2.051 & 2.896 \\
& T3 & 1.533 & -0.284 & 0.225 & 1.620 & 0.291 & 0.235 & 0.990 & 0.967 & 1.000 & 2.076 & 0.134 & 1.487 & 2.056 & 2.871 \\
& T4 & 1.136 & -0.234 & 0.202 & 1.426 & 0.257 & 0.213 & 1.000 & 0.987 & 1.000 & 2.428 & 0.523 & 1.635 & 2.186 & 4.476 \\
& T5 & 1.163 & -0.203 & 0.126 & 1.430 & 0.228 & 0.159 & 1.000 & 0.997 & 1.000 & 2.934 & 1.018 & 1.725 & 2.659 & 5.622 \\
& T6 & 1.201 & -0.240 & 0.202 & 1.463 & 0.260 & 0.214 & 0.997 & 0.993 & 0.997 & 2.364 & 0.536 & 1.541 & 2.146 & 4.479 \\
\addlinespace[2pt]
\cmidrule(lr){1-16}
\multicolumn{16}{@{}l@{}}{\textbf{Linear regression (OLS)}} \\
\multirow{6}{*}{\ParamBox{0.11\textwidth}{20}{300}{5}{0.04}{0.25}{0.6}{clean}} & T1 & -0.715 & -0.075 & 0.277 & 0.937 & 0.160 & 0.393 & 0.993 & 0.960 & 0.880 & 2.166 & 0.442 & 1.121 & 2.063 & 3.923 \\
& T2 & 0.048 & -0.053 & 0.085 & 0.818 & 0.147 & 0.310 & 1.000 & 0.987 & 0.963 & 2.628 & 0.419 & 1.438 & 2.614 & 4.018 \\
& T3 & -0.138 & -0.051 & 0.120 & 0.726 & 0.153 & 0.303 & 1.000 & 0.987 & 0.973 & 2.494 & 0.358 & 1.424 & 2.460 & 3.942 \\
& T4 & -0.338 & -0.007 & 0.082 & 0.778 & 0.118 & 0.266 & 1.000 & 1.000 & 0.973 & 2.712 & 0.414 & 1.688 & 2.697 & 4.064 \\
& T5 & 0.041 & -0.005 & 0.005 & 0.691 & 0.118 & 0.238 & 1.000 & 1.000 & 1.000 & 3.249 & 0.524 & 2.105 & 3.177 & 4.922 \\
& T6 & -0.376 & -0.007 & 0.086 & 0.744 & 0.120 & 0.277 & 1.000 & 0.993 & 0.973 & 2.597 & 0.446 & 1.442 & 2.530 & 4.117 \\
\addlinespace[2pt]
\multicolumn{16}{@{}l@{}}{\textbf{Sparse regression (EN)}} \\
\multirow{6}{*}{\ParamBox{0.11\textwidth}{20}{300}{5}{0.04}{0.25}{0.6}{{ext.}}} & T1 & 1.680 & -0.361 & 0.359 & 1.982 & 0.383 & 0.383 & 0.797 & 0.557 & 0.827 & 1.635 & 0.141 & 0.994 & 1.616 & 2.446 \\
& T2 & 1.635 & -0.289 & 0.217 & 1.916 & 0.311 & 0.247 & 0.960 & 0.970 & 0.990 & 2.225 & 0.310 & 1.411 & 2.129 & 3.520 \\
& T3 & 1.642 & -0.301 & 0.236 & 1.941 & 0.323 & 0.267 & 0.950 & 0.943 & 0.993 & 2.171 & 0.294 & 1.238 & 2.129 & 3.313 \\
& T4 & 0.583 & -0.155 & 0.151 & 1.503 & 0.236 & 0.208 & 0.997 & 1.000 & 0.997 & 3.258 & 1.419 & 1.657 & 2.974 & 6.326 \\
& T5 & 0.786 & -0.090 & -0.038 & 1.424 & 0.171 & 0.163 & 1.000 & 1.000 & 1.000 & 4.243 & 2.132 & 2.134 & 3.928 & 7.804 \\
& T6 & 0.867 & -0.197 & 0.183 & 1.580 & 0.259 & 0.230 & 0.987 & 0.980 & 0.990 & 2.847 & 1.196 & 1.506 & 2.580 & 5.603 \\
\bottomrule
\end{tabular*}
\noindent\parbox{\textwidth}{\raggedleft
  \Nchips{20}\\[-0.05em]
  \Nchips{40}%
}
\endgroup
\end{table}

\begin{table}[htbp]
\centering
\caption{Models coefficients bias, RMSE, 95\% coverage, and out-of-sample predictive MSE across MI methods for OLS (clean) and EN (contaminated) with \(\boldsymbol{n=20}\). Blocks: (i) \(n_{\text{sim}}=300\), \(\text{iter}=5\), \(\rho=0.6\), \(P_{\text{ext}}=0.15\), \(P_{\text{miss}}\in\{0.10,0.25\}\); (ii) \(n_{\text{sim}}=50\), \(\text{iter}=10\), \(\rho=0.6\), \(P_{\text{ext}}=0.05\), \(P_{\text{miss}}\in\{0.10,0.30\}\)}

\label{tab:n_20_40_graficos.6}

\begingroup
\scriptsize
\begin{tabular*}{\textwidth}{@{\extracolsep{\fill}} l l
ccc 
ccc 
ccc 
ccccc 
@{}}
\toprule
& & \multicolumn{3}{c}{Bias} & \multicolumn{3}{c}{RMSE} & \multicolumn{3}{c}{Coverage} & \multicolumn{5}{c}{Pred.~MSE (out-of-sample)} \\
\cmidrule(lr){3-5}\cmidrule(lr){6-8}\cmidrule(lr){9-11}\cmidrule(lr){12-16}
Setup & MI &
$\beta_0$ & $\beta_1$ & $\beta_2$ &
$\beta_0$ & $\beta_1$ & $\beta_2$ &
$\beta_0$ & $\beta_1$ & $\beta_2$ &
$\overline{X}$ & $\sigma^2$ & $Q_{2.5}$ & $Q_{50}$ & $Q_{97.5}$
 \\
\midrule
\multicolumn{16}{@{}l@{}}{\textbf{Linear regression (OLS)}} \\
\multirow{6}{*}{\ParamBox{0.11\textwidth}{20}{300}{5}{0.15}{0.10}{0.6}{clean}} & T1 & -0.424 & -0.043 & 0.158 & 0.528 & 0.097 & 0.246 & 1.000 & 0.990 & 0.983 & 2.480 & 0.461 & 1.606 & 2.322 & 4.545 \\
& T2 & -0.114 & -0.038 & 0.087 & 0.452 & 0.095 & 0.212 & 1.000 & 0.990 & 0.990 & 2.647 & 0.392 & 1.724 & 2.538 & 4.206 \\
& T3 & -0.211 & -0.032 & 0.097 & 0.413 & 0.093 & 0.214 & 1.000 & 0.990 & 0.990 & 2.576 & 0.367 & 1.743 & 2.445 & 4.219 \\
& T4 & -0.211 & -0.024 & 0.084 & 0.454 & 0.091 & 0.205 & 1.000 & 0.993 & 0.990 & 2.680 & 0.429 & 1.725 & 2.565 & 4.376 \\
& T5 & -0.124 & -0.019 & 0.060 & 0.449 & 0.097 & 0.191 & 1.000 & 0.997 & 0.993 & 2.909 & 0.569 & 2.005 & 2.756 & 4.661 \\
& T6 & -0.286 & -0.018 & 0.086 & 0.458 & 0.088 & 0.191 & 1.000 & 0.997 & 0.993 & 2.658 & 0.465 & 1.654 & 2.494 & 4.536 \\
\addlinespace[2pt]
\multicolumn{16}{@{}l@{}}{\textbf{Sparse regression (EN)}} \\
\multirow{6}{*}{\ParamBox{0.11\textwidth}{20}{300}{5}{0.15}{0.10}{0.6}{{ext.}}} & T1 & 0.959 & -0.251 & 0.244 & 1.004 & 0.256 & 0.252 & 0.997 & 0.943 & 0.997 & 1.447 & 0.055 & 1.051 & 1.439 & 1.971 \\
& T2 & 0.888 & -0.225 & 0.203 & 0.947 & 0.231 & 0.212 & 1.000 & 0.997 & 1.000 & 1.628 & 0.081 & 1.209 & 1.583 & 2.285 \\
& T3 & 0.905 & -0.230 & 0.209 & 0.965 & 0.236 & 0.220 & 1.000 & 0.993 & 1.000 & 1.599 & 0.069 & 1.187 & 1.592 & 2.080 \\
& T4 & 0.322 & -0.138 & 0.135 & 0.956 & 0.190 & 0.175 & 1.000 & 1.000 & 1.000 & 2.255 & 0.790 & 1.298 & 1.906 & 4.449 \\
& T5 & 0.311 & -0.096 & 0.046 & 0.929 & 0.156 & 0.130 & 1.000 & 1.000 & 1.000 & 2.760 & 1.133 & 1.475 & 2.512 & 5.358 \\
& T6 & 0.531 & -0.170 & 0.158 & 0.977 & 0.206 & 0.189 & 1.000 & 0.993 & 1.000 & 2.039 & 0.744 & 1.232 & 1.779 & 4.875 \\
\addlinespace[2pt]
\cmidrule(lr){1-16}
\multicolumn{16}{@{}l@{}}{\textbf{Linear regression (OLS)}} \\
\multirow{6}{*}{\ParamBox{0.11\textwidth}{20}{300}{5}{0.15}{0.25}{0.6}{clean}} & T1 & -0.725 & -0.075 & 0.277 & 0.944 & 0.160 & 0.393 & 0.993 & 0.960 & 0.880 & 2.159 & 0.424 & 1.071 & 2.081 & 3.692 \\
& T2 & 0.039 & -0.056 & 0.091 & 0.767 & 0.145 & 0.303 & 1.000 & 0.977 & 0.960 & 2.619 & 0.419 & 1.298 & 2.600 & 3.960 \\
& T3 & -0.142 & -0.052 & 0.121 & 0.739 & 0.150 & 0.305 & 1.000 & 0.977 & 0.963 & 2.484 & 0.341 & 1.406 & 2.464 & 3.794 \\
& T4 & -0.332 & -0.008 & 0.082 & 0.742 & 0.116 & 0.263 & 1.000 & 0.997 & 0.983 & 2.697 & 0.428 & 1.618 & 2.631 & 4.266 \\
& T5 & 0.000 & -0.003 & 0.010 & 0.748 & 0.122 & 0.244 & 1.000 & 1.000 & 1.000 & 3.224 & 0.441 & 2.186 & 3.132 & 4.888 \\
& T6 & -0.367 & -0.008 & 0.086 & 0.718 & 0.125 & 0.282 & 1.000 & 0.987 & 0.973 & 2.616 & 0.452 & 1.487 & 2.575 & 3.966 \\
\addlinespace[2pt]
\multicolumn{16}{@{}l@{}}{\textbf{Sparse regression (EN)}} \\
\multirow{6}{*}{\ParamBox{0.11\textwidth}{20}{300}{5}{0.15}{0.25}{0.6}{{ext.}}} & T1 & 1.137 & -0.308 & 0.327 & 1.290 & 0.327 & 0.354 & 0.893 & 0.590 & 0.817 & 1.266 & 0.092 & 0.756 & 1.244 & 1.989 \\
& T2 & 0.983 & -0.241 & 0.217 & 1.158 & 0.261 & 0.250 & 0.987 & 0.987 & 0.987 & 1.780 & 0.225 & 1.058 & 1.698 & 2.941 \\
& T3 & 1.016 & -0.253 & 0.233 & 1.194 & 0.274 & 0.268 & 0.970 & 0.977 & 0.963 & 1.716 & 0.186 & 0.990 & 1.662 & 2.666 \\
& T4 & -0.553 & 0.012 & -0.011 & 1.375 & 0.183 & 0.166 & 0.993 & 0.997 & 0.997 & 3.516 & 1.697 & 1.641 & 3.320 & 6.568 \\
& T5 & -0.454 & 0.090 & -0.199 & 1.271 & 0.186 & 0.264 & 0.993 & 1.000 & 1.000 & 4.462 & 1.769 & 2.250 & 4.403 & 7.146 \\
& T6 & -0.091 & -0.073 & 0.075 & 1.270 & 0.207 & 0.204 & 0.997 & 0.997 & 0.997 & 3.016 & 2.135 & 1.216 & 2.572 & 6.718 \\
\addlinespace[2pt]
\cmidrule(lr){1-16}
\multicolumn{16}{@{}l@{}}{\textbf{Linear regression (OLS)}} \\
\multirow{6}{*}{\ParamBox{0.11\textwidth}{20}{50}{10}{0.05}{0.10}{0.6}{clean}} & T1 & -0.411 & -0.025 & 0.117 & 0.533 & 0.067 & 0.175 & 1.000 & 1.000 & 0.980 & 2.283 & 0.243 & 1.521 & 2.221 & 3.317 \\
& T2 & -0.136 & -0.016 & 0.047 & 0.446 & 0.066 & 0.140 & 1.000 & 1.000 & 1.000 & 2.452 & 0.200 & 1.710 & 2.451 & 3.465 \\
& T3 & -0.247 & -0.012 & 0.062 & 0.412 & 0.061 & 0.144 & 1.000 & 1.000 & 1.000 & 2.411 & 0.241 & 1.739 & 2.353 & 3.649 \\
& T4 & -0.216 & -0.005 & 0.044 & 0.432 & 0.066 & 0.140 & 1.000 & 1.000 & 1.000 & 2.473 & 0.260 & 1.627 & 2.498 & 3.916 \\
& T5 & -0.090 & -0.004 & 0.020 & 0.334 & 0.063 & 0.126 & 1.000 & 1.000 & 1.000 & 2.678 & 0.201 & 1.988 & 2.580 & 3.521 \\
& T6 & -0.259 & 0.002 & 0.035 & 0.437 & 0.062 & 0.124 & 1.000 & 1.000 & 1.000 & 2.499 & 0.326 & 1.654 & 2.491 & 4.020 \\
\addlinespace[2pt]
\multicolumn{16}{@{}l@{}}{\textbf{Sparse regression (EN)}} \\
\multirow{6}{*}{\ParamBox{0.11\textwidth}{20}{50}{10}{0.05}{0.10}{0.6}{{ext.}}} & T1 & 1.650 & -0.314 & 0.267 & 1.778 & 0.323 & 0.274 & 0.900 & 0.880 & 0.980 & 1.866 & 0.095 & 1.349 & 1.849 & 2.415 \\
& T2 & 1.652 & -0.291 & 0.219 & 1.786 & 0.301 & 0.228 & 0.980 & 0.940 & 1.000 & 2.087 & 0.131 & 1.508 & 2.072 & 2.885 \\
& T3 & 1.638 & -0.293 & 0.223 & 1.743 & 0.301 & 0.232 & 0.980 & 0.980 & 1.000 & 2.083 & 0.132 & 1.518 & 2.049 & 2.951 \\
& T4 & 1.160 & -0.237 & 0.204 & 1.490 & 0.264 & 0.216 & 1.000 & 0.980 & 1.000 & 2.476 & 0.653 & 1.615 & 2.237 & 4.320 \\
& T5 & 1.192 & -0.201 & 0.117 & 1.459 & 0.224 & 0.138 & 1.000 & 1.000 & 1.000 & 3.013 & 1.118 & 1.821 & 2.653 & 5.465 \\
& T6 & 1.285 & -0.251 & 0.208 & 1.570 & 0.272 & 0.219 & 1.000 & 0.940 & 1.000 & 2.363 & 0.504 & 1.525 & 2.180 & 4.062 \\
\addlinespace[2pt]
\cmidrule(lr){1-16}
\multicolumn{16}{@{}l@{}}{\textbf{Linear regression (OLS)}} \\
\multirow{6}{*}{\ParamBox{0.11\textwidth}{20}{50}{10}{0.05}{0.30}{0.6}{clean}} & T1 & -0.640 & -0.112 & 0.331 & 1.063 & 0.214 & 0.470 & 0.980 & 0.920 & 0.800 & 1.980 & 0.459 & 0.837 & 1.956 & 3.319 \\
& T2 & 0.294 & -0.084 & 0.095 & 1.030 & 0.191 & 0.341 & 1.000 & 0.960 & 0.920 & 2.545 & 0.417 & 1.272 & 2.614 & 3.869 \\
& T3 & -0.008 & -0.075 & 0.138 & 0.920 & 0.192 & 0.347 & 1.000 & 0.960 & 0.920 & 2.394 & 0.354 & 1.156 & 2.491 & 3.456 \\
& T4 & -0.298 & -0.014 & 0.087 & 0.738 & 0.133 & 0.295 & 1.000 & 1.000 & 1.000 & 2.635 & 0.346 & 1.581 & 2.704 & 3.764 \\
& T5 & 0.191 & -0.019 & 0.004 & 0.643 & 0.137 & 0.266 & 1.000 & 1.000 & 1.000 & 3.224 & 0.234 & 2.276 & 3.390 & 3.946 \\
& T6 & -0.294 & -0.028 & 0.113 & 0.744 & 0.164 & 0.353 & 1.000 & 0.980 & 0.940 & 2.541 & 0.388 & 1.257 & 2.592 & 3.652 \\
\addlinespace[2pt]
\multicolumn{16}{@{}l@{}}{\textbf{Sparse regression (EN)}} \\
\multirow{6}{*}{\ParamBox{0.11\textwidth}{20}{50}{10}{0.05}{0.30}{0.6}{{ext.}}} & T1 & 1.748 & -0.396 & 0.423 & 2.124 & 0.429 & 0.459 & 0.740 & 0.560 & 0.660 & 1.594 & 0.165 & 1.046 & 1.608 & 2.656 \\
& T2 & 1.642 & -0.288 & 0.217 & 1.951 & 0.314 & 0.252 & 0.940 & 1.000 & 0.980 & 2.318 & 0.271 & 1.499 & 2.317 & 3.473 \\
& T3 & 1.746 & -0.319 & 0.257 & 2.067 & 0.350 & 0.295 & 0.920 & 0.980 & 0.940 & 2.270 & 0.317 & 1.402 & 2.179 & 3.470 \\
& T4 & 0.188 & -0.098 & 0.117 & 1.594 & 0.214 & 0.177 & 1.000 & 1.000 & 1.000 & 3.703 & 1.348 & 2.092 & 3.376 & 6.020 \\
& T5 & 0.518 & -0.007 & -0.148 & 1.323 & 0.144 & 0.252 & 1.000 & 1.000 & 1.000 & 5.068 & 2.619 & 2.927 & 4.846 & 8.192 \\
& T6 & 0.626 & -0.165 & 0.174 & 1.533 & 0.233 & 0.220 & 1.000 & 0.980 & 0.980 & 3.117 & 1.182 & 1.997 & 2.611 & 5.219 \\
\bottomrule
\end{tabular*}
\makebox[\textwidth][r]{\Nchips{20}} 
\endgroup
\end{table}

\begin{table}[htbp]
\centering
\caption{Models coefficients bias, RMSE, 95\% coverage, and out-of-sample predictive MSE across MI methods for OLS (clean) and EN (contaminated). First part: \(\boldsymbol{n=20}\), \(n_{\text{sim}}=50\), \(\text{iter}=10\), \(\rho=0.6\), \(P_{\text{ext}}=0.10\), \(P_{\text{miss}}\in\{0.10,0.30\}\). Second part: \(\boldsymbol{n=40}\), \(n_{\text{sim}}=50\), \(\text{iter}=5\), \(\rho=0\), \(P_{\text{ext}}=0.05\), \(P_{\text{miss}}\in\{0.10,0.30\}\)}

\label{tab:n_20_40_graficos.7}

\begingroup
\scriptsize
\begin{tabular*}{\textwidth}{@{\extracolsep{\fill}} l l
ccc 
ccc 
ccc 
ccccc 
@{}}
\toprule
& & \multicolumn{3}{c}{Bias} & \multicolumn{3}{c}{RMSE} & \multicolumn{3}{c}{Coverage} & \multicolumn{5}{c}{Pred.~MSE (out-of-sample)} \\
\cmidrule(lr){3-5}\cmidrule(lr){6-8}\cmidrule(lr){9-11}\cmidrule(lr){12-16}
Setup & MI &
$\beta_0$ & $\beta_1$ & $\beta_2$ &
$\beta_0$ & $\beta_1$ & $\beta_2$ &
$\beta_0$ & $\beta_1$ & $\beta_2$ &
$\overline{X}$ & $\sigma^2$ & $Q_{2.5}$ & $Q_{50}$ & $Q_{97.5}$
 \\
\midrule
\multicolumn{16}{@{}l@{}}{\textbf{Linear regression (OLS)}} \\
\multirow{6}{*}{\ParamBox{0.11\textwidth}{20}{50}{10}{0.10}{0.10}{0.6}{clean}} & T1 & -0.420 & -0.025 & 0.120 & 0.539 & 0.067 & 0.177 & 1.000 & 1.000 & 0.980 & 2.331 & 0.265 & 1.604 & 2.206 & 3.439 \\
& T2 & -0.141 & -0.018 & 0.054 & 0.443 & 0.064 & 0.144 & 1.000 & 1.000 & 1.000 & 2.491 & 0.205 & 1.758 & 2.420 & 3.404 \\
& T3 & -0.225 & -0.016 & 0.066 & 0.426 & 0.063 & 0.150 & 1.000 & 1.000 & 1.000 & 2.443 & 0.227 & 1.699 & 2.336 & 3.531 \\
& T4 & -0.241 & -0.002 & 0.043 & 0.409 & 0.057 & 0.134 & 1.000 & 1.000 & 1.000 & 2.501 & 0.240 & 1.668 & 2.433 & 3.536 \\
& T5 & -0.108 & 0.001 & 0.012 & 0.371 & 0.063 & 0.135 & 1.000 & 1.000 & 1.000 & 2.741 & 0.229 & 1.941 & 2.692 & 3.921 \\
& T6 & -0.258 & 0.001 & 0.037 & 0.412 & 0.060 & 0.122 & 1.000 & 1.000 & 1.000 & 2.494 & 0.259 & 1.688 & 2.444 & 3.743 \\
\addlinespace[2pt]
\multicolumn{16}{@{}l@{}}{\textbf{Sparse regression (EN)}} \\
\multirow{6}{*}{\ParamBox{0.11\textwidth}{20}{50}{10}{0.10}{0.10}{0.6}{{ext.}}} & T1 & 1.081 & -0.262 & 0.242 & 1.156 & 0.269 & 0.251 & 1.000 & 0.860 & 1.000 & 1.457 & 0.047 & 1.083 & 1.452 & 1.866 \\
& T2 & 1.021 & -0.237 & 0.202 & 1.102 & 0.244 & 0.211 & 1.000 & 1.000 & 1.000 & 1.629 & 0.064 & 1.268 & 1.599 & 2.207 \\
& T3 & 1.049 & -0.242 & 0.205 & 1.126 & 0.249 & 0.214 & 1.000 & 1.000 & 1.000 & 1.617 & 0.060 & 1.205 & 1.570 & 2.190 \\
& T4 & 0.433 & -0.159 & 0.153 & 1.111 & 0.207 & 0.179 & 1.000 & 1.000 & 1.000 & 2.256 & 0.823 & 1.263 & 1.863 & 4.260 \\
& T5 & 0.391 & -0.106 & 0.044 & 1.183 & 0.176 & 0.115 & 1.000 & 1.000 & 1.000 & 3.036 & 1.289 & 1.641 & 2.615 & 4.995 \\
& T6 & 0.497 & -0.174 & 0.171 & 1.140 & 0.217 & 0.193 & 1.000 & 0.980 & 1.000 & 2.146 & 0.761 & 1.238 & 1.815 & 3.951 \\
\addlinespace[2pt]
\cmidrule(lr){1-16}
\multicolumn{16}{@{}l@{}}{\textbf{Linear regression (OLS)}} \\
\multirow{6}{*}{\ParamBox{0.11\textwidth}{20}{50}{10}{0.10}{0.30}{0.6}{clean}} & T1 & -0.617 & -0.113 & 0.329 & 1.031 & 0.213 & 0.469 & 0.980 & 0.920 & 0.800 & 1.912 & 0.252 & 0.917 & 2.023 & 2.649 \\
& T2 & 0.256 & -0.083 & 0.101 & 0.911 & 0.186 & 0.363 & 1.000 & 0.980 & 0.940 & 2.484 & 0.292 & 1.298 & 2.470 & 3.473 \\
& T3 & 0.061 & -0.075 & 0.125 & 0.894 & 0.189 & 0.359 & 1.000 & 0.940 & 0.920 & 2.330 & 0.312 & 1.226 & 2.410 & 3.260 \\
& T4 & -0.266 & -0.010 & 0.075 & 0.666 & 0.139 & 0.314 & 1.000 & 1.000 & 0.960 & 2.661 & 0.246 & 1.603 & 2.707 & 3.442 \\
& T5 & 0.160 & -0.017 & 0.006 & 0.694 & 0.130 & 0.249 & 1.000 & 1.000 & 1.000 & 3.188 & 0.345 & 2.090 & 3.238 & 3.990 \\
& T6 & -0.274 & -0.033 & 0.117 & 0.672 & 0.153 & 0.339 & 1.000 & 0.980 & 0.960 & 2.541 & 0.235 & 1.425 & 2.666 & 3.256 \\
\addlinespace[2pt]
\multicolumn{16}{@{}l@{}}{\textbf{Sparse regression (EN)}} \\
\multirow{6}{*}{\ParamBox{0.11\textwidth}{20}{50}{10}{0.10}{0.30}{0.6}{{ext.}}} & T1 & 1.294 & -0.338 & 0.361 & 1.519 & 0.362 & 0.394 & 0.780 & 0.540 & 0.660 & 1.202 & 0.072 & 0.696 & 1.210 & 1.692 \\
& T2 & 1.196 & -0.262 & 0.220 & 1.417 & 0.290 & 0.266 & 0.960 & 0.960 & 0.960 & 1.832 & 0.174 & 1.103 & 1.827 & 2.685 \\
& T3 & 1.194 & -0.273 & 0.239 & 1.425 & 0.298 & 0.276 & 0.960 & 1.000 & 0.960 & 1.763 & 0.139 & 1.015 & 1.774 & 2.445 \\
& T4 & -0.506 & -0.006 & 0.023 & 1.413 & 0.162 & 0.124 & 1.000 & 1.000 & 1.000 & 3.583 & 1.237 & 2.003 & 3.358 & 5.400 \\
& T5 & -0.398 & 0.095 & -0.221 & 1.229 & 0.164 & 0.261 & 1.000 & 1.000 & 1.000 & 4.850 & 1.351 & 2.527 & 5.008 & 6.669 \\
& T6 & -0.166 & -0.068 & 0.092 & 1.304 & 0.193 & 0.192 & 1.000 & 1.000 & 1.000 & 3.180 & 2.296 & 1.401 & 2.891 & 6.975 \\
\addlinespace[2pt]
\cmidrule(lr){1-16}
\multicolumn{16}{@{}l@{}}{\textbf{Linear regression (OLS)}} \\
\multirow{6}{*}{\ParamBox{0.11\textwidth}{40}{50}{5}{0.05}{0.10}{0.0}{clean}} & T1 & 2.253 & -0.203 & -0.121 & 2.277 & 0.205 & 0.127 & 0.740 & 0.620 & 1.000 & 2.081 & 0.049 & 1.582 & 2.110 & 2.468 \\
& T2 & 2.490 & -0.199 & -0.177 & 2.512 & 0.201 & 0.181 & 0.680 & 0.780 & 1.000 & 2.242 & 0.053 & 1.763 & 2.284 & 2.636 \\
& T3 & 2.449 & -0.200 & -0.167 & 2.474 & 0.201 & 0.172 & 0.700 & 0.820 & 1.000 & 2.236 & 0.060 & 1.671 & 2.267 & 2.636 \\
& T4 & 2.367 & -0.191 & -0.167 & 2.399 & 0.194 & 0.173 & 0.840 & 0.900 & 1.000 & 2.336 & 0.072 & 1.931 & 2.346 & 2.806 \\
& T5 & 2.538 & -0.190 & -0.204 & 2.579 & 0.195 & 0.211 & 0.800 & 0.920 & 1.000 & 2.547 & 0.212 & 2.012 & 2.479 & 3.874 \\
& T6 & 2.405 & -0.193 & -0.172 & 2.434 & 0.196 & 0.178 & 0.800 & 0.900 & 1.000 & 2.320 & 0.078 & 1.831 & 2.354 & 2.758 \\
\addlinespace[2pt]
\multicolumn{16}{@{}l@{}}{\textbf{Sparse regression (EN)}} \\
\multirow{6}{*}{\ParamBox{0.11\textwidth}{40}{50}{5}{0.05}{0.10}{0.0}{{ext.}}} & T1 & 3.436 & -0.238 & -0.279 & 3.455 & 0.241 & 0.282 & 0.020 & 0.160 & 0.920 & 2.566 & 0.091 & 1.897 & 2.606 & 2.999 \\
& T2 & 3.652 & -0.234 & -0.329 & 3.669 & 0.238 & 0.333 & 0.020 & 0.360 & 0.460 & 2.736 & 0.113 & 2.097 & 2.797 & 3.347 \\
& T3 & 3.580 & -0.225 & -0.334 & 3.595 & 0.229 & 0.337 & 0.020 & 0.540 & 0.460 & 2.746 & 0.115 & 1.994 & 2.753 & 3.215 \\
& T4 & 3.547 & -0.221 & -0.338 & 3.572 & 0.226 & 0.342 & 0.060 & 0.460 & 0.480 & 2.854 & 0.187 & 2.049 & 2.905 & 3.654 \\
& T5 & 3.581 & -0.209 & -0.366 & 3.617 & 0.218 & 0.369 & 0.140 & 0.560 & 0.500 & 3.045 & 0.246 & 2.278 & 3.042 & 3.965 \\
& T6 & 3.531 & -0.219 & -0.337 & 3.561 & 0.225 & 0.341 & 0.060 & 0.520 & 0.480 & 2.826 & 0.160 & 2.049 & 2.837 & 3.532 \\
\addlinespace[2pt]
\cmidrule(lr){1-16}
\multicolumn{16}{@{}l@{}}{\textbf{Linear regression (OLS)}} \\
\multirow{6}{*}{\ParamBox{0.11\textwidth}{40}{50}{5}{0.05}{0.30}{0.0}{clean}} & T1 & 1.884 & -0.223 & -0.009 & 1.958 & 0.229 & 0.065 & 0.720 & 0.340 & 1.000 & 1.725 & 0.100 & 1.101 & 1.794 & 2.271 \\
& T2 & 2.733 & -0.215 & -0.190 & 2.798 & 0.223 & 0.203 & 0.660 & 0.840 & 1.000 & 2.325 & 0.176 & 1.476 & 2.351 & 3.098 \\
& T3 & 2.577 & -0.214 & -0.163 & 2.631 & 0.220 & 0.176 & 0.680 & 0.800 & 1.000 & 2.239 & 0.193 & 1.392 & 2.264 & 3.102 \\
& T4 & 2.422 & -0.196 & -0.169 & 2.486 & 0.203 & 0.185 & 0.900 & 0.920 & 1.000 & 2.605 & 0.312 & 1.651 & 2.589 & 3.660 \\
& T5 & 3.070 & -0.205 & -0.276 & 3.142 & 0.214 & 0.290 & 0.740 & 0.920 & 1.000 & 3.193 & 0.403 & 2.234 & 3.140 & 4.428 \\
& T6 & 2.333 & -0.192 & -0.161 & 2.400 & 0.198 & 0.182 & 0.820 & 0.880 & 1.000 & 2.459 & 0.277 & 1.472 & 2.415 & 3.614 \\
\addlinespace[2pt]
\multicolumn{16}{@{}l@{}}{\textbf{Sparse regression (EN)}} \\
\multirow{6}{*}{\ParamBox{0.11\textwidth}{40}{50}{5}{0.05}{0.30}{0.0}{{ext.}}} & T1 & 2.939 & -0.250 & -0.149 & 3.038 & 0.261 & 0.171 & 0.140 & 0.260 & 1.000 & 2.225 & 0.139 & 1.561 & 2.281 & 2.872 \\
& T2 & 3.599 & -0.223 & -0.337 & 3.661 & 0.234 & 0.349 & 0.220 & 0.620 & 0.720 & 2.840 & 0.189 & 2.195 & 2.832 & 3.573 \\
& T3 & 3.442 & -0.214 & -0.323 & 3.528 & 0.229 & 0.334 & 0.200 & 0.700 & 0.720 & 2.803 & 0.178 & 1.948 & 2.856 & 3.550 \\
& T4 & 3.226 & -0.185 & -0.337 & 3.316 & 0.205 & 0.350 & 0.360 & 0.720 & 0.680 & 3.181 & 0.334 & 2.183 & 3.123 & 4.371 \\
& T5 & 3.456 & -0.175 & -0.403 & 3.573 & 0.199 & 0.412 & 0.360 & 0.820 & 0.660 & 3.515 & 0.315 & 2.565 & 3.590 & 4.534 \\
& T6 & 3.317 & -0.192 & -0.342 & 3.393 & 0.208 & 0.355 & 0.320 & 0.820 & 0.780 & 3.079 & 0.277 & 2.153 & 2.997 & 3.976 \\
\bottomrule
\end{tabular*}
\noindent\parbox{\textwidth}{\raggedleft
  \Nchips{20}\\[-0.05em]
  \Nchips{40}%
}
\endgroup
\end{table}

\begin{table}[htbp]
\centering
\caption{Models coefficients bias, RMSE, 95\% coverage, and out-of-sample predictive MSE across MI methods for OLS (clean) and EN (contaminated). \(\boldsymbol{n=40}\), \(n_{\text{sim}}=50\), \(\text{iter}=5\), \(P_{\text{ext}}=0.10\), \(P_{\text{miss}}\in\{0.10,0.30\}\); followed by an additional block with \(\boldsymbol{n=20}\), \(n_{\text{sim}}=1000\), \(\text{iter}=10\),\(P_{\text{ext}}=0.30\),$P_{\text{miss}}$ $\in\{0.10,0.30\}$}

\label{tab:n_20_40_graficos.8}

\begingroup
\scriptsize
\begin{tabular*}{\textwidth}{@{\extracolsep{\fill}} l l
ccc 
ccc 
ccc 
ccccc 
@{}}
\toprule
& & \multicolumn{3}{c}{Bias} & \multicolumn{3}{c}{RMSE} & \multicolumn{3}{c}{Coverage} & \multicolumn{5}{c}{Pred.~MSE (out-of-sample)} \\
\cmidrule(lr){3-5}\cmidrule(lr){6-8}\cmidrule(lr){9-11}\cmidrule(lr){12-16}
Setup & MI &
$\beta_0$ & $\beta_1$ & $\beta_2$ &
$\beta_0$ & $\beta_1$ & $\beta_2$ &
$\beta_0$ & $\beta_1$ & $\beta_2$ &
$\overline{X}$ & $\sigma^2$ & $Q_{2.5}$ & $Q_{50}$ & $Q_{97.5}$
 \\
\midrule
\multicolumn{16}{@{}l@{}}{\textbf{Linear regression (OLS)}} \\
\multirow{6}{*}{\ParamBox{0.11\textwidth}{40}{50}{5}{0.10}{0.10}{0.0}{clean}} & T1 & 2.253 & -0.203 & -0.122 & 2.279 & 0.205 & 0.128 & 0.720 & 0.620 & 1.000 & 2.077 & 0.043 & 1.582 & 2.116 & 2.389 \\
& T2 & 2.503 & -0.201 & -0.176 & 2.526 & 0.203 & 0.181 & 0.680 & 0.800 & 1.000 & 2.251 & 0.050 & 1.763 & 2.283 & 2.623 \\
& T3 & 2.449 & -0.199 & -0.168 & 2.475 & 0.201 & 0.173 & 0.720 & 0.860 & 1.000 & 2.239 & 0.055 & 1.671 & 2.280 & 2.624 \\
& T4 & 2.374 & -0.189 & -0.172 & 2.408 & 0.193 & 0.177 & 0.860 & 0.900 & 1.000 & 2.351 & 0.068 & 1.931 & 2.346 & 2.852 \\
& T5 & 2.574 & -0.191 & -0.207 & 2.619 & 0.197 & 0.215 & 0.720 & 0.900 & 1.000 & 2.560 & 0.213 & 2.012 & 2.479 & 3.891 \\
& T6 & 2.380 & -0.191 & -0.173 & 2.415 & 0.194 & 0.178 & 0.800 & 0.880 & 1.000 & 2.314 & 0.070 & 1.831 & 2.299 & 2.718 \\
\addlinespace[2pt]
\multicolumn{16}{@{}l@{}}{\textbf{Sparse regression (EN)}} \\
\multirow{6}{*}{\ParamBox{0.11\textwidth}{40}{50}{5}{0.10}{0.10}{0.0}{{ext.}}} & T1 & 3.277 & -0.239 & -0.245 & 3.286 & 0.241 & 0.248 & 0.000 & 0.040 & 1.000 & 2.505 & 0.084 & 1.839 & 2.574 & 2.927 \\
& T2 & 3.428 & -0.229 & -0.296 & 3.436 & 0.232 & 0.300 & 0.000 & 0.240 & 0.820 & 2.698 & 0.106 & 2.017 & 2.735 & 3.205 \\
& T3 & 3.337 & -0.220 & -0.296 & 3.347 & 0.223 & 0.300 & 0.000 & 0.380 & 0.800 & 2.708 & 0.116 & 1.934 & 2.751 & 3.153 \\
& T4 & 3.322 & -0.214 & -0.304 & 3.337 & 0.219 & 0.308 & 0.020 & 0.520 & 0.900 & 2.842 & 0.199 & 2.016 & 2.843 & 3.644 \\
& T5 & 3.371 & -0.212 & -0.318 & 3.384 & 0.217 & 0.322 & 0.020 & 0.580 & 0.860 & 2.906 & 0.185 & 2.020 & 2.973 & 3.575 \\
& T6 & 3.367 & -0.221 & -0.298 & 3.375 & 0.225 & 0.303 & 0.000 & 0.460 & 0.840 & 2.743 & 0.136 & 1.922 & 2.794 & 3.367 \\
\addlinespace[2pt]
\cmidrule(lr){1-16}
\multicolumn{16}{@{}l@{}}{\textbf{Linear regression (OLS)}} \\
\multirow{6}{*}{\ParamBox{0.11\textwidth}{40}{50}{5}{0.10}{0.30}{0.0}{clean}} & T1 & 1.920 & -0.227 & -0.009 & 1.984 & 0.233 & 0.069 & 0.680 & 0.320 & 1.000 & 1.728 & 0.093 & 1.103 & 1.757 & 2.272 \\
& T2 & 2.766 & -0.218 & -0.190 & 2.830 & 0.226 & 0.205 & 0.660 & 0.820 & 1.000 & 2.334 & 0.167 & 1.517 & 2.318 & 3.133 \\
& T3 & 2.609 & -0.217 & -0.163 & 2.651 & 0.222 & 0.175 & 0.720 & 0.820 & 1.000 & 2.254 & 0.177 & 1.398 & 2.239 & 3.102 \\
& T4 & 2.477 & -0.202 & -0.166 & 2.540 & 0.210 & 0.185 & 0.880 & 0.880 & 1.000 & 2.605 & 0.274 & 1.763 & 2.563 & 3.660 \\
& T5 & 3.087 & -0.206 & -0.276 & 3.163 & 0.217 & 0.290 & 0.720 & 0.900 & 1.000 & 3.201 & 0.376 & 2.243 & 3.089 & 4.428 \\
& T6 & 2.357 & -0.193 & -0.161 & 2.424 & 0.201 & 0.182 & 0.800 & 0.860 & 1.000 & 2.467 & 0.221 & 1.595 & 2.418 & 3.614 \\
\addlinespace[2pt]
\multicolumn{16}{@{}l@{}}{\textbf{Sparse regression (EN)}} \\
\multirow{6}{*}{\ParamBox{0.11\textwidth}{40}{50}{5}{0.10}{0.30}{0.0}{{ext.}}} & T1 & 2.936 & -0.262 & -0.125 & 2.997 & 0.272 & 0.151 & 0.060 & 0.100 & 1.000 & 2.162 & 0.155 & 1.434 & 2.177 & 2.828 \\
& T2 & 3.401 & -0.221 & -0.300 & 3.436 & 0.230 & 0.314 & 0.060 & 0.580 & 0.780 & 2.809 & 0.217 & 2.010 & 2.779 & 3.627 \\
& T3 & 3.239 & -0.212 & -0.287 & 3.295 & 0.225 & 0.299 & 0.180 & 0.580 & 0.860 & 2.755 & 0.201 & 1.941 & 2.818 & 3.626 \\
& T4 & 2.939 & -0.169 & -0.314 & 3.015 & 0.192 & 0.329 & 0.240 & 0.780 & 0.780 & 3.252 & 0.299 & 2.344 & 3.264 & 4.220 \\
& T5 & 3.114 & -0.151 & -0.383 & 3.190 & 0.183 & 0.402 & 0.260 & 0.860 & 0.760 & 3.560 & 0.505 & 2.273 & 3.530 & 5.010 \\
& T6 & 3.178 & -0.193 & -0.314 & 3.248 & 0.213 & 0.332 & 0.220 & 0.740 & 0.880 & 3.072 & 0.322 & 2.001 & 2.986 & 4.296 \\
\addlinespace[2pt]
\cmidrule(lr){1-16}
\multicolumn{16}{@{}l@{}}{\textbf{Linear regression (OLS)}} \\
\multirow{6}{*}{\ParamBox{0.11\textwidth}{20}{1000}{10}{0.30}{0.10}{0.6}{clean}} & T1 & -0.413 & -0.036 & 0.143 & 0.518 & 0.085 & 0.217 & 1.000 & 0.996 & 0.991 & 2.524 & 0.419 & 1.607 & 2.368 & 4.287 \\
& T2 & -0.097 & -0.031 & 0.071 & 0.404 & 0.084 & 0.182 & 1.000 & 0.997 & 0.996 & 2.690 & 0.345 & 1.820 & 2.576 & 4.237 \\
& T3 & -0.208 & -0.025 & 0.083 & 0.402 & 0.081 & 0.183 & 1.000 & 0.996 & 0.996 & 2.620 & 0.335 & 1.804 & 2.500 & 4.145 \\
& T4 & -0.221 & -0.016 & 0.070 & 0.415 & 0.075 & 0.170 & 1.000 & 0.998 & 0.996 & 2.693 & 0.381 & 1.768 & 2.578 & 4.347 \\
& T5 & -0.140 & -0.014 & 0.052 & 0.401 & 0.083 & 0.162 & 1.000 & 0.998 & 0.998 & 2.883 & 0.397 & 2.007 & 2.749 & 4.577 \\
& T6 & -0.264 & -0.012 & 0.069 & 0.436 & 0.077 & 0.162 & 1.000 & 0.997 & 0.996 & 2.672 & 0.405 & 1.760 & 2.547 & 4.476 \\
\addlinespace[2pt]
\multicolumn{16}{@{}l@{}}{\textbf{Sparse regression (EN)}} \\
\multirow{6}{*}{\ParamBox{0.11\textwidth}{20}{1000}{10}{0.30}{0.10}{0.6}{{ext.}}} & T1 & 1.170 & -0.356 & 0.399 & 1.214 & 0.361 & 0.405 & 0.900 & 0.980 & 0.902 & 1.441 & 0.038 & 1.045 & 1.438 & 1.865 \\
& T2 & 1.033 & -0.305 & 0.323 & 1.107 & 0.315 & 0.339 & 0.921 & 0.997 & 0.921 & 1.577 & 0.046 & 1.183 & 1.566 & 2.025 \\
& T3 & 1.049 & -0.313 & 0.337 & 1.119 & 0.323 & 0.350 & 0.913 & 0.999 & 0.913 & 1.558 & 0.045 & 1.150 & 1.549 & 2.018 \\
& T4 & 0.081 & -0.047 & -0.007 & 0.786 & 0.201 & 0.251 & 1.000 & 1.000 & 1.000 & 2.337 & 0.648 & 1.290 & 2.168 & 4.096 \\
& T5 & 0.271 & -0.070 & 0.002 & 0.558 & 0.149 & 0.179 & 1.000 & 1.000 & 1.000 & 2.318 & 0.302 & 1.518 & 2.245 & 3.555 \\
& T6 & 0.797 & -0.238 & 0.237 & 0.929 & 0.266 & 0.283 & 0.950 & 1.000 & 0.950 & 1.846 & 0.289 & 1.206 & 1.700 & 3.117 \\
\addlinespace[2pt]
\cmidrule(lr){1-16}
\multicolumn{16}{@{}l@{}}{\textbf{Linear regression (OLS)}} \\
\multirow{6}{*}{\ParamBox{0.11\textwidth}{20}{1000}{10}{0.10}{0.30}{0.6}{clean}} & T1 & -0.875 & -0.087 & 0.328 & 1.151 & 0.181 & 0.452 & 0.989 & 0.903 & 0.799 & 2.045 & 0.496 & 0.877 & 1.995 & 3.702 \\
& T2 & 0.046 & -0.058 & 0.093 & 0.851 & 0.156 & 0.333 & 0.999 & 0.979 & 0.950 & 2.585 & 0.402 & 1.283 & 2.593 & 3.818 \\
& T3 & -0.211 & -0.051 & 0.133 & 0.819 & 0.160 & 0.336 & 0.999 & 0.973 & 0.937 & 2.453 & 0.397 & 1.228 & 2.451 & 3.692 \\
& T4 & -0.406 & -0.004 & 0.088 & 0.837 & 0.125 & 0.297 & 1.000 & 0.998 & 0.968 & 2.710 & 0.409 & 1.544 & 2.717 & 3.955 \\
& T5 & 0.074 & 0.003 & -0.016 & 0.738 & 0.124 & 0.240 & 1.000 & 1.000 & 1.000 & 3.355 & 0.389 & 2.264 & 3.312 & 4.753 \\
& T6 & -0.402 & -0.006 & 0.088 & 0.807 & 0.135 & 0.316 & 1.000 & 0.988 & 0.952 & 2.614 & 0.501 & 1.399 & 2.594 & 4.158 \\
\addlinespace[2pt]
\multicolumn{16}{@{}l@{}}{\textbf{Sparse regression (EN)}} \\
\multirow{6}{*}{\ParamBox{0.11\textwidth}{20}{1000}{10}{0.10}{0.30}{0.6}{{ext.}}} & T1 & 1.198 & -0.320 & 0.341 & 1.411 & 0.340 & 0.369 & 0.832 & 0.428 & 0.700 & 1.180 & 0.106 & 0.634 & 1.146 & 1.882 \\
& T2 & 1.069 & -0.248 & 0.211 & 1.281 & 0.266 & 0.242 & 0.994 & 0.982 & 0.991 & 1.793 & 0.228 & 1.005 & 1.715 & 2.804 \\
& T3 & 1.078 & -0.255 & 0.226 & 1.306 & 0.276 & 0.258 & 0.985 & 0.962 & 0.984 & 1.703 & 0.215 & 0.917 & 1.655 & 2.714 \\
& T4 & -0.612 & 0.001 & 0.022 & 1.632 & 0.188 & 0.147 & 0.978 & 0.999 & 1.000 & 3.608 & 1.734 & 1.741 & 3.410 & 6.421 \\
& T5 & -0.476 & 0.102 & -0.228 & 1.485 & 0.186 & 0.266 & 0.996 & 1.000 & 1.000 & 4.961 & 1.837 & 2.758 & 4.861 & 8.013 \\
& T6 & -0.306 & -0.059 & 0.090 & 1.540 & 0.204 & 0.183 & 0.990 & 0.993 & 0.999 & 3.197 & 2.359 & 1.254 & 2.828 & 6.813 \\
\bottomrule
\end{tabular*}
\noindent\parbox{\textwidth}{\raggedleft
  \Nchips{20}\\[-0.05em]
  \Nchips{40}%
}
\endgroup
\end{table}


\begin{table}[htbp]
\centering
\caption{Models coefficients bias, RMSE, 95\% coverage, and out-of-sample predictive MSE across MI methods for OLS (clean) and elastic net (EN, contaminated), with \(\boldsymbol{n=80}\), \(n_{\text{sim}}=50\), \(\text{iter}=5\), \(\rho=0.6\), contamination \(P_{\text{ext}}\in\{0.03,0.05\}\) and missingness \(P_{\text{miss}}\in\{0.05,0.10,0.30\}\)}
\label{tab:n_80_200_graficos.1}

\begingroup
\scriptsize
\begin{tabular*}{\textwidth}{@{\extracolsep{\fill}} l l
ccc 
ccc 
ccc 
ccccc 
@{}}
\toprule
& & \multicolumn{3}{c}{Bias} & \multicolumn{3}{c}{RMSE} & \multicolumn{3}{c}{Coverage} & \multicolumn{5}{c}{Pred.~MSE (out-of-sample)} \\
\cmidrule(lr){3-5}\cmidrule(lr){6-8}\cmidrule(lr){9-11}\cmidrule(lr){12-16}
Setup & MI &
$\beta_0$ & $\beta_1$ & $\beta_2$ &
$\beta_0$ & $\beta_1$ & $\beta_2$ &
$\beta_0$ & $\beta_1$ & $\beta_2$ &
$\overline{X}$ & $\sigma^2$ & $Q_{2.5}$ & $Q_{50}$ & $Q_{97.5}$
 \\
\midrule
\multicolumn{16}{@{}l@{}}{\textbf{Linear regression (OLS)}} \\
\multirow{6}{*}{\ParamBox{0.11\textwidth}{80}{50}{5}{0.03}{0.05}{0.6}{clean}} & T1 & 0.282 & 0.029 & -0.159 & 0.311 & 0.032 & 0.161 & 1.000 & 1.000 & 1.000 & 2.488 & 0.011 & 2.330 & 2.473 & 2.682 \\
& T2 & 0.389 & 0.035 & -0.192 & 0.417 & 0.037 & 0.194 & 1.000 & 1.000 & 1.000 & 2.557 & 0.010 & 2.413 & 2.555 & 2.724 \\
& T3 & 0.371 & 0.036 & -0.192 & 0.397 & 0.038 & 0.194 & 1.000 & 1.000 & 1.000 & 2.551 & 0.012 & 2.406 & 2.540 & 2.747 \\
& T4 & 0.387 & 0.034 & -0.191 & 0.415 & 0.037 & 0.194 & 1.000 & 1.000 & 1.000 & 2.538 & 0.011 & 2.369 & 2.522 & 2.750 \\
& T5 & 0.425 & 0.036 & -0.202 & 0.453 & 0.039 & 0.205 & 1.000 & 1.000 & 1.000 & 2.592 & 0.014 & 2.396 & 2.581 & 2.855 \\
& T6 & 0.401 & 0.036 & -0.197 & 0.428 & 0.038 & 0.199 & 1.000 & 1.000 & 1.000 & 2.539 & 0.010 & 2.368 & 2.527 & 2.723 \\
\addlinespace[2pt]
\multicolumn{16}{@{}l@{}}{\textbf{Sparse regression (EN)}} \\
\multirow{6}{*}{\ParamBox{0.11\textwidth}{80}{50}{5}{0.03}{0.05}{0.6}{{ext.}}} & T1 & -0.250 & 0.110 & -0.170 & 0.280 & 0.111 & 0.171 & 1.000 & 1.000 & 1.000 & 1.919 & 0.006 & 1.781 & 1.912 & 2.073 \\
& T2 & -0.215 & 0.121 & -0.199 & 0.253 & 0.122 & 0.199 & 1.000 & 1.000 & 1.000 & 1.984 & 0.006 & 1.875 & 1.972 & 2.153 \\
& T3 & -0.230 & 0.122 & -0.197 & 0.259 & 0.123 & 0.198 & 1.000 & 1.000 & 0.980 & 1.979 & 0.006 & 1.843 & 1.983 & 2.118 \\
& T4 & -0.284 & 0.128 & -0.199 & 0.336 & 0.129 & 0.200 & 1.000 & 0.980 & 1.000 & 2.029 & 0.014 & 1.849 & 2.012 & 2.281 \\
& T5 & -0.279 & 0.132 & -0.208 & 0.336 & 0.135 & 0.210 & 1.000 & 1.000 & 1.000 & 2.080 & 0.018 & 1.844 & 2.062 & 2.340 \\
& T6 & -0.240 & 0.123 & -0.196 & 0.280 & 0.124 & 0.197 & 1.000 & 1.000 & 1.000 & 1.994 & 0.008 & 1.868 & 1.976 & 2.210 \\
\addlinespace[2pt]
\cmidrule(lr){1-16}
\multicolumn{16}{@{}l@{}}{\textbf{Linear regression (OLS)}} \\
\multirow{6}{*}{\ParamBox{0.11\textwidth}{80}{50}{5}{0.03}{0.10}{0.6}{clean}} & T1 & 0.154 & 0.021 & -0.116 & 0.306 & 0.034 & 0.124 & 1.000 & 1.000 & 1.000 & 2.412 & 0.014 & 2.185 & 2.432 & 2.577 \\
& T2 & 0.383 & 0.035 & -0.192 & 0.474 & 0.046 & 0.198 & 1.000 & 1.000 & 1.000 & 2.565 & 0.015 & 2.313 & 2.583 & 2.762 \\
& T3 & 0.380 & 0.032 & -0.184 & 0.456 & 0.040 & 0.190 & 1.000 & 1.000 & 1.000 & 2.553 & 0.020 & 2.277 & 2.555 & 2.775 \\
& T4 & 0.383 & 0.032 & -0.184 & 0.466 & 0.041 & 0.189 & 1.000 & 1.000 & 1.000 & 2.554 & 0.015 & 2.343 & 2.571 & 2.761 \\
& T5 & 0.437 & 0.035 & -0.203 & 0.517 & 0.044 & 0.211 & 1.000 & 1.000 & 1.000 & 2.634 & 0.026 & 2.361 & 2.623 & 2.952 \\
& T6 & 0.441 & 0.026 & -0.184 & 0.504 & 0.037 & 0.190 & 1.000 & 1.000 & 1.000 & 2.538 & 0.016 & 2.316 & 2.542 & 2.790 \\
\addlinespace[2pt]
\multicolumn{16}{@{}l@{}}{\textbf{Sparse regression (EN)}} \\
\multirow{6}{*}{\ParamBox{0.11\textwidth}{80}{50}{5}{0.03}{0.10}{0.6}{{ext.}}} & T1 & -0.230 & 0.095 & -0.144 & 0.314 & 0.098 & 0.145 & 1.000 & 1.000 & 1.000 & 1.859 & 0.010 & 1.633 & 1.852 & 2.055 \\
& T2 & -0.206 & 0.121 & -0.200 & 0.322 & 0.125 & 0.202 & 1.000 & 0.980 & 0.980 & 2.002 & 0.013 & 1.787 & 2.002 & 2.198 \\
& T3 & -0.166 & 0.114 & -0.194 & 0.276 & 0.117 & 0.196 & 1.000 & 1.000 & 1.000 & 1.979 & 0.013 & 1.725 & 1.971 & 2.232 \\
& T4 & -0.294 & 0.129 & -0.200 & 0.423 & 0.134 & 0.201 & 1.000 & 0.920 & 0.960 & 2.058 & 0.024 & 1.794 & 2.041 & 2.349 \\
& T5 & -0.161 & 0.131 & -0.229 & 0.337 & 0.136 & 0.232 & 1.000 & 1.000 & 0.880 & 2.206 & 0.034 & 1.930 & 2.192 & 2.559 \\
& T6 & -0.232 & 0.126 & -0.204 & 0.330 & 0.129 & 0.206 & 1.000 & 1.000 & 0.940 & 2.035 & 0.015 & 1.786 & 2.008 & 2.275 \\
\addlinespace[2pt]
\cmidrule(lr){1-16}
\multicolumn{16}{@{}l@{}}{\textbf{Linear regression (OLS)}} \\
\multirow{6}{*}{\ParamBox{0.11\textwidth}{80}{50}{5}{0.03}{0.30}{0.6}{clean}} & T1 & -0.353 & -0.002 & 0.039 & 0.634 & 0.041 & 0.109 & 1.000 & 1.000 & 0.980 & 2.100 & 0.043 & 1.653 & 2.095 & 2.460 \\
& T2 & 0.418 & 0.036 & -0.200 & 0.650 & 0.052 & 0.225 & 1.000 & 1.000 & 0.960 & 2.565 & 0.051 & 2.058 & 2.549 & 2.938 \\
& T3 & 0.351 & 0.039 & -0.192 & 0.637 & 0.057 & 0.215 & 1.000 & 1.000 & 0.900 & 2.527 & 0.059 & 2.060 & 2.526 & 2.973 \\
& T4 & 0.373 & 0.042 & -0.207 & 0.580 & 0.055 & 0.232 & 1.000 & 1.000 & 0.900 & 2.628 & 0.058 & 2.171 & 2.625 & 3.022 \\
& T5 & 0.536 & 0.049 & -0.254 & 0.729 & 0.058 & 0.277 & 1.000 & 1.000 & 0.860 & 2.893 & 0.070 & 2.284 & 2.941 & 3.295 \\
& T6 & 0.375 & 0.034 & -0.188 & 0.559 & 0.050 & 0.211 & 1.000 & 1.000 & 0.960 & 2.552 & 0.057 & 1.993 & 2.549 & 2.943 \\
\addlinespace[2pt]
\multicolumn{16}{@{}l@{}}{\textbf{Sparse regression (EN)}} \\
\multirow{6}{*}{\ParamBox{0.11\textwidth}{80}{50}{5}{0.03}{0.30}{0.6}{{ext.}}} & T1 & -0.506 & 0.057 & -0.014 & 0.707 & 0.078 & 0.055 & 0.980 & 0.940 & 1.000 & 1.576 & 0.033 & 1.250 & 1.573 & 1.906 \\
& T2 & -0.283 & 0.128 & -0.196 & 0.545 & 0.139 & 0.204 & 1.000 & 0.960 & 0.940 & 2.047 & 0.059 & 1.687 & 2.007 & 2.514 \\
& T3 & -0.378 & 0.130 & -0.182 & 0.584 & 0.140 & 0.189 & 1.000 & 0.960 & 0.920 & 2.010 & 0.048 & 1.645 & 2.007 & 2.391 \\
& T4 & -0.692 & 0.172 & -0.206 & 0.936 & 0.183 & 0.216 & 0.980 & 0.760 & 0.820 & 2.237 & 0.085 & 1.769 & 2.240 & 2.757 \\
& T5 & -0.437 & 0.194 & -0.299 & 0.733 & 0.204 & 0.306 & 1.000 & 0.820 & 0.780 & 2.639 & 0.116 & 1.894 & 2.680 & 3.179 \\
& T6 & -0.513 & 0.151 & -0.199 & 0.779 & 0.162 & 0.207 & 0.960 & 0.900 & 0.920 & 2.149 & 0.099 & 1.671 & 2.137 & 2.831 \\
\addlinespace[2pt]
\cmidrule(lr){1-16}
\multicolumn{16}{@{}l@{}}{\textbf{Linear regression (OLS)}} \\
\multirow{6}{*}{\ParamBox{0.11\textwidth}{80}{50}{5}{0.05}{0.05}{0.6}{clean}} & T1 & 0.272 & 0.030 & -0.159 & 0.301 & 0.033 & 0.161 & 1.000 & 1.000 & 1.000 & 2.489 & 0.010 & 2.301 & 2.486 & 2.661 \\
& T2 & 0.374 & 0.036 & -0.192 & 0.399 & 0.038 & 0.194 & 1.000 & 1.000 & 1.000 & 2.543 & 0.011 & 2.342 & 2.543 & 2.717 \\
& T3 & 0.362 & 0.037 & -0.192 & 0.391 & 0.040 & 0.193 & 1.000 & 1.000 & 1.000 & 2.544 & 0.012 & 2.349 & 2.547 & 2.738 \\
& T4 & 0.393 & 0.034 & -0.193 & 0.420 & 0.037 & 0.195 & 1.000 & 1.000 & 1.000 & 2.557 & 0.011 & 2.346 & 2.553 & 2.738 \\
& T5 & 0.404 & 0.037 & -0.201 & 0.440 & 0.040 & 0.205 & 1.000 & 1.000 & 1.000 & 2.592 & 0.019 & 2.393 & 2.569 & 2.860 \\
& T6 & 0.384 & 0.036 & -0.194 & 0.411 & 0.038 & 0.196 & 1.000 & 1.000 & 1.000 & 2.544 & 0.010 & 2.356 & 2.543 & 2.711 \\
\addlinespace[2pt]
\multicolumn{16}{@{}l@{}}{\textbf{Sparse regression (EN)}} \\
\multirow{6}{*}{\ParamBox{0.11\textwidth}{80}{50}{5}{0.05}{0.05}{0.6}{{ext.}}} & T1 & -0.226 & 0.119 & -0.198 & 0.254 & 0.120 & 0.198 & 1.000 & 1.000 & 1.000 & 1.861 & 0.006 & 1.707 & 1.858 & 1.982 \\
& T2 & -0.227 & 0.133 & -0.224 & 0.265 & 0.134 & 0.225 & 1.000 & 1.000 & 0.560 & 1.926 & 0.006 & 1.768 & 1.923 & 2.073 \\
& T3 & -0.235 & 0.133 & -0.223 & 0.270 & 0.134 & 0.224 & 1.000 & 1.000 & 0.560 & 1.922 & 0.008 & 1.769 & 1.921 & 2.056 \\
& T4 & -0.260 & 0.137 & -0.227 & 0.299 & 0.139 & 0.228 & 1.000 & 0.980 & 0.540 & 1.962 & 0.012 & 1.773 & 1.950 & 2.181 \\
& T5 & -0.246 & 0.140 & -0.234 & 0.283 & 0.142 & 0.236 & 1.000 & 1.000 & 0.640 & 2.007 & 0.012 & 1.827 & 2.006 & 2.200 \\
& T6 & -0.243 & 0.135 & -0.225 & 0.268 & 0.136 & 0.226 & 1.000 & 1.000 & 0.500 & 1.937 & 0.008 & 1.769 & 1.938 & 2.114 \\
\bottomrule
\end{tabular*}
\noindent\parbox{\textwidth}{\raggedleft
  \Nchips{80}
}
\endgroup
\end{table}

\begin{table}[htbp]
\centering
\caption{Models coefficients bias, RMSE, 95\% coverage, and out-of-sample predictive MSE across MI methods for OLS (clean) and EN (contaminated), with \(\boldsymbol{n=80}\), \(n_{\text{sim}}=50\), \(\text{iter}=5\), \(\rho=0.6\), contamination \(P_{\text{ext}}\in\{0.05,0.10\}\) and missingness \(P_{\text{miss}}\in\{0.05,0.10,0.30\}\)}

\label{tab:n_80_200_graficos.2}

\begingroup
\scriptsize
\begin{tabular*}{\textwidth}{@{\extracolsep{\fill}} l l
ccc 
ccc 
ccc 
ccccc 
@{}}
\toprule
& & \multicolumn{3}{c}{Bias} & \multicolumn{3}{c}{RMSE} & \multicolumn{3}{c}{Coverage} & \multicolumn{5}{c}{Pred.~MSE (out-of-sample)} \\
\cmidrule(lr){3-5}\cmidrule(lr){6-8}\cmidrule(lr){9-11}\cmidrule(lr){12-16}
Setup & MI &
$\beta_0$ & $\beta_1$ & $\beta_2$ &
$\beta_0$ & $\beta_1$ & $\beta_2$ &
$\beta_0$ & $\beta_1$ & $\beta_2$ &
$\overline{X}$ & $\sigma^2$ & $Q_{2.5}$ & $Q_{50}$ & $Q_{97.5}$
 \\
\midrule
\multicolumn{16}{@{}l@{}}{\textbf{Linear regression (OLS)}} \\
\multirow{6}{*}{\ParamBox{0.11\textwidth}{80}{50}{5}{0.05}{0.10}{0.6}{clean}} & T1 & 0.159 & 0.022 & -0.118 & 0.309 & 0.033 & 0.125 & 1.000 & 1.000 & 1.000 & 2.416 & 0.012 & 2.191 & 2.417 & 2.565 \\
& T2 & 0.397 & 0.035 & -0.195 & 0.479 & 0.043 & 0.200 & 1.000 & 1.000 & 1.000 & 2.561 & 0.016 & 2.266 & 2.573 & 2.733 \\
& T3 & 0.367 & 0.034 & -0.187 & 0.443 & 0.041 & 0.192 & 1.000 & 1.000 & 1.000 & 2.561 & 0.016 & 2.309 & 2.566 & 2.768 \\
& T4 & 0.397 & 0.032 & -0.189 & 0.473 & 0.040 & 0.194 & 1.000 & 1.000 & 0.980 & 2.559 & 0.013 & 2.381 & 2.546 & 2.758 \\
& T5 & 0.459 & 0.035 & -0.210 & 0.530 & 0.043 & 0.217 & 1.000 & 1.000 & 1.000 & 2.660 & 0.030 & 2.368 & 2.657 & 2.994 \\
& T6 & 0.440 & 0.029 & -0.191 & 0.513 & 0.040 & 0.196 & 1.000 & 1.000 & 0.980 & 2.550 & 0.013 & 2.357 & 2.551 & 2.756 \\
\addlinespace[2pt]
\multicolumn{16}{@{}l@{}}{\textbf{Sparse regression (EN)}} \\
\multirow{6}{*}{\ParamBox{0.11\textwidth}{80}{50}{5}{0.05}{0.10}{0.6}{{ext.}}} & T1 & -0.181 & 0.101 & -0.171 & 0.263 & 0.104 & 0.172 & 1.000 & 1.000 & 1.000 & 1.798 & 0.010 & 1.581 & 1.795 & 1.985 \\
& T2 & -0.156 & 0.125 & -0.223 & 0.267 & 0.128 & 0.225 & 1.000 & 0.980 & 0.640 & 1.941 & 0.014 & 1.763 & 1.939 & 2.148 \\
& T3 & -0.166 & 0.125 & -0.221 & 0.265 & 0.128 & 0.222 & 1.000 & 1.000 & 0.580 & 1.916 & 0.012 & 1.675 & 1.913 & 2.151 \\
& T4 & -0.260 & 0.138 & -0.230 & 0.373 & 0.143 & 0.232 & 1.000 & 0.920 & 0.500 & 1.992 & 0.021 & 1.725 & 1.974 & 2.238 \\
& T5 & -0.196 & 0.144 & -0.252 & 0.317 & 0.148 & 0.255 & 1.000 & 0.960 & 0.620 & 2.110 & 0.033 & 1.832 & 2.072 & 2.497 \\
& T6 & -0.189 & 0.133 & -0.232 & 0.283 & 0.137 & 0.234 & 1.000 & 0.980 & 0.500 & 1.953 & 0.015 & 1.733 & 1.943 & 2.164 \\
\addlinespace[2pt]
\cmidrule(lr){1-16}
\multicolumn{16}{@{}l@{}}{\textbf{Linear regression (OLS)}} \\
\multirow{6}{*}{\ParamBox{0.11\textwidth}{80}{50}{5}{0.05}{0.30}{0.6}{clean}} & T1 & -0.359 & -0.002 & 0.040 & 0.650 & 0.045 & 0.113 & 1.000 & 1.000 & 0.980 & 2.093 & 0.045 & 1.653 & 2.103 & 2.461 \\
& T2 & 0.410 & 0.035 & -0.198 & 0.647 & 0.053 & 0.223 & 1.000 & 1.000 & 0.960 & 2.551 & 0.052 & 2.058 & 2.546 & 2.938 \\
& T3 & 0.357 & 0.038 & -0.192 & 0.647 & 0.059 & 0.217 & 1.000 & 1.000 & 0.900 & 2.521 & 0.061 & 2.060 & 2.514 & 2.973 \\
& T4 & 0.386 & 0.040 & -0.204 & 0.579 & 0.053 & 0.230 & 1.000 & 1.000 & 0.860 & 2.620 & 0.063 & 2.171 & 2.625 & 3.012 \\
& T5 & 0.540 & 0.049 & -0.255 & 0.787 & 0.062 & 0.280 & 1.000 & 1.000 & 0.840 & 2.900 & 0.072 & 2.284 & 2.934 & 3.339 \\
& T6 & 0.380 & 0.034 & -0.188 & 0.606 & 0.051 & 0.215 & 1.000 & 1.000 & 0.920 & 2.552 & 0.058 & 1.993 & 2.547 & 2.943 \\
\addlinespace[2pt]
\multicolumn{16}{@{}l@{}}{\textbf{Sparse regression (EN)}} \\
\multirow{6}{*}{\ParamBox{0.11\textwidth}{80}{50}{5}{0.05}{0.30}{0.6}{{ext.}}} & T1 & -0.328 & 0.052 & -0.044 & 0.563 & 0.075 & 0.069 & 0.980 & 0.980 & 1.000 & 1.533 & 0.032 & 1.236 & 1.535 & 1.853 \\
& T2 & -0.236 & 0.133 & -0.221 & 0.521 & 0.145 & 0.228 & 1.000 & 0.900 & 0.920 & 2.014 & 0.053 & 1.623 & 2.038 & 2.431 \\
& T3 & -0.344 & 0.138 & -0.209 & 0.559 & 0.148 & 0.215 & 1.000 & 0.920 & 0.840 & 1.955 & 0.045 & 1.585 & 1.961 & 2.340 \\
& T4 & -0.707 & 0.190 & -0.242 & 0.939 & 0.201 & 0.250 & 0.940 & 0.660 & 0.660 & 2.199 & 0.083 & 1.670 & 2.153 & 2.734 \\
& T5 & -0.495 & 0.215 & -0.334 & 0.807 & 0.228 & 0.342 & 0.960 & 0.660 & 0.540 & 2.592 & 0.097 & 2.059 & 2.604 & 3.260 \\
& T6 & -0.446 & 0.157 & -0.228 & 0.740 & 0.170 & 0.236 & 0.960 & 0.800 & 0.780 & 2.077 & 0.089 & 1.632 & 2.046 & 2.762 \\
\addlinespace[2pt]
\cmidrule(lr){1-16}
\multicolumn{16}{@{}l@{}}{\textbf{Linear regression (OLS)}} \\
\multirow{6}{*}{\ParamBox{0.11\textwidth}{80}{50}{5}{0.10}{0.05}{0.6}{clean}} & T1 & 0.273 & 0.030 & -0.159 & 0.302 & 0.032 & 0.161 & 1.000 & 1.000 & 1.000 & 2.493 & 0.009 & 2.339 & 2.490 & 2.673 \\
& T2 & 0.370 & 0.037 & -0.192 & 0.390 & 0.038 & 0.194 & 1.000 & 1.000 & 1.000 & 2.560 & 0.012 & 2.379 & 2.537 & 2.755 \\
& T3 & 0.373 & 0.036 & -0.192 & 0.397 & 0.038 & 0.193 & 1.000 & 1.000 & 1.000 & 2.564 & 0.010 & 2.396 & 2.576 & 2.755 \\
& T4 & 0.390 & 0.036 & -0.196 & 0.417 & 0.038 & 0.198 & 1.000 & 1.000 & 1.000 & 2.566 & 0.009 & 2.381 & 2.565 & 2.752 \\
& T5 & 0.395 & 0.036 & -0.198 & 0.426 & 0.039 & 0.200 & 1.000 & 1.000 & 1.000 & 2.589 & 0.013 & 2.405 & 2.587 & 2.846 \\
& T6 & 0.389 & 0.036 & -0.195 & 0.414 & 0.038 & 0.198 & 1.000 & 1.000 & 1.000 & 2.551 & 0.009 & 2.383 & 2.552 & 2.723 \\
\addlinespace[2pt]
\multicolumn{16}{@{}l@{}}{\textbf{Sparse regression (EN)}} \\
\multirow{6}{*}{\ParamBox{0.11\textwidth}{80}{50}{5}{0.10}{0.05}{0.6}{{ext.}}} & T1 & -0.205 & 0.114 & -0.191 & 0.222 & 0.115 & 0.192 & 1.000 & 1.000 & 0.960 & 1.755 & 0.006 & 1.645 & 1.743 & 1.911 \\
& T2 & -0.223 & 0.128 & -0.216 & 0.245 & 0.129 & 0.217 & 1.000 & 1.000 & 0.500 & 1.826 & 0.007 & 1.689 & 1.827 & 1.964 \\
& T3 & -0.219 & 0.128 & -0.216 & 0.238 & 0.129 & 0.216 & 1.000 & 1.000 & 0.560 & 1.817 & 0.007 & 1.704 & 1.811 & 1.998 \\
& T4 & -0.221 & 0.127 & -0.213 & 0.237 & 0.128 & 0.214 & 1.000 & 1.000 & 0.520 & 1.830 & 0.007 & 1.701 & 1.820 & 1.994 \\
& T5 & -0.227 & 0.132 & -0.222 & 0.249 & 0.134 & 0.223 & 1.000 & 1.000 & 0.580 & 1.896 & 0.012 & 1.745 & 1.885 & 2.193 \\
& T6 & -0.208 & 0.127 & -0.215 & 0.223 & 0.127 & 0.215 & 1.000 & 1.000 & 0.540 & 1.828 & 0.008 & 1.706 & 1.829 & 2.046 \\
\addlinespace[2pt]
\cmidrule(lr){1-16}
\multicolumn{16}{@{}l@{}}{\textbf{Linear regression (OLS)}} \\
\multirow{6}{*}{\ParamBox{0.11\textwidth}{80}{50}{5}{0.10}{0.10}{0.6}{clean}} & T1 & 0.147 & 0.022 & -0.115 & 0.301 & 0.036 & 0.125 & 1.000 & 1.000 & 1.000 & 2.397 & 0.017 & 2.171 & 2.403 & 2.667 \\
& T2 & 0.394 & 0.033 & -0.191 & 0.486 & 0.044 & 0.197 & 1.000 & 1.000 & 1.000 & 2.547 & 0.023 & 2.308 & 2.549 & 2.812 \\
& T3 & 0.383 & 0.032 & -0.185 & 0.449 & 0.040 & 0.191 & 1.000 & 1.000 & 1.000 & 2.523 & 0.021 & 2.253 & 2.518 & 2.770 \\
& T4 & 0.392 & 0.032 & -0.186 & 0.466 & 0.043 & 0.193 & 1.000 & 1.000 & 1.000 & 2.549 & 0.017 & 2.365 & 2.527 & 2.802 \\
& T5 & 0.421 & 0.035 & -0.199 & 0.516 & 0.045 & 0.209 & 1.000 & 1.000 & 1.000 & 2.615 & 0.029 & 2.367 & 2.599 & 3.067 \\
& T6 & 0.411 & 0.029 & -0.185 & 0.493 & 0.042 & 0.192 & 1.000 & 1.000 & 1.000 & 2.522 & 0.020 & 2.270 & 2.509 & 2.789 \\
\addlinespace[2pt]
\multicolumn{16}{@{}l@{}}{\textbf{Sparse regression (EN)}} \\
\multirow{6}{*}{\ParamBox{0.11\textwidth}{80}{50}{5}{0.10}{0.10}{0.6}{{ext.}}} & T1 & -0.183 & 0.099 & -0.166 & 0.248 & 0.102 & 0.168 & 1.000 & 1.000 & 1.000 & 1.698 & 0.013 & 1.455 & 1.687 & 1.901 \\
& T2 & -0.200 & 0.124 & -0.214 & 0.274 & 0.128 & 0.216 & 1.000 & 1.000 & 0.640 & 1.833 & 0.016 & 1.580 & 1.843 & 2.024 \\
& T3 & -0.234 & 0.128 & -0.214 & 0.288 & 0.131 & 0.216 & 1.000 & 1.000 & 0.660 & 1.817 & 0.015 & 1.561 & 1.813 & 2.019 \\
& T4 & -0.231 & 0.131 & -0.220 & 0.291 & 0.135 & 0.222 & 1.000 & 0.940 & 0.480 & 1.855 & 0.015 & 1.629 & 1.865 & 2.082 \\
& T5 & -0.232 & 0.142 & -0.241 & 0.284 & 0.144 & 0.243 & 1.000 & 1.000 & 0.460 & 1.982 & 0.022 & 1.727 & 1.993 & 2.237 \\
& T6 & -0.181 & 0.126 & -0.220 & 0.232 & 0.129 & 0.222 & 1.000 & 0.980 & 0.560 & 1.844 & 0.011 & 1.653 & 1.828 & 2.024 \\
\bottomrule
\end{tabular*}
\noindent\parbox{\textwidth}{\raggedleft
  \Nchips{80}
}
\endgroup
\end{table}

\begin{table}[htbp]
\centering
\caption{Models coefficients bias, RMSE, 95\% coverage, and out-of-sample predictive MSE across MI methods for OLS (clean) and EN (contaminated). Top block: \(\boldsymbol{n=80}\), \(n_{\text{sim}}=50\), \(\text{iter}=5\), \(\rho=0.6\), with \(P_{\text{ext}}=0.10\) and \(P_{\text{miss}}=0.30\); followed by an additional block with \(\boldsymbol{n=200}\), \(n_{\text{sim}}=50\), \(\text{iter}=5\), \(\rho=0.6\), \(P_{\text{ext}}=0.03\) and \(P_{\text{miss}}\in\{0.05,0.10,0.30\}\)}

\label{tab:n_80_200_graficos.3}

\begingroup
\scriptsize
\begin{tabular*}{\textwidth}{@{\extracolsep{\fill}} l l
ccc 
ccc 
ccc 
ccccc 
@{}}
\toprule
& & \multicolumn{3}{c}{Bias} & \multicolumn{3}{c}{RMSE} & \multicolumn{3}{c}{Coverage} & \multicolumn{5}{c}{Pred.~MSE (out-of-sample)} \\
\cmidrule(lr){3-5}\cmidrule(lr){6-8}\cmidrule(lr){9-11}\cmidrule(lr){12-16}
Setup & MI &
$\beta_0$ & $\beta_1$ & $\beta_2$ &
$\beta_0$ & $\beta_1$ & $\beta_2$ &
$\beta_0$ & $\beta_1$ & $\beta_2$ &
$\overline{X}$ & $\sigma^2$ & $Q_{2.5}$ & $Q_{50}$ & $Q_{97.5}$
 \\
\midrule
\multicolumn{16}{@{}l@{}}{\textbf{Linear regression (OLS)}} \\
\multirow{6}{*}{\ParamBox{0.11\textwidth}{80}{50}{5}{0.10}{0.30}{0.6}{clean}} & T1 & -0.368 & -0.004 & 0.047 & 0.672 & 0.046 & 0.101 & 0.980 & 1.000 & 1.000 & 2.112 & 0.036 & 1.694 & 2.126 & 2.410 \\
& T2 & 0.381 & 0.037 & -0.194 & 0.683 & 0.058 & 0.216 & 1.000 & 1.000 & 0.900 & 2.582 & 0.046 & 2.162 & 2.601 & 2.954 \\
& T3 & 0.328 & 0.037 & -0.183 & 0.577 & 0.054 & 0.202 & 1.000 & 1.000 & 0.960 & 2.540 & 0.064 & 1.964 & 2.561 & 2.953 \\
& T4 & 0.357 & 0.039 & -0.196 & 0.584 & 0.055 & 0.218 & 1.000 & 1.000 & 0.900 & 2.652 & 0.065 & 2.171 & 2.659 & 3.194 \\
& T5 & 0.560 & 0.044 & -0.249 & 0.753 & 0.056 & 0.269 & 1.000 & 1.000 & 0.900 & 2.896 & 0.074 & 2.494 & 2.876 & 3.415 \\
& T6 & 0.385 & 0.034 & -0.191 & 0.623 & 0.053 & 0.213 & 1.000 & 1.000 & 0.900 & 2.586 & 0.053 & 2.141 & 2.627 & 2.938 \\
\addlinespace[2pt]
\multicolumn{16}{@{}l@{}}{\textbf{Sparse regression (EN)}} \\
\multirow{6}{*}{\ParamBox{0.11\textwidth}{80}{50}{5}{0.10}{0.30}{0.6}{{ext.}}} & T1 & -0.193 & 0.040 & -0.047 & 0.391 & 0.064 & 0.071 & 0.980 & 0.980 & 1.000 & 1.452 & 0.030 & 1.154 & 1.443 & 1.791 \\
& T2 & -0.203 & 0.126 & -0.212 & 0.419 & 0.137 & 0.219 & 1.000 & 0.900 & 0.780 & 1.907 & 0.049 & 1.503 & 1.911 & 2.272 \\
& T3 & -0.305 & 0.131 & -0.202 & 0.467 & 0.140 & 0.209 & 1.000 & 0.880 & 0.800 & 1.864 & 0.045 & 1.462 & 1.846 & 2.226 \\
& T4 & -0.598 & 0.179 & -0.239 & 0.778 & 0.189 & 0.247 & 0.960 & 0.740 & 0.640 & 2.069 & 0.088 & 1.628 & 2.026 & 2.678 \\
& T5 & -0.442 & 0.201 & -0.312 & 0.671 & 0.211 & 0.319 & 0.960 & 0.860 & 0.620 & 2.418 & 0.090 & 1.895 & 2.385 & 3.030 \\
& T6 & -0.299 & 0.140 & -0.220 & 0.552 & 0.151 & 0.227 & 0.960 & 0.920 & 0.760 & 1.930 & 0.062 & 1.523 & 1.957 & 2.369 \\
\addlinespace[2pt]
\cmidrule(lr){1-16}
\multicolumn{16}{@{}l@{}}{\textbf{Linear regression (OLS)}} \\
\multirow{6}{*}{\ParamBox{0.11\textwidth}{200}{50}{5}{0.03}{0.05}{0.6}{clean}} & T1 & -0.050 & -0.011 & 0.034 & 0.128 & 0.015 & 0.036 & 1.000 & 1.000 & 1.000 & 2.089 & 0.004 & 1.930 & 2.100 & 2.179 \\
& T2 & 0.023 & -0.004 & 0.006 & 0.128 & 0.012 & 0.015 & 1.000 & 1.000 & 1.000 & 2.158 & 0.005 & 1.986 & 2.167 & 2.254 \\
& T3 & 0.019 & -0.004 & 0.005 & 0.132 & 0.012 & 0.016 & 1.000 & 1.000 & 1.000 & 2.161 & 0.005 & 2.001 & 2.169 & 2.251 \\
& T4 & 0.003 & -0.003 & 0.007 & 0.127 & 0.011 & 0.016 & 1.000 & 1.000 & 1.000 & 2.156 & 0.004 & 2.021 & 2.167 & 2.271 \\
& T5 & 0.057 & -0.005 & -0.001 & 0.163 & 0.015 & 0.015 & 1.000 & 1.000 & 1.000 & 2.198 & 0.005 & 2.045 & 2.201 & 2.317 \\
& T6 & 0.005 & -0.003 & 0.006 & 0.132 & 0.011 & 0.016 & 1.000 & 1.000 & 1.000 & 2.156 & 0.004 & 2.021 & 2.160 & 2.255 \\
\addlinespace[2pt]
\multicolumn{16}{@{}l@{}}{\textbf{Sparse regression (EN)}} \\
\multirow{6}{*}{\ParamBox{0.11\textwidth}{200}{50}{5}{0.03}{0.05}{0.6}{{ext.}}} & T1 & 0.874 & -0.047 & -0.027 & 0.878 & 0.049 & 0.030 & 0.960 & 1.000 & 1.000 & 2.129 & 0.003 & 2.046 & 2.128 & 2.224 \\
& T2 & 0.931 & -0.037 & -0.058 & 0.935 & 0.039 & 0.060 & 0.920 & 1.000 & 1.000 & 2.194 & 0.003 & 2.120 & 2.196 & 2.282 \\
& T3 & 0.909 & -0.035 & -0.058 & 0.914 & 0.037 & 0.060 & 0.940 & 1.000 & 1.000 & 2.192 & 0.003 & 2.112 & 2.185 & 2.287 \\
& T4 & 0.910 & -0.034 & -0.059 & 0.915 & 0.036 & 0.061 & 0.960 & 1.000 & 1.000 & 2.203 & 0.003 & 2.088 & 2.198 & 2.296 \\
& T5 & 0.948 & -0.035 & -0.065 & 0.954 & 0.037 & 0.067 & 0.920 & 1.000 & 1.000 & 2.219 & 0.004 & 2.116 & 2.215 & 2.338 \\
& T6 & 0.931 & -0.036 & -0.059 & 0.935 & 0.038 & 0.061 & 0.960 & 1.000 & 1.000 & 2.202 & 0.004 & 2.077 & 2.198 & 2.296 \\
\addlinespace[2pt]
\cmidrule(lr){1-16}
\multicolumn{16}{@{}l@{}}{\textbf{Linear regression (OLS)}} \\
\multirow{6}{*}{\ParamBox{0.11\textwidth}{200}{50}{5}{0.03}{0.10}{0.6}{clean}} & T1 & -0.143 & -0.018 & 0.065 & 0.224 & 0.022 & 0.068 & 1.000 & 1.000 & 1.000 & 2.010 & 0.006 & 1.833 & 2.022 & 2.142 \\
& T2 & 0.019 & -0.004 & 0.004 & 0.187 & 0.014 & 0.025 & 1.000 & 1.000 & 1.000 & 2.148 & 0.007 & 1.959 & 2.162 & 2.299 \\
& T3 & 0.013 & -0.004 & 0.006 & 0.166 & 0.013 & 0.022 & 1.000 & 1.000 & 1.000 & 2.150 & 0.007 & 1.977 & 2.166 & 2.290 \\
& T4 & -0.016 & -0.002 & 0.008 & 0.187 & 0.014 & 0.023 & 1.000 & 1.000 & 1.000 & 2.164 & 0.010 & 1.957 & 2.180 & 2.342 \\
& T5 & 0.052 & -0.003 & -0.004 & 0.213 & 0.016 & 0.025 & 1.000 & 1.000 & 1.000 & 2.229 & 0.009 & 2.097 & 2.227 & 2.405 \\
& T6 & -0.018 & -0.002 & 0.008 & 0.191 & 0.014 & 0.025 & 1.000 & 1.000 & 1.000 & 2.153 & 0.009 & 1.953 & 2.159 & 2.353 \\
\addlinespace[2pt]
\multicolumn{16}{@{}l@{}}{\textbf{Sparse regression (EN)}} \\
\multirow{6}{*}{\ParamBox{0.11\textwidth}{200}{50}{5}{0.03}{0.10}{0.6}{{ext.}}} & T1 & 0.872 & -0.064 & 0.007 & 0.884 & 0.066 & 0.019 & 0.840 & 1.000 & 1.000 & 2.054 & 0.009 & 1.886 & 2.056 & 2.249 \\
& T2 & 0.981 & -0.041 & -0.060 & 0.992 & 0.043 & 0.063 & 0.740 & 1.000 & 1.000 & 2.193 & 0.011 & 1.968 & 2.219 & 2.377 \\
& T3 & 0.949 & -0.039 & -0.057 & 0.961 & 0.042 & 0.060 & 0.820 & 1.000 & 1.000 & 2.187 & 0.011 & 2.005 & 2.198 & 2.375 \\
& T4 & 0.932 & -0.036 & -0.060 & 0.945 & 0.040 & 0.063 & 0.860 & 1.000 & 1.000 & 2.203 & 0.011 & 2.022 & 2.201 & 2.392 \\
& T5 & 0.991 & -0.037 & -0.068 & 1.008 & 0.041 & 0.072 & 0.680 & 1.000 & 1.000 & 2.246 & 0.012 & 2.041 & 2.242 & 2.414 \\
& T6 & 0.955 & -0.039 & -0.058 & 0.968 & 0.043 & 0.062 & 0.760 & 1.000 & 1.000 & 2.197 & 0.013 & 2.025 & 2.203 & 2.389 \\
\addlinespace[2pt]
\cmidrule(lr){1-16}
\multicolumn{16}{@{}l@{}}{\textbf{Linear regression (OLS)}} \\
\multirow{6}{*}{\ParamBox{0.11\textwidth}{200}{50}{5}{0.03}{0.30}{0.6}{clean}} & T1 & -0.514 & -0.045 & 0.194 & 0.628 & 0.055 & 0.198 & 0.840 & 0.940 & 0.100 & 1.707 & 0.015 & 1.515 & 1.702 & 1.987 \\
& T2 & -0.004 & 0.003 & -0.001 & 0.336 & 0.030 & 0.038 & 1.000 & 1.000 & 1.000 & 2.170 & 0.020 & 1.954 & 2.185 & 2.462 \\
& T3 & -0.009 & 0.002 & 0.002 & 0.372 & 0.031 & 0.044 & 1.000 & 1.000 & 1.000 & 2.162 & 0.021 & 1.961 & 2.140 & 2.491 \\
& T4 & -0.146 & 0.013 & 0.008 & 0.385 & 0.034 & 0.044 & 1.000 & 1.000 & 1.000 & 2.203 & 0.023 & 1.993 & 2.170 & 2.575 \\
& T5 & 0.096 & 0.013 & -0.043 & 0.391 & 0.038 & 0.063 & 1.000 & 1.000 & 1.000 & 2.454 & 0.021 & 2.240 & 2.437 & 2.759 \\
& T6 & -0.133 & 0.011 & 0.008 & 0.388 & 0.034 & 0.047 & 1.000 & 1.000 & 1.000 & 2.189 & 0.024 & 1.993 & 2.164 & 2.523 \\
\addlinespace[2pt]
\multicolumn{16}{@{}l@{}}{\textbf{Sparse regression (EN)}} \\
\multirow{6}{*}{\ParamBox{0.11\textwidth}{200}{50}{5}{0.03}{0.30}{0.6}{{ext.}}} & T1 & 0.654 & -0.119 & 0.151 & 0.735 & 0.123 & 0.155 & 0.800 & 0.440 & 0.700 & 1.761 & 0.022 & 1.513 & 1.763 & 2.056 \\
& T2 & 0.962 & -0.040 & -0.062 & 1.012 & 0.050 & 0.075 & 0.800 & 1.000 & 1.000 & 2.217 & 0.025 & 1.903 & 2.216 & 2.526 \\
& T3 & 0.943 & -0.038 & -0.061 & 0.987 & 0.048 & 0.070 & 0.860 & 1.000 & 1.000 & 2.193 & 0.027 & 1.880 & 2.201 & 2.518 \\
& T4 & 0.790 & -0.021 & -0.066 & 0.872 & 0.042 & 0.076 & 0.880 & 1.000 & 1.000 & 2.287 & 0.034 & 1.962 & 2.275 & 2.597 \\
& T5 & 0.939 & -0.017 & -0.101 & 1.003 & 0.040 & 0.109 & 0.840 & 1.000 & 1.000 & 2.418 & 0.028 & 2.172 & 2.424 & 2.657 \\
& T6 & 0.910 & -0.031 & -0.067 & 0.980 & 0.047 & 0.078 & 0.840 & 1.000 & 1.000 & 2.251 & 0.034 & 1.922 & 2.265 & 2.540 \\
\bottomrule
\end{tabular*}
\noindent\parbox{\textwidth}{\raggedleft
  \Nchips{80}\\[-0.05em]
  \Nchips{200}%
}
\endgroup
\end{table}

\begin{table}[htbp]
\centering
\caption{Models coefficients bias, RMSE, 95\% coverage, and out-of-sample predictive MSE across MI methods for OLS (clean) and EN (contaminated), with \(\boldsymbol{n=200}\), \(n_{\text{sim}}=50\), \(\text{iter}=5\), \(\rho=0.6\), contamination \(P_{\text{ext}}\in\{0.05,0.10\}\) and missingness \(P_{\text{miss}}\in\{0.05,0.10,0.30\}\)}

\label{tab:n_80_200_graficos.4}

\begingroup
\scriptsize
\begin{tabular*}{\textwidth}{@{\extracolsep{\fill}} l l
ccc 
ccc 
ccc 
ccccc 
@{}}
\toprule
& & \multicolumn{3}{c}{Bias} & \multicolumn{3}{c}{RMSE} & \multicolumn{3}{c}{Coverage} & \multicolumn{5}{c}{Pred.~MSE (out-of-sample)} \\
\cmidrule(lr){3-5}\cmidrule(lr){6-8}\cmidrule(lr){9-11}\cmidrule(lr){12-16}
Setup & MI &
$\beta_0$ & $\beta_1$ & $\beta_2$ &
$\beta_0$ & $\beta_1$ & $\beta_2$ &
$\beta_0$ & $\beta_1$ & $\beta_2$ &
$\overline{X}$ & $\sigma^2$ & $Q_{2.5}$ & $Q_{50}$ & $Q_{97.5}$
 \\
\midrule
\multicolumn{16}{@{}l@{}}{\textbf{Linear regression (OLS)}} \\
\multirow{6}{*}{\ParamBox{0.11\textwidth}{200}{50}{5}{0.05}{0.05}{0.6}{clean}} & T1 & -0.051 & -0.011 & 0.034 & 0.141 & 0.015 & 0.036 & 1.000 & 1.000 & 1.000 & 2.088 & 0.004 & 1.951 & 2.103 & 2.182 \\
& T2 & 0.020 & -0.004 & 0.006 & 0.136 & 0.011 & 0.015 & 1.000 & 1.000 & 1.000 & 2.156 & 0.004 & 2.012 & 2.172 & 2.241 \\
& T3 & 0.018 & -0.003 & 0.004 & 0.143 & 0.012 & 0.015 & 1.000 & 1.000 & 1.000 & 2.159 & 0.005 & 2.014 & 2.168 & 2.269 \\
& T4 & 0.007 & -0.003 & 0.006 & 0.134 & 0.010 & 0.015 & 1.000 & 1.000 & 1.000 & 2.161 & 0.004 & 2.028 & 2.176 & 2.278 \\
& T5 & 0.053 & -0.005 & 0.000 & 0.160 & 0.014 & 0.014 & 1.000 & 1.000 & 1.000 & 2.198 & 0.006 & 2.075 & 2.204 & 2.365 \\
& T6 & 0.009 & -0.003 & 0.004 & 0.151 & 0.012 & 0.016 & 1.000 & 1.000 & 1.000 & 2.155 & 0.005 & 2.027 & 2.160 & 2.284 \\
\addlinespace[2pt]
\multicolumn{16}{@{}l@{}}{\textbf{Sparse regression (EN)}} \\
\multirow{6}{*}{\ParamBox{0.11\textwidth}{200}{50}{5}{0.05}{0.05}{0.6}{{ext.}}} & T1 & 0.753 & -0.030 & -0.034 & 0.757 & 0.031 & 0.036 & 1.000 & 1.000 & 1.000 & 2.087 & 0.003 & 1.989 & 2.087 & 2.176 \\
& T2 & 0.792 & -0.018 & -0.064 & 0.795 & 0.021 & 0.066 & 1.000 & 1.000 & 1.000 & 2.157 & 0.003 & 2.064 & 2.162 & 2.262 \\
& T3 & 0.773 & -0.016 & -0.064 & 0.779 & 0.020 & 0.065 & 1.000 & 1.000 & 1.000 & 2.154 & 0.002 & 2.049 & 2.158 & 2.234 \\
& T4 & 0.786 & -0.017 & -0.065 & 0.790 & 0.020 & 0.066 & 1.000 & 1.000 & 1.000 & 2.159 & 0.004 & 2.035 & 2.166 & 2.260 \\
& T5 & 0.814 & -0.017 & -0.070 & 0.819 & 0.020 & 0.071 & 0.940 & 1.000 & 1.000 & 2.178 & 0.004 & 2.083 & 2.180 & 2.313 \\
& T6 & 0.800 & -0.018 & -0.066 & 0.803 & 0.020 & 0.067 & 1.000 & 1.000 & 1.000 & 2.158 & 0.003 & 2.033 & 2.166 & 2.248 \\
\addlinespace[2pt]
\cmidrule(lr){1-16}
\multicolumn{16}{@{}l@{}}{\textbf{Linear regression (OLS)}} \\
\multirow{6}{*}{\ParamBox{0.11\textwidth}{200}{50}{5}{0.05}{0.10}{0.6}{clean}} & T1 & -0.155 & -0.018 & 0.067 & 0.239 & 0.022 & 0.071 & 1.000 & 1.000 & 1.000 & 2.005 & 0.006 & 1.832 & 2.021 & 2.119 \\
& T2 & 0.005 & -0.003 & 0.006 & 0.191 & 0.014 & 0.025 & 1.000 & 1.000 & 1.000 & 2.141 & 0.008 & 1.957 & 2.149 & 2.274 \\
& T3 & 0.012 & -0.004 & 0.007 & 0.187 & 0.015 & 0.025 & 1.000 & 1.000 & 1.000 & 2.148 & 0.008 & 1.937 & 2.159 & 2.280 \\
& T4 & -0.033 & -0.001 & 0.010 & 0.207 & 0.015 & 0.028 & 1.000 & 1.000 & 1.000 & 2.159 & 0.009 & 1.955 & 2.176 & 2.312 \\
& T5 & 0.053 & -0.004 & -0.002 & 0.202 & 0.016 & 0.025 & 1.000 & 1.000 & 1.000 & 2.231 & 0.011 & 2.005 & 2.220 & 2.412 \\
& T6 & -0.037 & -0.001 & 0.010 & 0.208 & 0.014 & 0.027 & 1.000 & 1.000 & 1.000 & 2.144 & 0.009 & 1.940 & 2.152 & 2.306 \\
\addlinespace[2pt]
\multicolumn{16}{@{}l@{}}{\textbf{Sparse regression (EN)}} \\
\multirow{6}{*}{\ParamBox{0.11\textwidth}{200}{50}{5}{0.05}{0.10}{0.6}{{ext.}}} & T1 & 0.755 & -0.047 & -0.000 & 0.764 & 0.049 & 0.017 & 0.960 & 1.000 & 1.000 & 2.014 & 0.009 & 1.839 & 2.017 & 2.208 \\
& T2 & 0.830 & -0.021 & -0.065 & 0.838 & 0.025 & 0.068 & 0.960 & 1.000 & 1.000 & 2.152 & 0.011 & 1.933 & 2.172 & 2.336 \\
& T3 & 0.800 & -0.019 & -0.062 & 0.810 & 0.024 & 0.065 & 0.960 & 1.000 & 1.000 & 2.146 & 0.011 & 1.963 & 2.155 & 2.339 \\
& T4 & 0.810 & -0.019 & -0.064 & 0.819 & 0.024 & 0.067 & 0.980 & 1.000 & 1.000 & 2.165 & 0.010 & 1.962 & 2.170 & 2.317 \\
& T5 & 0.867 & -0.020 & -0.074 & 0.879 & 0.025 & 0.076 & 0.860 & 1.000 & 1.000 & 2.214 & 0.010 & 2.018 & 2.187 & 2.412 \\
& T6 & 0.837 & -0.022 & -0.063 & 0.845 & 0.027 & 0.066 & 0.920 & 1.000 & 1.000 & 2.149 & 0.011 & 1.984 & 2.164 & 2.321 \\
\addlinespace[2pt]
\cmidrule(lr){1-16}
\multicolumn{16}{@{}l@{}}{\textbf{Linear regression (OLS)}} \\
\multirow{6}{*}{\ParamBox{0.11\textwidth}{200}{50}{5}{0.05}{0.30}{0.6}{clean}} & T1 & -0.518 & -0.045 & 0.196 & 0.645 & 0.055 & 0.200 & 0.840 & 0.940 & 0.140 & 1.698 & 0.016 & 1.502 & 1.707 & 1.972 \\
& T2 & -0.015 & 0.003 & 0.001 & 0.370 & 0.030 & 0.043 & 1.000 & 1.000 & 1.000 & 2.155 & 0.022 & 1.920 & 2.177 & 2.462 \\
& T3 & -0.031 & 0.002 & 0.006 & 0.396 & 0.030 & 0.048 & 1.000 & 1.000 & 1.000 & 2.153 & 0.025 & 1.943 & 2.126 & 2.491 \\
& T4 & -0.159 & 0.013 & 0.011 & 0.418 & 0.033 & 0.045 & 1.000 & 1.000 & 1.000 & 2.188 & 0.024 & 1.949 & 2.165 & 2.554 \\
& T5 & 0.103 & 0.010 & -0.038 & 0.418 & 0.036 & 0.064 & 1.000 & 1.000 & 1.000 & 2.442 & 0.023 & 2.205 & 2.433 & 2.742 \\
& T6 & -0.152 & 0.011 & 0.015 & 0.415 & 0.033 & 0.051 & 1.000 & 1.000 & 1.000 & 2.173 & 0.027 & 1.962 & 2.153 & 2.486 \\
\addlinespace[2pt]
\multicolumn{16}{@{}l@{}}{\textbf{Sparse regression (EN)}} \\
\multirow{6}{*}{\ParamBox{0.11\textwidth}{200}{50}{5}{0.05}{0.30}{0.6}{{ext.}}} & T1 & 0.604 & -0.106 & 0.141 & 0.665 & 0.111 & 0.145 & 0.800 & 0.560 & 0.820 & 1.724 & 0.022 & 1.467 & 1.710 & 2.030 \\
& T2 & 0.806 & -0.019 & -0.068 & 0.856 & 0.036 & 0.079 & 0.860 & 1.000 & 1.000 & 2.178 & 0.027 & 1.867 & 2.164 & 2.500 \\
& T3 & 0.784 & -0.017 & -0.067 & 0.823 & 0.033 & 0.074 & 0.900 & 1.000 & 1.000 & 2.155 & 0.027 & 1.854 & 2.154 & 2.467 \\
& T4 & 0.715 & -0.008 & -0.073 & 0.775 & 0.035 & 0.081 & 0.940 & 1.000 & 1.000 & 2.245 & 0.042 & 1.882 & 2.227 & 2.641 \\
& T5 & 0.879 & -0.004 & -0.112 & 0.922 & 0.031 & 0.117 & 0.900 & 1.000 & 1.000 & 2.383 & 0.027 & 2.051 & 2.383 & 2.660 \\
& T6 & 0.836 & -0.021 & -0.069 & 0.884 & 0.039 & 0.079 & 0.820 & 1.000 & 1.000 & 2.202 & 0.031 & 1.897 & 2.192 & 2.531 \\
\addlinespace[2pt]
\cmidrule(lr){1-16}
\multicolumn{16}{@{}l@{}}{\textbf{Linear regression (OLS)}} \\
\multirow{6}{*}{\ParamBox{0.11\textwidth}{200}{50}{5}{0.10}{0.05}{0.6}{clean}} & T1 & -0.057 & -0.011 & 0.034 & 0.145 & 0.015 & 0.036 & 1.000 & 1.000 & 1.000 & 2.086 & 0.003 & 1.988 & 2.084 & 2.193 \\
& T2 & 0.018 & -0.004 & 0.006 & 0.143 & 0.012 & 0.016 & 1.000 & 1.000 & 1.000 & 2.156 & 0.004 & 2.064 & 2.150 & 2.271 \\
& T3 & 0.011 & -0.003 & 0.005 & 0.147 & 0.013 & 0.016 & 1.000 & 1.000 & 1.000 & 2.155 & 0.004 & 2.038 & 2.155 & 2.285 \\
& T4 & 0.007 & -0.003 & 0.006 & 0.144 & 0.011 & 0.016 & 1.000 & 1.000 & 1.000 & 2.157 & 0.003 & 2.063 & 2.153 & 2.262 \\
& T5 & 0.052 & -0.003 & -0.003 & 0.178 & 0.014 & 0.017 & 1.000 & 1.000 & 1.000 & 2.204 & 0.006 & 2.089 & 2.196 & 2.376 \\
& T6 & -0.014 & -0.001 & 0.006 & 0.134 & 0.011 & 0.015 & 1.000 & 1.000 & 1.000 & 2.155 & 0.004 & 2.054 & 2.144 & 2.296 \\
\addlinespace[2pt]
\multicolumn{16}{@{}l@{}}{\textbf{Sparse regression (EN)}} \\
\multirow{6}{*}{\ParamBox{0.11\textwidth}{200}{50}{5}{0.10}{0.05}{0.6}{{ext.}}} & T1 & 0.663 & -0.004 & -0.074 & 0.666 & 0.010 & 0.075 & 0.980 & 1.000 & 1.000 & 1.936 & 0.002 & 1.850 & 1.937 & 2.011 \\
& T2 & 0.670 & 0.011 & -0.104 & 0.672 & 0.015 & 0.105 & 1.000 & 1.000 & 1.000 & 1.999 & 0.002 & 1.915 & 2.001 & 2.092 \\
& T3 & 0.671 & 0.010 & -0.103 & 0.674 & 0.014 & 0.104 & 1.000 & 1.000 & 1.000 & 1.996 & 0.002 & 1.907 & 1.995 & 2.075 \\
& T4 & 0.685 & 0.010 & -0.105 & 0.687 & 0.013 & 0.105 & 1.000 & 1.000 & 1.000 & 2.008 & 0.002 & 1.903 & 2.005 & 2.104 \\
& T5 & 0.694 & 0.011 & -0.108 & 0.697 & 0.015 & 0.109 & 0.960 & 1.000 & 1.000 & 2.014 & 0.003 & 1.925 & 2.012 & 2.103 \\
& T6 & 0.682 & 0.011 & -0.106 & 0.684 & 0.014 & 0.106 & 1.000 & 1.000 & 1.000 & 2.008 & 0.002 & 1.890 & 2.006 & 2.081 \\
\bottomrule
\end{tabular*}
\noindent\parbox{\textwidth}{\raggedleft
  \Nchips{200}%
}
\endgroup
\end{table}

\begin{table}[htbp]
\centering
\caption{Models coefficients bias, RMSE, 95\% coverage, and out-of-sample predictive MSE across MI methods for OLS (clean) and EN (contaminated). First block: \(\boldsymbol{n=200}\), \(n_{\text{sim}}=50\), \(\text{iter}=5\), \(\rho=0.6\), with \(P_{\text{ext}}=0.10\) and \(P_{\text{miss}}\in\{0.10,0.30\}\); followed by an additional block with \(\boldsymbol{n=80}\), \(n_{\text{sim}}=300\), \(\text{iter}=5\), \(\rho=0.6\), \(P_{\text{ext}}=0.04\) and \(P_{\text{miss}}\in\{0.10,0.25\}\)}

\label{tab:n_80_200_graficos.5}

\begingroup
\scriptsize
\begin{tabular*}{\textwidth}{@{\extracolsep{\fill}} l l
ccc 
ccc 
ccc 
ccccc 
@{}}
\toprule
& & \multicolumn{3}{c}{Bias} & \multicolumn{3}{c}{RMSE} & \multicolumn{3}{c}{Coverage} & \multicolumn{5}{c}{Pred.~MSE (out-of-sample)} \\
\cmidrule(lr){3-5}\cmidrule(lr){6-8}\cmidrule(lr){9-11}\cmidrule(lr){12-16}
Setup & MI &
$\beta_0$ & $\beta_1$ & $\beta_2$ &
$\beta_0$ & $\beta_1$ & $\beta_2$ &
$\beta_0$ & $\beta_1$ & $\beta_2$ &
$\overline{X}$ & $\sigma^2$ & $Q_{2.5}$ & $Q_{50}$ & $Q_{97.5}$
 \\
\midrule
\multicolumn{16}{@{}l@{}}{\textbf{Linear regression (OLS)}} \\
\multirow{6}{*}{\ParamBox{0.11\textwidth}{200}{50}{5}{0.10}{0.10}{0.6}{clean}} & T1 & -0.127 & -0.018 & 0.062 & 0.202 & 0.022 & 0.065 & 1.000 & 1.000 & 1.000 & 2.008 & 0.006 & 1.840 & 2.028 & 2.153 \\
& T2 & 0.022 & -0.004 & 0.003 & 0.160 & 0.013 & 0.021 & 1.000 & 1.000 & 1.000 & 2.159 & 0.008 & 2.016 & 2.167 & 2.333 \\
& T3 & 0.034 & -0.004 & 0.003 & 0.175 & 0.014 & 0.021 & 1.000 & 1.000 & 1.000 & 2.146 & 0.008 & 1.976 & 2.157 & 2.285 \\
& T4 & 0.010 & -0.003 & 0.006 & 0.177 & 0.015 & 0.023 & 1.000 & 1.000 & 1.000 & 2.152 & 0.007 & 1.975 & 2.158 & 2.321 \\
& T5 & 0.072 & -0.004 & -0.006 & 0.189 & 0.017 & 0.026 & 1.000 & 1.000 & 1.000 & 2.221 & 0.010 & 2.053 & 2.217 & 2.381 \\
& T6 & 0.011 & -0.002 & 0.004 & 0.174 & 0.014 & 0.023 & 1.000 & 1.000 & 1.000 & 2.150 & 0.009 & 1.989 & 2.155 & 2.340 \\
\addlinespace[2pt]
\multicolumn{16}{@{}l@{}}{\textbf{Sparse regression (EN)}} \\
\multirow{6}{*}{\ParamBox{0.11\textwidth}{200}{50}{5}{0.10}{0.10}{0.6}{{ext.}}} & T1 & 0.676 & -0.022 & -0.041 & 0.684 & 0.026 & 0.044 & 0.840 & 1.000 & 1.000 & 1.871 & 0.007 & 1.677 & 1.878 & 2.015 \\
& T2 & 0.701 & 0.007 & -0.103 & 0.711 & 0.017 & 0.104 & 0.880 & 1.000 & 1.000 & 2.003 & 0.008 & 1.830 & 1.998 & 2.141 \\
& T3 & 0.690 & 0.007 & -0.100 & 0.698 & 0.017 & 0.102 & 0.920 & 1.000 & 1.000 & 1.991 & 0.009 & 1.802 & 2.006 & 2.153 \\
& T4 & 0.715 & 0.005 & -0.100 & 0.721 & 0.015 & 0.102 & 0.900 & 1.000 & 1.000 & 2.004 & 0.008 & 1.849 & 2.005 & 2.160 \\
& T5 & 0.730 & 0.007 & -0.108 & 0.738 & 0.018 & 0.110 & 0.880 & 1.000 & 1.000 & 2.052 & 0.008 & 1.871 & 2.048 & 2.201 \\
& T6 & 0.707 & 0.006 & -0.101 & 0.713 & 0.016 & 0.102 & 0.920 & 1.000 & 1.000 & 2.008 & 0.007 & 1.853 & 2.006 & 2.155 \\
\addlinespace[2pt]
\cmidrule(lr){1-16}
\multicolumn{16}{@{}l@{}}{\textbf{Linear regression (OLS)}} \\
\multirow{6}{*}{\ParamBox{0.11\textwidth}{200}{50}{5}{0.10}{0.30}{0.6}{clean}} & T1 & -0.567 & -0.041 & 0.196 & 0.693 & 0.052 & 0.201 & 0.780 & 0.960 & 0.180 & 1.683 & 0.017 & 1.464 & 1.679 & 1.959 \\
& T2 & -0.050 & 0.006 & 0.001 & 0.384 & 0.029 & 0.043 & 1.000 & 1.000 & 1.000 & 2.131 & 0.022 & 1.831 & 2.147 & 2.436 \\
& T3 & -0.104 & 0.007 & 0.009 & 0.405 & 0.029 & 0.053 & 1.000 & 1.000 & 1.000 & 2.132 & 0.025 & 1.878 & 2.113 & 2.480 \\
& T4 & -0.143 & 0.013 & 0.007 & 0.442 & 0.033 & 0.045 & 1.000 & 1.000 & 1.000 & 2.174 & 0.023 & 1.940 & 2.143 & 2.468 \\
& T5 & 0.119 & 0.008 & -0.036 & 0.431 & 0.034 & 0.064 & 1.000 & 1.000 & 1.000 & 2.433 & 0.033 & 2.149 & 2.427 & 2.797 \\
& T6 & -0.175 & 0.015 & 0.008 & 0.448 & 0.036 & 0.048 & 1.000 & 1.000 & 1.000 & 2.152 & 0.026 & 1.878 & 2.171 & 2.457 \\
\addlinespace[2pt]
\multicolumn{16}{@{}l@{}}{\textbf{Sparse regression (EN)}} \\
\multirow{6}{*}{\ParamBox{0.11\textwidth}{200}{50}{5}{0.10}{0.30}{0.6}{{ext.}}} & T1 & 0.637 & -0.088 & 0.092 & 0.665 & 0.092 & 0.098 & 0.640 & 0.720 & 0.980 & 1.602 & 0.019 & 1.365 & 1.604 & 1.844 \\
& T2 & 0.655 & 0.013 & -0.110 & 0.685 & 0.030 & 0.117 & 0.900 & 1.000 & 0.960 & 2.027 & 0.024 & 1.714 & 2.022 & 2.313 \\
& T3 & 0.667 & 0.011 & -0.107 & 0.693 & 0.027 & 0.112 & 0.900 & 1.000 & 1.000 & 2.001 & 0.025 & 1.752 & 1.981 & 2.314 \\
& T4 & 0.751 & 0.002 & -0.106 & 0.771 & 0.024 & 0.111 & 0.860 & 1.000 & 0.980 & 2.052 & 0.027 & 1.759 & 2.059 & 2.335 \\
& T5 & 0.764 & 0.018 & -0.141 & 0.787 & 0.032 & 0.146 & 0.800 & 1.000 & 0.940 & 2.190 & 0.022 & 1.963 & 2.180 & 2.465 \\
& T6 & 0.731 & 0.006 & -0.109 & 0.751 & 0.030 & 0.116 & 0.840 & 1.000 & 0.980 & 2.059 & 0.030 & 1.801 & 2.066 & 2.374 \\
\addlinespace[2pt]
\cmidrule(lr){1-16}
\multicolumn{16}{@{}l@{}}{\textbf{Linear regression (OLS)}} \\
\multirow{6}{*}{\ParamBox{0.11\textwidth}{80}{300}{5}{0.04}{0.10}{0.6}{clean}} & T1 & 0.200 & 0.021 & -0.125 & 0.305 & 0.030 & 0.131 & 1.000 & 1.000 & 1.000 & 2.438 & 0.018 & 2.188 & 2.439 & 2.725 \\
& T2 & 0.421 & 0.034 & -0.199 & 0.484 & 0.041 & 0.203 & 1.000 & 1.000 & 1.000 & 2.580 & 0.020 & 2.321 & 2.578 & 2.869 \\
& T3 & 0.406 & 0.033 & -0.193 & 0.472 & 0.039 & 0.197 & 1.000 & 1.000 & 1.000 & 2.565 & 0.021 & 2.286 & 2.558 & 2.862 \\
& T4 & 0.407 & 0.033 & -0.193 & 0.463 & 0.039 & 0.198 & 1.000 & 1.000 & 1.000 & 2.584 & 0.018 & 2.353 & 2.577 & 2.871 \\
& T5 & 0.475 & 0.035 & -0.212 & 0.533 & 0.041 & 0.217 & 1.000 & 1.000 & 1.000 & 2.666 & 0.025 & 2.367 & 2.656 & 3.024 \\
& T6 & 0.414 & 0.033 & -0.195 & 0.465 & 0.039 & 0.201 & 1.000 & 1.000 & 1.000 & 2.575 & 0.018 & 2.350 & 2.574 & 2.877 \\
\addlinespace[2pt]
\multicolumn{16}{@{}l@{}}{\textbf{Sparse regression (EN)}} \\
\multirow{6}{*}{\ParamBox{0.11\textwidth}{80}{300}{5}{0.04}{0.10}{0.6}{{ext.}}} & T1 & -0.285 & 0.113 & -0.173 & 0.346 & 0.115 & 0.174 & 1.000 & 1.000 & 0.997 & 1.790 & 0.012 & 1.549 & 1.791 & 1.992 \\
& T2 & -0.244 & 0.136 & -0.227 & 0.319 & 0.139 & 0.228 & 1.000 & 0.983 & 0.610 & 1.926 & 0.016 & 1.680 & 1.925 & 2.194 \\
& T3 & -0.266 & 0.136 & -0.223 & 0.336 & 0.138 & 0.224 & 1.000 & 0.990 & 0.643 & 1.918 & 0.014 & 1.668 & 1.921 & 2.144 \\
& T4 & -0.352 & 0.149 & -0.231 & 0.437 & 0.152 & 0.232 & 1.000 & 0.933 & 0.563 & 1.992 & 0.021 & 1.730 & 1.979 & 2.279 \\
& T5 & -0.301 & 0.155 & -0.252 & 0.416 & 0.158 & 0.254 & 1.000 & 0.923 & 0.553 & 2.100 & 0.027 & 1.818 & 2.096 & 2.418 \\
& T6 & -0.287 & 0.141 & -0.228 & 0.355 & 0.143 & 0.229 & 1.000 & 0.967 & 0.567 & 1.945 & 0.015 & 1.691 & 1.946 & 2.203 \\
\addlinespace[2pt]
\cmidrule(lr){1-16}
\multicolumn{16}{@{}l@{}}{\textbf{Linear regression (OLS)}} \\
\multirow{6}{*}{\ParamBox{0.11\textwidth}{80}{300}{5}{0.04}{0.25}{0.6}{clean}} & T1 & -0.192 & 0.001 & -0.000 & 0.492 & 0.045 & 0.077 & 0.997 & 0.993 & 1.000 & 2.189 & 0.038 & 1.801 & 2.191 & 2.554 \\
& T2 & 0.405 & 0.033 & -0.193 & 0.607 & 0.056 & 0.209 & 1.000 & 1.000 & 0.977 & 2.562 & 0.049 & 2.130 & 2.557 & 2.983 \\
& T3 & 0.359 & 0.033 & -0.183 & 0.573 & 0.054 & 0.198 & 1.000 & 0.997 & 0.977 & 2.528 & 0.047 & 2.103 & 2.521 & 2.987 \\
& T4 & 0.369 & 0.036 & -0.193 & 0.527 & 0.051 & 0.209 & 1.000 & 1.000 & 0.950 & 2.600 & 0.049 & 2.225 & 2.582 & 3.031 \\
& T5 & 0.568 & 0.038 & -0.238 & 0.706 & 0.054 & 0.253 & 1.000 & 1.000 & 0.907 & 2.811 & 0.049 & 2.368 & 2.820 & 3.268 \\
& T6 & 0.381 & 0.035 & -0.192 & 0.548 & 0.051 & 0.209 & 1.000 & 1.000 & 0.947 & 2.574 & 0.043 & 2.193 & 2.568 & 2.979 \\
\addlinespace[2pt]
\multicolumn{16}{@{}l@{}}{\textbf{Sparse regression (EN)}} \\
\multirow{6}{*}{\ParamBox{0.11\textwidth}{80}{300}{5}{0.04}{0.25}{0.6}{{ext.}}} & T1 & -0.340 & 0.069 & -0.077 & 0.516 & 0.083 & 0.090 & 0.990 & 0.977 & 0.997 & 1.578 & 0.026 & 1.241 & 1.586 & 1.869 \\
& T2 & -0.251 & 0.134 & -0.221 & 0.479 & 0.142 & 0.226 & 0.997 & 0.910 & 0.693 & 1.948 & 0.047 & 1.505 & 1.953 & 2.339 \\
& T3 & -0.290 & 0.134 & -0.214 & 0.470 & 0.141 & 0.219 & 0.997 & 0.920 & 0.747 & 1.912 & 0.038 & 1.521 & 1.921 & 2.267 \\
& T4 & -0.570 & 0.171 & -0.234 & 0.743 & 0.180 & 0.240 & 0.997 & 0.733 & 0.663 & 2.122 & 0.066 & 1.644 & 2.108 & 2.678 \\
& T5 & -0.400 & 0.188 & -0.299 & 0.635 & 0.197 & 0.305 & 1.000 & 0.800 & 0.597 & 2.427 & 0.086 & 1.922 & 2.412 & 3.031 \\
& T6 & -0.403 & 0.152 & -0.227 & 0.581 & 0.160 & 0.233 & 1.000 & 0.903 & 0.720 & 2.008 & 0.053 & 1.600 & 2.000 & 2.498 \\
\bottomrule
\end{tabular*}
\noindent\parbox{\textwidth}{\raggedleft
  \Nchips{80}\\[-0.05em]
  \Nchips{200}%
}
\endgroup
\end{table}

\begin{table}[htbp]
\centering
\caption{Models coefficients bias, RMSE, 95\% coverage, and out-of-sample predictive MSE across MI methods for OLS (clean) and EN (contaminated). Top block: \(\boldsymbol{n=80}\), \(n_{\text{sim}}=300\), \(\text{iter}=5\), \(\rho=0.6\), with \(P_{\text{ext}}=0.15\) and \(P_{\text{miss}}\in\{0.10,0.25\}\); followed by an additional block with \(\boldsymbol{n=80}\), \(n_{\text{sim}}=50\), \(\text{iter}=10\), \(\rho=0.6\), \(P_{\text{ext}}=0.05\) and \(P_{\text{miss}}\in\{0.10,0.30\}\)}

\label{tab:n_80_200_graficos.6}

\begingroup
\scriptsize
\begin{tabular*}{\textwidth}{@{\extracolsep{\fill}} l l
ccc 
ccc 
ccc 
ccccc 
@{}}
\toprule
& & \multicolumn{3}{c}{Bias} & \multicolumn{3}{c}{RMSE} & \multicolumn{3}{c}{Coverage} & \multicolumn{5}{c}{Pred.~MSE (out-of-sample)} \\
\cmidrule(lr){3-5}\cmidrule(lr){6-8}\cmidrule(lr){9-11}\cmidrule(lr){12-16}
Setup & MI &
$\beta_0$ & $\beta_1$ & $\beta_2$ &
$\beta_0$ & $\beta_1$ & $\beta_2$ &
$\beta_0$ & $\beta_1$ & $\beta_2$ &
$\overline{X}$ & $\sigma^2$ & $Q_{2.5}$ & $Q_{50}$ & $Q_{97.5}$
 \\
\midrule
\multicolumn{16}{@{}l@{}}{\textbf{Linear regression (OLS)}} \\
\multirow{6}{*}{\ParamBox{0.11\textwidth}{80}{300}{5}{0.15}{0.10}{0.6}{clean}} & T1 & 0.200 & 0.021 & -0.126 & 0.306 & 0.030 & 0.131 & 1.000 & 1.000 & 1.000 & 2.442 & 0.019 & 2.174 & 2.436 & 2.752 \\
& T2 & 0.425 & 0.034 & -0.199 & 0.487 & 0.040 & 0.203 & 1.000 & 1.000 & 1.000 & 2.578 & 0.022 & 2.293 & 2.570 & 2.910 \\
& T3 & 0.405 & 0.033 & -0.193 & 0.472 & 0.040 & 0.198 & 1.000 & 1.000 & 1.000 & 2.568 & 0.021 & 2.271 & 2.566 & 2.871 \\
& T4 & 0.404 & 0.034 & -0.195 & 0.460 & 0.040 & 0.200 & 1.000 & 1.000 & 1.000 & 2.585 & 0.019 & 2.325 & 2.584 & 2.874 \\
& T5 & 0.475 & 0.035 & -0.213 & 0.531 & 0.042 & 0.218 & 1.000 & 1.000 & 0.987 & 2.673 & 0.027 & 2.406 & 2.660 & 3.022 \\
& T6 & 0.418 & 0.033 & -0.196 & 0.477 & 0.039 & 0.201 & 1.000 & 1.000 & 0.997 & 2.580 & 0.020 & 2.332 & 2.567 & 2.893 \\
\addlinespace[2pt]
\multicolumn{16}{@{}l@{}}{\textbf{Sparse regression (EN)}} \\
\multirow{6}{*}{\ParamBox{0.11\textwidth}{80}{300}{5}{0.15}{0.10}{0.6}{{ext.}}} & T1 & -0.031 & 0.086 & -0.164 & 0.143 & 0.089 & 0.166 & 1.000 & 1.000 & 0.960 & 1.566 & 0.009 & 1.366 & 1.575 & 1.739 \\
& T2 & -0.067 & 0.113 & -0.210 & 0.170 & 0.116 & 0.212 & 1.000 & 0.997 & 0.547 & 1.699 & 0.013 & 1.477 & 1.700 & 1.903 \\
& T3 & -0.081 & 0.114 & -0.209 & 0.175 & 0.116 & 0.211 & 1.000 & 0.997 & 0.537 & 1.680 & 0.011 & 1.439 & 1.684 & 1.867 \\
& T4 & -0.055 & 0.114 & -0.214 & 0.140 & 0.116 & 0.216 & 1.000 & 1.000 & 0.467 & 1.712 & 0.012 & 1.502 & 1.709 & 1.942 \\
& T5 & -0.074 & 0.122 & -0.227 & 0.160 & 0.125 & 0.229 & 1.000 & 0.997 & 0.547 & 1.809 & 0.022 & 1.559 & 1.786 & 2.150 \\
& T6 & -0.049 & 0.112 & -0.210 & 0.129 & 0.114 & 0.212 & 1.000 & 1.000 & 0.520 & 1.694 & 0.011 & 1.507 & 1.692 & 1.901 \\
\addlinespace[2pt]
\cmidrule(lr){1-16}
\multicolumn{16}{@{}l@{}}{\textbf{Linear regression (OLS)}} \\
\multirow{6}{*}{\ParamBox{0.11\textwidth}{80}{300}{5}{0.15}{0.25}{0.6}{clean}} & T1 & -0.183 & 0.000 & -0.002 & 0.485 & 0.046 & 0.075 & 0.997 & 0.993 & 1.000 & 2.187 & 0.041 & 1.805 & 2.173 & 2.572 \\
& T2 & 0.405 & 0.034 & -0.196 & 0.608 & 0.058 & 0.210 & 1.000 & 1.000 & 0.960 & 2.559 & 0.049 & 2.130 & 2.557 & 2.983 \\
& T3 & 0.353 & 0.034 & -0.184 & 0.581 & 0.057 & 0.200 & 1.000 & 0.997 & 0.963 & 2.527 & 0.049 & 2.077 & 2.522 & 2.939 \\
& T4 & 0.389 & 0.034 & -0.194 & 0.556 & 0.052 & 0.210 & 1.000 & 1.000 & 0.937 & 2.598 & 0.056 & 2.146 & 2.595 & 3.087 \\
& T5 & 0.564 & 0.039 & -0.238 & 0.705 & 0.056 & 0.252 & 1.000 & 1.000 & 0.930 & 2.815 & 0.052 & 2.391 & 2.807 & 3.284 \\
& T6 & 0.411 & 0.032 & -0.193 & 0.565 & 0.051 & 0.210 & 1.000 & 1.000 & 0.953 & 2.570 & 0.047 & 2.156 & 2.568 & 2.971 \\
\addlinespace[2pt]
\multicolumn{16}{@{}l@{}}{\textbf{Sparse regression (EN)}} \\
\multirow{6}{*}{\ParamBox{0.11\textwidth}{80}{300}{5}{0.15}{0.25}{0.6}{{ext.}}} & T1 & 0.039 & 0.037 & -0.082 & 0.261 & 0.057 & 0.095 & 1.000 & 1.000 & 0.993 & 1.369 & 0.023 & 1.062 & 1.381 & 1.650 \\
& T2 & -0.052 & 0.108 & -0.205 & 0.272 & 0.117 & 0.211 & 1.000 & 0.950 & 0.673 & 1.726 & 0.040 & 1.313 & 1.751 & 2.081 \\
& T3 & -0.098 & 0.111 & -0.201 & 0.293 & 0.120 & 0.207 & 1.000 & 0.957 & 0.680 & 1.679 & 0.035 & 1.309 & 1.689 & 2.040 \\
& T4 & -0.114 & 0.124 & -0.223 & 0.309 & 0.132 & 0.229 & 0.997 & 0.937 & 0.513 & 1.772 & 0.044 & 1.378 & 1.767 & 2.160 \\
& T5 & -0.141 & 0.148 & -0.266 & 0.324 & 0.157 & 0.272 & 1.000 & 0.920 & 0.633 & 2.080 & 0.058 & 1.628 & 2.075 & 2.590 \\
& T6 & -0.016 & 0.108 & -0.210 & 0.233 & 0.116 & 0.216 & 1.000 & 0.967 & 0.607 & 1.716 & 0.036 & 1.327 & 1.732 & 2.079 \\
\addlinespace[2pt]
\cmidrule(lr){1-16}
\multicolumn{16}{@{}l@{}}{\textbf{Linear regression (OLS)}} \\
\multirow{6}{*}{\ParamBox{0.11\textwidth}{80}{50}{10}{0.05}{0.10}{0.6}{clean}} & T1 & 0.159 & 0.022 & -0.118 & 0.309 & 0.033 & 0.125 & 1.000 & 1.000 & 1.000 & 2.416 & 0.012 & 2.191 & 2.417 & 2.565 \\
& T2 & 0.381 & 0.034 & -0.190 & 0.456 & 0.041 & 0.195 & 1.000 & 1.000 & 1.000 & 2.556 & 0.015 & 2.334 & 2.551 & 2.752 \\
& T3 & 0.382 & 0.032 & -0.186 & 0.466 & 0.041 & 0.190 & 1.000 & 1.000 & 1.000 & 2.556 & 0.014 & 2.339 & 2.548 & 2.736 \\
& T4 & 0.392 & 0.032 & -0.188 & 0.467 & 0.041 & 0.192 & 1.000 & 1.000 & 1.000 & 2.561 & 0.010 & 2.390 & 2.579 & 2.730 \\
& T5 & 0.448 & 0.034 & -0.204 & 0.515 & 0.042 & 0.209 & 1.000 & 1.000 & 1.000 & 2.632 & 0.017 & 2.397 & 2.627 & 2.914 \\
& T6 & 0.402 & 0.032 & -0.188 & 0.485 & 0.041 & 0.194 & 1.000 & 1.000 & 1.000 & 2.536 & 0.017 & 2.332 & 2.552 & 2.772 \\
\addlinespace[2pt]
\multicolumn{16}{@{}l@{}}{\textbf{Sparse regression (EN)}} \\
\multirow{6}{*}{\ParamBox{0.11\textwidth}{80}{50}{10}{0.05}{0.10}{0.6}{{ext.}}} & T1 & -0.181 & 0.101 & -0.171 & 0.263 & 0.104 & 0.172 & 1.000 & 1.000 & 1.000 & 1.798 & 0.010 & 1.581 & 1.795 & 1.985 \\
& T2 & -0.161 & 0.127 & -0.226 & 0.263 & 0.130 & 0.227 & 1.000 & 1.000 & 0.640 & 1.944 & 0.015 & 1.657 & 1.949 & 2.150 \\
& T3 & -0.173 & 0.127 & -0.223 & 0.250 & 0.129 & 0.225 & 1.000 & 1.000 & 0.620 & 1.922 & 0.013 & 1.692 & 1.943 & 2.087 \\
& T4 & -0.251 & 0.138 & -0.231 & 0.353 & 0.142 & 0.232 & 1.000 & 0.980 & 0.480 & 1.993 & 0.018 & 1.772 & 1.977 & 2.226 \\
& T5 & -0.192 & 0.143 & -0.252 & 0.288 & 0.147 & 0.254 & 1.000 & 0.980 & 0.460 & 2.106 & 0.032 & 1.782 & 2.092 & 2.471 \\
& T6 & -0.179 & 0.131 & -0.230 & 0.256 & 0.134 & 0.232 & 1.000 & 1.000 & 0.480 & 1.956 & 0.011 & 1.764 & 1.934 & 2.157 \\
\addlinespace[2pt]
\cmidrule(lr){1-16}
\multicolumn{16}{@{}l@{}}{\textbf{Linear regression (OLS)}} \\
\multirow{6}{*}{\ParamBox{0.11\textwidth}{80}{50}{10}{0.05}{0.30}{0.6}{clean}} & T1 & -0.359 & -0.002 & 0.040 & 0.650 & 0.045 & 0.113 & 1.000 & 1.000 & 0.980 & 2.093 & 0.045 & 1.653 & 2.103 & 2.461 \\
& T2 & 0.417 & 0.039 & -0.207 & 0.649 & 0.057 & 0.230 & 1.000 & 1.000 & 0.900 & 2.552 & 0.050 & 2.080 & 2.537 & 2.932 \\
& T3 & 0.322 & 0.038 & -0.184 & 0.584 & 0.057 & 0.208 & 1.000 & 1.000 & 0.940 & 2.499 & 0.052 & 2.035 & 2.519 & 2.886 \\
& T4 & 0.366 & 0.039 & -0.199 & 0.609 & 0.056 & 0.222 & 1.000 & 1.000 & 0.900 & 2.626 & 0.058 & 2.125 & 2.655 & 2.992 \\
& T5 & 0.613 & 0.044 & -0.259 & 0.796 & 0.057 & 0.278 & 1.000 & 1.000 & 0.840 & 2.899 & 0.053 & 2.417 & 2.927 & 3.273 \\
& T6 & 0.410 & 0.034 & -0.194 & 0.610 & 0.050 & 0.218 & 1.000 & 1.000 & 0.960 & 2.568 & 0.059 & 2.036 & 2.576 & 2.980 \\
\addlinespace[2pt]
\multicolumn{16}{@{}l@{}}{\textbf{Sparse regression (EN)}} \\
\multirow{6}{*}{\ParamBox{0.11\textwidth}{80}{50}{10}{0.05}{0.30}{0.6}{{ext.}}} & T1 & -0.328 & 0.052 & -0.044 & 0.563 & 0.075 & 0.069 & 0.980 & 0.980 & 1.000 & 1.533 & 0.032 & 1.236 & 1.535 & 1.853 \\
& T2 & -0.232 & 0.133 & -0.221 & 0.519 & 0.143 & 0.228 & 1.000 & 0.900 & 0.800 & 2.002 & 0.046 & 1.640 & 1.981 & 2.362 \\
& T3 & -0.304 & 0.137 & -0.214 & 0.513 & 0.144 & 0.219 & 1.000 & 0.900 & 0.800 & 1.941 & 0.043 & 1.609 & 1.948 & 2.308 \\
& T4 & -0.703 & 0.189 & -0.243 & 0.931 & 0.199 & 0.249 & 0.920 & 0.680 & 0.720 & 2.206 & 0.073 & 1.797 & 2.187 & 2.817 \\
& T5 & -0.472 & 0.212 & -0.331 & 0.772 & 0.224 & 0.337 & 1.000 & 0.680 & 0.380 & 2.576 & 0.088 & 2.010 & 2.563 & 3.229 \\
& T6 & -0.481 & 0.164 & -0.233 & 0.734 & 0.175 & 0.240 & 0.960 & 0.820 & 0.740 & 2.071 & 0.081 & 1.681 & 2.016 & 2.623 \\
\bottomrule
\end{tabular*}
\noindent\parbox{\textwidth}{\raggedleft
  \Nchips{80}
}
\endgroup
\end{table}

\begin{table}[htbp]
\centering
\caption{Models coefficients bias, RMSE, 95\% coverage, and out-of-sample predictive MSE across MI methods for OLS (clean) and EN (contaminated). Top block: \(\boldsymbol{n=80}\), \(n_{\text{sim}}=50\), \(\text{iter}=10\), \(\rho=0.6\), with \(P_{\text{ext}}=0.10\) and \(P_{\text{miss}}\in\{0.10,0.30\}\); followed by an additional block with \(\boldsymbol{n=200}\), \(n_{\text{sim}}=50\), \(\text{iter}=5\), \(\rho=0\), \(P_{\text{ext}}=0.05\) and \(P_{\text{miss}}\in\{0.10,0.30\}\)}

\label{tab:n_80_200_graficos.7}

\begingroup
\scriptsize
\begin{tabular*}{\textwidth}{@{\extracolsep{\fill}} l l
ccc 
ccc 
ccc 
ccccc 
@{}}
\toprule
& & \multicolumn{3}{c}{Bias} & \multicolumn{3}{c}{RMSE} & \multicolumn{3}{c}{Coverage} & \multicolumn{5}{c}{Pred.~MSE (out-of-sample)} \\
\cmidrule(lr){3-5}\cmidrule(lr){6-8}\cmidrule(lr){9-11}\cmidrule(lr){12-16}
Setup & MI &
$\beta_0$ & $\beta_1$ & $\beta_2$ &
$\beta_0$ & $\beta_1$ & $\beta_2$ &
$\beta_0$ & $\beta_1$ & $\beta_2$ &
$\overline{X}$ & $\sigma^2$ & $Q_{2.5}$ & $Q_{50}$ & $Q_{97.5}$
 \\
\midrule
\multicolumn{16}{@{}l@{}}{\textbf{Linear regression (OLS)}} \\
\multirow{6}{*}{\ParamBox{0.11\textwidth}{80}{50}{10}{0.10}{0.10}{0.6}{clean}} & T1 & 0.147 & 0.022 & -0.115 & 0.301 & 0.036 & 0.125 & 1.000 & 1.000 & 1.000 & 2.397 & 0.017 & 2.171 & 2.403 & 2.667 \\
& T2 & 0.366 & 0.035 & -0.188 & 0.452 & 0.045 & 0.194 & 1.000 & 1.000 & 1.000 & 2.536 & 0.023 & 2.247 & 2.549 & 2.807 \\
& T3 & 0.353 & 0.034 & -0.184 & 0.436 & 0.044 & 0.190 & 1.000 & 1.000 & 1.000 & 2.533 & 0.020 & 2.294 & 2.522 & 2.839 \\
& T4 & 0.359 & 0.033 & -0.183 & 0.442 & 0.043 & 0.190 & 1.000 & 1.000 & 1.000 & 2.534 & 0.018 & 2.318 & 2.522 & 2.805 \\
& T5 & 0.426 & 0.033 & -0.196 & 0.517 & 0.043 & 0.204 & 1.000 & 1.000 & 1.000 & 2.611 & 0.021 & 2.389 & 2.597 & 2.888 \\
& T6 & 0.399 & 0.032 & -0.187 & 0.461 & 0.041 & 0.194 & 1.000 & 1.000 & 1.000 & 2.531 & 0.019 & 2.311 & 2.517 & 2.799 \\
\addlinespace[2pt]
\multicolumn{16}{@{}l@{}}{\textbf{Sparse regression (EN)}} \\
\multirow{6}{*}{\ParamBox{0.11\textwidth}{80}{50}{10}{0.10}{0.10}{0.6}{{ext.}}} & T1 & -0.183 & 0.099 & -0.166 & 0.248 & 0.102 & 0.168 & 1.000 & 1.000 & 1.000 & 1.698 & 0.013 & 1.455 & 1.687 & 1.901 \\
& T2 & -0.203 & 0.127 & -0.217 & 0.266 & 0.129 & 0.219 & 1.000 & 0.980 & 0.540 & 1.837 & 0.017 & 1.517 & 1.832 & 2.049 \\
& T3 & -0.214 & 0.126 & -0.215 & 0.278 & 0.129 & 0.216 & 1.000 & 0.980 & 0.620 & 1.824 & 0.015 & 1.615 & 1.825 & 2.050 \\
& T4 & -0.238 & 0.132 & -0.220 & 0.297 & 0.135 & 0.222 & 1.000 & 0.940 & 0.400 & 1.852 & 0.016 & 1.656 & 1.824 & 2.111 \\
& T5 & -0.223 & 0.141 & -0.241 & 0.277 & 0.143 & 0.243 & 1.000 & 1.000 & 0.480 & 1.985 & 0.024 & 1.722 & 1.960 & 2.297 \\
& T6 & -0.171 & 0.125 & -0.218 & 0.228 & 0.127 & 0.220 & 1.000 & 1.000 & 0.460 & 1.842 & 0.012 & 1.644 & 1.836 & 2.035 \\
\addlinespace[2pt]
\cmidrule(lr){1-16}
\multicolumn{16}{@{}l@{}}{\textbf{Linear regression (OLS)}} \\
\multirow{6}{*}{\ParamBox{0.11\textwidth}{80}{50}{10}{0.10}{0.30}{0.6}{clean}} & T1 & -0.368 & -0.004 & 0.047 & 0.672 & 0.046 & 0.101 & 0.980 & 1.000 & 1.000 & 2.112 & 0.036 & 1.694 & 2.126 & 2.410 \\
& T2 & 0.403 & 0.039 & -0.202 & 0.645 & 0.057 & 0.219 & 1.000 & 1.000 & 0.960 & 2.600 & 0.053 & 2.220 & 2.596 & 3.013 \\
& T3 & 0.311 & 0.036 & -0.176 & 0.612 & 0.056 & 0.197 & 1.000 & 1.000 & 0.920 & 2.532 & 0.045 & 2.176 & 2.529 & 2.956 \\
& T4 & 0.354 & 0.039 & -0.194 & 0.565 & 0.055 & 0.215 & 1.000 & 1.000 & 0.940 & 2.637 & 0.065 & 2.105 & 2.646 & 3.053 \\
& T5 & 0.632 & 0.040 & -0.255 & 0.779 & 0.053 & 0.273 & 1.000 & 1.000 & 0.880 & 2.905 & 0.056 & 2.451 & 2.914 & 3.310 \\
& T6 & 0.368 & 0.037 & -0.193 & 0.592 & 0.055 & 0.216 & 1.000 & 1.000 & 0.920 & 2.600 & 0.048 & 2.134 & 2.669 & 2.953 \\
\addlinespace[2pt]
\multicolumn{16}{@{}l@{}}{\textbf{Sparse regression (EN)}} \\
\multirow{6}{*}{\ParamBox{0.11\textwidth}{80}{50}{10}{0.10}{0.30}{0.6}{{ext.}}} & T1 & -0.193 & 0.040 & -0.047 & 0.391 & 0.064 & 0.071 & 0.980 & 0.980 & 1.000 & 1.452 & 0.030 & 1.154 & 1.443 & 1.791 \\
& T2 & -0.200 & 0.127 & -0.213 & 0.380 & 0.135 & 0.220 & 1.000 & 0.940 & 0.740 & 1.918 & 0.043 & 1.531 & 1.929 & 2.246 \\
& T3 & -0.283 & 0.132 & -0.208 & 0.421 & 0.139 & 0.213 & 1.000 & 0.900 & 0.740 & 1.851 & 0.042 & 1.507 & 1.862 & 2.252 \\
& T4 & -0.580 & 0.176 & -0.240 & 0.766 & 0.186 & 0.246 & 0.960 & 0.780 & 0.540 & 2.083 & 0.063 & 1.653 & 2.065 & 2.577 \\
& T5 & -0.428 & 0.199 & -0.312 & 0.599 & 0.206 & 0.317 & 0.980 & 0.840 & 0.500 & 2.435 & 0.063 & 1.965 & 2.429 & 2.883 \\
& T6 & -0.293 & 0.140 & -0.220 & 0.503 & 0.149 & 0.227 & 0.960 & 0.920 & 0.740 & 1.924 & 0.049 & 1.569 & 1.919 & 2.362 \\
\addlinespace[2pt]
\cmidrule(lr){1-16}
\multicolumn{16}{@{}l@{}}{\textbf{Linear regression (OLS)}} \\
\multirow{6}{*}{\ParamBox{0.11\textwidth}{200}{50}{5}{0.05}{0.10}{0.0}{clean}} & T1 & -0.268 & 0.003 & 0.049 & 0.350 & 0.016 & 0.053 & 1.000 & 1.000 & 1.000 & 1.983 & 0.007 & 1.793 & 2.000 & 2.099 \\
& T2 & -0.007 & -0.001 & 0.004 & 0.238 & 0.017 & 0.021 & 1.000 & 1.000 & 1.000 & 2.141 & 0.008 & 1.938 & 2.144 & 2.280 \\
& T3 & -0.005 & -0.002 & 0.006 & 0.229 & 0.017 & 0.021 & 1.000 & 1.000 & 1.000 & 2.147 & 0.009 & 1.922 & 2.158 & 2.269 \\
& T4 & -0.033 & -0.000 & 0.009 & 0.261 & 0.019 & 0.024 & 1.000 & 1.000 & 1.000 & 2.142 & 0.008 & 1.980 & 2.144 & 2.293 \\
& T5 & 0.120 & -0.008 & -0.007 & 0.295 & 0.021 & 0.023 & 1.000 & 1.000 & 1.000 & 2.284 & 0.016 & 2.027 & 2.302 & 2.470 \\
& T6 & -0.047 & 0.001 & 0.007 & 0.272 & 0.018 & 0.025 & 1.000 & 1.000 & 1.000 & 2.148 & 0.009 & 1.927 & 2.158 & 2.291 \\
\addlinespace[2pt]
\multicolumn{16}{@{}l@{}}{\textbf{Sparse regression (EN)}} \\
\multirow{6}{*}{\ParamBox{0.11\textwidth}{200}{50}{5}{0.05}{0.10}{0.0}{{ext.}}} & T1 & 1.478 & -0.090 & -0.057 & 1.485 & 0.091 & 0.058 & 0.000 & 0.620 & 1.000 & 2.000 & 0.010 & 1.824 & 2.003 & 2.188 \\
& T2 & 1.587 & -0.079 & -0.099 & 1.595 & 0.081 & 0.100 & 0.000 & 0.940 & 1.000 & 2.163 & 0.011 & 1.949 & 2.175 & 2.368 \\
& T3 & 1.577 & -0.078 & -0.099 & 1.585 & 0.080 & 0.099 & 0.000 & 0.940 & 1.000 & 2.160 & 0.011 & 1.963 & 2.159 & 2.361 \\
& T4 & 1.664 & -0.088 & -0.096 & 1.670 & 0.089 & 0.097 & 0.000 & 0.880 & 0.980 & 2.175 & 0.011 & 1.982 & 2.180 & 2.365 \\
& T5 & 1.671 & -0.082 & -0.107 & 1.683 & 0.084 & 0.108 & 0.000 & 0.940 & 0.960 & 2.252 & 0.013 & 1.978 & 2.254 & 2.431 \\
& T6 & 1.698 & -0.090 & -0.100 & 1.701 & 0.090 & 0.101 & 0.000 & 0.860 & 0.980 & 2.170 & 0.012 & 1.965 & 2.175 & 2.383 \\
\addlinespace[2pt]
\cmidrule(lr){1-16}
\multicolumn{16}{@{}l@{}}{\textbf{Linear regression (OLS)}} \\
\multirow{6}{*}{\ParamBox{0.11\textwidth}{200}{50}{5}{0.05}{0.30}{0.0}{clean}} & T1 & -0.819 & 0.013 & 0.140 & 0.935 & 0.037 & 0.145 & 0.740 & 1.000 & 0.240 & 1.642 & 0.015 & 1.467 & 1.650 & 1.902 \\
& T2 & -0.019 & 0.004 & 0.001 & 0.453 & 0.035 & 0.036 & 1.000 & 1.000 & 1.000 & 2.155 & 0.023 & 1.928 & 2.169 & 2.473 \\
& T3 & -0.045 & 0.004 & 0.004 & 0.487 & 0.036 & 0.039 & 1.000 & 1.000 & 1.000 & 2.153 & 0.027 & 1.931 & 2.120 & 2.496 \\
& T4 & -0.066 & 0.004 & 0.009 & 0.493 & 0.036 & 0.042 & 1.000 & 1.000 & 1.000 & 2.186 & 0.029 & 1.935 & 2.184 & 2.549 \\
& T5 & 0.473 & -0.020 & -0.051 & 0.656 & 0.040 & 0.064 & 1.000 & 1.000 & 1.000 & 2.648 & 0.042 & 2.286 & 2.610 & 3.009 \\
& T6 & -0.050 & 0.005 & 0.005 & 0.426 & 0.034 & 0.037 & 1.000 & 1.000 & 1.000 & 2.156 & 0.032 & 1.897 & 2.176 & 2.592 \\
\addlinespace[2pt]
\multicolumn{16}{@{}l@{}}{\textbf{Sparse regression (EN)}} \\
\multirow{6}{*}{\ParamBox{0.11\textwidth}{200}{50}{5}{0.05}{0.30}{0.0}{{ext.}}} & T1 & 1.219 & -0.113 & 0.034 & 1.265 & 0.117 & 0.046 & 0.200 & 0.220 & 1.000 & 1.660 & 0.020 & 1.414 & 1.648 & 1.982 \\
& T2 & 1.572 & -0.078 & -0.103 & 1.603 & 0.084 & 0.107 & 0.220 & 0.940 & 0.880 & 2.174 & 0.034 & 1.856 & 2.140 & 2.543 \\
& T3 & 1.558 & -0.077 & -0.099 & 1.591 & 0.083 & 0.104 & 0.300 & 0.840 & 0.960 & 2.172 & 0.029 & 1.867 & 2.157 & 2.556 \\
& T4 & 1.638 & -0.090 & -0.091 & 1.668 & 0.095 & 0.096 & 0.100 & 0.820 & 0.880 & 2.233 & 0.043 & 1.875 & 2.229 & 2.613 \\
& T5 & 1.835 & -0.088 & -0.134 & 1.871 & 0.095 & 0.137 & 0.140 & 0.840 & 0.760 & 2.506 & 0.049 & 2.154 & 2.504 & 2.858 \\
& T6 & 1.843 & -0.106 & -0.098 & 1.865 & 0.111 & 0.102 & 0.080 & 0.600 & 0.860 & 2.194 & 0.034 & 1.848 & 2.205 & 2.547 \\
\bottomrule
\end{tabular*}
\noindent\parbox{\textwidth}{\raggedleft
  \Nchips{80}\\[-0.05em]
  \Nchips{200}%
}
\endgroup
\end{table}

\begin{table}[htbp]
\centering
\caption{Models coefficients bias, RMSE, 95\% coverage, and out-of-sample predictive MSE across MI methods for OLS (clean) and EN (contaminated). Top block: \(\boldsymbol{n=200}\), \(n_{\text{sim}}=50\), \(\text{iter}=5\), \(\rho=0\), with \(P_{\text{ext}}=0.10\) and \(P_{\text{miss}}\in\{0.10,0.30\}\); followed by an additional block with \(\boldsymbol{n=80}\), \(n_{\text{sim}}=1000\), \(\text{iter}=10\), \(\rho=0.6\), \(P_{\text{ext}}=0.30\) and \(P_{\text{miss}}\in\{0.10,0.30\}\)}

\label{tab:n_80_200_graficos.8}

\begingroup
\scriptsize
\begin{tabular*}{\textwidth}{@{\extracolsep{\fill}} l l
ccc 
ccc 
ccc 
ccccc 
@{}}
\toprule
& & \multicolumn{3}{c}{Bias} & \multicolumn{3}{c}{RMSE} & \multicolumn{3}{c}{Coverage} & \multicolumn{5}{c}{Pred.~MSE (out-of-sample)} \\
\cmidrule(lr){3-5}\cmidrule(lr){6-8}\cmidrule(lr){9-11}\cmidrule(lr){12-16}
Setup & MI &
$\beta_0$ & $\beta_1$ & $\beta_2$ &
$\beta_0$ & $\beta_1$ & $\beta_2$ &
$\beta_0$ & $\beta_1$ & $\beta_2$ &
$\overline{X}$ & $\sigma^2$ & $Q_{2.5}$ & $Q_{50}$ & $Q_{97.5}$
 \\
\midrule
\multicolumn{16}{@{}l@{}}{\textbf{Linear regression (OLS)}} \\
\multirow{6}{*}{\ParamBox{0.11\textwidth}{200}{50}{5}{0.10}{0.10}{0.0}{clean}} & T1 & -0.230 & 0.000 & 0.045 & 0.298 & 0.015 & 0.048 & 1.000 & 1.000 & 1.000 & 1.986 & 0.006 & 1.816 & 2.003 & 2.127 \\
& T2 & 0.009 & -0.002 & 0.003 & 0.199 & 0.015 & 0.018 & 1.000 & 1.000 & 1.000 & 2.158 & 0.008 & 2.012 & 2.175 & 2.326 \\
& T3 & 0.027 & -0.003 & 0.002 & 0.209 & 0.016 & 0.017 & 1.000 & 1.000 & 1.000 & 2.146 & 0.008 & 1.971 & 2.161 & 2.287 \\
& T4 & 0.002 & -0.001 & 0.002 & 0.229 & 0.018 & 0.019 & 1.000 & 1.000 & 1.000 & 2.156 & 0.010 & 1.946 & 2.159 & 2.339 \\
& T5 & 0.144 & -0.009 & -0.010 & 0.295 & 0.022 & 0.023 & 1.000 & 1.000 & 1.000 & 2.274 & 0.012 & 2.118 & 2.255 & 2.500 \\
& T6 & -0.006 & -0.001 & 0.003 & 0.218 & 0.017 & 0.017 & 1.000 & 1.000 & 1.000 & 2.150 & 0.009 & 1.947 & 2.150 & 2.312 \\
\addlinespace[2pt]
\multicolumn{16}{@{}l@{}}{\textbf{Sparse regression (EN)}} \\
\multirow{6}{*}{\ParamBox{0.11\textwidth}{200}{50}{5}{0.10}{0.10}{0.0}{{ext.}}} & T1 & 1.683 & -0.089 & -0.105 & 1.688 & 0.091 & 0.106 & 0.000 & 0.560 & 0.720 & 1.856 & 0.008 & 1.659 & 1.867 & 2.019 \\
& T2 & 1.738 & -0.074 & -0.144 & 1.743 & 0.076 & 0.145 & 0.000 & 0.940 & 0.100 & 2.006 & 0.011 & 1.758 & 2.005 & 2.191 \\
& T3 & 1.735 & -0.075 & -0.143 & 1.740 & 0.076 & 0.144 & 0.000 & 0.920 & 0.100 & 2.006 & 0.009 & 1.807 & 2.014 & 2.179 \\
& T4 & 1.796 & -0.082 & -0.140 & 1.799 & 0.083 & 0.141 & 0.000 & 0.840 & 0.080 & 2.018 & 0.009 & 1.831 & 2.020 & 2.163 \\
& T5 & 1.820 & -0.079 & -0.151 & 1.827 & 0.080 & 0.152 & 0.000 & 0.880 & 0.040 & 2.094 & 0.014 & 1.854 & 2.085 & 2.333 \\
& T6 & 1.830 & -0.085 & -0.140 & 1.833 & 0.087 & 0.141 & 0.000 & 0.800 & 0.100 & 2.028 & 0.010 & 1.843 & 2.041 & 2.189 \\
\addlinespace[2pt]
\cmidrule(lr){1-16}
\multicolumn{16}{@{}l@{}}{\textbf{Linear regression (OLS)}} \\
\multirow{6}{*}{\ParamBox{0.11\textwidth}{200}{50}{5}{0.10}{0.30}{0.0}{clean}} & T1 & -0.872 & 0.018 & 0.140 & 0.990 & 0.040 & 0.145 & 0.680 & 1.000 & 0.280 & 1.628 & 0.016 & 1.424 & 1.624 & 1.888 \\
& T2 & -0.053 & 0.007 & -0.000 & 0.472 & 0.035 & 0.035 & 1.000 & 1.000 & 1.000 & 2.134 & 0.024 & 1.820 & 2.140 & 2.469 \\
& T3 & -0.131 & 0.011 & 0.007 & 0.505 & 0.035 & 0.043 & 1.000 & 1.000 & 1.000 & 2.132 & 0.027 & 1.881 & 2.092 & 2.498 \\
& T4 & -0.080 & 0.006 & 0.007 & 0.504 & 0.036 & 0.040 & 1.000 & 1.000 & 1.000 & 2.168 & 0.027 & 1.906 & 2.182 & 2.489 \\
& T5 & 0.395 & -0.015 & -0.047 & 0.643 & 0.042 & 0.063 & 1.000 & 1.000 & 1.000 & 2.624 & 0.035 & 2.255 & 2.621 & 2.980 \\
& T6 & -0.112 & 0.010 & 0.007 & 0.484 & 0.038 & 0.037 & 1.000 & 1.000 & 1.000 & 2.125 & 0.028 & 1.859 & 2.102 & 2.406 \\
\addlinespace[2pt]
\multicolumn{16}{@{}l@{}}{\textbf{Sparse regression (EN)}} \\
\multirow{6}{*}{\ParamBox{0.11\textwidth}{200}{50}{5}{0.10}{0.30}{0.0}{{ext.}}} & T1 & 1.572 & -0.122 & -0.023 & 1.586 & 0.125 & 0.037 & 0.000 & 0.000 & 1.000 & 1.536 & 0.017 & 1.306 & 1.531 & 1.765 \\
& T2 & 1.705 & -0.071 & -0.148 & 1.720 & 0.076 & 0.152 & 0.000 & 0.960 & 0.380 & 2.017 & 0.027 & 1.773 & 1.994 & 2.352 \\
& T3 & 1.717 & -0.073 & -0.146 & 1.733 & 0.077 & 0.148 & 0.020 & 0.920 & 0.420 & 2.014 & 0.027 & 1.730 & 1.988 & 2.374 \\
& T4 & 1.911 & -0.100 & -0.133 & 1.918 & 0.102 & 0.135 & 0.000 & 0.620 & 0.420 & 2.047 & 0.024 & 1.806 & 2.034 & 2.323 \\
& T5 & 1.948 & -0.082 & -0.175 & 1.960 & 0.086 & 0.177 & 0.000 & 0.860 & 0.180 & 2.311 & 0.028 & 2.039 & 2.328 & 2.604 \\
& T6 & 1.976 & -0.104 & -0.136 & 1.983 & 0.107 & 0.139 & 0.000 & 0.500 & 0.460 & 2.072 & 0.032 & 1.747 & 2.065 & 2.432 \\
\addlinespace[2pt]
\cmidrule(lr){1-16}
\multicolumn{16}{@{}l@{}}{\textbf{Linear regression (OLS)}} \\
\multirow{6}{*}{\ParamBox{0.11\textwidth}{80}{1000}{10}{0.30}{0.10}{0.6}{clean}} & T1 & 0.185 & 0.022 & -0.125 & 0.301 & 0.032 & 0.131 & 1.000 & 1.000 & 1.000 & 2.435 & 0.019 & 2.172 & 2.431 & 2.713 \\
& T2 & 0.403 & 0.035 & -0.198 & 0.468 & 0.042 & 0.202 & 1.000 & 1.000 & 1.000 & 2.572 & 0.020 & 2.300 & 2.569 & 2.842 \\
& T3 & 0.391 & 0.034 & -0.193 & 0.460 & 0.041 & 0.198 & 1.000 & 1.000 & 1.000 & 2.562 & 0.019 & 2.289 & 2.560 & 2.828 \\
& T4 & 0.393 & 0.034 & -0.194 & 0.447 & 0.040 & 0.198 & 1.000 & 1.000 & 1.000 & 2.575 & 0.019 & 2.331 & 2.569 & 2.846 \\
& T5 & 0.460 & 0.036 & -0.211 & 0.515 & 0.042 & 0.216 & 1.000 & 1.000 & 0.999 & 2.657 & 0.022 & 2.400 & 2.653 & 2.953 \\
& T6 & 0.406 & 0.034 & -0.197 & 0.462 & 0.040 & 0.202 & 1.000 & 1.000 & 1.000 & 2.568 & 0.019 & 2.321 & 2.565 & 2.856 \\
\addlinespace[2pt]
\multicolumn{16}{@{}l@{}}{\textbf{Sparse regression (EN)}} \\
\multirow{6}{*}{\ParamBox{0.11\textwidth}{80}{1000}{10}{0.30}{0.10}{0.6}{{ext.}}} & T1 & 0.047 & 0.083 & -0.181 & 0.114 & 0.086 & 0.183 & 1.000 & 0.998 & 0.476 & 1.141 & 0.006 & 0.974 & 1.140 & 1.296 \\
& T2 & -0.025 & 0.110 & -0.220 & 0.118 & 0.112 & 0.222 & 1.000 & 0.991 & 0.228 & 1.249 & 0.008 & 1.071 & 1.250 & 1.421 \\
& T3 & -0.032 & 0.111 & -0.220 & 0.113 & 0.113 & 0.222 & 1.000 & 0.991 & 0.188 & 1.231 & 0.007 & 1.063 & 1.234 & 1.389 \\
& T4 & -0.015 & 0.111 & -0.225 & 0.100 & 0.114 & 0.226 & 1.000 & 0.993 & 0.170 & 1.246 & 0.008 & 1.070 & 1.247 & 1.430 \\
& T5 & -0.062 & 0.125 & -0.242 & 0.130 & 0.127 & 0.244 & 1.000 & 0.987 & 0.229 & 1.363 & 0.017 & 1.143 & 1.343 & 1.681 \\
& T6 & -0.010 & 0.109 & -0.221 & 0.106 & 0.112 & 0.223 & 1.000 & 0.988 & 0.191 & 1.240 & 0.009 & 1.056 & 1.238 & 1.431 \\
\addlinespace[2pt]
\cmidrule(lr){1-16}
\multicolumn{16}{@{}l@{}}{\textbf{Linear regression (OLS)}} \\
\multirow{6}{*}{\ParamBox{0.11\textwidth}{80}{1000}{10}{0.10}{0.30}{0.6}{clean}} & T1 & -0.304 & -0.007 & 0.037 & 0.611 & 0.050 & 0.088 & 0.996 & 0.997 & 1.000 & 2.115 & 0.047 & 1.698 & 2.114 & 2.525 \\
& T2 & 0.444 & 0.034 & -0.206 & 0.670 & 0.058 & 0.220 & 1.000 & 1.000 & 0.949 & 2.584 & 0.056 & 2.125 & 2.583 & 3.021 \\
& T3 & 0.377 & 0.034 & -0.191 & 0.614 & 0.057 & 0.206 & 1.000 & 1.000 & 0.962 & 2.539 & 0.053 & 2.130 & 2.541 & 2.984 \\
& T4 & 0.399 & 0.035 & -0.201 & 0.588 & 0.054 & 0.218 & 1.000 & 1.000 & 0.931 & 2.617 & 0.060 & 2.154 & 2.617 & 3.155 \\
& T5 & 0.630 & 0.040 & -0.257 & 0.773 & 0.057 & 0.271 & 1.000 & 1.000 & 0.876 & 2.872 & 0.055 & 2.437 & 2.864 & 3.371 \\
& T6 & 0.422 & 0.033 & -0.200 & 0.615 & 0.052 & 0.217 & 1.000 & 1.000 & 0.930 & 2.581 & 0.049 & 2.154 & 2.584 & 3.027 \\
\addlinespace[2pt]
\multicolumn{16}{@{}l@{}}{\textbf{Sparse regression (EN)}} \\
\multirow{6}{*}{\ParamBox{0.11\textwidth}{80}{1000}{10}{0.10}{0.30}{0.6}{{ext.}}} & T1 & -0.137 & 0.039 & -0.056 & 0.384 & 0.063 & 0.075 & 0.994 & 0.992 & 0.998 & 1.414 & 0.030 & 1.082 & 1.418 & 1.750 \\
& T2 & -0.180 & 0.125 & -0.216 & 0.394 & 0.134 & 0.222 & 0.999 & 0.921 & 0.726 & 1.871 & 0.049 & 1.426 & 1.876 & 2.279 \\
& T3 & -0.240 & 0.128 & -0.211 & 0.415 & 0.137 & 0.217 & 0.999 & 0.915 & 0.710 & 1.810 & 0.045 & 1.388 & 1.811 & 2.248 \\
& T4 & -0.495 & 0.170 & -0.245 & 0.645 & 0.179 & 0.250 & 0.989 & 0.736 & 0.495 & 2.021 & 0.061 & 1.565 & 2.026 & 2.502 \\
& T5 & -0.355 & 0.193 & -0.315 & 0.527 & 0.201 & 0.320 & 0.994 & 0.831 & 0.479 & 2.399 & 0.076 & 1.918 & 2.375 & 2.996 \\
& T6 & -0.213 & 0.134 & -0.226 & 0.365 & 0.142 & 0.232 & 1.000 & 0.941 & 0.607 & 1.884 & 0.049 & 1.450 & 1.872 & 2.304 \\
\bottomrule
\end{tabular*}
\noindent\parbox{\textwidth}{\raggedleft
  \Nchips{80}\\[-0.05em]
  \Nchips{200}%
}
\endgroup
\end{table}


\begin{table}[htbp]
\centering
\caption{Models coefficients bias, RMSE, 95\% coverage, and out-of-sample predictive MSE across MI methods for OLS (clean) and elastic net (EN, contaminated), with \(\boldsymbol{n=500}\), \(n_{\text{sim}}=50\), \(\text{iter}=5\), \(\rho=0.6\), contamination \(P_{\text{ext}}\in\{0.03,0.05\}\) and missingness \(P_{\text{miss}}\in\{0.05,0.10,0.30\}\)}
\label{tab:n_500_graficos.1}

\begingroup
\scriptsize
\begin{tabular*}{\textwidth}{@{\extracolsep{\fill}} l l
ccc 
ccc 
ccc 
ccccc 
@{}}
\toprule
& & \multicolumn{3}{c}{Bias} & \multicolumn{3}{c}{RMSE} & \multicolumn{3}{c}{Coverage} & \multicolumn{5}{c}{Pred.~MSE (out-of-sample)} \\
\cmidrule(lr){3-5}\cmidrule(lr){6-8}\cmidrule(lr){9-11}\cmidrule(lr){12-16}
Setup & MI &
$\beta_0$ & $\beta_1$ & $\beta_2$ &
$\beta_0$ & $\beta_1$ & $\beta_2$ &
$\beta_0$ & $\beta_1$ & $\beta_2$ &
$\overline{X}$ & $\sigma^2$ & $Q_{2.5}$ & $Q_{50}$ & $Q_{97.5}$
 \\
\midrule
\multicolumn{16}{@{}l@{}}{\textbf{Linear regression (OLS)}} \\
\multirow{6}{*}{\ParamBox{0.11\textwidth}{500}{50}{5}{0.03}{0.05}{0.6}{clean}} & T1 & 0.117 & 0.005 & -0.023 & 0.130 & 0.008 & 0.025 & 1.000 & 1.000 & 1.000 & 2.152 & 0.001 & 2.110 & 2.147 & 2.203 \\
& T2 & 0.176 & 0.016 & -0.057 & 0.185 & 0.018 & 0.058 & 1.000 & 1.000 & 1.000 & 2.214 & 0.001 & 2.157 & 2.216 & 2.278 \\
& T3 & 0.167 & 0.017 & -0.056 & 0.179 & 0.018 & 0.057 & 1.000 & 1.000 & 1.000 & 2.213 & 0.001 & 2.162 & 2.208 & 2.289 \\
& T4 & 0.167 & 0.017 & -0.056 & 0.179 & 0.018 & 0.057 & 1.000 & 1.000 & 1.000 & 2.215 & 0.001 & 2.158 & 2.211 & 2.282 \\
& T5 & 0.177 & 0.017 & -0.059 & 0.189 & 0.019 & 0.060 & 1.000 & 1.000 & 1.000 & 2.224 & 0.001 & 2.162 & 2.221 & 2.282 \\
& T6 & 0.176 & 0.017 & -0.057 & 0.187 & 0.018 & 0.058 & 1.000 & 1.000 & 1.000 & 2.215 & 0.001 & 2.154 & 2.213 & 2.279 \\
\addlinespace[2pt]
\multicolumn{16}{@{}l@{}}{\textbf{Sparse regression (EN)}} \\
\multirow{6}{*}{\ParamBox{0.11\textwidth}{500}{50}{5}{0.03}{0.05}{0.6}{{ext.}}} & T1 & 0.165 & -0.077 & 0.115 & 0.172 & 0.077 & 0.116 & 1.000 & 0.060 & 0.200 & 2.080 & 0.001 & 2.018 & 2.083 & 2.133 \\
& T2 & 0.203 & -0.066 & 0.086 & 0.210 & 0.066 & 0.087 & 1.000 & 0.860 & 0.980 & 2.154 & 0.001 & 2.073 & 2.158 & 2.211 \\
& T3 & 0.194 & -0.065 & 0.086 & 0.200 & 0.065 & 0.087 & 1.000 & 0.860 & 0.960 & 2.151 & 0.001 & 2.081 & 2.149 & 2.205 \\
& T4 & 0.205 & -0.065 & 0.085 & 0.210 & 0.065 & 0.085 & 1.000 & 0.900 & 1.000 & 2.155 & 0.001 & 2.083 & 2.160 & 2.225 \\
& T5 & 0.204 & -0.063 & 0.082 & 0.212 & 0.064 & 0.082 & 1.000 & 0.900 & 0.980 & 2.179 & 0.001 & 2.118 & 2.184 & 2.242 \\
& T6 & 0.202 & -0.065 & 0.085 & 0.207 & 0.066 & 0.086 & 1.000 & 0.720 & 0.940 & 2.150 & 0.002 & 2.084 & 2.156 & 2.220 \\
\addlinespace[2pt]
\cmidrule(lr){1-16}
\multicolumn{16}{@{}l@{}}{\textbf{Linear regression (OLS)}} \\
\multirow{6}{*}{\ParamBox{0.11\textwidth}{500}{50}{5}{0.03}{0.10}{0.6}{clean}} & T1 & 0.058 & -0.007 & 0.013 & 0.114 & 0.012 & 0.019 & 1.000 & 1.000 & 1.000 & 2.082 & 0.002 & 1.985 & 2.082 & 2.163 \\
& T2 & 0.165 & 0.017 & -0.055 & 0.197 & 0.019 & 0.057 & 1.000 & 1.000 & 1.000 & 2.213 & 0.003 & 2.127 & 2.213 & 2.300 \\
& T3 & 0.162 & 0.017 & -0.054 & 0.191 & 0.020 & 0.057 & 1.000 & 1.000 & 1.000 & 2.213 & 0.003 & 2.101 & 2.213 & 2.292 \\
& T4 & 0.161 & 0.016 & -0.053 & 0.195 & 0.020 & 0.055 & 1.000 & 1.000 & 1.000 & 2.211 & 0.003 & 2.107 & 2.204 & 2.311 \\
& T5 & 0.185 & 0.019 & -0.062 & 0.213 & 0.021 & 0.064 & 1.000 & 1.000 & 1.000 & 2.245 & 0.003 & 2.165 & 2.238 & 2.322 \\
& T6 & 0.164 & 0.017 & -0.055 & 0.195 & 0.020 & 0.057 & 1.000 & 1.000 & 1.000 & 2.209 & 0.003 & 2.103 & 2.213 & 2.294 \\
\addlinespace[2pt]
\multicolumn{16}{@{}l@{}}{\textbf{Sparse regression (EN)}} \\
\multirow{6}{*}{\ParamBox{0.11\textwidth}{500}{50}{5}{0.03}{0.10}{0.6}{{ext.}}} & T1 & 0.140 & -0.091 & 0.150 & 0.169 & 0.092 & 0.150 & 1.000 & 0.040 & 0.000 & 2.000 & 0.003 & 1.903 & 1.995 & 2.088 \\
& T2 & 0.215 & -0.067 & 0.088 & 0.237 & 0.068 & 0.089 & 1.000 & 0.740 & 0.860 & 2.151 & 0.003 & 2.048 & 2.147 & 2.243 \\
& T3 & 0.205 & -0.067 & 0.089 & 0.230 & 0.068 & 0.091 & 1.000 & 0.680 & 0.880 & 2.144 & 0.003 & 2.049 & 2.149 & 2.231 \\
& T4 & 0.205 & -0.067 & 0.090 & 0.222 & 0.068 & 0.091 & 1.000 & 0.660 & 0.840 & 2.146 & 0.004 & 2.051 & 2.140 & 2.284 \\
& T5 & 0.214 & -0.064 & 0.081 & 0.236 & 0.065 & 0.082 & 1.000 & 0.840 & 0.940 & 2.203 & 0.004 & 2.065 & 2.213 & 2.294 \\
& T6 & 0.207 & -0.067 & 0.089 & 0.226 & 0.068 & 0.091 & 1.000 & 0.700 & 0.880 & 2.146 & 0.004 & 2.039 & 2.152 & 2.256 \\
\addlinespace[2pt]
\cmidrule(lr){1-16}
\multicolumn{16}{@{}l@{}}{\textbf{Linear regression (OLS)}} \\
\multirow{6}{*}{\ParamBox{0.11\textwidth}{500}{50}{5}{0.03}{0.30}{0.6}{clean}} & T1 & -0.186 & -0.063 & 0.170 & 0.269 & 0.066 & 0.172 & 0.980 & 0.620 & 0.000 & 1.786 & 0.006 & 1.664 & 1.791 & 1.929 \\
& T2 & 0.206 & 0.012 & -0.055 & 0.279 & 0.025 & 0.062 & 1.000 & 1.000 & 1.000 & 2.216 & 0.008 & 2.051 & 2.220 & 2.372 \\
& T3 & 0.192 & 0.014 & -0.055 & 0.266 & 0.026 & 0.062 & 1.000 & 1.000 & 1.000 & 2.208 & 0.008 & 2.066 & 2.210 & 2.391 \\
& T4 & 0.158 & 0.017 & -0.055 & 0.238 & 0.027 & 0.060 & 1.000 & 1.000 & 1.000 & 2.221 & 0.007 & 2.076 & 2.226 & 2.375 \\
& T5 & 0.225 & 0.025 & -0.083 & 0.298 & 0.033 & 0.088 & 1.000 & 0.980 & 0.980 & 2.329 & 0.007 & 2.177 & 2.314 & 2.476 \\
& T6 & 0.164 & 0.018 & -0.057 & 0.249 & 0.031 & 0.066 & 1.000 & 1.000 & 1.000 & 2.221 & 0.010 & 2.013 & 2.232 & 2.408 \\
\addlinespace[2pt]
\multicolumn{16}{@{}l@{}}{\textbf{Sparse regression (EN)}} \\
\multirow{6}{*}{\ParamBox{0.11\textwidth}{500}{50}{5}{0.03}{0.30}{0.6}{{ext.}}} & T1 & 0.007 & -0.144 & 0.280 & 0.150 & 0.145 & 0.281 & 1.000 & 0.000 & 0.000 & 1.694 & 0.006 & 1.534 & 1.694 & 1.862 \\
& T2 & 0.262 & -0.071 & 0.085 & 0.307 & 0.073 & 0.090 & 1.000 & 0.700 & 0.880 & 2.172 & 0.009 & 1.964 & 2.165 & 2.359 \\
& T3 & 0.223 & -0.066 & 0.082 & 0.268 & 0.068 & 0.087 & 1.000 & 0.840 & 0.920 & 2.153 & 0.009 & 1.972 & 2.143 & 2.351 \\
& T4 & 0.206 & -0.066 & 0.086 & 0.236 & 0.067 & 0.089 & 1.000 & 0.740 & 0.820 & 2.160 & 0.010 & 1.968 & 2.151 & 2.325 \\
& T5 & 0.292 & -0.055 & 0.049 & 0.324 & 0.058 & 0.057 & 1.000 & 0.940 & 1.000 & 2.354 & 0.015 & 2.115 & 2.340 & 2.544 \\
& T6 & 0.232 & -0.067 & 0.084 & 0.274 & 0.069 & 0.088 & 0.980 & 0.820 & 0.840 & 2.147 & 0.009 & 1.971 & 2.149 & 2.305 \\
\addlinespace[2pt]
\cmidrule(lr){1-16}
\multicolumn{16}{@{}l@{}}{\textbf{Linear regression (OLS)}} \\
\multirow{6}{*}{\ParamBox{0.11\textwidth}{500}{50}{5}{0.05}{0.05}{0.6}{clean}} & T1 & 0.112 & 0.006 & -0.023 & 0.126 & 0.008 & 0.025 & 1.000 & 1.000 & 1.000 & 2.152 & 0.001 & 2.106 & 2.153 & 2.203 \\
& T2 & 0.169 & 0.018 & -0.058 & 0.180 & 0.019 & 0.058 & 1.000 & 1.000 & 1.000 & 2.218 & 0.001 & 2.160 & 2.214 & 2.287 \\
& T3 & 0.172 & 0.017 & -0.056 & 0.181 & 0.018 & 0.057 & 1.000 & 1.000 & 1.000 & 2.213 & 0.001 & 2.157 & 2.207 & 2.265 \\
& T4 & 0.170 & 0.017 & -0.056 & 0.180 & 0.018 & 0.057 & 1.000 & 1.000 & 1.000 & 2.216 & 0.001 & 2.158 & 2.221 & 2.270 \\
& T5 & 0.175 & 0.018 & -0.060 & 0.186 & 0.019 & 0.061 & 1.000 & 1.000 & 1.000 & 2.230 & 0.001 & 2.166 & 2.231 & 2.306 \\
& T6 & 0.172 & 0.017 & -0.058 & 0.184 & 0.019 & 0.059 & 1.000 & 1.000 & 1.000 & 2.214 & 0.001 & 2.154 & 2.211 & 2.267 \\
\addlinespace[2pt]
\multicolumn{16}{@{}l@{}}{\textbf{Sparse regression (EN)}} \\
\multirow{6}{*}{\ParamBox{0.11\textwidth}{500}{50}{5}{0.05}{0.05}{0.6}{{ext.}}} & T1 & 0.236 & -0.081 & 0.109 & 0.240 & 0.082 & 0.109 & 1.000 & 0.000 & 0.200 & 1.968 & 0.001 & 1.926 & 1.968 & 2.010 \\
& T2 & 0.269 & -0.071 & 0.081 & 0.274 & 0.071 & 0.081 & 1.000 & 0.460 & 1.000 & 2.039 & 0.001 & 1.985 & 2.043 & 2.082 \\
& T3 & 0.255 & -0.069 & 0.080 & 0.261 & 0.069 & 0.081 & 1.000 & 0.480 & 1.000 & 2.036 & 0.001 & 1.996 & 2.035 & 2.077 \\
& T4 & 0.268 & -0.070 & 0.080 & 0.271 & 0.070 & 0.081 & 1.000 & 0.440 & 1.000 & 2.041 & 0.001 & 1.991 & 2.033 & 2.119 \\
& T5 & 0.270 & -0.068 & 0.076 & 0.274 & 0.068 & 0.076 & 1.000 & 0.700 & 1.000 & 2.067 & 0.002 & 1.995 & 2.063 & 2.148 \\
& T6 & 0.268 & -0.070 & 0.079 & 0.272 & 0.070 & 0.080 & 1.000 & 0.540 & 0.980 & 2.041 & 0.001 & 1.983 & 2.039 & 2.101 \\
\bottomrule
\end{tabular*}
\noindent\parbox{\textwidth}{\raggedleft
  \Nchips{500}
}
\endgroup
\end{table}

\begin{table}[htbp]
\centering
\caption{Models coefficients bias, RMSE, 95\% coverage, and out-of-sample predictive MSE across MI methods for OLS (clean) and EN (contaminated), with \(\boldsymbol{n=500}\), \(n_{\text{sim}}=50\), \(\text{iter}=5\), \(\rho=0.6\). The first two blocks report \(P_{\text{ext}}=0.05\) with \(P_{\text{miss}}\in\{0.10,0.30\}\); the last two blocks report \(P_{\text{ext}}=0.10\) with \(P_{\text{miss}}\in\{0.05,0.10\}\)}

\label{tab:n_500_graficos.2}

\begingroup
\scriptsize
\begin{tabular*}{\textwidth}{@{\extracolsep{\fill}} l l
ccc 
ccc 
ccc 
ccccc 
@{}}
\toprule
& & \multicolumn{3}{c}{Bias} & \multicolumn{3}{c}{RMSE} & \multicolumn{3}{c}{Coverage} & \multicolumn{5}{c}{Pred.~MSE (out-of-sample)} \\
\cmidrule(lr){3-5}\cmidrule(lr){6-8}\cmidrule(lr){9-11}\cmidrule(lr){12-16}
Setup & MI &
$\beta_0$ & $\beta_1$ & $\beta_2$ &
$\beta_0$ & $\beta_1$ & $\beta_2$ &
$\beta_0$ & $\beta_1$ & $\beta_2$ &
$\overline{X}$ & $\sigma^2$ & $Q_{2.5}$ & $Q_{50}$ & $Q_{97.5}$
 \\
\midrule
\multicolumn{16}{@{}l@{}}{\textbf{Linear regression (OLS)}} \\
\multirow{6}{*}{\ParamBox{0.11\textwidth}{500}{50}{5}{0.05}{0.10}{0.6}{clean}} & T1 & 0.065 & -0.007 & 0.012 & 0.113 & 0.011 & 0.018 & 1.000 & 1.000 & 1.000 & 2.081 & 0.002 & 2.013 & 2.079 & 2.167 \\
& T2 & 0.177 & 0.016 & -0.056 & 0.206 & 0.019 & 0.058 & 1.000 & 1.000 & 1.000 & 2.208 & 0.002 & 2.135 & 2.202 & 2.319 \\
& T3 & 0.175 & 0.016 & -0.055 & 0.196 & 0.019 & 0.057 & 1.000 & 1.000 & 1.000 & 2.211 & 0.002 & 2.123 & 2.206 & 2.313 \\
& T4 & 0.174 & 0.016 & -0.055 & 0.201 & 0.019 & 0.057 & 1.000 & 1.000 & 1.000 & 2.212 & 0.002 & 2.133 & 2.211 & 2.308 \\
& T5 & 0.188 & 0.017 & -0.060 & 0.215 & 0.020 & 0.062 & 1.000 & 1.000 & 1.000 & 2.240 & 0.002 & 2.163 & 2.237 & 2.326 \\
& T6 & 0.179 & 0.016 & -0.057 & 0.201 & 0.019 & 0.059 & 1.000 & 1.000 & 1.000 & 2.212 & 0.003 & 2.123 & 2.214 & 2.319 \\
\addlinespace[2pt]
\multicolumn{16}{@{}l@{}}{\textbf{Sparse regression (EN)}} \\
\multirow{6}{*}{\ParamBox{0.11\textwidth}{500}{50}{5}{0.05}{0.10}{0.6}{{ext.}}} & T1 & 0.218 & -0.096 & 0.141 & 0.234 & 0.096 & 0.142 & 1.000 & 0.000 & 0.000 & 1.888 & 0.002 & 1.817 & 1.888 & 1.955 \\
& T2 & 0.270 & -0.071 & 0.083 & 0.285 & 0.072 & 0.085 & 1.000 & 0.440 & 0.900 & 2.035 & 0.002 & 1.946 & 2.036 & 2.113 \\
& T3 & 0.261 & -0.070 & 0.084 & 0.275 & 0.071 & 0.085 & 1.000 & 0.520 & 0.920 & 2.027 & 0.002 & 1.939 & 2.026 & 2.098 \\
& T4 & 0.266 & -0.072 & 0.085 & 0.278 & 0.073 & 0.086 & 1.000 & 0.440 & 0.960 & 2.026 & 0.002 & 1.950 & 2.020 & 2.128 \\
& T5 & 0.289 & -0.069 & 0.076 & 0.300 & 0.070 & 0.077 & 1.000 & 0.560 & 1.000 & 2.088 & 0.003 & 1.974 & 2.086 & 2.203 \\
& T6 & 0.266 & -0.071 & 0.084 & 0.276 & 0.072 & 0.085 & 1.000 & 0.500 & 0.960 & 2.031 & 0.003 & 1.947 & 2.027 & 2.139 \\
\addlinespace[2pt]
\cmidrule(lr){1-16}
\multicolumn{16}{@{}l@{}}{\textbf{Linear regression (OLS)}} \\
\multirow{6}{*}{\ParamBox{0.11\textwidth}{500}{50}{5}{0.05}{0.30}{0.6}{clean}} & T1 & -0.189 & -0.063 & 0.169 & 0.286 & 0.066 & 0.171 & 0.960 & 0.600 & 0.000 & 1.797 & 0.005 & 1.685 & 1.786 & 1.948 \\
& T2 & 0.199 & 0.015 & -0.061 & 0.294 & 0.028 & 0.068 & 1.000 & 1.000 & 1.000 & 2.230 & 0.007 & 2.069 & 2.222 & 2.402 \\
& T3 & 0.191 & 0.016 & -0.059 & 0.271 & 0.025 & 0.065 & 1.000 & 1.000 & 1.000 & 2.221 & 0.007 & 2.087 & 2.214 & 2.381 \\
& T4 & 0.138 & 0.020 & -0.058 & 0.228 & 0.028 & 0.064 & 1.000 & 1.000 & 1.000 & 2.231 & 0.005 & 2.083 & 2.246 & 2.362 \\
& T5 & 0.220 & 0.024 & -0.081 & 0.299 & 0.032 & 0.086 & 1.000 & 1.000 & 0.960 & 2.330 & 0.006 & 2.187 & 2.340 & 2.464 \\
& T6 & 0.154 & 0.020 & -0.061 & 0.260 & 0.028 & 0.066 & 1.000 & 1.000 & 1.000 & 2.237 & 0.006 & 2.121 & 2.233 & 2.409 \\
\addlinespace[2pt]
\multicolumn{16}{@{}l@{}}{\textbf{Sparse regression (EN)}} \\
\multirow{6}{*}{\ParamBox{0.11\textwidth}{500}{50}{5}{0.05}{0.30}{0.6}{{ext.}}} & T1 & 0.127 & -0.148 & 0.263 & 0.195 & 0.149 & 0.264 & 0.980 & 0.000 & 0.000 & 1.602 & 0.006 & 1.485 & 1.597 & 1.734 \\
& T2 & 0.294 & -0.070 & 0.077 & 0.334 & 0.073 & 0.079 & 0.980 & 0.600 & 0.980 & 2.063 & 0.007 & 1.907 & 2.068 & 2.210 \\
& T3 & 0.283 & -0.069 & 0.076 & 0.316 & 0.071 & 0.079 & 1.000 & 0.700 & 0.960 & 2.047 & 0.009 & 1.921 & 2.043 & 2.232 \\
& T4 & 0.301 & -0.071 & 0.078 & 0.323 & 0.073 & 0.081 & 1.000 & 0.620 & 0.960 & 2.039 & 0.007 & 1.882 & 2.048 & 2.200 \\
& T5 & 0.367 & -0.062 & 0.046 & 0.385 & 0.063 & 0.050 & 0.980 & 0.840 & 1.000 & 2.229 & 0.010 & 2.029 & 2.235 & 2.415 \\
& T6 & 0.301 & -0.071 & 0.077 & 0.322 & 0.073 & 0.079 & 0.980 & 0.720 & 0.980 & 2.036 & 0.008 & 1.861 & 2.023 & 2.215 \\
\addlinespace[2pt]
\cmidrule(lr){1-16}
\multicolumn{16}{@{}l@{}}{\textbf{Linear regression (OLS)}} \\
\multirow{6}{*}{\ParamBox{0.11\textwidth}{500}{50}{5}{0.10}{0.05}{0.6}{clean}} & T1 & 0.112 & 0.006 & -0.023 & 0.126 & 0.008 & 0.025 & 1.000 & 1.000 & 1.000 & 2.153 & 0.001 & 2.112 & 2.154 & 2.192 \\
& T2 & 0.167 & 0.017 & -0.057 & 0.179 & 0.019 & 0.058 & 1.000 & 1.000 & 1.000 & 2.215 & 0.001 & 2.166 & 2.214 & 2.261 \\
& T3 & 0.165 & 0.017 & -0.056 & 0.175 & 0.019 & 0.057 & 1.000 & 1.000 & 1.000 & 2.213 & 0.001 & 2.166 & 2.210 & 2.256 \\
& T4 & 0.173 & 0.017 & -0.057 & 0.184 & 0.018 & 0.057 & 1.000 & 1.000 & 1.000 & 2.217 & 0.001 & 2.161 & 2.221 & 2.259 \\
& T5 & 0.176 & 0.019 & -0.061 & 0.187 & 0.020 & 0.062 & 1.000 & 1.000 & 1.000 & 2.231 & 0.001 & 2.166 & 2.232 & 2.298 \\
& T6 & 0.172 & 0.017 & -0.057 & 0.184 & 0.018 & 0.058 & 1.000 & 1.000 & 1.000 & 2.215 & 0.001 & 2.156 & 2.215 & 2.272 \\
\addlinespace[2pt]
\multicolumn{16}{@{}l@{}}{\textbf{Sparse regression (EN)}} \\
\multirow{6}{*}{\ParamBox{0.11\textwidth}{500}{50}{5}{0.10}{0.05}{0.6}{{ext.}}} & T1 & 0.253 & -0.093 & 0.132 & 0.255 & 0.094 & 0.132 & 1.000 & 0.000 & 0.000 & 1.866 & 0.001 & 1.828 & 1.871 & 1.899 \\
& T2 & 0.263 & -0.081 & 0.106 & 0.266 & 0.081 & 0.106 & 1.000 & 0.000 & 0.360 & 1.932 & 0.001 & 1.876 & 1.939 & 1.976 \\
& T3 & 0.253 & -0.080 & 0.105 & 0.256 & 0.080 & 0.105 & 1.000 & 0.000 & 0.400 & 1.932 & 0.001 & 1.881 & 1.933 & 1.971 \\
& T4 & 0.264 & -0.080 & 0.105 & 0.266 & 0.081 & 0.105 & 1.000 & 0.040 & 0.300 & 1.938 & 0.001 & 1.883 & 1.940 & 1.990 \\
& T5 & 0.271 & -0.079 & 0.101 & 0.274 & 0.079 & 0.101 & 1.000 & 0.100 & 0.520 & 1.963 & 0.002 & 1.898 & 1.956 & 2.080 \\
& T6 & 0.261 & -0.080 & 0.105 & 0.263 & 0.080 & 0.105 & 1.000 & 0.000 & 0.360 & 1.935 & 0.001 & 1.875 & 1.938 & 1.989 \\
\addlinespace[2pt]
\cmidrule(lr){1-16}
\multicolumn{16}{@{}l@{}}{\textbf{Linear regression (OLS)}} \\
\multirow{6}{*}{\ParamBox{0.11\textwidth}{500}{50}{5}{0.10}{0.10}{0.6}{clean}} & T1 & 0.057 & -0.007 & 0.013 & 0.113 & 0.012 & 0.017 & 1.000 & 1.000 & 1.000 & 2.081 & 0.002 & 2.008 & 2.086 & 2.146 \\
& T2 & 0.165 & 0.017 & -0.056 & 0.195 & 0.020 & 0.057 & 1.000 & 1.000 & 1.000 & 2.208 & 0.002 & 2.115 & 2.213 & 2.289 \\
& T3 & 0.167 & 0.016 & -0.055 & 0.199 & 0.020 & 0.056 & 1.000 & 1.000 & 1.000 & 2.209 & 0.002 & 2.124 & 2.208 & 2.290 \\
& T4 & 0.158 & 0.018 & -0.055 & 0.188 & 0.020 & 0.057 & 1.000 & 1.000 & 1.000 & 2.213 & 0.002 & 2.140 & 2.217 & 2.289 \\
& T5 & 0.182 & 0.018 & -0.062 & 0.209 & 0.022 & 0.063 & 1.000 & 1.000 & 1.000 & 2.244 & 0.002 & 2.155 & 2.235 & 2.345 \\
& T6 & 0.170 & 0.016 & -0.055 & 0.198 & 0.020 & 0.056 & 1.000 & 1.000 & 1.000 & 2.207 & 0.002 & 2.122 & 2.214 & 2.279 \\
\addlinespace[2pt]
\multicolumn{16}{@{}l@{}}{\textbf{Sparse regression (EN)}} \\
\multirow{6}{*}{\ParamBox{0.11\textwidth}{500}{50}{5}{0.10}{0.10}{0.6}{{ext.}}} & T1 & 0.251 & -0.108 & 0.162 & 0.258 & 0.108 & 0.163 & 1.000 & 0.000 & 0.000 & 1.790 & 0.002 & 1.715 & 1.789 & 1.856 \\
& T2 & 0.265 & -0.081 & 0.106 & 0.272 & 0.081 & 0.107 & 1.000 & 0.080 & 0.400 & 1.932 & 0.003 & 1.839 & 1.924 & 2.017 \\
& T3 & 0.257 & -0.080 & 0.107 & 0.265 & 0.081 & 0.108 & 1.000 & 0.140 & 0.400 & 1.925 & 0.002 & 1.835 & 1.927 & 2.013 \\
& T4 & 0.265 & -0.083 & 0.109 & 0.270 & 0.083 & 0.110 & 1.000 & 0.080 & 0.320 & 1.928 & 0.003 & 1.820 & 1.935 & 2.010 \\
& T5 & 0.281 & -0.078 & 0.097 & 0.289 & 0.078 & 0.098 & 1.000 & 0.280 & 0.760 & 1.998 & 0.003 & 1.901 & 1.995 & 2.104 \\
& T6 & 0.269 & -0.083 & 0.109 & 0.275 & 0.084 & 0.110 & 1.000 & 0.080 & 0.340 & 1.931 & 0.003 & 1.838 & 1.930 & 2.034 \\
\bottomrule
\end{tabular*}
\noindent\parbox{\textwidth}{\raggedleft
  \Nchips{500}
}
\endgroup
\end{table}

\begin{table}[htbp]
\centering
\caption{Models coefficients bias, RMSE, 95\% coverage, and out-of-sample predictive MSE across MI methods for OLS (clean) and EN (contaminated). Top block: \(\boldsymbol{n=500}\), \(n_{\text{sim}}=50\), \(\text{iter}=5\), \(\rho=0.6\), \(P_{\text{ext}}=0.10\), \(P_{\text{miss}}=0.30\). Remaining blocks: \(\boldsymbol{n=500}\), \(n_{\text{sim}}=3000\), \(\text{iter}=5\), \(\rho=0.6\), with \(P_{\text{ext}}\in\{0.10,0.30\}\) and \(P_{\text{miss}}\in\{0.10,0.30\}\)}

\label{tab:n_500_graficos.3}

\begingroup
\scriptsize
\begin{tabular*}{\textwidth}{@{\extracolsep{\fill}} l l
ccc 
ccc 
ccc 
ccccc 
@{}}
\toprule
& & \multicolumn{3}{c}{Bias} & \multicolumn{3}{c}{RMSE} & \multicolumn{3}{c}{Coverage} & \multicolumn{5}{c}{Pred.~MSE (out-of-sample)} \\
\cmidrule(lr){3-5}\cmidrule(lr){6-8}\cmidrule(lr){9-11}\cmidrule(lr){12-16}
Setup & MI &
$\beta_0$ & $\beta_1$ & $\beta_2$ &
$\beta_0$ & $\beta_1$ & $\beta_2$ &
$\beta_0$ & $\beta_1$ & $\beta_2$ &
$\overline{X}$ & $\sigma^2$ & $Q_{2.5}$ & $Q_{50}$ & $Q_{97.5}$
 \\
\midrule
\multicolumn{16}{@{}l@{}}{\textbf{Linear regression (OLS)}} \\
\multirow{6}{*}{\ParamBox{0.11\textwidth}{500}{50}{5}{0.10}{0.30}{0.6}{clean}} & T1 & -0.192 & -0.064 & 0.174 & 0.273 & 0.067 & 0.176 & 1.000 & 0.580 & 0.000 & 1.784 & 0.008 & 1.593 & 1.786 & 1.954 \\
& T2 & 0.206 & 0.012 & -0.053 & 0.280 & 0.024 & 0.061 & 1.000 & 1.000 & 1.000 & 2.211 & 0.011 & 2.027 & 2.213 & 2.402 \\
& T3 & 0.195 & 0.013 & -0.053 & 0.268 & 0.024 & 0.060 & 1.000 & 1.000 & 1.000 & 2.210 & 0.012 & 1.979 & 2.194 & 2.406 \\
& T4 & 0.150 & 0.018 & -0.056 & 0.222 & 0.025 & 0.063 & 1.000 & 1.000 & 1.000 & 2.235 & 0.011 & 2.040 & 2.236 & 2.393 \\
& T5 & 0.227 & 0.022 & -0.078 & 0.279 & 0.028 & 0.082 & 1.000 & 1.000 & 1.000 & 2.323 & 0.010 & 2.112 & 2.325 & 2.467 \\
& T6 & 0.163 & 0.017 & -0.055 & 0.242 & 0.025 & 0.063 & 1.000 & 1.000 & 0.980 & 2.222 & 0.012 & 2.025 & 2.235 & 2.424 \\
\addlinespace[2pt]
\multicolumn{16}{@{}l@{}}{\textbf{Sparse regression (EN)}} \\
\multirow{6}{*}{\ParamBox{0.11\textwidth}{500}{50}{5}{0.10}{0.30}{0.6}{{ext.}}} & T1 & 0.225 & -0.167 & 0.282 & 0.257 & 0.168 & 0.283 & 0.920 & 0.000 & 0.000 & 1.497 & 0.005 & 1.371 & 1.498 & 1.636 \\
& T2 & 0.275 & -0.082 & 0.105 & 0.301 & 0.084 & 0.108 & 0.980 & 0.360 & 0.600 & 1.948 & 0.006 & 1.825 & 1.942 & 2.093 \\
& T3 & 0.255 & -0.079 & 0.103 & 0.289 & 0.081 & 0.105 & 1.000 & 0.520 & 0.720 & 1.931 & 0.007 & 1.775 & 1.924 & 2.083 \\
& T4 & 0.287 & -0.086 & 0.112 & 0.305 & 0.088 & 0.115 & 0.940 & 0.280 & 0.420 & 1.931 & 0.008 & 1.788 & 1.934 & 2.101 \\
& T5 & 0.324 & -0.069 & 0.071 & 0.347 & 0.072 & 0.075 & 0.920 & 0.760 & 0.980 & 2.136 & 0.008 & 1.982 & 2.127 & 2.328 \\
& T6 & 0.277 & -0.081 & 0.103 & 0.298 & 0.083 & 0.106 & 0.980 & 0.340 & 0.600 & 1.936 & 0.008 & 1.772 & 1.931 & 2.074 \\
\addlinespace[2pt]
\cmidrule(lr){1-16}
\multicolumn{16}{@{}l@{}}{\textbf{Linear regression (OLS)}} \\
\multirow{6}{*}{\ParamBox{0.11\textwidth}{500}{3000}{5}{0.10}{0.10}{0.6}{clean}} & T1 & 0.060 & -0.006 & 0.011 & 0.109 & 0.011 & 0.018 & 1.000 & 1.000 & 1.000 & 2.093 & 0.002 & 2.007 & 2.094 & 2.173 \\
& T2 & 0.175 & 0.017 & -0.058 & 0.201 & 0.020 & 0.059 & 1.000 & 1.000 & 1.000 & 2.222 & 0.002 & 2.125 & 2.222 & 2.316 \\
& T3 & 0.172 & 0.017 & -0.057 & 0.197 & 0.020 & 0.059 & 1.000 & 1.000 & 1.000 & 2.220 & 0.002 & 2.122 & 2.220 & 2.306 \\
& T4 & 0.165 & 0.018 & -0.057 & 0.190 & 0.021 & 0.059 & 1.000 & 1.000 & 1.000 & 2.224 & 0.002 & 2.127 & 2.226 & 2.314 \\
& T5 & 0.185 & 0.019 & -0.064 & 0.211 & 0.022 & 0.066 & 1.000 & 1.000 & 1.000 & 2.255 & 0.003 & 2.156 & 2.255 & 2.357 \\
& T6 & 0.171 & 0.018 & -0.058 & 0.198 & 0.021 & 0.060 & 1.000 & 1.000 & 1.000 & 2.225 & 0.002 & 2.124 & 2.225 & 2.321 \\
\addlinespace[2pt]
\multicolumn{16}{@{}l@{}}{\textbf{Sparse regression (EN)}} \\
\multirow{6}{*}{\ParamBox{0.11\textwidth}{500}{3000}{5}{0.10}{0.10}{0.6}{{ext.}}} & T1 & 0.247 & -0.108 & 0.162 & 0.253 & 0.108 & 0.162 & 1.000 & 0.000 & 0.000 & 1.794 & 0.002 & 1.707 & 1.796 & 1.869 \\
& T2 & 0.261 & -0.081 & 0.107 & 0.268 & 0.082 & 0.108 & 1.000 & 0.094 & 0.409 & 1.932 & 0.002 & 1.840 & 1.933 & 2.018 \\
& T3 & 0.254 & -0.080 & 0.107 & 0.260 & 0.081 & 0.108 & 1.000 & 0.108 & 0.424 & 1.928 & 0.002 & 1.830 & 1.929 & 2.016 \\
& T4 & 0.264 & -0.082 & 0.108 & 0.269 & 0.083 & 0.109 & 1.000 & 0.069 & 0.370 & 1.932 & 0.002 & 1.839 & 1.932 & 2.022 \\
& T5 & 0.279 & -0.079 & 0.098 & 0.286 & 0.079 & 0.099 & 0.997 & 0.215 & 0.694 & 1.986 & 0.004 & 1.877 & 1.981 & 2.123 \\
& T6 & 0.263 & -0.082 & 0.107 & 0.268 & 0.082 & 0.108 & 1.000 & 0.083 & 0.400 & 1.931 & 0.002 & 1.832 & 1.932 & 2.027 \\
\addlinespace[2pt]
\cmidrule(lr){1-16}
\multicolumn{16}{@{}l@{}}{\textbf{Linear regression (OLS)}} \\
\multirow{6}{*}{\ParamBox{0.11\textwidth}{500}{3000}{5}{0.10}{0.30}{0.6}{clean}} & T1 & -0.200 & -0.061 & 0.170 & 0.280 & 0.065 & 0.172 & 0.984 & 0.682 & 0.007 & 1.795 & 0.006 & 1.635 & 1.796 & 1.943 \\
& T2 & 0.192 & 0.016 & -0.058 & 0.273 & 0.026 & 0.065 & 0.999 & 1.000 & 0.994 & 2.228 & 0.009 & 2.039 & 2.229 & 2.415 \\
& T3 & 0.169 & 0.017 & -0.056 & 0.251 & 0.026 & 0.063 & 0.999 & 1.000 & 0.994 & 2.216 & 0.008 & 2.032 & 2.220 & 2.392 \\
& T4 & 0.139 & 0.021 & -0.057 & 0.235 & 0.029 & 0.064 & 1.000 & 1.000 & 0.993 & 2.234 & 0.008 & 2.053 & 2.235 & 2.418 \\
& T5 & 0.208 & 0.026 & -0.081 & 0.283 & 0.033 & 0.086 & 0.997 & 0.999 & 0.970 & 2.335 & 0.009 & 2.156 & 2.336 & 2.519 \\
& T6 & 0.154 & 0.019 & -0.058 & 0.252 & 0.029 & 0.066 & 0.999 & 0.999 & 0.988 & 2.232 & 0.009 & 2.050 & 2.232 & 2.414 \\
\addlinespace[2pt]
\multicolumn{16}{@{}l@{}}{\textbf{Sparse regression (EN)}} \\
\multirow{6}{*}{\ParamBox{0.11\textwidth}{500}{3000}{5}{0.10}{0.30}{0.6}{{ext.}}} & T1 & 0.227 & -0.168 & 0.284 & 0.258 & 0.169 & 0.285 & 0.912 & 0.000 & 0.000 & 1.494 & 0.005 & 1.351 & 1.495 & 1.635 \\
& T2 & 0.272 & -0.083 & 0.108 & 0.296 & 0.085 & 0.111 & 0.979 & 0.370 & 0.584 & 1.939 & 0.007 & 1.773 & 1.940 & 2.113 \\
& T3 & 0.252 & -0.081 & 0.107 & 0.280 & 0.083 & 0.111 & 0.981 & 0.442 & 0.606 & 1.924 & 0.008 & 1.753 & 1.923 & 2.106 \\
& T4 & 0.281 & -0.086 & 0.111 & 0.298 & 0.087 & 0.114 & 0.972 & 0.272 & 0.506 & 1.941 & 0.008 & 1.760 & 1.942 & 2.127 \\
& T5 & 0.324 & -0.072 & 0.077 & 0.347 & 0.075 & 0.082 & 0.943 & 0.665 & 0.922 & 2.124 & 0.011 & 1.924 & 2.122 & 2.340 \\
& T6 & 0.282 & -0.085 & 0.109 & 0.300 & 0.087 & 0.113 & 0.971 & 0.323 & 0.564 & 1.936 & 0.009 & 1.749 & 1.934 & 2.134 \\
\addlinespace[2pt]
\cmidrule(lr){1-16}
\multicolumn{16}{@{}l@{}}{\textbf{Linear regression (OLS)}} \\
\multirow{6}{*}{\ParamBox{0.11\textwidth}{500}{3000}{5}{0.30}{0.10}{0.6}{clean}} & T1 & 0.060 & -0.006 & 0.011 & 0.109 & 0.012 & 0.018 & 1.000 & 1.000 & 1.000 & 2.093 & 0.002 & 2.007 & 2.094 & 2.169 \\
& T2 & 0.176 & 0.017 & -0.058 & 0.201 & 0.020 & 0.059 & 1.000 & 1.000 & 1.000 & 2.222 & 0.002 & 2.128 & 2.222 & 2.312 \\
& T3 & 0.173 & 0.017 & -0.057 & 0.197 & 0.020 & 0.059 & 1.000 & 1.000 & 1.000 & 2.220 & 0.002 & 2.126 & 2.221 & 2.310 \\
& T4 & 0.165 & 0.018 & -0.058 & 0.191 & 0.021 & 0.060 & 1.000 & 1.000 & 1.000 & 2.224 & 0.002 & 2.132 & 2.225 & 2.314 \\
& T5 & 0.185 & 0.019 & -0.064 & 0.211 & 0.022 & 0.066 & 1.000 & 1.000 & 1.000 & 2.254 & 0.003 & 2.154 & 2.255 & 2.353 \\
& T6 & 0.171 & 0.018 & -0.058 & 0.198 & 0.021 & 0.060 & 1.000 & 1.000 & 1.000 & 2.225 & 0.002 & 2.128 & 2.226 & 2.320 \\
\addlinespace[2pt]
\multicolumn{16}{@{}l@{}}{\textbf{Sparse regression (EN)}} \\
\multirow{6}{*}{\ParamBox{0.11\textwidth}{500}{3000}{5}{0.30}{0.10}{0.6}{{ext.}}} & T1 & 0.125 & -0.093 & 0.150 & 0.131 & 0.094 & 0.151 & 1.000 & 0.000 & 0.000 & 1.342 & 0.001 & 1.263 & 1.343 & 1.407 \\
& T2 & 0.076 & -0.060 & 0.095 & 0.087 & 0.061 & 0.097 & 1.000 & 0.743 & 0.652 & 1.443 & 0.002 & 1.358 & 1.446 & 1.519 \\
& T3 & 0.077 & -0.061 & 0.096 & 0.087 & 0.062 & 0.097 & 1.000 & 0.740 & 0.649 & 1.442 & 0.002 & 1.355 & 1.444 & 1.516 \\
& T4 & 0.086 & -0.063 & 0.098 & 0.093 & 0.064 & 0.100 & 1.000 & 0.710 & 0.626 & 1.450 & 0.002 & 1.362 & 1.450 & 1.530 \\
& T5 & 0.082 & -0.056 & 0.087 & 0.092 & 0.057 & 0.088 & 1.000 & 0.884 & 0.855 & 1.499 & 0.004 & 1.393 & 1.491 & 1.651 \\
& T6 & 0.085 & -0.062 & 0.098 & 0.092 & 0.063 & 0.099 & 1.000 & 0.699 & 0.631 & 1.453 & 0.002 & 1.367 & 1.454 & 1.535 \\
\bottomrule
\end{tabular*}
\noindent\parbox{\textwidth}{\raggedleft
  \Nchips{500}
}
\endgroup
\end{table}

\end{document}